\documentclass[twoside,12pt]{article}
\pdfoutput=1

\usepackage{epsfig}

\usepackage{graphicx}
\usepackage{amsmath}
\usepackage{amssymb}
\usepackage{array}
\usepackage{wrapfig}
\usepackage{units}
\usepackage{color}

\usepackage{braket}
\usepackage{mathtools}

\newcommand{\be}{\begin{equation}}
\newcommand{\ee}{\end{equation}}
\newcommand{\bea}{\begin{eqnarray}}
\newcommand{\eea}{\end{eqnarray}}

\newcommand{\ba}{\begin{eqnarray}}
\newcommand{\ea}{\end{eqnarray}}

\newcommand{\diracslash}[1]{\not\!\! #1}

\newcommand{\nc}{\newcommand}

\nc{\newsection}[1]{\section{#1}\setcounter{equation}{0}}
\nc{\newappendix}[1]{\section*{#1}\setcounter{equation}{0}}
\nc{\scm}{\scriptscriptstyle\mathrm}
\nc{\f}{\frac}
\nc{\baa}{\begin{array}}      \nc{\eaa}{\end{array}}
\nc{\bit}{\begin{itemize}}    \nc{\eit}{\end{itemize}}
\nc{\ben}{\begin{enumerate}}  \nc{\een}{\end{enumerate}}
\nc{\bce}{\begin{center}}     \nc{\ece}{\end{center}}
\nc{\bfl}{\begin{flushright}} \nc{\efl}{\end{flushright}}
\nc{\btb}{\begin{tabular}}    \nc{\etb}{\end{tabular}}
\nc{\eps}{\varepsilon}
\nc{\vp}{\varphi}
\nc{\tvp}{\widetilde{\varphi}}
\nc{\D}{\mbox{$\not\!\!D$}}
\nc{\Db}{\mbox{${\raisebox{2mm}{\boldmath ${}^\leftarrow$}\hspace{-4mm} D}$}}
\nc{\Dfb}{\mbox{$\raisebox{2mm}{\boldmath ${}^\leftrightarrow$}\hspace{-4mm} D$}}
\nc{\vpj }{\mbox{${\vp^\dag i\,\raisebox{2mm}{\boldmath ${}^\leftrightarrow$}\hspace{-4mm} D_\mu\,\vp}$}}
\nc{\vpjt}{\mbox{${\vp^\dag i\,\raisebox{2mm}{\boldmath ${}^\leftrightarrow$}\hspace{-4mm} D_\mu^{\,I}\,\vp}$}}

\def\wt{\widetilde}
\def\gpbz{{\bar g}_\pi^{(0)}}
\def\gpbo{{\bar g}_\pi^{(1)}}
\def\gpbt{{\bar g}_\pi^{(2)}}
\def\gpbi{{\bar g}_\pi^{(i)}}

\def\gebz{{\bar g}_\eta^{(0)}}
\def\gebo{{\bar g}_\eta^{(1)}}

\def\grbz{{\bar g}_\rho^{(0)}}
\def\grbo{{\bar g}_\rho^{(1)}}
\def\grbt{{\bar g}_\rho^{(2)}}

\def\gobz{{\bar g}_\omega^{(0)}}
\def\gobo{{\bar g}_\omega^{(1)}}

\newcommand\slurp[1]{#1}
\newcommand\sslurp[2]{#1{#2}}
{\catcode`/=\active \expandafter}%
\slurp{\newcommand/}{/\penalty1000\hskip0pt\relax}
{\catcode`:=\active \expandafter}%
\slurp{\newcommand:}{:\penalty1000\hskip0pt\relax}

\newcommand\addspace{\ifcat\nextchar a\spacefactor999. \else.\fi}
{\catcode`\.=\active \expandafter}%
\slurp{\newcommand.}{\futurelet\nextchar\addspace}

\usepackage[unicode]{hyperref} 
\ifx\href\undefined\def\href#1{}\fi
\ifx\texorpdfstring\undefined\def\texorpdfstring#1#2{#1}\fi

\newcommand\myslash{/} \newcommand\mycolon{:}
\newcommand\doi{{\catcode`/=\active \catcode`:=\active \expandafter}\sslurp\realdoi}
{\catcode`/=\active \catcode`:=\active \expandafter}%
\slurp{\newcommand\realdoi[1]{{\let/=\myslash \let:=\mycolon
                               \edef\raw{{http://dx.doi.org/#1}}\expandafter}%
                               \expandafter\href\raw{doi:#1}}}
\newcommand\eprint[2]{{\escapechar-1%
                       \edef\a{\expandafter\string\csname arXiv\endcsname}%
                       \edef\b{\expandafter\string\csname #1\endcsname}%
                       \edef\c{\expandafter\string\csname #2\endcsname}%
                       \edef\d{\noexpand\href{http://arXiv.org/abs/\c}}%
                       \ifx\a\b\expandafter\d\fi{\tt #1:#2}}}

\newcommand{\lsim}{\buildrel < \over {_\sim}}
\newcommand{\gsim}{\buildrel > \over {_\sim}}

\newcommand{\simge}{\hspace*{0.2em}\raisebox{0.5ex}{$>$}
     \hspace{-0.8em}\raisebox{-0.3em}{$\sim$}\hspace*{0.2em}}

\newcommand{\slashPTsub}{TVPV}

\newcommand{\dslash}[1]{#1 \llap{/\kern-0.5pt}}
\newcommand{\Dslash}[1]{#1 \llap{/\kern+1.2pt}}
\newcommand{\DDslash}[1]{#1 \llap{/\kern+2.3pt}}
\newcommand{\dslashh}[1]{#1 \llap{/\kern+1pt}}
\newcommand{\boldtau}{\mbox{\boldmath $\tau$}}
\newcommand{\boldpi}{\mbox{\boldmath $\pi$}}
\newcommand{\boldtheta}{\mbox{\boldmath $\theta$}}
\newcommand{\boldV}{\mbox{\boldmath $V$}}

\newcommand{\boldrho}{\mbox{\boldmath $\rho$}}

\newcommand{\lamchi}{\Lambda_\chi}

\newcommand{\svec}[1]{\ensuremath\vec{#1}}

\begin{document}

\title{ \vspace{1cm} 
Electric Dipole Moments of Nucleons, Nuclei, and Atoms: \\
The Standard Model and Beyond }
\author{ Jonathan Engel$^{1}$, Michael J. Ramsey-Musolf$^{2,3}$, and U. van
Kolck$^{4,5}$ \\
\\
$^1$ Department of Physics and Astronomy, CB3255, University of North Carolina,\\
Chapel Hill, NC 27599-3255, USA\\
$^2$ Department of Physics, University of Wisconsin-Madison,\\
 1150 University Ave., Madison, WI 53706, USA \\
$^3$ California Institute of Technology,\\
Pasadena, CA 91125, USA \\
$^4$ Institut de Physique Nucl\'eaire, Universit\'e Paris-Sud, CNRS/IN2P3,\\
91406 Orsay, France \\
$^5$ Department of Physics, University of Arizona,\\
Tucson, AZ 85721, USA }

\maketitle
\begin{abstract} 
Searches for the permanent electric dipole moments (EDMs) of molecules, atoms,
nucleons and nuclei provide powerful probes of CP violation both within and
beyond the Standard Model (BSM). The interpretation of experimental EDM limits
requires careful delineation of physics at a wide range of distance scales,
from the long-range atomic and molecular scales to the short-distance dynamics
of physics at or beyond the Fermi scale. In this review, we provide a framework
for disentangling contributions from physics at these disparate scales,
building out from the set of dimension four and six effective operators that
embody CP violation at the Fermi scale. We survey existing computations of
hadronic and nuclear matrix elements associated with Fermi-scale CP violation
in systems of experimental interest, and quantify the present level of
theoretical uncertainty in these calculations.  Using representative BSM
scenarios of current interest, we illustrate how the interplay of physics at
various scales generates EDMs at a potentially observable level. 

\end{abstract}

\newpage

\section{Introduction} 
\label{sec:intro}

Nuclear physics tests of fundamental symmetries have played a vital role in the
development of the Standard Model (SM) and provide powerful probes of what may
lie beyond it.  As described elsewhere in this issue, these tests have
uncovered the left-handed nature of the charged current weak interaction,
helped single out the SM structure of the weak neutral current from various
alternatives, revealed the phenomena of quark-mixing via the slight deviation
from exact lepton-quark universality in weak decays, and provided stringent
upper bounds on the neutrino mass scale. In this article, we focus on two
symmetries for which nuclear physics studies have a long and illustrious
history: time reversal invariance (T) and invariance under the combination of
change conjugation (C) and parity (P). It is well-known, of course, that CP is
not conserved in flavor-changing weak interactions, a phenomena now associated
with the complex phase in the Cabibbo-Kobayashi-Maskawa (CKM) matrix. A
consistent phenomenology of CKM CP violation (CPV) has emerged from extensive
studies of $K$- and $B$-meson properties and interactions. As a local quantum
field theory satisfying the postulates of the CPT theorem, the SM thus also
admits time-reversal violation, as the combined operation of CPT leaves the SM
interactions unchanged. 

It is likely, however, that the SM picture of CP and T violation is incomplete.
Numerous scenarios for physics beyond the SM (BSM) readily admit new sources of
CPV. Given that the SM is likely embedded in a more complete theory of 
fundamental interactions, it is reasonable to expect novel signatures of CPV to
appear along with other manifestations of new physics. Cosmology provides an 
additional compelling motivation for BSM CPV. Assuming the Universe was 
matter-antimatter symmetric at its birth or at the end of the inflationary 
epoch, additional sources of CPV are needed to explain the presently observed 
cosmic matter-antimatter asymmetry (for reviews and extensive references, see 
Refs.~\cite{Morrissey:2012db,Dine:2003ax,Riotto:1999yt}). From a perhaps even 
more speculative standpoint, the generation of the matter-antimatter asymmetry 
could also entail the violation of CPT invariance, while various exotic BSM 
frameworks also incorporate such a violation. While exploring the possibilities
for BSM CPV and even CPT violation, one should bear in mind that there remains 
within the SM itself one as yet unobserved source of CP and T violation: the 
dimension-four QCD \lq\lq $\theta$" term, whose dimensionless coefficient,
the vacuum angle ${\bar\theta}$, is now constrained to be no larger than 
$\sim 10^{-10}$ by the non-observation of permanent electric dipole moments 
(EDMs) of the $^{199}$Hg atom and neutron. This exceedingly small upper limit 
-- and the associated \lq\lq strong CP problem" -- has motivated the idea of 
an additional symmetry, the \lq\lq Peccei-Quinn" (PQ) symmetry, whose 
spontaneous breakdown would imply the existence of the axion that has also not 
yet been observed. (For a recent review, see Ref.~\cite{Kim:2008hd}.)

In the quest to discover both BSM CPV as well as CPV generated by the SM strong
interaction, EDM searches have generally provided by far the most powerful
probes. In contrast to the CPV observed in the $K$- and $B$-meson sectors, the
existence of an EDM of an elementary particle or quantum system requires no
flavor-changing interactions. The situation is more complicated in the SM,
however, since CKM CPV requires the participation of three generations of
quarks and, thus, flavor-changing interactions at the loop level. As a result,
the EDMs of light quark and lepton systems generated by CKM CPV are highly
suppressed. The individual quark EDMs vanish at two-loop order
\cite{Shabalin:1978rs,Shabalin:1982sg}.  The lowest-order contribution to the
neutron EDM, then, arises not from the individual quark EDMs but from a
two-loop hadronic interaction involving two $\Delta S=1$ weak interactions
between quarks.  The CP-conserving interaction appears at tree-level, while the
CPV interaction is generated from the one-loop $d\to s$ \lq\lq penguin"
operator that contains a sum over all three flavors of positive-charge quarks.
The electron EDM first appears at four-loop order \cite{Bernreuther:1990jx},
suppressing it by several orders of magnitude with respect to the neutron EDM. 

Given the present and prospective EDM search sensitivities (see Table
\ref{tab:edmexp}), one may consider CKM CPV to be something of a negligible
\lq\lq background", making these searches primarily probes of either SM strong
or BSM CPV (together with P violation). Moreover, if the latter is
flavor-diagonal as one encounters in many (but not all) BSM scenarios, then the
sensitivity of EDMs generally exceeds that of other possible tests, such as
CP-odd observables in a high-energy collider experiment. For these reasons, the
emphasis in this article on nuclear physics tests of CP and T will fall on EDMs.

Unraveling the implications of EDM searches is a multi-faceted problem, 
entailing physics at a variety of length scales. The experiments themselves are
extraordinarily challenging, requiring exquisite control over a number of 
possible effects that could mimic an EDM.
Still, significant improvements are expected for the traditional
searches on neutral systems (Table \ref{tab:edmexp}) 
and proposals have been been made to measure the EDMs of charged
particles in storage rings at similar levels.
(For discussions of the experimental 
status and prospects, see, {\em e.g.} Ref. \cite{Chupp:2008zz} as well as a 
forthcoming companion article to this review \cite{Chupp:2013}.) 
In what 
follows, we concentrate on the theoretical problem, seeking to provide a 
framework for interpreting experimental results that delineates the physics at 
different length scales that one must consider: the atomic, nuclear, and 
hadronic scales, wherein one contends with the complications of 
non-perturbative strong interactions and many-body physics; the Fermi scale, 
associated with the various effective operators outlined in the introductory 
article \cite{PPNPIntro}; 
the scale of BSM physics, $1/\Lambda$, at which one encounters explicit 
new degrees of freedom whose interactions give rise to the effective-operator 
Wilson coefficients; the short-distance scale of high-energy collider 
experiments that may produce these new degrees of freedom directly; and the 
scales associated with early Universe cosmology that may be responsible for the
generation of the matter-antimatter asymmetry. 

\begin{figure}[tb]
\begin{minipage}[t]{8 cm}
\epsfig{file=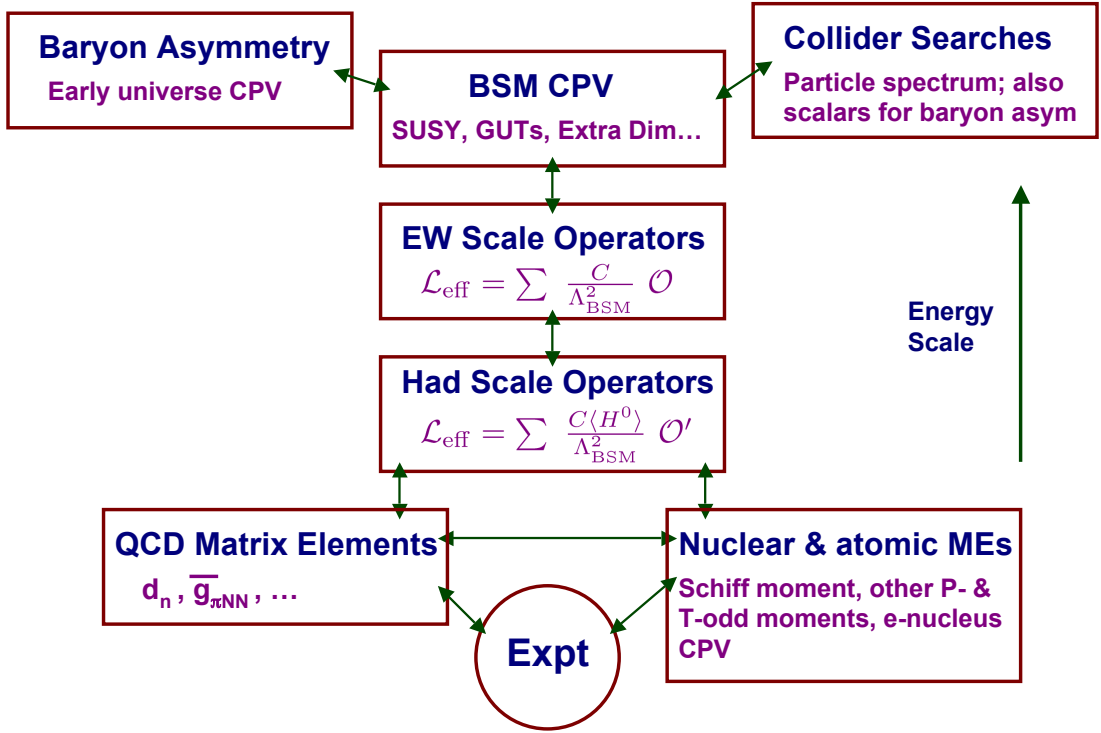,scale=0.6}
\end{minipage}
\begin{center}
\begin{minipage}[t]{16.5 cm}
\caption{Electric dipole moments and the interplay of various scales. For purposes of illustration, only the impact of dimension six CPV operators is shown. Below the weak scale, some operators, such as the fermion EDMs and quark chromo EDMs are effectively dimension five, carrying an explicit factor of the Higgs vacuum expectation value $\langle H^0\rangle$. A summary of the operators of interest to this article appears in Table \ref{tab:effopsum}. See text for a full discussion. 
\label{fig:edmscales}}
\end{minipage}
\end{center}
\end{figure}

The interplay of these different scales is illustrated in
Fig.~\ref{fig:edmscales}. For purposes of this article, we assume the
underlying dynamics of BSM CPV are associated with an energy scale $\Lambda$
that lies well above the electroweak or Fermi scale\footnote{It is possible
that new CPV interactions are generated by new light degrees of freedom, a
possibility that we do not treat extensively in this article.}.  If
$\Lambda\lsim 10$ TeV, high energy collider searches may discover the
elementary particles responsible for BSM CPV, determine their masses, and
provide information about the nature of their CP-conserving interactions. The
new CPV interactions may also provide one of the ingredients needed for
successful electroweak baryogenesis, though additional scalar degrees of
freedom would also be expected in order to obtain a first order electroweak
phase transition.  In principle, collider searches could also observe the
latter and measure their relevant properties\cite{Morrissey:2012db}. At the
low-end of the energy scale, EDM searches look for the CPV \lq\lq footprints"
of these new interactions. In this energy regime, the extent to which the
underlying CPV interactions become manifest depends on their interplay with the
many-body and strong interaction dynamics of the hadronic, nuclear, atomic, and
molecular systems of interest. The quantities that one extracts most directly
from EDM searches, then, are not the underlying CPV interactions, but the
hadronic, nuclear, and atomic matrix elements that they induce, such as the
neutron EDM, time-reversal-violating and parity-violating (TVPV) $\pi NN$
interaction, nuclear Schiff moment, and TVPV effective electron-nucleus
interaction\footnote{Henceforth, we will use \lq\lq CPV" when referring to the underlying 
elementary particle interactions and \lq\lq TVPV" when referring to the resulting effects induced 
at the hadronic, nuclear, atomic, and molecular levels.}.  

The bridge between these matrix elements and the underlying CPV dynamics is
provided by a set of effective operators, whose coefficients are governed by an
appropriate power of $1/\Lambda$ and a dimensionless Wilson coefficient that
depends on the details of the underlying dynamics. At energies between
$\Lambda$ and the weak scale, these operators contain all of the Standard Model
fields and respect the SU(3)$_C\times$SU(2)$_L\times$U(1)$_Y$ gauge symmetry of
the theory. Below the weak scale, the heavy SM degrees of freedom are \lq\lq
integrated out", leaving a set of effective operators that respect the residual
SU(3)$_C\times$U(1)$_\mathrm{EM}$ symmetry and that may have reduced mass
dimension. Nevertheless, the Wilson coefficients of the hadronic scale
operators derive from those that enter at the weak scale, wherein the full
gauge symmetry of the SM enforces certain relations between them. 

As we discuss below, it is in principle possible to use a combination of
experimental results and theoretical hadronic, nuclear, and atomic computations
to determine, or at least constrain, the Wilson coefficients without making any
assumptions about the details of the underlying BSM CPV other than that it is
associated with a high energy scale. In this sense, the theoretical effort
associated with the bottom half of Fig.~\ref{fig:edmscales} is
model-independent. More broadly, however, the goal of the \lq\lq EDM program"
is to derive as much information as possible from EDM searches, in conjunction
with other precision tests and high-energy experiments, for both BSM and strong
CPV as well as for the origin of matter. Doing so requires running operators
from one scale to the next; matching the interactions at the boundaries
between neighboring scales; and identifying and quantifying where possible the
attendant theoretical uncertainties. One may then ask whether the emerging
picture is consistent with any existing model for the underlying BSM CPV,
precludes others, or perhaps points to one not yet invented. 

In what follows, we lay out the overall framework for this program. In doing
so, we attempt to address a question that has been somewhat underemphasized in
previous work, namely, the level of theoretical uncertainty associated with
various steps in the interpretive chain of \lq\lq running and matching". While
the absence of theoretically robust hadronic, nuclear and atomic/molecular
computations would not detract from the significance of the observation of an
EDM, the level of theoretical uncertainty does affect one's ability to utilize
present and prospective experimental results to pinpoint the underlying CPV
mechanism or rule out various possibilities. In response to this issue, we
provide a set of benchmark theoretical error bars associated with various
quantities of interest, recognizing that this effort remains a work in progress
and alerting the reader to the website 
where updated
information will appear. 

The focus of this framework is on the $\theta$-term as well as Wilson
coefficients $ C_k$ for the dimension-six CPV operators indicated in the center
of Fig.~\ref{fig:edmscales}. These operators, which break P as well as T,
include the elementary fermion EDMs, the quark \lq\lq chomo-electric dipole
moments" (CEDMs), Weinberg three-gluon operator, and various four-fermion CPV
operators (both semileptonic and non-leptonic). We provide a general set of
expressions relating these operator coefficients to the hadronic, nuclear,
atomic and molecular quantities of interest, given in
Eqs.~(\ref{eq:dNhad}-\ref{eq:etadef}, \ref{eq:coefs},
\ref{eq:datom}-\ref{eq:rhoArel}) and Table \ref{tab:lightEDMs}. 
We take into account the chirality-flipping nature of the elementary fermion
EDMs and quark CEDMs, writing the corresponding Wilson coefficients as the
product of the fermion Yukawa couplings and a BSM scenario-dependent factor
($\delta_f$ or ${\tilde\delta}_q$). Doing so allows us to place the EDM and
CEDM operator coefficients on the same footing as those for operators that do
not flip chirality, such as the CPV electron-quark operators. A summary of the
dimension-four and -six operators for light flavors (electron, up- and
down-quarks, gluons) is given in Table \ref{tab:effopsum}.  Note that one
encounters thirteen quantities at this order, though some combinations have a
more significant impact that others on the systems of experimental interest.
One could, of course, expand the list to include the muon EDM as well as
heavier flavors of quarks. In some cases the manifestation of the latter may in
light quark systems may be non-negligible. Nonetheless, for purposes of this
review we will concentrate largely on the already sizeable set of operators
involving only the light flavors\footnote{One should also bear in mind that not
all possible sources of CPV naturally fit within the effective operator
framework. If a new CPV interaction is mediated by a very light weakly coupled
boson, the latter must be retained as an explicit degree of freedom}.

The dependence of various 
hadronic, atomic, and molecular quantities on ${\bar\theta}$, 
$\mathrm{Im} C_k$, $\delta_f$, and ${\tilde\delta}_q$ is then governed by the 
physics at the relevant scales. We compile the existing set of corresponding 
matrix-element calculations and give a set of benchmark values and theoretical 
ranges that can be used when extracting limits on ${\bar\theta}$, 
$\mathrm{Im} C_k$, $\delta_f$, and ${\tilde\delta}_q$ from experimental results.
For the discussion of hadronic matrix elements, we rely heavily on 
considerations of chiral symmetry as an overall guide, though we also quote 
results from lattice QCD, QCD sum rules, and quark models as well. An important
conclusion from this survey is the need for a concerted future effort on the 
hadronic and nuclear matrix elements. While the literature on computations of 
the $\mathrm{Im}\, C_k$, $\delta_f$, and ${\tilde\delta}_q$ in various BSM 
scenarios is deep, the corresponding set of results for hadronic and nuclear 
matrix elements is relatively thin. Given the level of effort and resources 
devoted to the experimental measurements of EDMs, a commensurate attack on the 
theoretical side is clearly in order.

\begin{table}[t] 
\centering \renewcommand{\arraystretch}{1.5} \btb{||c|c|c||c||} 
\hline
Wilson Coefficient & Operator (dimension)  & Number & Systems\\
\hline \hline
${\bar\theta}$  & theta term (4) & 1 & hadronic \&  \\
& & & diamagnetic atoms\\
\hline
$\delta_e$ & electron EDM (6) & 1 & paramagenetic atoms \\
$\mathrm{Im}\, C_{\ell e q u}^{(1,3)}$, $\mathrm{Im}\, C_{\ell e q d}$ & semi-leptonic (6) & 3 & \& molecules \\
\hline
$\delta_q$ & quark EDM (6) & 2 & hadronic \&\\
${\tilde\delta}_q$ & quark chromo EDM (6) & 2 & diamagnetic atoms \\
$C_{\tilde G}$ & three-gluon (6)  & 1 & \\
$\mathrm{Im}\, C_{quqd}^{(1,8)}$ & four-quark (6) & 2 & \\
$\mathrm{Im}\, C_{\varphi ud}$ & induced four-quark (6) & 1& \\
\hline
total & & 13 & \\
\hline \hline
\etb
\caption[ . ]{Dimension four and dimension six CPV operator coefficients for
light flavors. First column gives dimensionless Wilson coefficient (see Sec.
\ref{sec:def}), followed by operator name and mass dimension (second column)
and number of operators (third column). Final column indicates type of system
in which a given operator will have its most significant impact.  }
\label{tab:effopsum}
\end{table}

Our discussion of this theoretical framework is organized in the remainder of 
the article as follows. In Section \ref{sec:def}, we briefly review the 
conventions and definitions, drawing on the notation of the introductory 
article. Section \ref{sec:had} contains a discussion of physics at the hadronic
scale, including the running of the weak-scale operators to the hadronic scale,
the various hadronic interactions cast in the context of chiral symmetry, and a
summary of sensitivities of these hadronic quantities to the weak-scale 
operator coefficients. In Section \ref{sec:nuc}, we review the status and open 
questions related to computations at the nuclear and atomic scales, including 
P- and T-odd nuclear moments such as the Schiff moment. We follow this 
discussion with an illustrative overview of the high-scale physics that may 
give rise to the weak-scale operators in Section \ref{sec:bsm}. 
A discussion and outlook appears in Section \ref{sec:concl}. Throughout the article, we refer to other recent 
reviews \cite{Dzuba:2012bh,Ellis:2008zy,Ginges:2003qt,Pospelov:2005pr} when 
appropriate, endeavoring to avoid excessively duplicating material that is 
amply covered elsewhere but updating when necessary. We also do not discuss 
other tests of CP and T violation, given the limitations of space for this 
review (for a discussion of T violation in neutron and nuclear $\beta$-decay, 
see the companion article 
in this issue on charged current processes).

\section{Conventions and Definitions}
\label{sec:def}

The starting point for our analysis is the weak scale operators defined in the
introductory article \cite{PPNPIntro}.  We concentrate on three sources of CPV,
\be
\label{eq:LCPV1}
\mathcal{L}_\mathrm{CPV} = \mathcal{L}_\mathrm{CKM}+\mathcal{L}_{\bar\theta}
+\mathcal{L}_\mathrm{BSM}^\mathrm{eff}\ .  \ee Here the CPV SM CKM
\cite{Kobayashi:1973fv} and QCD
\cite{'tHooft:1976up,Jackiw:1976pf,Callan:1976je} interactions are 
\bea
\mathcal{L}_\mathrm{CKM} &=& -\frac{ig_2}{\sqrt{2}} V_\mathrm{CKM}^{pq} {\bar
U}_L^p \diracslash{W}^+ D_L^q +\mathrm{h.c.}\ , \\
\mathcal{L}_{\bar\theta} &=& -\frac{g_3^2}{16\pi^2} {\bar\theta} \,
\mathrm{Tr}\left(G^{\mu\nu}{\tilde G}_{\mu\nu}\right) \ , 
\label{thetaterm}
\eea 
where $g_2$ and $g_3$ are the weak and strong coupling constants,
respectively, $U_L^p$ ($D_L^p$) is a generation-$p$ left-handed up-type
(down-type) quark field, $V_\mathrm{CKM}^{pq}$ denotes a CKM matrix element,
$W_\mu^{\pm}$ are the charged weak gauge fields, and ${\tilde G}_{\mu\nu}
=\epsilon_{\mu\nu\alpha\beta}G^{\alpha\beta}/2$
($\epsilon_{0123}=1$ \footnote{Note that our sign convention for
$\epsilon_{\mu\nu\alpha\beta}$, which follows that of
Ref.~\cite{Grzadkowski:2010es}, is opposite to what is used in
Ref.~\cite{Pospelov:2005pr} and elsewhere.  Consequently,
$\mathcal{L}_{\bar\theta}$ carries an overall $-1$ compared to what frequently
appears in the literature.}) is the dual to the gluon field strength
$G^{\mu\nu}$.  In addition, 
\be
\label{eq:LCPV2}
\mathcal{L}_\mathrm{BSM}^\mathrm{eff}= \frac{1}{\Lambda^2}\ \sum_i
\alpha^{(n)}_i \, O_i^{(6)} \ , 
\ee 
gives the set of dimension-six CPV
operators at the weak scale $v=246$ GeV generated by BSM physics at a scale
$\Lambda>v$.  These operators \cite{Grzadkowski:2010es} are listed in Tables
\ref{tab:cpvdim6-1} and \ref{tab:cpvdim6-2}. Note that the operators containing
fermions are not CPV in and of themselves. Rather CPV effects arise when the
corresponding coefficients $\alpha^{(n)}_i$ are complex, as discussed below.

\begin{table}[t] 
\centering \renewcommand{\arraystretch}{1.5} \btb{||c|c||c|c||c|c||} 
\hline \hline
\multicolumn{2}{||c||}{Pure Gauge} & \multicolumn{2}{|c||}{Gauge-Higgs} &
\multicolumn{2}{|c||}{Gauge-Higgs-Fermion}\\
\hline
$Q_{\wt G}$ & $f^{ABC} \wt G_\mu^{A\nu} G_\nu^{B\rho} G_\rho^{C\mu} $ &
$Q_{\vp\wt G}$ & $\vp^\dag \vp\, \wt G^A_{\mu\nu} G^{A\mu\nu}$ & $Q_{u
G}$       & $(\bar Q \sigma^{\mu\nu} T^A u_R) \tvp\, G_{\mu\nu}^A$ \\
%
%
$Q_{\wt W}$ & $\eps^{IJK} \wt W_\mu^{I\nu} W_\nu^{J\rho} W_\rho^{K\mu}$ &
$Q_{\vp\wt W}$ & $\vp^\dag \vp\, \wt W^I_{\mu\nu} W^{I\mu\nu}$ & $Q_{d
G}$        & $(\bar Q \sigma^{\mu\nu} T^A d_R) \vp\, G_{\mu\nu}^A$ \\
%
&& $Q_{\vp\wt B}$ & $\vp^\dag \vp\, \wt B_{\mu\nu} B^{\mu\nu}$ & $Q_{f
W}$         & $(\bar F \sigma^{\mu\nu} f_R) \tau^I \Phi\, W_{\mu\nu}^I$ \\
%
&& $Q_{\vp\wt WB}$ & $\vp^\dag \tau^I \vp\, \wt W^I_{\mu\nu} B^{\mu\nu}$ &
$Q_{f B}$               & $(\bar F \sigma^{\mu\nu} f_R) \Phi\, B_{\mu\nu}$ \\
%
\hline \hline
\etb
\caption[ . ]{Dimension-six CPV operators involving gauge and/or Higgs degrees
of freedom. Notation largely follows that of introductory article
\cite{PPNPIntro} with the following modifications: $q_p\to Q$ denotes a
left-handed quark doublet; $F$ denotes a left-handed fermion doublet; $f_R$
denotes a right-handed SU(2) singlet fermion; and $\Phi=\tvp$ ($\vp$) for $f_R$
being an up-type (down-type) fermion.  For simplicity, generation indices have
been omitted.  }
\label{tab:cpvdim6-1}
\end{table}


\begin{table}[t]
\centering \renewcommand{\arraystretch}{1.5}
\begin{tabular}{||c|c||}
\hline\hline
\multicolumn{2}{||c||}{$(\bar LR)(\bar RL)$ and $(\bar LR)(\bar LR)$} \\\hline
$Q_{ledq}$ & $(\bar L^j e_R)(\bar d_R Q^j)$ \\
$Q_{quqd}^{(1)}$ & $(\bar Q^j u_R) \epsilon_{jk} (\bar Q^k d_R)$ \\
$Q_{quqd}^{(8)}$ & $(\bar Q^j T^A u_R) \epsilon_{jk} (\bar Q^k T^A d_R)$ \\
$Q_{lequ}^{(1)}$ & $(\bar L^j e_R) \epsilon_{jk} (\bar Q^k u_R)$ \\
$Q_{lequ}^{(3)}$ & $(\bar L^j \sigma_{\mu\nu} e_R) \epsilon_{jk} (\bar Q^k
\sigma^{\mu\nu} u_R)$ \\
\hline\hline
\end{tabular}
\caption{Dimension-six CPV operators involving four fermions. Notation as in
Table \ref{tab:cpvdim6-1}, with the additional modification $l_p\to L$ with
respect to the introductory article \cite{PPNPIntro}. 
\label{tab:cpvdim6-2}}
\end{table}

In this review we are mostly interested in the atomic/moleclar, hadronic, and
nuclear aspects of CPV. We will therefore concentrate on the two lightest
quarks, up and down, but will occasionally also point out effects of other
quarks, especially strange.

\subsection{CPV at Dimension Four}
\label{sec:dim4}

CPV from the $\theta$-term in Eq.~(\ref{thetaterm})
is intimately connected with the quark masses.
The \lq\lq bar" notation indicates 
that this dimensionless quantity is a linear combination of a bare 
$\theta$-parameter and argument of the quark Yukawa couplings:
\be
\label{eq:thetabardef}
{\bar\theta} = \theta+\mathrm{arg}\, \mathrm{det} (\lambda_u\lambda_d)\ ,
\ee
where the second term arises after redefining the phases of all the quark 
fields. 
Alternatively, the $\theta$-term can be eliminated,
thanks to the axial anomaly, through a chiral rotation 
\cite{Baluni:1978rf}.
Enforcing vacuum stability
to first order in the quark masses, all CPV is then in the quark bilinear 
\be
\label{eq:cpvmass}
\mathcal{L}_{\bar\theta}\leftrightarrow \mathcal{L}_\mathrm{CPV}^\mathrm{QCD} =
-m^\ast \bar{\theta} \; \bar{q}i\gamma_5 q\ ,
\ee
where $\bar \theta\ll 1$ was used. Here
\be
m^\ast = \frac{m_u m_d}{m_u+m_d} = \frac{\bar{m}}{2}\left(1-\epsilon^2\right)\ .
\ee
in terms of the average light quark mass $\bar{m}=(m_u+m_d)/2$
and relative splitting $\epsilon=(m_d-m_u)/2\bar{m}$.
Below we use $m_u=2.3^{+0.7}_{-0.5}$ MeV and $m_d=4.8^{+0.7}_{-0.3}$ MeV 
\cite{Beringer:1900zz}.
In the same notation the CP-even quark mass operator is given by
\be
\label{eq:Lcpmass}
\mathcal{L}_\mathrm{mass}^\mathrm{quark} = -{\bar m} {\bar q} q 
+{\epsilon\bar m} {\bar q}\tau_3 q\ .
\ee
Equation \eqref{eq:cpvmass} will be the starting point 
in Section \ref{sec:had} for chiral
considerations that impact TVPV observables in nuclear physics.

\subsection{CPV at Dimension Six}
\label{sec:dim6}

We now focus on a subset of the operators in Tables \ref{tab:cpvdim6-1} and 
\ref{tab:cpvdim6-2} that have been the the objects of most scrutiny,
because they are expected to give the largest contributions at low energies
\cite{deVries:2012ab}:
$Q_{q G}$ ($q=u,d$), $Q_{f W}$,  $Q_{f B}$, $Q_{\wt G}$, and various 
four-fermion operators. 

After electroweak 
symmetry breaking (EWSB) wherein $\phi^T\to(0, v/\sqrt{2})$, $Q_{q G}$ gives 
rise to the quark chromo-electric dipole moment (CEDM) interaction:
\be
\label{eq:cedmdef}
\mathcal{L}^\mathrm{CEDM} = -i\sum_q\ \frac{g_3 {\tilde d}_q}{2}\ 
{\bar q} \sigma^{\mu\nu} T^A\gamma_5 q\ G_{\mu\nu}^A \ ,
\ee
where $T^A$ ($A=1, \ldots, 8$) are the generators of the color group.
Analogously, 
$Q_{f W}$ and $Q_{f B}$ generate the elementary fermion EDM interactions,
\be
\label{eq:edmdef}
\mathcal{L}^\mathrm{EDM} = -i\sum_f\ \frac{ d_f}{2} 
{\bar f}\sigma^{\mu\nu} \gamma_5 f\ F_{\mu\nu} \ ,
\ee
where $F_{\mu\nu}$ is the electromagnetic field strength.
In the non-relativistic limit, Eq.~(\ref{eq:edmdef}) contains 
the CPV interactions with the electric field ${\vec E}$,
\be
\mathcal{L}^\mathrm{EDM}\rightarrow \sum_f {d_f}\ 
\chi^\dag_f {\vec\sigma}\chi_f \cdot {\vec E} \ ,
\ee
where $\chi_f$ is the Pauli spinor for fermion $f$ and ${\vec\sigma}$ is the 
vector of Pauli matrices. Thus, $d_f$ gives the EDM typically quoted units 
of $e$ cm or $e$ fm. Letting 
\be
\alpha^{(6)}_{f V_k}\equiv g_k C_{f V_k} \ ,
\ee
where $V_k=$ $B$, $W$, and $G$ for $k=1,2,3$ respectively,
the relationships between the ${\tilde d}_q$ and $d_f$ and the $C_{f V_k}$ are 
\bea
\label{eq:pcdef}
{\tilde d_q} & =& - \frac{\sqrt{2}}{v} \left(\frac{ v}{\Lambda}\right)^2\ 
\mathrm{Im} \ C_{q G} \ ,
\\
\label{eq:prdef}
d_f & = &  - \frac{\sqrt{2} e}{v}\ \left(\frac{ v}{\Lambda}\right)^2\ 
\mathrm{Im}\ C_{f \gamma} \ ,
\eea
where
\be
\label{eq:Cfgammadef}
\mathrm{Im}\ C_{f \gamma} \equiv 
\mathrm{Im}\ C_{f B}+ 2I_3^f \; \mathrm{Im}\ C_{f W} \ ,
\ee
and $I_3^f$ is the third component of weak isospin for fermion $f$.  Here, we 
have expressed $d_f$ and ${\tilde d}_q$ in terms of the  Fermi scale 
$1/v$, a 
dimensionless ratio involving the BSM scale $\Lambda$ and $v$, and the dimensionless 
Wilson coefficients. Expressing these quantities in units of fm one has
\bea
{\tilde d_q} & = &  - (1.13 \times 10^{-3}\ \mathrm{fm}) 
\left(\frac{ v}{\Lambda}\right)^2\ \mathrm{Im} \ C_{q G} \ ,\\
d_f & = & - (1.13 \times 10^{-3}\ e \, \mathrm{fm})
\left(\frac{ v}{\Lambda}\right)^2\ \mathrm{Im}\ C_{f \gamma} \ .
\eea

It is useful to observe that the EDM and CEDM operator coefficients are 
typically proportional to the corresponding fermion masses, as the operators 
that generate them above the weak scale 
($Q_{q \wt G}$, $Q_{f \wt W}$, $Q_{f \wt B}$) contain explicit factors of the Higgs 
field dictated by electroweak gauge invariance. 
More physically, the
EDM and CEDM operators -- like the fermion magnetic moment -- induce a flip of 
chirality and, thus, are naturally proportional to the fermion mass 
$m_f=Y_f v/\sqrt{2}$. 
Broadly speaking, then, one expects the Wilson coefficients to contain a factor
of the fermion Yukawa coupling $Y_f$. 
In the Minimial Supersymmetric Standard Model (MSSM), for example, a one-loop 
contribution to $\mathrm{Im}\ C_{q\wt G}$ from squark-gluino loop (see Fig.~\ref{fig:susy1})
has the magnitude \cite{Ibrahim:1997gj,Ibrahim:1998je}
\be
\mathrm{Im}\ C_{q G} = \frac{g_3^2}{16 \pi^2}\, Y_q\,
\sin[\mathrm{Arg}(\mu M_3 b^\ast)]\, F({\tilde m_j})\ ,
\ee
where $\mu$ is the supersymmetric Higgs-Higgsino mass parameter, $M_3$ is the 
soft SUSY-breaking gluino mass,  $b$ gives the Higgs SUSY-breaking mass 
parameter, and $F({\tilde m_j})$ is a loop function that depends on the various
superpartner masses ${\tilde m}_j$. In this case, the scale $\Lambda$ would be 
the largest value of ${\tilde m}_j$ entering the loop 
\footnote{Note that we have not 
included a similar contribution involving the relative phases of $M_3$ and the 
squark triscalar terms (see Section \ref{sec:susy}).}.
It is convenient to define two quantities ${\tilde\delta}_q$ and $\delta_f$ 
that embody all of the model-specific dynamics responsible for the EDM and CEDM
apart from Yukawa insertion:
\bea
\mathrm{Im}\ C_{q G} & \equiv & Y_q\, {\tilde\delta}_q 
\rightarrow {\tilde d}_q = -\frac{2m_q}{v^2}\, 
\left(\frac{v}{\Lambda}\right)^2\, {\tilde\delta}_q \ ,
\label{eq:nda1prime}
\\
\mathrm{Im}\ C_{{f\gamma }} & \equiv & Y_f\, {\delta}_f 
\rightarrow d_f =  -e\frac{2m_f}{v^2} \, 
\left(\frac{v}{\Lambda}\right)^2\, {\delta}_f \ .
\label{eq:nda1}
\eea
While one often finds bounds on the elementary fermion EDM and CEDMs quoted in 
terms of $d_f$ and ${\tilde d}_q$, the quantities $\delta_f$ and 
${\tilde\delta}_q$ are more appropriate when comparing with the Wilson 
coefficients of other dimension-six CPV operators, such as the three-gluon or 
semileptonic four-fermion interactions, that do not generally carry the 
Yukawa suppression. In what follows, we will provide expressions in terms of 
the $\mathrm{Im}\, C_{fV}$, ($d_f$, ${\tilde d}_q$), and 
($\delta_f$, ${\tilde\delta}_q$). In doing so, we will neglect the light-quark mass splitting and replace
\be
Y_u,\ Y_d \rightarrow Y_q \equiv \frac{\sqrt{2}{\bar m}}{v}
\ee
with ${\bar m}$ being the average light quark mass.

The extraction of the CPV three-gluon and low-energy, flavor-diagonal 
CPV four-fermion operators
from $\mathcal{L}_\mathrm{BSM}^\mathrm{eff}$ is generally more straightforward. 
Making the identifications 
\be
\label{eq:alphas}
\alpha^{(6)}_{\tilde G} \equiv g_3\, C_{\tilde G}\, \qquad
\alpha^{(6)}_{\ell e d q}\equiv C_{\ell e d q}\ , \qquad 
\alpha^{(6)}_{\ell e q u (1,3)}\equiv C_{\ell e d u}^{(1,3)}\ , \qquad 
\alpha^{(6)}_{q u q d (1,8)}\equiv g_3^2\, C_{q u q d}^{(1,8)} 
\ee
gives
the so-called 
Weinberg three-gluon operator \cite{Weinberg:1989dx} 
\be
\label{eq:weinbop}
\mathcal{L}^\mathrm{\tilde G}_\mathrm{CPV} =
\frac{g_3\, C_{\tilde G}}{\Lambda^2} 
f^{ABC} \wt G_\mu^{A\nu} G_\nu^{B\rho} G_\rho^{C\mu} \ ,
\ee
the CPV semileptonic interaction  
\bea
\label{eq:lsemilept1}
\mathcal{L}^\mathrm{eq}_\mathrm{CPV} & = & 
i \frac{\mathrm{Im} C_{\ell e d q}}{2\Lambda^2} 
\left[ {\bar e}\gamma_5 e\ {\bar d}d - {\bar e} e\ {\bar d}\gamma_5d\right]
-i\frac{\mathrm{Im} C_{\ell e q u}^{(1)}}{2\Lambda^2}
\left[ {\bar e}\gamma_5 e\ {\bar u}u + {\bar e} e\ {\bar u}\gamma_5u\right]
\\
\nonumber
&& - \frac{\mathrm{Im} C_{\ell e q u}^{(3)}}{2\Lambda^2}\ 
\epsilon_{\mu\nu\alpha\beta}\  
{\bar e}\sigma^{\mu\nu} e\ {\bar u} \sigma^{\alpha\beta} u  \ ,
\eea
and the CPV hadronic interaction \cite{RamseyMusolf:2006vr}
\bea
\label{eq:hadcpveff}
\mathcal{L}^\mathrm{qq}_\mathrm{CPV} & = & 
i \frac{g_3^2\,  \mathrm{Im} C_{q u q d}^{(1)}}{2\Lambda^2}
\left[{\bar u}\gamma_5 u \ {\bar d}d +{\bar u} u\  {\bar d}\gamma_5 d  
-{\bar d}\gamma_5 u\ {\bar u} d-{\bar d} u\ {\bar u}\gamma_5 d\right]\\
\nonumber
&&+ i \frac{g_3^2\, \mathrm{Im} C_{q u q d}^{(8)}}{2\Lambda^2}
\left[{\bar u}\gamma_5 T^A u\  {\bar d}T^A d +{\bar u} T^A u\  
{\bar d}\gamma_5 T^A d  -{\bar d}\gamma_5 T^A u\  {\bar u} T^A d
-{\bar d} T^A u\  {\bar u}\gamma_5 T^A d\right]\ .
\eea
Note that in contrast to the all other CPV $d=6$ operators of interest here, the coefficient of
the three-gluon operator (\ref{eq:weinbop}) does not require the imaginary part. 

In addition to these four-fermion operators, the operator 
\be
Q_{\varphi u d} = i \left({\tilde\varphi}^\dag D_\mu \varphi\right) 
{\bar u}_R \gamma^\mu d_R
\ee
with ${\tilde\varphi}=i\sigma_2\phi^\ast$
can also give rise to a four-fermion operator through exchange of the $W$ boson
contained in the covariant derivative. After EWSB, one has
\be
\label{eq:rhcurrent}
Q_{\varphi u d} \rightarrow 
\frac{g v^2}{2\sqrt{2}}\ {\bar u}_R \gamma^\mu d_R\, W_\mu^+ \ .
\ee
Exchange of the $W^+$ between the right-handed current in 
Eq.~(\ref{eq:rhcurrent}) and the left-handed current of the SM leads to 
an effective left-right (LR) Lagrangian
with the CPV part given by
\cite{Herczeg:1997ei,Zhang:2007da,Xu:2009nt,Ng:2011ui,deVries:2012ab}
\be
\label{eq:lreff0}
\mathcal{L}^\mathrm{eff}_\mathrm{LR,\, CPV} = 
-i \frac{\mathrm{Im}\, C_{\varphi u d}}{\Lambda^2} 
\left[{\bar d}_L\gamma^\mu u_L\, {\bar u}_R\gamma_\mu d_R
-{\bar u}_L\gamma^\mu d_L\, {\bar d}_R\gamma_\mu u_R\right] \ .
\ee
After a Fierz transformation, one then obtains
\be
\label{eq:lreff}
\mathcal{L}^\mathrm{eff}_\mathrm{LR,\, CPV} = 
i \frac{\mathrm{Im}\, C_{\varphi u d}}{3\Lambda^2} 
\left\{
{\bar u} u \, {\bar d}\gamma_5 d 
- {\bar u}\gamma_5 u \, {\bar d} d 
+ 3 \left[ 
{\bar u}T^A u \, {\bar d}\gamma_5 T^Ad 
- {\bar u}\gamma_5 T^Au \, 
{\bar d} T^A d 
\right]
\right\}\ .
\ee
Although the RHS of Eq.~(\ref{eq:lreff}) has the form of a product of scalar 
and pseudoscalar bilinears, it has a different flavor structure from the 
similar spacetime structures appearing in Eq.~(\ref{eq:hadcpveff}). As we 
discuss in Section \ref{sec:bsm} below, the interaction (\ref{eq:lreff}) 
is naturally generated in left-right symmetric theories. 
We also note that the operator in Eq.~(\ref{eq:lreff0}) will mix \cite{Dekens13}
with 
an operator of the form 
\be
{\bar d}_L\gamma^\mu T^A u_L\, {\bar u}_R\gamma_\mu T^A d_R
-{\bar u}_L\gamma^\mu T^A d_L\, {\bar d}_R\gamma_\mu T^A u_R
\ , 
\label{eq:mix}
\ee
generating
the corresponding scalar $\otimes$ pseudoscalar structures in 
Eq.~(\ref{eq:lreff}),
when running from the weak scale to the hadronic scale.

\subsection{Naturalness, Peccei-Quinn, and an Induced Vacuum Angle}
\label{sec:PQ}

It is well known that null results for the neutron and $^{199}$Hg EDMs imply 
that the coefficient of the dimension four operator in Eq.~(\ref{thetaterm}) 
is tiny: ${\bar\theta}\lsim 10^{-10}$. 
In general, one would expect both terms of the right side of 
Eq.~(\ref{eq:thetabardef}) to be $\mathcal{O}(1)$. Obtaining a value that is 
ten or more orders of magnitude smaller would require a highly unnatural 
degree of 
fine-tuning to obtain a cancellation between the two terms. 
Note that in the limit of one massless quark, $m^\ast\to 0$, 
CPV from Eq. \eqref{eq:cpvmass} disappears entirely. 
However, such a possibility seems to be excluded on phenomenological grounds
\cite{Leutwyler:2009jg}.

Alternatively, one may construct a mechanism that would generate a tiny 
${\bar\theta}$ at a more fundamental level,  through imposition of a symmetry 
or \lq\lq geography" (see, {\it e.g.}, Ref.~\cite{Cheung:2007bu}). 
The most well-known example of a symmetry argument is the Peccei-Quinn (PQ) 
mechanism. In brief, one starts from the anomalous axial U(1) symmetry of the SM 
in the limit of massless quarks, adding  one or more additional scalar 
fields to the SM whose interactions with the quarks preserves 
the tree-level axial U(1) symmetry. The enlarged symmetry, denoted U(1)$_{PQ}$, 
is spontaneously broken at a high scale, leading to a pseudoscalar Goldstone 
boson $a$, the axion. The corresponding axion Lagrangian is 
\be
\label{eq:Laxion}
\mathcal{L}_\mathrm{axion} = \frac{1}{2}\partial^\mu a \partial_\mu a - V(a) 
- \frac{a(x)}{f_a}\ \frac{g_3^2}{16\pi^2}\ 
\mathrm{Tr}\left(G^{\mu\nu}{\tilde G}_{\mu\nu}\right)\ ,
\ee
where the axion potential
\be
\label{eq:axionpot1}
V(a) = \frac{1}{2} \chi(0) \left( {\bar\theta}+\frac{a}{f_a}\right)^2+\cdots
\ee
is given in terms of the topological susceptibility $\chi(0)$ as well as 
the axion decay decay constant $f_a$ whose value indicates the scale of 
spontaneous PQ-symmetry breaking. In two-flavor QCD, one finds
\be
\chi(0) = - m^\ast\langle {\bar q} q\rangle \ ,
\ee
with
$\langle {\bar q} q\rangle\approx -(225\, \mathrm{MeV})^3$ 
\cite{Pospelov:2005pr}. Physical observables depend on the combination 
${\bar\theta}+\langle a\rangle/f_a$ rather than on ${\bar\theta}$, 
where $\langle a \rangle$ is the axion vacuum expectation value (vev). 
Minimization of $V(a)$ then implies that this combination vanishes, leading 
to a vanishing contribution to EDMs. 
The fluctuations about $\langle a \rangle$ correspond to the physical axion 
particle, whose mass is set by the ratio of $\sqrt{\chi(0)}$ and $f_a$. 

Within the SM as well as in BSM scenarios, CPV radiative corrections to the 
quark masses (or Yukawa interactions) can generate a non-vanishing 
contribution to $\mathrm{arg}\mathrm{det} (\lambda_u\lambda_d)$, 
re-introducing a possibly unacceptably large magnitude for ${\bar\theta}$. 
If the given CPV scenario does not suppress these contributions, 
the constraints on the underlying source of CPV can be quite severe. 
Invoking the PQ mechanism can alleviate these constraints.


As emphasized in Ref.~\cite{Pospelov:2005pr}, the presence of higher dimension 
CPV operators $\mathcal{O}_\mathrm{CPV} = \mathrm{Im}\, C\, Q_\mathrm{CPV}/\Lambda^2$ 
can lead to an induced $\theta$-term. 
The operator $Q_{\vp\wt G}$ in Table \ref{tab:cpvdim6-1}
gives a tree-level shift in $\bar\theta$,
which can still be removed through the PQ mechanism.
More importantly, there is
a shift in the minimum of the axion potential, 
which now reads \cite{Bigi:1991rh}  
\be
\label{eq:axionpot2}
V(a) = \chi(0)_{\mathcal{O}_\mathrm{CPV}}
\left({\bar\theta}-\frac{a}{f_a} \right) +\frac{1}{2} \chi(0) 
\left( {\bar\theta}+\frac{a}{f_a}\right)^2+\cdots \ ,
\ee
where
\be
\chi(0)_{\mathcal{O}_\mathrm{CPV}} = -i\, \lim_{k\to 0} 
\int\, d^4x\, e^{ix\cdot k} 
\bra{0} T\{ G{\tilde G}(x), \mathcal{O}_\mathrm{CPV}(0)\}\ket{0}\ .
\ee
As a result the minimum of the potential occurs for 
\be
\bar{\theta}+\frac{a}{f_a} = 
\frac{\chi(0)_{\mathcal{O}_\mathrm{CPV}} }{\chi(0)} 
\equiv \theta_\mathrm{ind}\ ,
\ee
a so-called \lq\lq induced" $\theta$-term
\footnote{Note that our definition gives an opposite sign to 
${\theta}_\mathrm{ind}$ compared to Ref.~\cite{Pospelov:2005pr}.}. 
Thus, use of the PQ mechanism to eliminate the contribution of 
${\bar\theta}$ to an EDM will introduce an additional contribution linear 
in the coefficient of a higher-dimensional CPV operator, 
$\mathrm{Im}\, C/\Lambda^2$. In the case of the CEDM operator, 
for example, one has
\be
\theta_\mathrm{ind} = \frac{m_0^2}{2}
\sum_{q=u,d,s}\, \frac{{\tilde d}_q}{m_q} \ ,
\ee
where $m_0^2$ characterizes the strength of the quark-gluon condensate
 $\langle {\bar q} \sigma^{\mu\nu} T^A G_{\mu\nu}^A q\rangle$.
In discussing the contributions of the dimension-six CPV operators to various 
P- and T-odd hadronic quantities, we will include the contributions from 
$\theta_\mathrm{ind}$ wherever they have been explicitly computed.

\section{CP and T at the Hadronic Scale}
\label{sec:had}

In order to relate the interactions defined in Section \ref{sec:def} to P- and
T-odd (TVPV)observables at the hadronic, nuclear, and atomic levels, we first
introduce the most relevant hadronic quantities in the context of heavy baryon
chiral perturbation theory (HB$\chi$PT) \cite{Bernard:2007zu}.  From the
standpoint of effective field theory (EFT), HB$\chi$PT provides the natural and
model-independent framework -- consistent with the approximate chiral symmetry
of QCD -- in which to parameterize one's ignorance about presently incalculable
non-perturbative strong-interaction matrix elements of the various CPV
operators appearing in $\mathcal{L}_\mathrm{CPV}$. For both the hadronic scale
analysis as well as the Fermi scale effective operator formulation embodied in
Eqs.~(\ref{eq:LCPV1}), and (\ref{eq:LCPV2}), the EFT philosophy entails
expressing the physical impact of unknown physics (BSM CPV or non-perturbative
QCD) in terms of an infinite tower of operators having successively higher-mass
dimension that carry appropriate inverse powers of the relevant mass scale:
$\Lambda$ in the case of BSM CPV and the QCD mass scale (or chiral
symmetry-breaking scale) $\lamchi$ $\sim 1$ GeV in the case of the low-energy
hadronic interaction. Doing so affords a systematic expansion of CPV
observables in scale ratios, such as $\lamchi/\Lambda$ or $P/\lamchi$ where $P$
denotes a soft momentum or pion mass. After truncation at a given order in
these ratios, one has a reasonable estimate of the error incurred through
omission of higher-order terms. 

Below $\lamchi$ all meson fields besides the pions can be 
\lq\lq integrated out", their effects being captured by short-range 
interactions.
Pions are light because they are the pseudo-Goldstone bosons of
chiral symmetry, which plays an important role in determining the 
relative importance of the effective interactions.
The term \lq\lq heavy" in HB$\chi$PT indicates that one is only interested in 
dynamics where the nucleon is non-relativistic, having momentum 
$p^\mu = m_N v^\mu+k^\mu$ with $v^\mu$ being its velocity and $|k^\mu| \ll m_N$,
the nucleon mass. 
The nucleon is, then, described by a two-component field 
$N_v(k)$ associated with a given velocity rather than a four-component Dirac 
field $\psi_N$. The anti-nucleon degrees of freedom are effectively 
also integrated out in terms of operators containing only $N_v(k)$, 
its derivatives, and the pion field. Dropping the subscript \lq\lq $v$" for 
notational simplicity, we give 
some representative terms in the resulting 
 T-violating and P-violating (TVPV) Lagrangian 
\cite{Mereghetti:2010tp,Maekawa:2011vs,deVries:2012ab}: 
\begin{eqnarray}
{\cal L}_{N\pi}^\mathrm{TVPV}&=&
-2 \bar{N} \left(\bar{d}_0+ \bar{d}_1\tau_3\right)S_\mu N \; v_\nu F^{\mu\nu}
+\bar{N}\left[\gpbz{\boldtau}\cdot{\boldpi} +\gpbo \pi^0 
+\gpbt\left( 3\tau_3\pi^0- {\boldtau}\cdot{\boldpi} \right)\right]N
\nonumber\\
\nonumber \\
&&
+\bar{C}_1 \bar{N}N \;\partial_\mu \left(\bar{N} S^\mu N \right) 
+\bar{C}_2 \bar{N}\boldtau N \cdot  
\partial_\mu \left(\bar{N} S^\mu\boldtau N \right)
+\cdots.
\label{chiPTTV}
\end{eqnarray}
Here, $\boldtau$ and $\boldpi$ denote the isovectors of Pauli matrices and 
pion fields, respectively, while $S^\mu$ and $v^\mu$ denote the spin and 
velocity of the nucleon that take on values in the nucleon rest frame:
$S^\mu\rightarrow (0, \vec \sigma/2)$ when $v^\mu\rightarrow (1, {\vec 0})$.

The first term in Eq.~(\ref{chiPTTV}) defines the isoscalar ($\bar{d}_0$) and 
isovector ($\bar{d}_1$) \lq\lq short-range" contributions to the nucleon EDM 
interaction:
\be
{\cal H}_\mathrm{EDM}^\mathrm{eff} = - \left(\bar{d}_0\pm\bar{d}_1\right) 
\chi^\dag{\vec\sigma}\chi \cdot{\vec E} \ ,
\ee
where the upper (lower) sign correspond to the proton (neutron) EDM 
interaction.

The second term is the T- and P-odd pion-nucleon non-derivative interaction 
\cite{Barton:1961eg}, consisting of isoscalar ($\gpbz$), isovector ($\gpbo$) 
and isotensor ($\gpbt$) pieces. 
These interactions have formally the same form when written in terms of 
a Dirac spinor $\psi_N$.
Note, however, that various authors follow differing notation for the
pion interactions. 
From the standpoint of a non-linear realization of chiral symmetry, 
it is more natural to build the Lagrangian from functions of ${\boldpi}/F_\pi$,
where $F_\pi=185$ MeV is the pion decay constant. 
The resulting P- and T-odd 
$\pi NN$ couplings would then have dimension of mass. 
Moreover, in the absence of any breaking of chiral symmetry,
the best choices of pion field are such that 
pion interactions involve derivatives, in which case the leading
P-, T-even pion-nucleon interaction is of the pseudovector form.
Yet, frequently a pseudoscalar form in terms of $\psi_N$ is used.
Chiral symmetry is then only ensured if additional 
$\boldpi^2 \bar{\psi}_N \psi_N$ 
interactions are included. Fortunately, in most of the instances
we are concerned with here, these additional interactions
are irrelevant, and pseudoscalar and pseudovector interactions
give the same result, once the corresponding couplings are related.
A summary of notation 
used by various authors is given in Table \ref{tab:pinnconv}.


\begin{table}[t]
\centering
\renewcommand{\arraystretch}{1.5}
\begin{tabular}{||c||c|c|c|c|c|c||}
\hline\hline
Interaction &  This work & Herczeg &  Pospelov  & Engel 
& Dmitriev & Mereghetti 
\\
& & & \& Ritz & {\it et al. }& {\it et al.} & {\it et al.} \\
& & \cite{Herczeg:1997ei} & \cite{Pospelov:2005pr} 
& \cite{jesus05,dobaczewski05} \ldots & \cite{dmitriev03,dmitriev05}
&\cite{Mereghetti:2010tp,Maekawa:2011vs,deVries:2012ab}
\\
\hline
$\bar{N}{\boldtau}\cdot{\boldpi}N$ & $\gpbz$ & ${\bar g}_{\pi NN}^{(0)\,\prime}$ 
& ${\bar g}_{\pi NN}^{(0)}$ & ${\bar g}_0$  & $-g_0$
&$\displaystyle{-\frac{({\bar g}_{0}+{\bar g}_2/3)}{F_\pi}}$ 
\\
$\pi^0 \bar{N}N$ &  $\gpbo$ & ${\bar g}_{\pi NN}^{(1)\,\prime}$ & 
${\bar g}_{\pi NN}^{(1)}$ & ${\bar g}_1$ & $g_1$  &$-{\bar g}_{1}/F_\pi$ 
\\
$\bar{N}\left( 3\tau_3\pi^0- {\boldtau}\cdot{\boldpi} \right)N$ & 
$\gpbt$ & ${\bar g}_{\pi NN}^{(2)\,\prime}$ & $-{\bar g}_{\pi NN}^{(2)}$  & 
${\bar g}_2$ & $g_2$ & $ -{\bar g}_2/3 F_\pi$ 
\\
\hline
$(\partial_\mu\boldpi) \cdot \bar N\boldtau S^\mu N$ 
& $-2g_A/F_\pi$ 
& $-g_{\pi NN}/m_N$ 
& $-g_{\pi NN}/m_N$ 
& $-g/m_N$ 
& $-g/m_N$
& $-2g_A/F_\pi$
\\
\hline\hline
\end{tabular}
\caption{ Conventions for 
$\pi NN$ couplings: 
first three rows give TVPV
non-derivative interactions while the last row gives the leading-order strong 
interaction. 
Note that $N$ denotes a heavy nucleon field,
so the pseudovector interaction in the
last row corresponds to the pseudoscalar coupling 
${\boldpi}\cdot {\bar \psi}_N i\gamma_5{\boldtau} \psi_N$
(plus an additional two-pion interaction) 
in terms of a relativistic field $\psi_N$.
In the chiral limit $g_A\simeq 1.26$ and $g=g_{\pi NN}\simeq 12.6$,
while accounting for the Goldberger-Treiman discrepancy
$g=g_{\pi NN}\simeq 13.5$ \cite{Stoks:1992ja,vanKolck:1996rm} 
and $g_A\simeq 1.33$.
Here, $F_\pi = 185$ MeV. 
\label{tab:pinnconv}}
\end{table}


The third and fourth terms in Eq.~(\ref{chiPTTV}) contain T- and P-odd
two-nucleon contact interactions, which represent all dynamics of range $\sim
1/\lamchi$, such as vector meson ($\eta$, $\rho$, $\omega$, {\em etc.})
exchange.  As we discuss below, these are expected to be the most significant
short-range TVPV interactions among nucleons.  The ``$\cdots$'' subsume an
infinite number of other TVPV interactions: terms related to the above
by chiral symmetry (see below) as well as interactions involving larger number
of derivatives and nucleon fields and/or more powers of small parameters. For
purposes of the present analysis we will not draw on these additional
interactions explicitly.  The reader should be warned that in general the
Lagrangian \eqref{chiPTTV} contains pion tadpoles
\cite{Mereghetti:2010tp,deVries:2012ab}, as no spacetime symmetry forbids a
$\pi^0$ term (accompanied by its chiral partners with an odd number of pions)
representing the disappearance of the neutral pion into vacuum.  Tadpoles can
be eliminated by field redefinitions, but for the left-right four-quark
operator \eqref{eq:lreff0} a multi-pion vertex survives at leading order
\cite{deVries:2012ab}.  Although usually this is of no consequence, it might
give rise to a significant TVPV three-nucleon force.

The various hadronic interactions in Eq.~(\ref{chiPTTV}) can be generated 
through the $\theta$-term or any of the dimension-six CPV operators introduced 
above that contains only quarks and/or gluons. The semileptonic four-fermion 
operators $Q_{\ell e dq}$ and $Q_{\ell e du}^{(1,3)}$ will give rise to effective 
electron-hadron interactions.  Concentrating on the electron-nucleon sector, 
we follow roughly the convention of Ref.~\cite{Ginges:2003qt} to write 
\bea
\label{eq:eNcpv}
\mathcal{L}^\mathrm{eff}_{eN} & = &- \frac{G_F}{\sqrt{2}}
\Bigl\{{\bar e}i\gamma_5 e\ {\bar \psi_N} \left[ C_S^{(0)} 
+C_S^{(1)}\tau_3\right] \psi_N
+ {\bar e} e\ {\bar \psi_N} i\gamma_5 \left[ C_P^{(0)} 
+C_P^{(1)}\tau_3\right] \psi_N
\\
\nonumber
&&- \epsilon_{\mu\nu\alpha\beta}\, {\bar e} \sigma^{\mu\nu} e\
{\bar \psi_N} \sigma^{\alpha\beta} \left[ C_T^{(0)} 
+C_T^{(1)}\tau_3\right] \psi_N\Bigr\}
+\cdots 
\eea
in terms of a relativistic nucleon field $\psi_N$. Normalizing to the 
$G_F=1/(\sqrt{2} v^2)$ allows us to make a straightforward comparison with 
limits quoted in the atomic EDM literature. A conversion to the operators 
normalized to $\Lambda$ appears in Section \ref{sec:semi} below.
This interaction simplifies for a heavy nucleon field,
\bea
\label{eq:eheavyNcpv}
\mathcal{L}^\mathrm{eff}_{eN} & = & -\frac{G_F}{\sqrt{2}}
\Bigl\{{\bar e}i\gamma_5 e\ {\bar N} \left[ C_S^{(0)} +C_S^{(1)}\tau_3\right] N
-8\, {\bar e} \sigma_{\mu\nu} e\ v^\nu
{\bar N} \left[ C_T^{(0)} +C_T^{(1)}\tau_3\right] S^\mu N\Bigr\}
+\cdots \ .
\eea
Here again, we neglect higher-derivative terms, operators containing more 
than two nucleon fields, and terms containing 
explicit factors of the pion field as implied by chiral symmetry. 
Note that the electron scalar $\otimes$ nucleon pseudoscalar operators
vanish at lowest order in the heavy baryon expansion\footnote{Note that we have introduced an overall minus sign
on the right hand sides of Eqs.~(\ref{eq:eNcpv},\ref{eq:eheavyNcpv}) to match the convention in 
Ref.~\cite{Ginges:2003qt} and elsewhere, where
the corresponding coefficients are defined for the Hamiltonian rather than the Lagrangian. Note also that an
explicit $-1$ appears in front of the tensor interactions in order to facilitate comparison with other work in which 
an opposite sign convention is used for $\epsilon_{\mu\nu\alpha\beta}$.}.

\subsection{Hadronic Matrix Elements}
\label{sub:hadme}

In order to determine the dependence of the hadronic couplings defined above on
the underlying sources of CPV, one must compute matrix elements of the various 
CPV operators introduced in Section \ref{sec:def}. The result will be a set of 
expressions of the form
\bea
\label{eq:dNhad}
d_N & = & \alpha_N\ {\bar\theta} +\left(\frac{v}{\Lambda}\right)^2\ 
\sum_k\beta_N^{(k)} \ \mathrm{Im}\, C_k \ , 
\\
\label{eq:gpbhad}
\gpbi & = & \lambda_{(i)}\ {\bar\theta} + \left(\frac{v}{\Lambda}\right)^2\ 
\sum_k \gamma_{(i)}^{(k)} \ \mathrm{Im}\, C_k \ , 
\\
\label{eq:Cihad}
{\bar C}_i & = & \kappa_{i}\ {\bar\theta} + \left(\frac{v}{\Lambda}\right)^2\ 
\sum_k \delta_{i}^{(k)} \ \mathrm{Im}\, C_k \ , 
\eea
where $C_k$ denotes the Wilson coefficients for operator $Q_k$,
as appropriate, and 
the coefficients $\alpha_N$ {\em etc.} embody the results of the hadronic
matrix-element computation. 
Note that for the three-gluon operator \eqref{eq:weinbop}, here and in 
the rest of this review $\mathrm{Im}\, C_k $ stands for $C_{\tilde G}$.
The coefficients $\alpha_N$ and 
$\beta_N^{(k)}$ have the units of electric charge times length, and we will 
express all results as $e$ fm. The coefficients $\lambda_{(i)}$ and 
$\gamma_{(i)}^{(k)}$ are dimensionless, while $\kappa_{i}$ and 
$\delta_{i}^{(k)}$ have dimensions of fm$^3$. We note that presently
very little is known about ${\bar C}_{1,2}$.

For future purposes, it will 
be convenient to define the sensitivity of the 
other hadronic quantities to 
either ${\tilde\delta}_q$ and $\delta_q$ 
or $d_q$ and ${\tilde d}_q$ {\em via}
\bea
\label{eq:zetadef}
\left(\frac{v}{\Lambda}\right)^2\ 
\left[\beta_N^{q G}\ \mathrm{Im}\, C_{q G} 
+ \beta_N^{q \gamma}\ \mathrm{Im}\, C_{q\gamma}\right]&=& 
e\,  {\tilde\rho}_N^{q}\, {\tilde d}_q 
+ \rho_N^{q}\, d_q 
= \left(\frac{v}{\Lambda}\right)^2\ 
\left[ e\, {\tilde\zeta}_N^{q}\, {\tilde \delta}_q 
+e\,  \zeta_N^{q}\, \delta_q \right] \ ,
\\
\label{eq:etadef}
\left(\frac{v}{\Lambda}\right)^2\ 
\left[ \gamma_{(i)}^{q G}\ \mathrm{Im}\, C_{q G}
+ \gamma_{(i)}^{q \gamma}\ \mathrm{Im}\, C_{q\gamma}\right]
&=&{\tilde\omega}_{(i)}^{q}\, {\tilde d}_q 
+ \omega_{(i)}^{q}\, d_q 
=\left(\frac{v}{\Lambda}\right)^2\ 
\left[ {\tilde\eta}_{(i)}^{q}\, {\tilde \delta}_q 
+ \eta_{(i)}^{q}\, \delta_q \right]\ .
\eea

Similarly, for the semileptonic interactions, we use $G_F=1/{\sqrt{2} v^2}$;
define $g_{S,P,T}^{(i)}$ as the isoscalar and isovector form factors in the limit
of isospsin symmetry
\bea
\label{eq:ffdef}
\frac{1}{2} \bra{N} \left[{\bar u} \Gamma u + {\bar d}\Gamma d\right]\ket{N} 
&\equiv& g_\Gamma^{(0)} {\bar \psi_N} \Gamma \psi_N\ ,\\
\frac{1}{2} \bra{N} \left[{\bar u} \Gamma u - {\bar d}\Gamma d\right]\ket{N} 
&\equiv& g_\Gamma^{(1)} {\bar \psi_N} \Gamma \tau_3 \psi_N\ ,
\eea
where $\Gamma = 1$, $\gamma_5$, $\sigma_{\mu\nu}$;
and write for $C_{S,P,T}^{(0,1)}$,
\begin{align}
C_S^{(0)}  &=& -g_S^{(0)}\, \left(\frac{v}{\Lambda}\right)^2\,  
\mathrm{Im}\, C_{eq}^{(-)}
&\qquad\mathrm{and} &
C_S^{(1)}  &=&  g_S^{(1)}\, \left(\frac{v}{\Lambda}\right)^2\,  
\mathrm{Im}\, C_{eq}^{(+)}
\label{eq:CSi}
\\
C_P^{(0)}  &= & g_P^{(0)}\, \left(\frac{v}{\Lambda}\right)^2\,  
\mathrm{Im}\, C_{eq}^{(+)}
&\qquad\mathrm{and} &
C_P^{(1)}  &= & -g_P^{(1)}\, \left(\frac{v}{\Lambda}\right)^2\,  
\mathrm{Im}\, C_{eq}^{(-)}
\label{eq:CPi}
\\
C_T^{(0)}  &= & -g_T^{(0)}\, \left(\frac{v}{\Lambda}\right)^2\,  
\mathrm{Im}\, C_{\ell e q u}^{(3)} 
&\qquad\mathrm{and} &
C_T^{(1)}  &= & -g_T^{(1)}\, \left(\frac{v}{\Lambda}\right)^2\,  
\mathrm{Im}\, C_{\ell e q u}^{(3)}  \ .
\label{eq:CTi}
\end{align}
where  we  define the combinations
\be
\label{eq:Ceqdef}
C_{eq}^{(\pm)}= C_{\ell e dq} \pm C_{\ell e q u}^{(1)} \ \ \ .
\ee


For the dimension-six operators generated by BSM CPV, performing the hadronic 
computation entails two successive steps of running and matching. 
\begin{itemize}
\item[(i)] One must first  run the operators perturbatively from the BSM scale $\Lambda$ to the weak scale. After integrating out the heavy SM degrees of freedom with appropriate matching, one then continues the running 
from the weak scale to the hadronic scale. The quantities $C_{q G}$, 
${\tilde d}_q$, ${\tilde \delta}$, {\em etc.} are then defined at the hadronic scale 
$\lamchi\sim 1$ GeV where nucleon matrix elements are then taken. They 
can be related to the quantities at the BSM scale $\Lambda$ through an 
appropriate \lq\lq $K$-factor", as in
\bea
\label{eq:Kfactordef}
\mathrm{Im}\, \left[g_3\, C_{q G}\right] (\lamchi) &=& K_{q G}\ \mathrm{Im}\, \left[g_3\, C_{q G}\right](\Lambda)\ ,
\ 
[g_3\, {\tilde d}_q] (\lamchi) = K_{q G}[g_3\, {\tilde d}_q](\Lambda)\ , \\
\nonumber
[g_3\, {\tilde \delta}_q] (\lamchi)  &=& K_{q G}\ [g_3\, {\tilde \delta}_q] (\Lambda)\ ,
\eea
where we follow the convention in the literature and bundle the strong coupling with the
Wilson coefficients $\mathrm{Im}\, C_k$ {\em etc.} The $K$-factors then relate the product of $g_3$ and the
Wilson coefficients at the two scales $\Lambda$ and $\lamchi$.


\item[(ii)]Second, one must compute the relevant matrix element at the hadronic scale 
utilizing non-perturbative methods. For the QCD $\theta$-term, only the second 
step is required. 
\end{itemize}

Carrying out the perturbative running is generally straightforward. In general, one must account for
mixing among various operators. The full anomalous dimension matrix that takes into account
the EDM, CEDM, three-gluon and four quark operators has recently been obtained in Ref.~\cite{Hisano:2012cc}. Prior
to this work, efforts concentrated largely on the evolution of the EDM, CEDM, and three-gluon operators\cite{Morozov:1985ef,Chang:1990jv,Arnowitt:1990eh,Braaten:1990gq}. Within this limited subset
of operators, only the three-gluon operator is multiplicatively renormalized. The resulting \lq\lq $K$-factor", obtained after taking into account two-loop running and threshold corrections, is given in the first line of Table \ref{tab:pQCD}. The three-gluon operator, however, will mix into the CEDM while the latter will mix into the EDM. Consequently it is not generally possible to quote a single $K$-factor for the latter two operators. Since the work of Ref.~\cite{Arnowitt:1990eh}, however, it has often been the practice to do so in the literature. The reason is that that in the MSSM, $C_{\tilde G}$ arises at two-loop order, whereas the CEDM first occurs at one-loop. Thus, the mixing of $Q_{\tilde G}$ into $Q_{qG}$ is effectively higher loop order. In the more general case, one must consider the full effects of operator mixing. Nevertheless, for illustrative purposes we quote a $K$-factor for the CEDM to illustrate the magnitude impact made by QCD evolution from the weak to hadronic scales. Under similar assumptions, the authors of Ref.~\cite{Arnowitt:1990eh} obtained the $K$-factor given in Table \ref{tab:pQCD}. 

\begin{table}[t]
\centering
\renewcommand{\arraystretch}{1.5}
\begin{tabular}{||c|c|c|c||}
\hline\hline
Operator &  $K_Q$ & Reference & Remarks\\
\hline
$Q_{\tilde G}$ &3.30 & \cite{Morozov:1985ef,Chang:1990jv,Arnowitt:1990eh,Braaten:1990gq} & Mult Renorm\\
$Q_{q G}$ &3.30 & \cite{Arnowitt:1990eh} & Mixing neglected \\
$Q_{q V}$,\ $V=B,W$ & 1.53&\cite{Arnowitt:1990eh} & Mixing neglected \\
$Q_{quqd}^{(1,8)}$ & matrix & \cite{Hisano:2012cc} &  \\
 \hline\hline
\end{tabular}
\caption{Illustrative perturbative renormalization factors for dimension-six CPV operators, 
$Q$. As in Eq.~(\ref{eq:Kfactordef}) the $K$-factors apply to the product of $g_3$ with the $\mathrm{Im}\, C_k$, {\em etc}. In general, only $Q_{\tilde G}$ is multiplicatively renormalized, with the renormalization
$K$-factor given in the first row. For all other operators, one must take into account mixing. Under the
assumptions made in Ref.~\cite{Arnowitt:1990eh} for the MSSM, approximate $K$-factor for the EDM
and CEDM operators have been obtained as quoted above. No analogous approximation
has been made for the four-quark operators, so we do not list corresponding entries. For a recent determination of the anomalous dimension matrix, 
see Ref.~\cite{Hisano:2012cc}.
\label{tab:pQCD}}
\end{table}

Performing non-perturbative computations is a more challenging task. Before 
reviewing the status of such calculations, it is useful to delineate 
expectations for the hadronic matrix elements based on the chiral symmetry 
properties of the operators, following the framework developed in 
Refs.~\cite{Mereghetti:2010tp,deVries:2012ab}. For a parallel treatment 
in the context of chiral SU(3),  see Refs.~\cite{Bsaisou:2012rg,Guo:2012vf}. We follow this discussion with 
a review of explicit computations utilizing various approaches.

\subsection{Chiral Symmetry 
and Na\"ive Dimensional Analysis}
\label{sect:chiralcons}

In the limit of vanishing quark masses, the QCD Lagrangian is invariant under 
separate rotations of the right- and left-handed fields. Specializing to the 
two lightest flavors, these rotations are given by
\be
\label{eq:chiral1}
q\to\mathrm{exp}\left[i{\boldtau}\cdot
\left(\boldtheta_R P_R+\boldtheta_L P_L\right)\right]\ q
\ee
where $P_{R(L)}$ denote right- (left-) handed projection operators and 
$\boldtheta_{R(L)}$ are three-component vectors of arbitrary real numbers. 
For future reference, it is useful to re-express Eq.~(\ref{eq:chiral1}) in 
terms of vector and axial rotations:
\be
\label{eq:chiral2}
q\to\mathrm{exp}\left[i{\boldtau}\cdot
\left(\boldtheta_V +\boldtheta_A \gamma_5\right)\right]\ q  \ .
\ee

The chiral SU(2$)_R\times$SU(2$)_L$ transformation embodied in Eqs. 
(\ref{eq:chiral1}, \ref{eq:chiral2})
are isomorphic to those of SO(4), and for 
present purposes it is convenient to consider objects that have definite 
SO(4) transformation properties. 
For example,
four-component SO(4) vectors 
\be
V=\left(
\begin{array}{c}
{\boldV}\\
V_4
\end{array}
\right) 
\ee
change, under an infinitesimal transform, by
\be
\delta V = 
\left(
\begin{array}{c}
{\boldV}\times{\boldtheta}_V+V_4\, {\boldtheta}_A\\
-{\boldtheta}_A\cdot{\boldV}
\end{array}
\right) \ ,
\ee
where ${\boldtheta}_{V,A}$ are presumed to be tiny. 

Terms in the effective Lagrangian just above the hadronic scale, 
and in particular the CPV
interactions in Eq.~(\ref{eq:LCPV1}), break chiral
symmetry in specific ways. 
In order to reproduce the corresponding S matrix,
the effective Lagrangian written in terms of hadronic fields
has interactions that break the symmetries in the same way.
For example, instead of a component of a chiral four-vector $V[q]$
built out of quark fields, there will be 
a hadronic chiral four-vector $V[\boldpi, N]$ built from nucleon 
and pion fields.
For a particular choice of pion fields, the latter 
can be related to one having no pions, $V[0,N]$, by
\bea
\label{eq:Vhad}
{\boldV}[\boldpi, N] & = & {\boldV}[0,N]
-\frac{2\boldpi}{D F_\pi}
\left(\frac{\boldpi}{F_\pi} \cdot{\boldV}[0,N]-V_4[0,N]\right) \ ,
\\
V_4[\boldpi, N]  & = & V_4[0,N]
-\frac{2\boldpi}{D F_\pi}\cdot
\left(\frac{\boldpi}{F_\pi} V_4[0,N] + {\boldV}[0,N] \right)\ , 
\label{eq:V4had}
\eea
where $D=1+\boldpi^2/F_\pi^2$.

The proportionality constant between 
the hadronic interaction strength and the interaction
strength above the hadronic scale is the hadronic matrix element.
Hadronic interactions obtained from different components of
the same object share the same matrix element.
When the matrix element is not known, it can be estimated
using naive dimensional analysis (NDA) \cite{Manohar:1983md, Weinberg:1989dx}.
Because the short-distance physics incorporated in operators in the 
Lagrangian
cannot be separated from quantum-mechanical
effects represented by loops in Feynman diagrams,
one assumes that the natural size of the operator coefficients 
is given by loop cutoff changes of ${\cal O}(1)$.
If $M$ denotes the scale of breakdown of the EFT,
the dimensionless ``reduced'' coefficient $(4\pi)^{2-N} M^{D-4} g$
of an operator 
of canonical dimension $D$ involving $N$ fields 
is assumed to be ${\cal O}(1)$ times the appropriate powers
of the reduced couplings of the underlying theory.
When applied to chiral-symmetric operators,
which are characterized only by the reduced QCD coupling $g_3/4\pi$,
but to any power,
consistency requires that we take $g_3\sim 4\pi$ in matrix-element estimates. 




Before proceeding with detailed applications, we consider two simple illustrations. First, when matching the CEDM operators onto the nucleon EDMs, we note that both operators posses the same canonical dimension. In this case $N=3$ and $D=5$ (as we are below the weak scale). Since ${\tilde d}_q$ and $d_N$ have dimension $M^{-1}$, we need only focus on the powers $g_3$ and $4\pi$. In this case, $N=3$ implies that   
\be
{\tilde\rho}_N^\mathrm{NDA}  =  {g_3}/{4\pi}\sim 1\ \ \ . 
\ee
On the other hand for the contribution of the three-gluon operator, which is $D=6$ even below the weak scale, we require one additional power of the hadronic scale $\lamchi$ as in 
\be
\beta_N^{\tilde G}  = {g_3 \lamchi}/{4\pi}\ \sim \lamchi.
\ee


A second simple example is provided by the quark mass effects on
the proton and neutron masses. 
The 
quark mass operators in Eq. \eqref{eq:Lcpmass} are components of 
two SO(4) vectors
whose fourth component transform as a scalar 
or pseudoscalar 
under parity, 
\be
\label{eq:SO4SP}
S[q]=\left(
\begin{array}{c}
-i{\bar q}{\boldtau} \gamma_5 q\\
{\bar q} q
\end{array}
\right)
\qquad \mathrm{and} \qquad
P[q]=\left(
\begin{array}{c}
{\bar q}{\boldtau} q\\
i{\bar q}\gamma_5 q
\end{array}
\right)\ ,
\ee
respectively. 
Replacing the light quark doublet in Eq.~(\ref{eq:SO4SP}) by heavy nucleon 
fields and noting that the pseudoscalar operators vanish to lowest order
in the heavy nucleon limit, we obtain the corresponding nucleon SO(4) vectors:
\be
\label{eq:SP}
S[0,N]=\left(
\begin{array}{c}
\mbox{\boldmath $0$}\\
{\bar N}{N}
\end{array}
\right)
\qquad \mathrm{and} \qquad
P[0,N]=\left(
\begin{array}{c}
{\bar N} {\boldtau} N\\
0
\end{array}
\right) \ .
\ee
Using Eqs. (\ref{eq:Vhad}, \ref{eq:V4had})
one finds that the fourth component of $S[\boldpi,N]$ and third component of 
$P[\boldpi,N]$
give contributions from the average quark mass and mass splitting to 
the average nucleon mass, 
$({\bar m}_N)_q$,
and nucleon mass difference, 
$(\Delta m_N)_q\equiv (m_n-m_p)_q$,
respectively: 
\be
\mathcal{L}_\mathrm{mass}^{N}  = 
-\left( {\bar m}_N\right)_q {\bar N} N 
+
\frac{\left( \Delta m_N\right)_q}{2} {\bar N}\tau_3 N\ .
\label{eq:nucleonmassL}
\ee
The reduced coefficients are $({\bar m}_N)_q/\Lambda_\chi$ 
and $(\Delta m_N)_q/\Lambda_\chi$, which should be linear
in $\bar{m}/\Lambda_\chi$ and $\epsilon \bar{m}/\Lambda_\chi$,
respectively, so that from NDA one expects 
$({\bar m}_N)_q \sim \bar{m}$ and 
$(\Delta m_N)_q\sim \epsilon \bar{m}$.



These terms are linked by chiral symmetry
to others that contain an even number of pion fields, which 
contribute to pionic processes such as pion-nucleon scattering
and pion production in nucleon-nucleon collisions.
The two terms in Eq. \ref{eq:nucleonmassL}
can be seen as the isospin-symmetric and
breaking components of the sigma term.
The corresponding coefficients have therefore been evaluated in lattice QCD
and also extracted
from data.
Results are, by and large, in agreement.
The extrapolation of lattice results
on octet baryon masses in 2+1 flavor
QCD \cite{Aoki:2008sm,Shanahan:2012wh}, for example,
gives $( \bar{m}_N)_q=45\pm 6$ MeV, which agrees
with the venerable value from Ref. \cite{Gasser:1990ce}.
Other extractions from data give similar values (see, {\it e.g.},
the compilation in Ref. \cite{Baru:2011bw}; see also \cite{Hoferichter:2012tu} ).
Similarly, the lattice value
$( {\Delta m}_N)_q=2.26\pm0.57\pm 0.42\pm 0.10$ MeV \cite{Beane:2006fk}
is consistent with other lattice
evaluations (see Ref. \cite{WalkerLoud:2010qq}),
with a determination of the electromagnetic 
splitting using dispersion relations \cite{WalkerLoud:2012bg},
and with an extraction from pion production \cite{Filin:2009yh}.



For future purposes, it is useful to relate ${\bar m}$ to the pion mass.  
Since the pion is the pseudo-Goldstone boson of spontaneously broken chiral 
symmetry, its mass 
vanishes in the limit ${\bar m}\to 0$.
Away from this limit, the pion mass term arises from a scalar SO(4)
vector as in Eq. \eqref{eq:SP}, but with $\bar{N}N\to 1$.
It is 
thus proportional to $\bar{m}$, and from NDA
$m_\pi^2\sim \bar{m} \lamchi$.
It is convenient, as we do below,
to estimate the coefficients of chiral-breaking operators
steming from the quark masses in terms of  $m_\pi^2/\lamchi$.


\subsubsection{Applications: $\bar\theta$-term}

We now use these classifications to identify the expected scaling of various 
hadronic operators as they are generated by underlying CPV interactions. We 
begin with the QCD ${\bar\theta}$-term, $\mathcal{L}_{\bar\theta}$ 
\eqref{thetaterm}, in the form of the quark bilinear \eqref{eq:cpvmass}.

This transformed $\theta$-term
is the fourth component
of the same chiral SO(4) pseudovector 
as the quark mass term, and it gives rise to the 
$-2{\bar N}{\boldtau}\cdot{\boldpi} N/F_\pi$ 
term in Eq. \eqref{chiPTTV} \cite{Mereghetti:2010tp}.
Thus, ${\bar N}\tau_3 N$ and $-2{\bar N}{\boldtau}\cdot{\boldpi} N/F_\pi$ 
(plus its chiral partners with more pion fields)
transform as the third and fourth components of the same SO(4) 
pseudovector at the hadronic level, and the coefficient of the latter must be 
given in terms of matrix elements of the former:
\be
C_\mathrm{had} P[\boldpi,N] = 
\langle{\mathrm{had}}\vert P \vert{\mathrm{had}}\rangle \ ,
\ee
where the state $\vert \mathrm{had}\rangle$ contains appropriate nucleon and 
pion modes, {\em viz.}
\be
\langle N \vert P_3 \vert{N}\rangle =  C_\mathrm{had}  {\bar N}\tau_3 N
\qquad\mathrm{and}\qquad
\langle N \vert P_4 \vert{N\pi}\rangle   =  
- 2C_\mathrm{had} {\bar N}{\boldtau}\cdot{\boldpi} N/F_\pi\ .
\ee
On the other hand, since 
\be
\epsilon {\bar m} \langle N \vert P_3 \vert{N}\rangle = 
\frac{\left( \Delta m_N\right)_q}{2} {\bar N}\tau_3 N \ ,
\ee
we have that 
\be
C_\mathrm{had} = \left( \Delta m_N\right)_q/2\epsilon {\bar m}\ .
\ee
Hence, the matrix element of the QCD ${\theta}$-term operator is
\be
\label{eq:intstepforg0}
-\frac{\bar{m}}{2}\left(1-\epsilon^2\right) \bar{\theta}\ 
\langle N \vert P_4 \vert{N\pi}\rangle =
\frac{1-\epsilon^2}{2\epsilon} \
\frac{\left( \Delta m_N\right)_q}{F_\pi} \ \bar\theta
{\bar N}{\boldtau}\cdot{\boldpi} N
\equiv\gpbz {\bar N}{\boldtau}\cdot{\boldpi} N\ .
\ee
Thus, we obtain the prediction
\be
\label{eq:gpbzchiral1}
\gpbz 
= \frac{1-\epsilon^2}{2\epsilon}\ \frac{\left( \Delta m_N\right)_q}{F_\pi} 
\ \bar\theta
\qquad\mathrm{or}\qquad
\lambda_{(0)}= \frac{1-\epsilon^2}{2\epsilon}\ 
\frac{\left( \Delta m_N\right)_q}{F_\pi} 
\ .
\ee
If the matrix element $(1-\epsilon^2)( \Delta m_N)_q/\epsilon$ is calculated,
$\gpbz/\bar\theta$ comes for free.
As a rough estimate, taking 
$\left( \Delta m_N\right)_q\sim 2\epsilon {\bar m}$ 
and ${\bar m}\sim F_\pi/20$ one would expect 
$\gpbz\sim 0.05 \, {\bar\theta}$, or $\lambda_{(0)}\sim 0.05$.
We may also express the relationships in Eq.~(\ref{eq:gpbzchiral1}) as
\be
\label{eq:gpbzchiral3}
\lambda_{(0)} \sim \frac{m_\pi^2}{\lamchi F_\pi} \ .
\ee
This expectation is given in Table \ref{tab:hadme1} along with predictions 
for the same quantity using other approaches. 
The lattice value for $( {\Delta m}_N)_q$ given above
implies
\be
\lambda_{(0)} = 0.017 \pm 0.005\ ,
\ee
where the error is obtained by adding the lattice uncertainties in $( {\Delta m}_N)_q$ in quadrature.




The foregoing reasoning leads \cite{Mereghetti:2010tp} to analogous 
expectations for the other $\lambda_{(i)}$ as well as the hadronic coefficients 
$\alpha_N$, $\beta_N^{(k)}$, and $\gamma_{(i)}^{(k)}$. 
For example, 
the simplest way to produce the isovector 
TVPV $\pi NN$ interaction
$\pi^0 {\bar N}N$ in Eq. \eqref{chiPTTV}
is from a tensor product of two pseudoscalar vectors, and as a consequence
\be
\label{eq:lam1theta}
\lambda_{(1)} \sim \frac{m_\pi^4}{\lamchi^3 F_\pi}\ ,
\ee
where we took $\epsilon \sim 1$.
The isotensor $\pi NN$ interaction in Eq. \eqref{chiPTTV}
is even more suppressed.

The analogous arguments for the short-range components of the nucleon
EDM are more complicated because one needs to account for the chiral
transformation properties of the interaction between quarks and
the photon field $A_\mu$,
\be
\label{eq:Lcpcharge}
\mathcal{L}_\mathrm{charge}^\mathrm{quark} = -\frac{e}{6} A_\mu \
{\bar q}\gamma^\mu \left(1+3\tau_3 \right) q \ .
\ee
While the first term is a chiral scalar, the second is the 3-4 component
of an antisymmetric tensor. Taking the tensor product with
the pseudoscalar vector $P$, they give rise, respectively,
to the isoscalar and isovector nucleon EDMs.
Thus, one expects 
\be
\label{eq:alphanchiral}
\bar{d}_{0,1}\sim e {\bar\theta}\ \frac{m_\pi^2}{\lamchi^3}
\qquad\mathrm{or}\qquad
\alpha_N\sim e\, \frac{m_\pi^2}{\lamchi^3}
\sim 0.2\, \frac{m_\pi^2}{\lamchi^2 }\ e \, \mathrm{fm}\ ,
\ee
where the additional factors of $\lamchi^{-2}$ are simply a consequence of
dimensional analysis.

\subsubsection{Dimension-six operators}

We now consider the dimension-six CPV operators appearing in 
Eq.~(\ref{eq:LCPV2}) and arising from BSM physics~\cite{deVries:2012ab}. 
The quark CEDMs can be embedded in SO(4) vectors and pseudovectors:
\be
S^{A\mu\nu} G_{\mu\nu}^A \equiv\left(
\begin{array}{c}
-i{\bar q} \sigma^{\mu\nu} \gamma_5 {\boldtau}\, T^A q
\\
{\bar q} \sigma^{\mu\nu} T^A q
\end{array}
\right) G_{\mu\nu}^A
\qquad \mathrm{and} \qquad
P^{A\mu\nu} G_{\mu\nu}^A \equiv \left(
\begin{array}{c}
{\bar q}\sigma^{\mu\nu}{\boldtau}\, T^A q
\\
i{\bar q}\sigma^{\mu\nu}\gamma_5 T^A q
\end{array}
\right) G_{\mu\nu}^A \ .
\label{eq:SAPA}
\ee
Thus, the isoscalar and isovector CEDM operators transform as the $P_4$ and 
$S_3$ components of an SO(4) 
pseudovector and vector, respectively. 
They contribute to $\gpbz$ and $\gpbo$ without any additional factors 
associated with chiral symmetry breaking. Thus, we expect these two couplings
to be comparable,
\be
{\bar g}_\pi^{(0,1)} \sim \frac{\lamchi^2}{vF_\pi} 
\left( \frac{v}{\Lambda}\right)^2\ 
\mathrm{Im} C_{q G}
\qquad\mathrm{or}\qquad
\gamma_{(0,1)}^{q G} \sim \frac{\lamchi^2}{vF_\pi} \ .
\ee 
Again $\gpbt$ is a higher-order effect.
However, since the EDM and CEDM 
Wilson coefficients carry an explicit 
factor of the quark Yukawa couplings 
$Y_q\sim m_q/v\sim m_\pi^2/ (v \lamchi)$,  it is useful to express the 
$\gpbi$ in terms of the quantity ${\tilde \delta}_q$ appearing in 
Eq.~(\ref{eq:nda1prime}) as well as the pion mass and QCD mass scales:
\be
{\bar g}_\pi^{(0,1)} \sim
\frac{m_\pi^2 \lamchi}{F_\pi\Lambda^2} \ {\tilde\delta}_q
\qquad\mathrm{or}\qquad
{\tilde\eta}_{(0,1)}^q \sim 
\frac{m_\pi^2 \lamchi}{F_\pi v^2} \ .
\ee
Similarly, we obtain for the sensitivity of ${\bar g}_\pi^{(0,1)}$ to the 
${\tilde d}_q$
\be
{\tilde\omega}_{(0,1)}^q \sim \frac{\Lambda_\chi^2}{F_\pi}\ .
\ee


From similar considerations, the CEDM contributions to the nucleon EDM requires
no additional chiral suppression but only an electromagnetic interaction, 
leading to
\be
\label{eq:EDMdelta}
\bar{d}_{(0,1)}\sim \frac{e}{v}
\left( \frac{v}{\Lambda}\right)^2\ \mathrm{Im}\ C_{q G}  
\qquad\mathrm{or}\qquad
\beta_N^{q G}\sim \frac{e}{v} \ ,
\ee
or, alternatively,
\be
\label{eq:EDMdeltaprime}
\bar{d}_{(0,1)}\sim  
\frac{e m_\pi^2}{\lamchi\Lambda^2}\  {\tilde\delta}_q
\qquad\mathrm{or}\qquad
{\tilde\zeta}_{0,1} \sim 
\frac{m_\pi^2}{\lamchi v^2}\ 
\ee
and 
\be
{\tilde\rho}_N^q = 1 \ .
\ee

For the effect of the quark EDM operators, the logic is similar.
The transformation properties of the isoscalar and isovector quark EDMs are 
obtained by replacing the $T^A\, G_{\mu\nu}^A$ in Eq. \eqref{eq:SAPA}
by $F^{\mu\nu}$.
Purely hadronic operators now require integrating out high-momentum
photons exchanged among quarks, which generates an additional factor of 
at least $\alpha/\pi$. In particular, for the $\gpbi$
we expect
\be
\gamma_{(i)}^{q \gamma} \sim \frac{\alpha}{\pi}    
\frac{\lamchi^2}{v F_\pi}\ ,
\qquad \omega_{(i)}^q \sim \frac{\alpha}{\pi} \frac{\Lambda_\chi^2}{e F_\pi}\ ,
\qquad\mathrm{and}\qquad 
{\eta}_{(i)}^q \sim \frac{\alpha}{\pi}
\frac{m_\pi^2 \lamchi}{F_\pi v^2} \ .
\ee
This suppression renders these operators irrelevant for most purposes.
In contrast, no such $\alpha/\pi$ factor is needed for the quark EDM 
contribution to the nucleon EDMs since the photon is external. In this case, 
we expect
\be
\beta_N^{q \gamma}\sim\beta_N^{q G} \ ,
\qquad \rho_N^q\sim {\tilde\rho}_N^q\ ,
\qquad\mathrm{and}\qquad 
\zeta_{(0,1)}\sim{\tilde\zeta}_{(0,1)} \ .
\ee

The situation for the three-gluon and four-quark operators operators is more 
subtle. Both the $Q_{quqd}^{(1,8)}$ and $Q_{\tilde G}$ are chiral-invariant 
pseudoscalars. 
Using $S$ and $P$ from Eq.~(\ref{eq:SO4SP}) we obtain 
\be
S\cdot P = -  {\bar q}{\boldtau}i\gamma_5 q \cdot {\bar q}{\boldtau} q 
+{\bar q} q \, {\bar q}i \gamma_5 q
= Q_{quqd}^{(1)}\ .
\ee
A similar result applies to the analogous definitions of $S$ and $P$ but with 
the SU(3$)_C$ generators $T^A$ included, yielding the operator $Q_{quqd}^{(8)}$. 
The CPV three-gluon operator is trivially a chiral pseudoscalar as the gluon 
fields do not transform under SU(2$)_R\times$SU(2$)_L$.
Consequently, the contributions from these operators to the $\gpbi$ require 
explicit factors of $m_\pi^2/\lamchi$ and $\epsilon m_\pi^2/\lamchi$
that reflect the chiral symmetry
breaking needed to generate components of an SO(4) vector. Letting 
$C_k$ denote the Wilson coefficient for any one of these 
three operators 
we then expect
\be
\label{gsforchiinv}
\bar{g}_\pi^{(0,1)} \sim \frac{m_\pi^2 \lamchi}{F_\pi {\Lambda}^2}\ 
\mathrm{Im}\, C_k
\qquad\mathrm{or}\qquad
\gamma_{(0,1)}^{(k)} \sim \left(\frac{m_\pi}{v}\right)^2
\frac{\lamchi}{F_\pi}\ ,
\ee
with $\bar{g}_\pi^{(2)}$ yet again at a higher order.
Note that for such chiral-invariant TVPV sources the contact interactions
in Eq. \eqref{chiPTTV}, which are chiral invariant, can be generated without
any suppression from $\bar m$:
\be
\bar{C}_{1,2} \sim \frac{\lamchi}{F_\pi^2 {\Lambda}^2}\ \mathrm{Im}\, C_k 
\qquad\mathrm{or}\qquad
\delta^{(k)}_{1,2}\sim \frac{\lamchi}{F_\pi^2 v^2}\ .
\ee
In nuclei they can be competitive with
one-pion exchange with one $\bar{g}_\pi^{(0,1)}$ interaction,
as the enhancement $\sim m_\pi^{-2}$ 
from a pion propagator is compensated by the
$m_\pi^2$ in Eq. \eqref{gsforchiinv}.

There is no suppression also when,
in combination with the quark electromagnetic interaction,
these sources
produce isoscalar and isovector components of the nucleon EDM
transforming as chiral scalar and antisymmetric tensor, respectively: 
\be
\label{eq:EDMchiinv}
\bar{d}_{(0,1)}\sim \frac{e \lamchi}{{\Lambda}^2}\ \mathrm{Im}\, C_k
\qquad\mathrm{or}\qquad
\beta_N^{k}\sim \frac{e \lamchi}{v^2} \ .
\ee

Matching the four-quark operator in Eq.~(\ref{eq:lreff}) onto hadronic 
operators, we first observe that it is the 3-4 component of a symmetric chiral 
tensor. Looking at the terms without color matrices $T^A$,
one may decompose them into two terms, each 
having the form of the product 
of the third and fourth components 
of either $S$ or $P$:
\be
\label{eq:sspp}
{\bar u} u \, {\bar d}i\gamma_5 d -{\bar u}i \gamma_5 u\, {\bar d} d =
\frac{1}{2}\left[ S_3 \otimes S_4 +P_4\otimes P_3\right]\ .
\ee
Similar relations hold for terms with $T^A$. 

The $S_3 \otimes S_4$ structure leads directly to a contribution to 
$\gpbo$. However, the presence
of $S_3$ signals vacuum instability, as it generates also
terms with an odd number of pions, including a neutral pion tadpole. 
Such a tadpole can be eliminated by a chiral
rotation, but it leaves behind a three-pion interaction that has
not been well studied and is neglected here.
After this rotation, one finds also contributions to 
both $\gpbz$ and $\gpbo$, and 
\be
\label{eq:piNLR}
\bar{g}_\pi^{(0,1)} \sim \frac{\lamchi^3}{F_\pi {\Lambda}^2}\
\frac{\mathrm{Im}\, C_{\varphi u d}}{(4\pi)^2 }
\qquad\mathrm{or}\qquad
\gamma_{(0,1)}^{(\varphi u d)} \sim \frac{\lamchi^3}{(4\pi)^2 v^2 F_\pi}\ .
\ee
As for CEDM, there is no suppression in the nucleon EDM,
\be
\label{eq:EDMLR}
\bar{d}_{(0,1)}\sim \frac{e \lamchi}{{\Lambda}^2}\
\frac{\mathrm{Im}\, C_{\varphi u d}}{(4\pi)^2 }
\qquad\mathrm{or}\qquad
\beta_N^{(\varphi u d)}\sim \frac{e \lamchi}{(4\pi)^2 v^2} \ .
\ee

\subsubsection{Semileptonic interactions}
\label{sec:semi}



For the semileptonic matrix elements in 
Eqs. (\ref{eq:CSi},\ref{eq:CPi},\ref{eq:CTi}), 
we can follow similar considerations.
Scalar and tensor interactions are most
important at low energies, Eq. \eqref{eq:eheavyNcpv}. 
The values of $g_S^{(i)}$ follow straightforwardly from the contributions of the
light quarks to the nucleon masses \eqref{eq:nucleonmassL}:
\be
\label{eq:gs01}
g_S^{(0)} = \frac{( \bar{m}_N)_q}{2\bar m}\ , 
\qquad 
g_S^{(1)} = \frac{( {\Delta m}_N)_q}{4\epsilon{\bar m}}\ .
\ee
From the empirical and lattice values for 
$( \bar{m}_N)_q$ and $( {\Delta m}_N)_q$ given in 
Section \ref{sect:chiralcons}, one obtains
\be
\label{eq:gs01b}
g_S^{(0)} = 6.3\pm 0.8 , 
\qquad 
g_S^{(1)} = 0.45 \pm 0.15 \ \ \ ,
\ee
where the errors do not include the range of values for the light quark masses.

Similarly, the tensor matrix elements are related to those of the quark EDM,
\be
\label{eq:gt01}
g_T^{(0)} = 4\, \rho^q_{(0)} \ , 
\qquad 
g_T^{(1)} = 4\, \rho^q_{(1)} \ .
\ee





In principle, we could adopt the chiral SO(4) approach to estimate the 
$g_{P}^{(i)}$. Instead, we find it useful to follow a variant of the arguments 
discussed in Refs.~\cite{Ellis:2008zy,Anselm:1985cf}. Taking the 
divergence of 
\be
\langle N\vert {\bar u} i\gamma_\mu\gamma_5 u 
- {\bar d} i\gamma_\mu\gamma_5 d \vert N\rangle = 
g_A {\bar \psi}_N \gamma_\mu\gamma_5\tau_3 \psi_N\ ,
\ee
employing partial conservation of the isovector axial current, 
using nucleon equations of motion, and observing that 
the divergence of the isovector axial current is anomaly-free as well as the 
definitions of the $g_P^{(k)}$ in Eq.~(\ref{eq:ffdef}) we obtain
\be
\label{eq:gp}
g_P^{(1)} = \frac{g_A  \bar{m}_N}{{\bar m}}\ ,
\ee
assuming exact isospin symmetry. 
Analogously,
we can define for matrix elements of 
the remaining flavors
\be
\label{eq:heavyffdef}
\bra{N} {\bar Q} \Gamma Q\ket{N} \equiv  
g_\Gamma^{Q} {\bar \psi_N} \Gamma \psi_N\ ,
\ee
with $Q$ denoting here $s$, $c$, $b$, or $t$. We will henceforth treat the 
strange quark separately and refer to $c$, $b$, and $t$ as the heavy flavors. 

The scalar form factors $g_S^s$ can be obtained from analyses of the 
$\sigma$ term. Defining
\be
\bra{N} m_s {\bar s} s\ket{N} =\sigma_s {\bar N} N
\ee
and 
\be
\kappa_s=\frac{ \sigma_s}{220\ \mathrm{MeV}}
\ee
as has been conventional in the recent literature (see, {\em e.g.} \cite{Ellis:2008zy}), one has
\be
\label{eq:gsstrange}
g_S^s = \kappa_s \frac{220\, \mathrm{MeV}}{m_s} 
\ee
with $\kappa_s = 21\pm 6$ MeV \cite{Shanahan:2012wh}. 
For heavier flavors a short derivation appears in Appendix \ref{sec:heavy} 
and we simply quote 
the result here: 
\be
\label{eq:gsheavy}
g_S^Q = \left(\frac{66\, \mathrm{MeV}}{m_Q} \right) 
\left(1-0.25\kappa_s\right)\ .
\ee
Note that the strange and heavy quark scalar form factors are isoscalar.

Again, the derivation for the pseudoscalar form factors
is given in Appendix \ref{sec:heavy} 
and leads to 
\be
\bra{N} m_Q {\bar Q} i\gamma_5 Q\ket{N} 
= {\bar \psi}_N \left[ g_P^{Q(0)} +g_P^{Q(1)}\tau_3 \right]i\gamma_5\psi_N\ ,
\ee
with
\bea
g_P^{Q(0)} & = & \frac{1}{4}\left[ g_A^{(0)}\left(\frac{m_N}{m_Q}\right) 
+ g_A\left(\frac{m_u+m_d}{m_u-m_d}\right)\, 
\left(\frac{\Delta m_N}{2 m_Q}\right)\right]
\\
g_P^{Q(1)} & = & \frac{1}{4} g_A\left(\frac{m_N}{2 m_Q}\right)\, 
\left(\frac{m_u-m_d}{m_u+m_d}\right)\ .
\label{eq:gPheavy}
\eea

With these considerations in mind, we now review explicit computations of the 
hadronic matrix elements, referring to the expectations based on considerations 
of chiral symmetry wherever possible. To date,  first-principles 
computations have been undertaken for the quantity $\alpha_n$; the scalar coupling $g_S^{(0,1)}$ that can be obtained
from $\sigma_{\pi N}$ and the contribution from the quark mass difference to the nucleon mass splitting as discussed above; direct computations of the isovector
form factors $g_\Gamma^{(1)}$ for $\Gamma=S,T$  using lattice 
methods\footnote{Note that the computations in Ref.~\cite{Bhattacharya:2011qm} apply to the charged current 
form factors that are related to those of interest here by an isospin rotation. Note also that the relative normalizations of the quantities here and in that work are given by  $g_{S,T}=2 g_{S,T}^{(1)}$ .}(for a compilation and recent results,
see Ref.~\cite{Bhattacharya:2011qm} );
and for the nucleon electric dipole moment form factors using HB$\chi$PT.
Indeed,  HB$\chi$PT can provide considerable insight into the dependence of 
matrix elements
on light quark masses while implementing a consistent expansion of QCD in 
scale ratios as discussed above. On the other
hand, knowledge of the low-energy constants requires additional input, 
either from a lattice computation, direct measurement,
or a model estimate. The present state of the art still relies heavily 
on the latter approach. As a result, 
there exists a considerable degree of model-dependent
uncertainty in the values of the $\beta_N^{(k)}$, {\em etc.}. One objective of 
this review is to provide a set of benchmark values and theoretical 
uncertainties for these parameters. A compilation of existing results drawn 
from various methods is given in Tables \ref{tab:hadme1} through 
\ref{tab:hadme4Q} appearing in Appendix \ref{sec:compile}. We now review 
explicit computations of the hadronic matrix 
elements, leading to the results quoted in these tables.


\subsection{Chiral Perturbation Theory}
\label{sect:chpt}

At low energies QCD reduces to HB$\chi$PT, where 
the symmetries of QCD, including the considerations of 
Section \ref{sect:chiralcons},
are naturally expressed. 
The isoscalar and isovector
electric dipole form factors (EDFF) of the nucleon are defined
through the TVPV electric current
\be
J_{\slashPTsub}^\mu(q,k)= 2 \left[ F_0(-q^2)+F_1(-q^2)\tau_3\right]
\left[ S^\mu v\cdot q -S\cdot q v^\mu 
+ \frac{1}{m_N} \left( S^\mu k\cdot q -S\cdot q k^\mu \right) 
+\cdots \right] \ ,
\ee
where the outgoing photon momentum $q=p-p^\prime$, 
and $k=(p+p^\prime)/2$. The EDFF 
\be
F_i(Q^2)= d_i - S^\prime_i Q^2 + \cdots
\ee
gives
the EDM at zero momentum transfer, $q^2=0$, and
the term linear in $q^2$ (the form-factor radius)
provides an electromagnetic contribution 
that cannot be separated from a short-range
electron-nucleon interaction.


HB$\chi$PT provides the momentum and pion mass dependence
of the EDFF in terms of the parameters appearing in the TVPV Lagrangian
\eqref{chiPTTV} and in the Lagrangian encoding P-even, T-even interactions.
The full nucleon EDFF has been calculated 
to leading order (LO) 
\cite{Crewther:1979pi,Thomas:1994wi,Hockings:2005cn,deVries:2010ah,
deVries:2012ab}
and next-to-leading order (NLO) 
\cite{Narison:2008jp, Ottnad:2009jw,deVries:2010ah,Mereghetti:2010kp,
deVries:2012ab}
in the $P/M_{QCD}$ expansion for all sources described above: 
the ${\theta}$-term and the dimension-six CPV operators.
(For related results in three-flavor $\chi$PT, see Refs. 
\cite{Cheng:1990pi,Pich:1991fq,Cho:1992rv,Borasoy:2000pq,Ottnad:2009jw,
Guo:2012vf}.) 
The resulting contributions fall into two classes: 
{\it i)} short-range contributions 
associated with momentum scales of order $\lamchi$, encoded
in the Lagrangian parameters and high-momentum part of pion loops; 
and {\it ii)} long-range contributions associated with scales of order $m_\pi$ 
and below, which appear in the low-momentum part of
loops and can be explicitly computed.
The arbitrary separation between Lagrangian parameters and high-momentum part 
of pion loops is controlled by the regularization scheme,
and after renormalization observables are regularization-scheme independent.

In this context, the short-range EDMs $\bar{d}_i$ appear at LO for all sources.
Long-range contributions resulting from pion loops introduce a dependence on 
$\gpbz$ and $\gpbo$. 
The specific chiral order
at which a long-range contribution arises depends on the chiral properties of 
the underlying source, as detailed above.
For the $\theta$-term, CEDM, and left-right four-fermion
operators, there are contributions
proportional to $\gpbz$ at both LO and NLO,
but at LO only to the isovector
nucleon EDM.
For the CEDM and left-right four-fermion
operators there are additional NLO terms proportional to 
$\gpbo$ as well.
The dependence of loops on the regulator, choice of renormalization scheme, 
and renormalization scale can be absorbed into a renormalization of the 
$\bar{d}_i$.
The momentum dependence, and in particular the contribution
to the Schiff moment,
is finite and set by $2m_\pi$.
For other sources, the relative suppression of pion-nucleon
couplings means smaller loop contributions. 
The momentum dependence becomes a higher-order effect and the scale of 
its variation is determined by $M_{QCD}$ rather than $2m_\pi$.

For the nucleon EDM one obtains to NLO:
\begin{equation}
d_{0}=\bar{d}_{0}
-\frac{eg_{A}}{16\pi F_{\pi}}
\left\{ \gpbz \left[\frac{3m_\pi}{m_N} -\frac{4(\Delta m_N)_q}{m_\pi} \right]
+\gpbo \frac{m_\pi}{m_N}\right\}
\label{d0}
\end{equation}
and
\begin{equation}
d_{1}=\bar{d}_{1}(\mu)+\delta\bar{d}_{1}(\mu)
-\frac{eg_{A}}{(2\pi)^{2} F_{\pi}}
\left\{\gpbz \left[\ln\frac{m_{N}^{2}}{m_{\pi}^{2}}+\frac{5\pi m_\pi}{4m_N}
-\frac{\Delta m_\pi^2}{m_\pi^2}
\right]-\gpbo\frac{\pi m_\pi}{4 m_N} \right\}
 \ ,
\label{d1}
\end{equation}
where $\Delta m_\pi^2=m_{\pi^\pm}^2-m_{\pi^0}^2$ is the
(mostly electromagnetic) pion mass splitting
and
\begin{equation}
\delta\bar{d}_{1}(\mu)\equiv-\frac{eg_{A}\gpbz}{(2\pi)^{2} F_{\pi}}
\left(\frac{2}{4-d}-\gamma_{E}+\ln\frac{4\pi\mu^{2}}{m_{N}^{2}}\right)\,
\end{equation}
in terms of the dimension of spacetime $d$, the renormalization scale $\mu$,
and the Euler-Mascheroni constant $\gamma_E\simeq 0.577$.

The $\mu$ dependence of $\delta\bar{d}_{1}(\mu)$ can be absorbed in 
$\bar{d}_{1}(\mu)$. One cannot separate
parameter and loop contributions in a model independent way, and
since the same combination of loops and parameters appears when
the nucleon is inserted in a nucleus, it is simplest
to redefine $\bar{d}_0$ and $\bar{d}_1$ to represent,
respectively, the 
full isoscalar and isovector nucleon EDMs.
However, one expects no cancellations between loop contributions that
are non-analytic in the quark masses and thus $m_\pi^2$,
and short-range pieces, which are analytic
\footnote{It is often conventional to retain only the non-analytic terms from 
loop computations and absorb all analytic terms into the parameters.}.
In this case the magnitude of the 
non-analytic contributions at a ``reasonable''
renormalization scale serves as a lower bound for the 
redefined $\bar{d}_i$. 
For $\mu=m_N$,
we expect for $\bar{\theta}$ \cite{Crewther:1979pi},
qCEDM \cite{Mereghetti:2010kp} and left-right four-fermion
operators \cite{deVries:2012ab},
\begin{equation}
|\bar d_0| \gtrsim 
0.01 
\left[\gpbz+ 0.3 \ \gpbo\right] 
e\,\textrm{fm} \ , 
\qquad
|\bar d_1 |\sim  
0.1 
\left[\gpbz+ 0.03 \ \gpbo\right]
e\,\textrm{fm} \ .
\label{d1est}
\end{equation}
Relying on the arguments leading to Eq.~(\ref{eq:gpbzchiral1}) for $\gpbz$,  
for the
$\theta$-term we obtain
$|d_N|\simge 2\cdot 10^{-3}\ \bar\theta \ e$ fm,
from which the current bound on $\bar\theta$ arises.
For the other sources one currently has to rely on NDA or model-dependent 
estimates, as we discuss below.



\subsection{Lattice QCD}
\label{sec:lattice}

Lattice QCD holds the promise of providing the various matrix elements for all
relevant CPV mechanisms.  However, to our knowledge, 
calculations have focused on the nucleon EDM for the ${\theta}$-term, the
isoscalar scalar couplings $g_S^{(0,1)}$, and the and tensor form factors.  In
the former case, the most recent published computations of the $\alpha_N$ date
back nearly five years or more
\cite{Aoki:1989rx,Shintani:2005xg,Berruto:2005hg,Shintani:2006xr,
Shintani:2008nt,Aoki:2008gv}. Generally, these computations have followed one
of two approaches: (a) computing the shift in the nucleon energy in the
presence of an electric field, or (b) computing the 
nucleon electric dipole form factor by expanding to leading non-trivial order
in ${\bar\theta}$.  These calculations are carried out at unphysical values of
the pion mass.  The nucleon EDFF discussed in Section \ref{sect:chpt} provides
in principle the tool to extrapolate results to smaller momentum and pion
mass.  Here, we summarize the most recently reported computations for each
approach. 

The most recent computation of the first method has been reported in 
Ref.~\cite{Shintani:2008nt}.  Using a two-flavor dynamical clover action, 
the authors considered the ratio of spin-up and spin-down nucleon propagators
\be
R_3(E,t;{\bar\theta}) = 
\frac{\langle N_1{\bar N}_1\rangle}{\langle N_2{\bar N}_2\rangle}
=\left[1+\mathcal{O}({\bar\theta})\right] \mathrm{exp}
\left[-\alpha_N{\bar\theta} E t\right]\ ,
\ee
where the subscript $\sigma$ on $N_\sigma$ denotes spin, $t$ denotes the time, 
and $E$ gives the magnitude of the electric field along the $z$-direction. 
The electric field is introduced {\em via} a replacement of the gauge link 
variables $U_k(x)$ in the Dirac-Wilson action as
\be
\label{eq:Elink}
U_k(x)\rightarrow e^{Q_q E_k t} U_k(x) \ ,
\ee
where $Q_q$ is the quark charge and $k$ labels the direction. 
For this computation, the replacement (\ref{eq:Elink}) was applied only 
to the valence quarks; so-called \lq\lq disconnected" insertions of the 
electric field on the sea quarks that enter through the quark determinant 
have not been included \footnote{In the limit of degenerate sea quarks in 
three-flavor QCD, the disconnected contribution is identically zero due 
to the vanishing trace over the quark charges \cite{Shintani:2008nt}.}. 
A corrected ratio $R_3^\mathrm{corr}$ was used to minimize the effect of 
insufficient statistics associated with  vanishing $E$ and/or ${\bar\theta}$. 
One then has
\be
2\alpha_N{\bar\theta}E = \ln\left[\frac{R_3^\mathrm{corr}(E,t-1;{\bar\theta})}
{R_3^\mathrm{corr}(E, t; {\bar\theta})}\right] \ .
\ee

\begin{figure}[tb]
\begin{center}
\begin{minipage}[t]{12 cm}
\epsfig{file=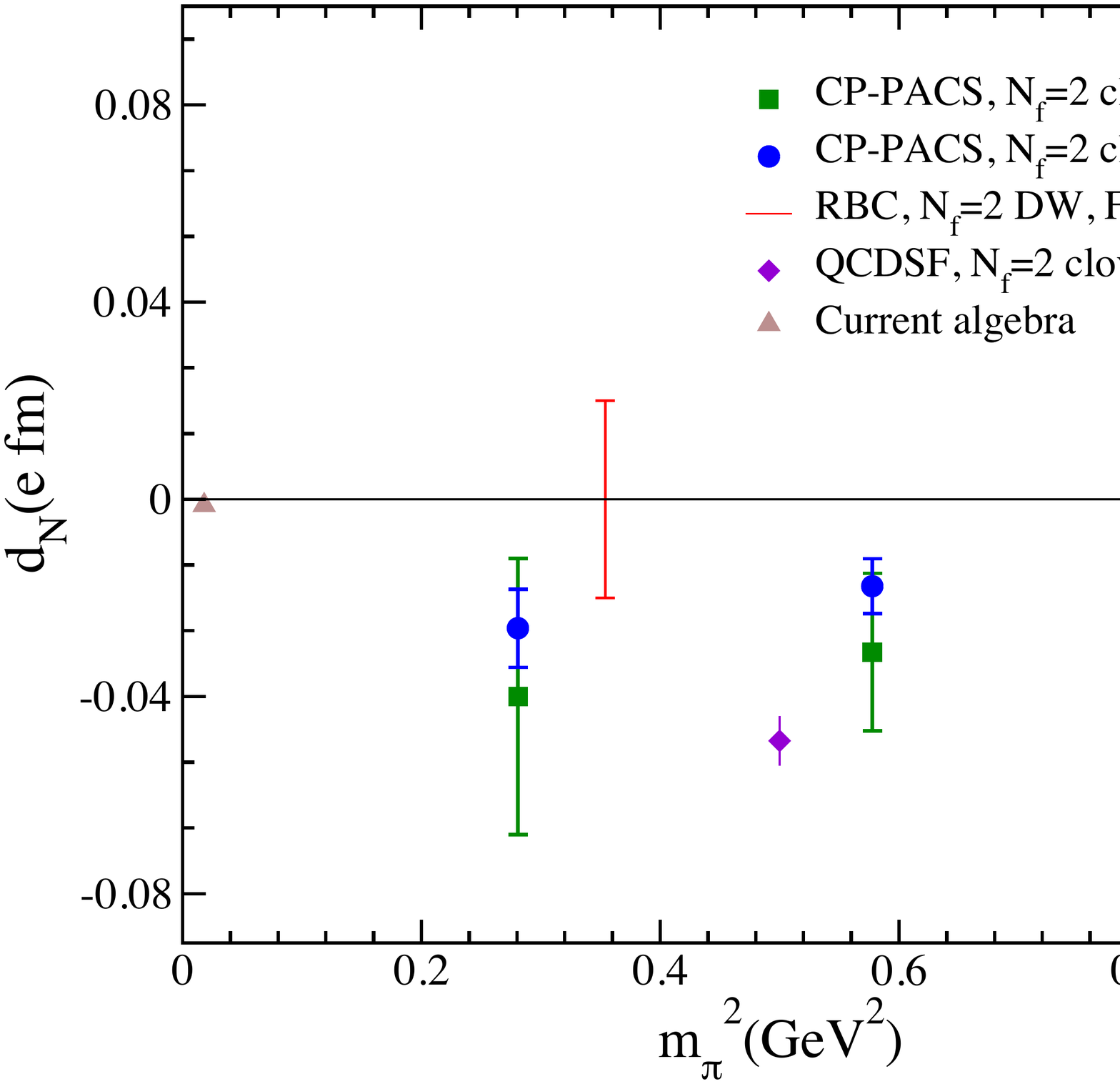,scale=0.3}
\epsfig{file=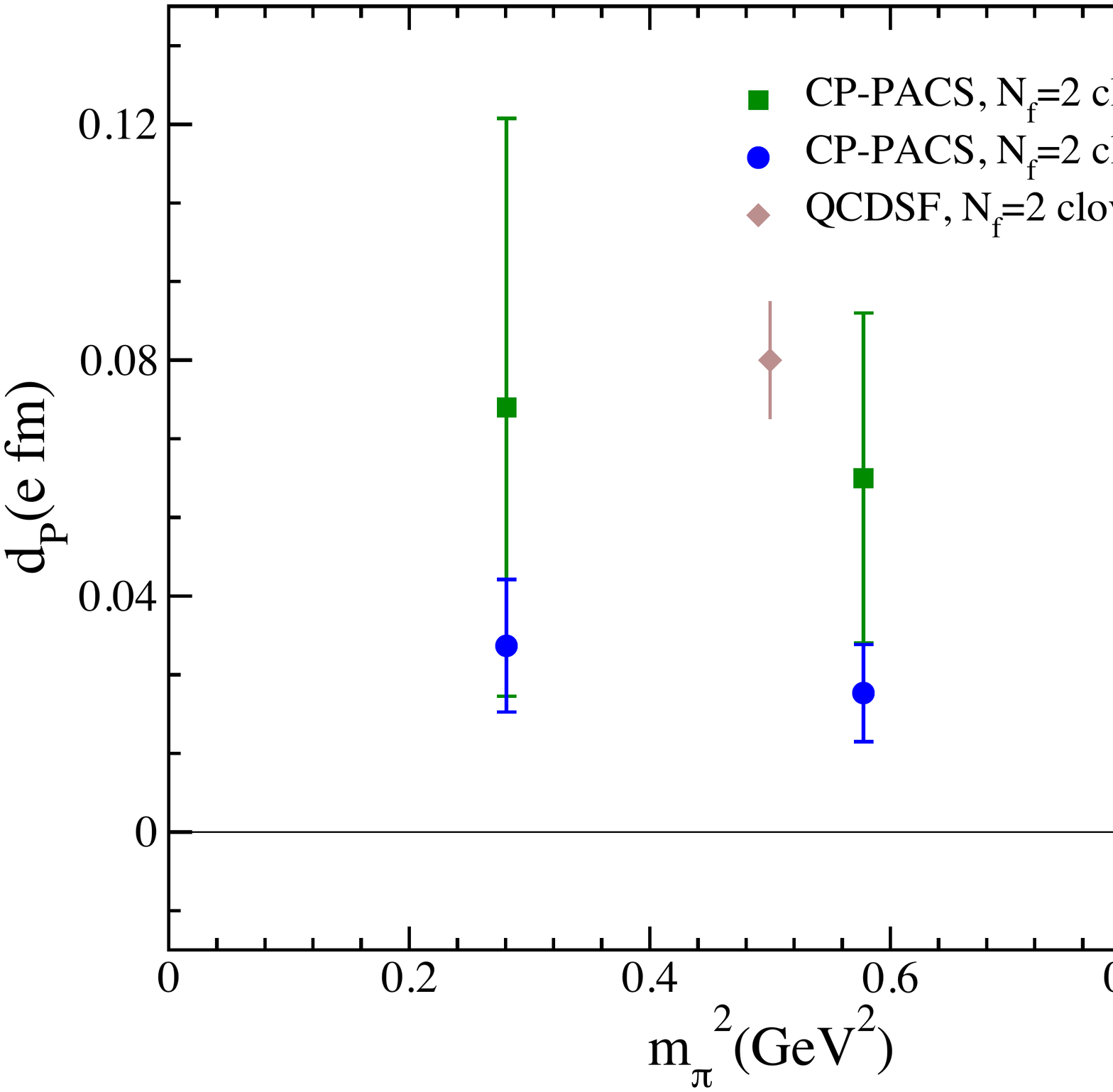,scale=0.3}
\end{minipage}
\begin{minipage}[t]{16.5 cm}
\caption{
Lattice computation of nucleon EDMs induced by the QCD $\theta$-term.
The pion mass squared dependence of $d_n$  (left)
and $d_p$ (right) obtained using various approaches.
Square symbols denote the results in external electric field
method in $N_f=2$ clover fermion \cite{Shintani:2008nt},
and circle symbols denote one in form factor method \cite{Shintani:2009}
with same gauge configurations.
Red bar denotes the bound of EDM in $N_f=2$ domain-wall fermion in
\cite{Berruto:2005hg}, and diamond is a result from EDM form factor
of imaginary $\theta$ method quoted in \cite{Aoki:2008gv}.
Note that the error bar of diamond symbol may be an underestimate due to
large systematic error associated chiral symmetry breaking of clover fermion.
The triangle symbol is model estimate in current algebra.
\label{fig:latt1} }
\end{minipage}
\end{center}
\end{figure}

A $24^3 \times 48$ lattice with $\beta=2.1$ and lattice spacing $a\approx 0.11$ 
fm was employed, where the latter is set by the $\rho$-meson mass 
$m_\rho=768.4$ GeV. Results for $\alpha_n$ obtained with a lightest quark mass 
corresponding to $m_\pi=0.53$ GeV are shown in Fig. \ref{fig:latt1}, 
using ${\bar\theta}= 0.025$ and $E=0.004/a^2$. 
The corresponding values are quoted in Table  \ref{tab:hadme1}.

The authors also studied the dependence of $\alpha_N$ on the light quark mass 
to determine if this coefficient vanishes in the chiral limit as required. 
Results were obtained at $m_\pi=1.13$, 0.93, 0.76 and 0.53 GeV. 
Results for the neutron are indicated in 
Fig. \ref{fig:latt1}. It is apparent that the computation does not exhibit 
the correct chiral behavior. The authors conclude that this situation is 
likely due to the explicit breaking of chiral symmetry by the Wilson-type 
quark action and the relatively large value of the lightest quark mass used. 
As the authors also emphasize, obtaining a significant, non-vanishing signal 
for the nucleon EDM does not appear to require the presence of appropriate 
chiral behavior.

The most recent computation utilizing the form factor method has been reported 
in Ref.~\cite{Aoki:2008gv}. The computation was performed by rotating 
${\bar\theta}$ into the quark mass matrix and taking it to have an imaginary 
value:
\be
\label{eq:imtheta}
{\bar\theta} = -i{\bar\theta}^I\ , 
\ee
with ${\bar\theta}^I$ being a real number. Simulations were performed using 
the Iwasaki gauge action and  two-flavors of dynamical clover fermions with 
$\beta=2.1$, $a \approx 0.11$ fm (again set by $m_\rho$), 
$m_\pi/m_\rho\approx 0.8$, and several values of the imaginary vacuum angle: 
${\bar\theta}^I=0$, $0.2$, $0.4$, $1.0$, and $1.5$. The EDM form factor $F_3$ 
was obtained from the ratio of three- and two-point correlators:
\be
R(t) = \frac{G_{NJ_\mu N}^{\theta\Gamma}
(t^\prime, t; {\vec p}^\prime, {\vec p})}
{\mathrm{Tr}\left[G_{NN}^\theta(t^\prime; {\vec p}^\prime)\Gamma_4\right]}\ ,
\ee
where $t$ denotes the time co-ordinate for the insertion of the vector current 
$J_\mu$, $t^\prime$ gives the time for the nucleon \lq\lq sink", and ${\vec p}$ 
(${\vec p}^\prime$) gives the nucleon momentum before (after) the vector curren
insertion. 

\begin{figure}[tb]
\begin{center}
\begin{minipage}[t]{8 cm}
\epsfig{file=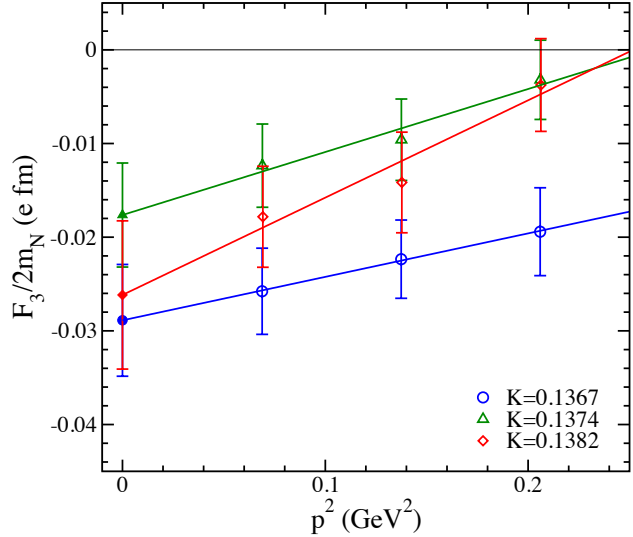,scale=0.4}
\end{minipage}
\begin{minipage}[t]{12 cm}
\caption{Lattice computation of ${\bar\theta}$-dependence of $d_n$ using the form factor method method\cite{Aoki:2008gv} for ${\bar\theta}^I=0.2$. 
Shown is the squared momentum transfer dependence  at three mass parameters $K=0.1382$--0.1367
which correspond to $m_\pi^2 = 0.3$--0.85 GeV$^2$.
These are results in $N_f=2$ clover fermion configurations.\label{fig:latt2}   }
\end{minipage}
\end{center}
\end{figure}

Results at vanishing momentum transfer were obtained using two different 
extrapolation methods: (a) employing a dipole {\em ansatz} for the 
$q^2$-dependence of the form factor and 
(b) assuming the EDM and Dirac form factors have the same $q^2$-dependence and 
utilizing the latter (see Fig. \ref{fig:latt2} ). 
Both methods give consistent values for the EDM. Taking
\be
d_N^\theta = \frac{\partial d_N^\theta}{\partial{\bar\theta}^I} 
+ c\left[ {\bar\theta}^I\right]^3
\ee
and using the coefficient of the linear term to define the EDM, the authors 
obtain the results indicated in Table \ref{tab:hadme1}. 
The results agree with those of Ref.~\cite{Shintani:2008nt} 
(electric field method) within error bars.

In addition to the direct computations of $d_N$, lattice QCD results provide 
input for the determination of $\lambda_{(0)}$ via Eq.~(\ref{eq:gpbzchiral1}) 
and for the $g_{S,T}^{(0,1)}$. 
As discussed above, values of $g_S^{(0,1)}$ may be inferred from lattice computations of $\sigma_{\pi N}$ and $(\Delta m_N)_q$. Alternately, one may obtain $g_{S,T}^{(1)}$ from direct computations of the charge changing scalar and tensor form factors\cite{Bhattacharya:2011qm} {\em via} isospin rotation. Taking into account the factor of two difference in normalization of these form factors, the preliminary lattice values quoted in Ref.~\cite{Bhattacharya:2011qm} imply
\bea
g_S^{(1)}(\overline{\mathrm{MS}}, \mu  =  2\, \mathrm{GeV} ) & = & 0.4(2)\\
g_T^{(1)}(\overline{\mathrm{MS}}, \mu = 2\, \mathrm{GeV} ) & = & 0.53(18)\ \ \ .
\eea
The computation of $g_S^{(1)}$ was obtained using two different gauge field ensembles with pion masses in the ranges $390< m_\pi < 780$ MeV and $350< m_\pi < 700$ MeV, respectively. A chiral extrapolation was performed assuming a linear dependence on $m_q$. The value for $g_T^{(1)}$ was derived by combining RBC/UKQCD and LHPC results, with a chiral extrapolation based on HB$\chi$PT results. A comparison of the value for $g_S^{(1)}$ with a result obtained using $(\Delta m_N)_q$ is given in Table \ref{tab:hadme4} below.

\subsection{QCD Sum Rules}

In recent times, the most widely quoted hadronic computations of $d_n$ and 
the ${\bar g}_\pi^{(k)}$ rely on the method of QCD Sum Rules (QCDSR). 
(For an extensive review in the context of EDMs, see 
Ref.~\cite{Pospelov:2005pr}; see Ref.~\cite{Hisano:2012sc} for a more recent 
discussion.) This approach entails computing hadronic correlators at 
large virtuality where the operator product expansion (OPE) can be rigorously 
applied and matching the result onto a phenomenological {\em ansatz} for 
the structure of the correlator at lower virtuality. The reliability of 
this matching is improved by performing a Borel transform to the OPE and 
phenomenological forms for the correlator. 

In the present instance, the relevant correlator $\Pi(Q^2)$ involves 
two nucleon sources $\eta_N$, 
\be
\Pi(Q^2) = i \int\, d^4x\, e^{iq\cdot x}\, 
\bra{0} T\left\{ \eta_N(x) {\bar\eta}_N(0)\right\}\ket{0}\ ,
\ee
where $\eta_N(x)$ contains combination(s) of quark field operators that carry
the nucleon quantum numbers, where $Q^2=-q^2$, and where the Dirac indices 
on $\Pi$ have been suppressed for simplicity. In general, one uses a linear 
combination of two sources,
\be
\eta_N= \eta_1 +\beta\eta_2 \ ,
\ee
where
\be
\eta_1  =  2\epsilon_{abc} d_a^T C\gamma_5 u_b d_c \qquad \mathrm{and}\qquad
\eta_2  =  2\epsilon_{abc} d_a^T C u_b \gamma_5 u_c
\ee
with the subscripts $a$ {\em etc.} denoting color. One computes $\Pi(Q^2)$ 
in a background that contains photon and pion fields as well as the TVPV 
interactions introduced above and identifies various Lorentz structures 
that are invariant under chiral rotations and that exhibit the appropriate 
spacetime symmetries associated with the EDM or $\pi NN$ interactions:
\bea
\nonumber
\mathrm{EDM} & \rightarrow &\left\{ \tilde F\cdot\sigma, \dslash{q}\right\}\ ,
\\
\pi NN & \rightarrow &  \dslash{q}\ ,
\eea
where ${\tilde F}\cdot\sigma \equiv {\tilde F}^{\mu\nu}\sigma_{\mu\nu}$. 
The corresponding phenomenological {\em ansatz} for the correlator 
at low virtuality is
\be
\label{eq:pipheno}
\Pi(Q^2)^\mathrm{pheno}  = - \frac{1}{2} f_d(Q^2) 
\left\{ {\tilde F}\cdot\sigma, \dslash{q}\right\} 
+\frac{1}{2} f_\pi(Q^2) \dslash{q} +\cdots \ ,
\ee
with 
\bea
\label{eq:pif}
f_d(Q^2) &=& \frac{\lambda^2 d_n m_N}{(Q^2+m_N^2)^2} 
- \frac{A_d (Q^2)}{Q^2+m_N^2} +B_d(Q^2)\ , 
\\ \nonumber
f_\pi(Q^2) &=& \frac{\lambda^2 {\bar g}_\pi^{(k)} m_N}{(Q^2+m_N^2)^2} 
- \frac{A_\pi (Q^2)}{Q^2+m_N^2} +B_\pi(Q^2) \ .
\eea
The dependence of the $f_k(Q^2)$ on the quantities of interest here 
($d_n$, ${\bar g}_\pi^{(k)}$) arises from the first (double pole) term 
on the RHS of Eq.~(\ref{eq:pif}). As we discuss shortly, it carries 
an important universal dependence on the parameter $\lambda$. 
The single pole term represents contributions associated with transitions 
between the neutron and excited states. Its strength, parameterized by 
the quantities $A_{d,\pi}$ is generally unknown, leading to one source of 
theoretical uncertainty. The continuum terms $B_{d,\pi}$ are generated by 
transitions between excited nucleon states and are also generally unknown. 

For the OPE evaluation, one performs all possible contractions of the $q$ 
and $\bar{q}$ fields in the sources $\eta_N$ and ${\bar\eta}_N$, leading to 
an expression in terms of the quark propagators evaluated in the presence of 
a photon and CPV background. The OPE gives the quark propagators in terms of 
various Wilson coefficients times condensates, {\em viz.}
\be
S(x)  =  S(x)^\mathrm{(0)} + \chi(x){\bar\chi}(0) 
+  S(x)^\mathrm{1\, photon}+S(x)^\mathrm{1\, gluon}+\cdots \ ,
\ee
where $S(x)^\mathrm{(0)}$ contains the free quark propagator, 
$S(x)^\mathrm{1\, photon}$ and $S(x)^\mathrm{1\, gluon}$ contain dependences 
on the photon and gluon field strength tensors, respectively, and 
the $\chi(x){\bar\chi}(0)$ carries dependences on the condensate 
$\langle {\bar q}q\rangle$ as well as tensor condensates in the presence 
of the photon background that can be related to $\langle {\bar q}q\rangle$ 
through various susceptibilities $\chi$, $\kappa$, and $\xi$:
\bea
\langle {\bar q} \sigma_{\mu\nu} q \rangle_F & =  & 
Q_q\chi\ F_{\mu\nu} \langle {\bar q}q\rangle \ ,
\\
g_s \langle {\bar q} G_{\mu\nu} q \rangle_F & =  & 
Q_q\kappa\ F_{\mu\nu} \langle {\bar q}q\rangle \ ,
\\
2 g_s \langle {\bar q} G_{\mu\nu}\gamma_5 q \rangle_F & =  & 
i Q_q\xi\ F_{\mu\nu} \langle {\bar q}q\rangle\ .
\eea
After including the CPV interactions in the background, one obtains 
the dependence of $\chi(x){\bar\chi}(0)$ on $\langle {\bar q}q\rangle$, 
the susceptibilities $\chi$, $\kappa$, and $\xi$, and the quantities 
${\bar\theta}$, $d_q$ and ${\tilde d}_q$. A detailed expression for 
the resulting correlator $\Pi(Q^2)^\mathrm{OPE}$ goes beyond the scope 
of this review but can be found in, {\em e.g.}, 
Refs.~\cite{Pospelov:1999mv,Pospelov:1999ha,Pospelov:2000bw,Hisano:2012sc}.

Applying the Borel transformation to both $\Pi(Q^2)^\mathrm{pheno}$ and 
$\Pi(Q^2)^\mathrm{OPE}$ and matching the coefficients of the relevant Lorentz 
structures then allows one to obtain $d_n$ and the ${\bar g}_\pi^{(k)}$ 
in terms of the CPV parameters, susceptibilities, $\langle {\bar q}q\rangle$ 
condensate, Borel mass $M$, and phenomenological parameters 
$\lambda$, $A_d$, {\em etc}. For example, for the neutron EDM one has 
\cite{Pospelov:2005pr,Hisano:2012sc}
\be
\lambda^2 d_n m_N - A M^2 = 
-\Theta\ \langle {\bar q}q\rangle\ \frac{M^4}{8\pi^2} e^{m_n^2/M^2}\ ,
\ee
with
\be
\Theta = \left(4 Q_d m_d\rho_d-Q_u m_u \rho_u\right)\chi{\hat\theta} 
+ \left(4d_d-d_u\right) 
+ \left(\kappa-\xi/2\right) \left(4 Q_d {\tilde d_d}-Q_u{\tilde d_u}\right)\ ,
\ee
and with the $\rho_q$ carrying a dependence on ratios of the quark masses. 
In the absence of a PQ mechanism, one has ${\hat\theta}={\bar\theta}$. 
As discussed above, in the presence of the PQ mechanism the other QCD 
CPV interactions, such as the CEDM, lead to a shift in the vacuum angle 
due to their effect on the axion potential. In this case, one must take 
${\hat\theta}=\theta_\mathrm{ind}$, with $\theta_\mathrm{ind}$ being the shift 
due to 
the additional axion potential contributions. 

The most recent results \cite{Pospelov:2005pr,Hisano:2012sc} for the 
dependence of $d_n$ and the ${\bar g}_\pi^{(k)}$ are indicated in 
Tables \ref{tab:hadme1} - \ref{tab:hadme3G}. We concentrate first on the 
dependence of $d_n$ on ${\bar\theta}$. Notably, the magnitude of $\alpha_n$ 
obtained from the QCDSR computations are one to two orders of magnitude 
smaller than those obtained using lattice calculations. Moreover, the most 
recent QCDSR determination of $\alpha_n$ \cite{Hisano:2012sc} is nearly 
a factor of six smaller in magnitude that the earlier work of 
Ref. \cite{Pospelov:2005pr}. This difference results, in part, from a 
different value of $\lambda$ used in the two analysis. The authors of 
Ref.~\cite{Pospelov:2005pr} utilized the smaller value for $\lambda$ obtained 
directly from QCDSR studies, while Ref.~\cite{Hisano:2012sc} employed a value 
obtained from a lattice computation of proton decay matrix elements and 
isospin symmetry, which is roughly two times larger. 

The foregoing technique also provides the dependence of $d_n$ on the quark EDM 
and chromo-EDMs. Again, the analyses in the two most recent computations
\cite{Pospelov:2005pr,Hisano:2012sc} are similar, differing primarily in their 
extraction of $\lambda$. Illustrative results are given in 
Tables \ref{tab:hadmeCEDMd} and \ref{tab:hadmeCEDMg} for the case when one 
assumes the presence of PQ symmetry. To our knowledge, the only QCDSR 
computations of the dependence of the $\gpbi$ on ${\bar\theta}$, 
${\tilde d}_q$, $C_{\tilde G}$ and four-quark operators have been reported in 
Refs.~\cite{Pospelov:2001ys,Pospelov:2005pr}. As discussed above, one may 
extract the ${\bar\theta}$-dependence of $\gpbz$ using chiral methods,  
so we do not quote an additional QCDSR result for this case. Moreover, 
the contribution of ${\bar\theta}$ to $\gpbo$ requires an additional power 
of $m_q$ (or $m_\pi^2$) as indicated in Eq.~(\ref{eq:lam1theta}), 
so its impact will generally be negligible except for systems in which 
the effect of $\gpbz$ vanishes. 
Consequently, we do not list any entry for $\lambda_{(1)}$.

For the contributions of other CPV operators, the situation is more subtle. 
In the case of  CEDM contributions to the $\gpbi$, the constraints of chiral symmetry imply the 
presence of two canceling contributions:
\be
\bra{N} \left({\tilde d}_q\ {\bar q} g_s  \sigma_{\mu\nu} G^{\mu\nu} q 
- m_0^2 {\bar q} q\right)\ket{N}
\ee
where the first term arises from the PCAC commutator term and the second 
is generated by the pion pole contribution. This cancellation renders 
the computation susceptible to theoretical uncertainties, particularly 
associated with the choice of $\beta$ in the nucleon sources. For example, 
taking $\beta=1$ yields a vanishing contribution through NLO under 
the assumption of pure valence quark dominance. On the other hand, 
choosing $\beta\not=1$ yields a non-vanishing result. Assuming that 
the double-pole term in Eq.~(\ref{eq:pif}) dominates, one obtains
\bea
\gpbz & \approx & \frac{3}{10}\ \frac{4\pi^2 |\langle{\bar q} q\rangle| m_0^2}
{m_N F_\pi M^2}\ F_0(\beta)\ \left( {\tilde d}_u+{\tilde d}_d\right)\ ,
\\
\gpbo & \approx & \frac{3}{2}\ \frac{4\pi^2 |\langle{\bar q} q\rangle| m_0^2}
{m_N F_\pi M^2}\ F_1(\beta)\ \left( {\tilde d}_u-{\tilde d}_d\right)\ ,
\eea
where at leading order the $F_k(\beta=1)=0$ and $F_k(\beta=0)=1$. 
For the latter choice, one obtains  a generically stronger CEDM sensitivity 
of $\gpbo$ compared to that of $\gpbz$. Going beyond LO, including 
uncertainties associated with additional condensates that consequently appear, 
the choice of Borel mass $M$ and variations with $\beta$, the analysis of
Ref.~\cite{Pospelov:2001ys} yields a \lq\lq best value" and range for $\gpbo$ 
as well a broad range but no best value for $\gpbz$ quoted in 
Table \ref{tab:hadmeCEDMg}.

Providing robust computations of the three-gluon and four-quark operators to 
the neutron EDM and the $\gpbi$ is even more challenging, as one encounters 
additional unknown condensates as well as the presence of infrared divergences 
at lower order in the OPE than for the other sources of CPV discussed thus far.
It is possible, however, to estimate the contributions to $d_n$ by relating the
EDM to the nucleon magnetic moment through a CPV rotation of the nucleon 
wavefunction. One then has, for example \cite{Demir:2002gg},
\be
\label{eq:3G}
d_n[C_{\tilde G}] \sim \mu_n\  \frac{9 g_s m_0^2}{32\pi^2}\ 
\ln\left(\frac{M^2}{\Lambda_{IR}^2} \right)\ 
\left(\frac{v}{\Lambda} \right)^2\ C_{\tilde G}\ ,
\ee
where $\mu_n$ is the neutron magnetic moment and where the other prefactors 
in Eq.~(\ref{eq:3G}) arise from a QCDSR evaluation of the correlator 
of two nucleon currents in the presence of the CPV three-gluon interaction. 
The latter evaluation is used to determine the CPV rotation needed for 
relating $d_n$ and $\mu_n$. Taking $M/\Lambda_{IR}=2$ and $g_s=2.1$ yields 
the estimate of the coefficient $\beta_n^{\tilde G}$ given 
in Table \ref{tab:hadme3G}.

\subsection{Quark Models}

The constituent quark model (CQM) has proven remarkably successful in 
accounting for a number of static properties of the lowest-lying baryons, 
most notably their magnetic moments. In the latter instance, one assumes 
each constituent quark posses a distinct magnetic moment that is proportional 
to its spin, 
\be
{\vec\mu}_Q = 2 \mu_Q {\vec S}_Q \ ,
\ee
and computes the nucleon magnetic moment using the appropriate 
spin-flavor-color-spatial nucleon wavefunction, resulting in 
\be
\mu_p = \frac{1}{3}\left[4\mu_U-\mu_D\right]
\qquad \mathrm{and}\qquad \mu_n = \frac{1}{3}\left[4\mu_D-\mu_U\right]\ .
\ee
To the extent that the constituent up- and down-quark magnetic moments differ 
only by the overall charge of the quark, one obtains for the ratio of nucleon
magnetic moments $\mu_p/\mu_n\simeq -3/2$, in close agreement with the 
experimental value. 

In the case of the EDM, the dimensionless Wilson coefficients $C_{q\tilde\gamma}$ and $\delta_q$
(or the equivalent dimensionful quark EDMs $d_q$) correspond to the EDMs of 
the current quarks of QCD rather than the constituent quarks of the 
quark model. Nevertheless, one may make the bold {\em ansatz} that 
the coefficients of the constituent quark operators are given by the 
corresponding coefficients for those of the current quarks after 
appropriate RG running from the scale $\Lambda$ to the hadronic scale 
$\lamchi$ that introduces the $K_q$ factor in Table \ref{tab:pQCD}:
\be
{\vec d}_Q = 2 d_Q {\vec S}_Q\qquad \mathrm{with}\qquad d_Q = d_q(\lamchi) 
= K_q d_q (\Lambda)+\cdots\ ,
\ee
where the last relation may just as easily be expressed in terms of 
$C_{q\tilde\gamma}$ or $\delta_q$ and where the $+\cdots$ indicate contributions
associated with operator mixing.
In this case, the computations of the nucleon EDMs proceed as in the 
case of the magnetic moments, leading to 
\be
\label{eq:QMedm}
d_p = \frac{1}{3}\left[4d_U-d_D\right]\qquad \mathrm{and}\qquad d_n 
= \frac{1}{3}\left[4d_D-d_U\right]\ .
\ee
Thus, one obtains
\be
\rho_n^d =  \rho_p^u = -4\rho_n^u =-4\rho_p^d = \frac{4}{3}\ .
\ee

An alternate approach, first proposed in Ref.~\cite{Ellis:1996dg}, is to 
retain the identity of the quarks as partonic degrees of freedom and relate 
the nucleon matrix elements of ${\bar q}\sigma_{\mu\nu} q {\tilde F}^{\mu\nu}$ 
to those of the quark axial vector currents ${\bar q}\gamma_\mu\gamma_5 q$ 
that contribute to the nucleon spin:
\bea
\frac{1}{2} \langle N\vert {\bar q}\gamma_\mu\gamma_5 q\vert N \rangle & = & 
\left(\Delta q\right)_N \, S_\mu \ ,
\\
\frac{1}{4} \langle N\vert {\bar q}\sigma_{\mu\nu}\gamma_5 q\vert N \rangle  
& = & \left(\Delta q\right)_N \bar{N}\sigma_{\mu\nu} N  \ ,
\eea
so that the nucleon EDM is given by
\be
d_N=\sum_{q=u,d,s} d_q \left(\Delta q\right)_N\ \ \ ,
\ee
leading to a \lq\lq parton quark model" (PQM) prediction
\be
\rho_N^q= \left(\Delta q\right)_N\ \ \ .
\ee
Information on the $ \left(\Delta q\right)_N$ can be obtained from a number 
of sources, including the determination of the axial vector coupling $g_A$ 
that enters neutron decay,
\be
g_A=\left(\Delta u\right)_p-\left(\Delta u\right)_n \ ,
\ee
and studies of both inclusive and semi-inclusive polarized, 
deep inelastic leptoproduction, 
\be
\left(\Delta u\right)_p = \left(\Delta d\right)_n = 0.746 \ ,
\qquad 
\left(\Delta d\right)_p = \left(\Delta u\right)_n = - 0.508 \ ,
\qquad 
\left(\Delta s\right)_p =\left(\Delta s\right)_n = -0.226\ .
\ee

The CQM and PQM fail to provide guidance for the quark CEDM or three-gluon 
operator contributions to the nucleon EDMs. They similarly offer no reliable 
means for estimating the contributions of these operators to the $\gpbi$. 
In these cases, one might complement quark model estimates
with the NDA discussed in Section \ref{sect:chiralcons}.


\subsection{Saturation Methods}
\label{subsec:saturation}

Over the years, estimates of four-quark matrix elements have often been 
obtained by assuming certain intermediate states dominate or \lq\lq saturate" 
the dynamics. Here, we illustrate the application of the saturation 
approximation to estimate two classes of matrix elements: 
(a) contributions to $\gpbo$ generated by the operator $Q_{\varphi ud}$ and 
(b) the values of the TVPV four-nucleon operator coefficients ${\bar C}_{1,2}$. 

Starting with $\gpbo$,
we recall that the first term in Eq.~(\ref{eq:sspp}) generates a contribution 
to $\gpbo$. When the scale $1/\Lambda^2$ is included, one expects 
$\gpbo\sim \mathrm{Im} C_{\varphi ud} \lamchi F_\pi/\Lambda^2$ as noted earlier. 
An explicit evaluation can be made using factorization and vacuum saturation, 
partial conservation of the axial current (PCAC), and the 
$\pi N$ $\sigma$-term. Vacuum saturation in this context amounts to first 
relating the $\bra{N \pi_2} S_3 \otimes S_4 \ket{N}$ to the crossed matrix
element $\bra{ \pi_2} S_3 \otimes S_4 \ket{N{\bar N}}$, inserting a complete 
set of states between the $S_3$ and $S_4$ bilinears, assuming the dominant 
contribution arises from the vacuum, and then uncrossing the ${\bar N}$. 
One then obtains
\be
\label{eq:lreff1}
\bra{N \pi_2} S_3 \otimes S_4 \ket{N} \sim 
\bra{N} S_4\ket{N}\ \bra{\pi_3} S_3 \ket{0}\ .
\ee
Now, the matrix element of $S_4$ is related to the pion-nucleon $\sigma$-term as
\be
\label{eq:lreff2}
\bra{N} S_4\ket{N} = \bra{N} {\bar u} u +{\bar d} d\ket{N} = 
\frac{\sigma_{\pi N}}{\bar m} \ ,
\ee
while the second matrix element can be evaluated by taking the divergence of
\be
\bra{\pi_3(p)} {\bar q}\frac{\tau_3}{2}\gamma^\mu\gamma_5 q\ket{0} = 
-i F_\pi p^\mu
\ee
and taking the pion on-shell, leading to
\be
\label{eq:lreff3}
\bra{\pi_3} S_3\ket{0} = \frac{F_\pi m_\pi^2}{{\bar m}}\ .
\ee
Using Eqs.~(\ref{eq:lreff1}-\ref{eq:lreff3}) and including the coefficient 
of the four-quark operators appearing in Eq.~(\ref{eq:lreff}) leads to
\be
\gpbo = \left(\mathrm{Im}\, C_{\varphi u d}\right) 
\left(\frac{m_\pi}{\Lambda}\right)^2 
\left(\frac{F_\pi\sigma_{\pi N}}{3{\bar m}^2}\right)\ ,
\ee
which is of order $\lamchi F_\pi/\Lambda^2$ as advertised. Using 
$\sigma_{\pi N}\approx 45\pm 6$ MeV\cite{Aoki:2008sm,Shanahan:2012wh}, ${\bar m} \approx 3.85$ MeV \cite{Beringer:1900zz}
 gives
\be
\label{eq:lreff4}
\gpbo = (3.3\times 10^{-5})\times\left(\mathrm{Im}\, C_{\varphi u d}\right) 
\left(\frac{v}{\Lambda}\right)^2\ .
\ee

The other approach is to assume the saturating states are the lowest
available single-meson states.
For example, a meson of mass $m_m\gg m_\pi$ that can be exchanged between
two nucleons gives rise to a potential of range $\sim 1/m_m$,
which is short
compared to typical nuclear distances $\sim 1/m_\pi \gg 1/m_m$.  
In an expansion in powers of  
$m_\pi/m_m$, such a potential can be replaced
by contact interactions with an increasing number of derivatives.
In first order in the relevant TVPV parameter, the meson couples
through a TVPV coupling to one nucleon and a P-, T-even coupling
to the other, resulting in TVPV contact interactions such as
the $\bar{C}_{1,2}$ terms in Eq. \eqref{chiPTTV}.

In this context, the mesons that have been considered are the lightest: the
pseudoscalar $\eta$ and the vector mesons $V=\rho, \omega$
\cite{GHM93,TH94,Tim+04},
with TVPV interactions given by
\begin{eqnarray}
{\cal L}_{N\pi}^\mathrm{TVPV}&=&
\eta \bar{N}\left(\gebz +\gebo \tau_3 \right)N
+
\frac{\omega_\mu }{2m_N}
\left[\bar{N}\left(\gobz+\gobo\tau_3\right)i\sigma^{\mu\nu}\gamma_5  
\partial_\nu N + \mathrm{H.c.}\right]
\nonumber \\
&&
+
\frac{1}{2m_N}
\bar{N}\left\{\left[\grbz {\boldtau}\cdot{\boldrho}_\mu 
+\grbo \rho_\mu^{0} 
+\grbt \left(3\tau_3 \rho_\mu^0 -{\boldtau}\cdot{\boldrho}_\mu \right)
\right]i\sigma^{\mu\nu}\gamma_5 \partial_\nu N + \mathrm{H.c.}\right\} \ .
\label{heavymesonTV}
\end{eqnarray}
They lead to \bea \bar C^{(\eta,\omega)}_{1} &= & \frac{1}{m_N}
\left(\frac{g_{\eta N N}\gebz}{m^2_{\eta}} -\frac{g_{\omega N
N}\gobz}{m^2_{\omega}}\right) \ ,
\label{etaomega}
\\
\bar C^{(\rho)}_{2} & =& - \frac{g_{\rho NN}}{m_N m_\rho^2}
\left(\grbz+\grbt\right) \ ,
\label{rhodelta}
\eea where $m_{\eta}$ and $m_{V}$ 
are meson masses, and $g_{\eta N N}$ and $g_{V N N}$ are P-, T-even
meson-nucleon couplings, respectively the axial coupling of the eta and the
vector coupling of the vector meson.  This type of meson exchange also produces
other contact interactions \cite{Tim+04}, which are, however, expected to be of
higher order for CPV sources of dimension up to six.

Since $m_{V} \sim \Lambda_\chi$ and there is no reason
for $\bar g_{V}^{(0)}/\gpbz$ to be particularly big or small, 
the size of vector-meson contributions is 
comparable to the NDA expectations,
with some suppression coming from the numerical smallness of the 
P-, T-even rho-nucleon vector coupling $g_{\rho NN} \simeq 3.2$ 
\cite{Stoks:1994wp,Machleidt:2000ge} compared to the analogous
pion-nucleon coupling $2m_Ng_A/F_\pi\simeq 13.5$. 
(In contrast the same ratio for the omega is close to 1 
\cite{Stoks:1994wp,Machleidt:2000ge}.)
For the eta meson, the enhancement due to the relatively light mass is 
offset by the relative smallness of $g_{\eta N N} \simeq 2.24$ 
\cite{Tiator:1994et}. 
Obviously the limitation of this type of saturation
is that there are no firmer estimates of the TVPV couplings
$\bar g_{\eta}^{(i)}$ and  $\bar g_{V}^{(i)}$ than for the 
$\bar{C}_{1,2}$.


\subsection{Hadronic Matrix Elements: Discussion}

While there exists a solid body of work devoted to matching the $\theta$-term and dimension six operators onto hadronic quantities, there clearly exists 
considerable need for further advances. In what follows, we attempt to provide a sense of the present range of theoretical uncertainty in sensitivity of various hadronic quantities to the underlying operator coefficients. To that end, for each sensitivity coefficient ($\alpha_n$, $\lambda_{(0)}$, ${\tilde\zeta}_n^q$, {\em etc.}) we provide a \lq\lq best value" and \lq\lq reasonable range". The importance of attempting to quantify the theoretical uncertainty is two-fold. First, when using EDM search limits to derive bounds on the underlying parameters such as ${\bar\theta}$ or ${\tilde\delta}_q$, previous studies have often included only the experimental uncertainty while relying on a single hadronic computational framework. As a result, the quoted bounds may be unrealistically stringent. Second, we anticipate that hadronic structure theorists involved in the relevant computations will find our benchmarks helpful in setting goals for future refinements. In short, it is useful to know where and to what level reductions in theoretical uncertainties are called for.

Before proceeding, we make an important caveat. In comparison with the analysis of experimental uncertainties, the task of assigning theoretical error bars entails a greater degree of subjectivity. Consequently, the benchmarks we provide below should be taken somewhat impressionistically rather than as quantitatively robust. Nevertheless, we believe they offer reasonable guidance as to the present level of uncertainty as well as quantitative targets for further refinements. 

With these considerations in mind, we now discuss specifics. In setting the best values and ranges, we will use the considerations based on chiral symmetry/NDA as well as the spread of current theoretical computations as guides. Chiral symmetry and NDA is particularly helpful in determining if the results of specific computation are anomalously large. While one expects the chiral arguments to be realistic up to a factor of a few, a computation that yields a result an order of magnitude larger would likely be open to question. On the other hand, it is quite possible that specific dynamics can suppress a given quantity by more than a factor of a few.  In general, our best values will be somewhere close to the mid-point of the range of explicit computations and close to the magnitude indicated by chiral symmetry/NDA.
When assessing the range of explicit computations in a given framework, we will be rather inclusive, except when a given computation seems to be particularly anomalous with respect to the chiral/NDA expectations. Hence, our reasonable ranges will be roughly consistent with the spread of explicit computations and the factors of a few variation one might expect with respect to chiral/NDA arguments. 

Table \ref{tab:best}, then, gives these benchmarks. Generally speaking, we see that the quantities with the most narrow ranges are: 
\begin{itemize}
\item[(a)] $\alpha_n$, the sensitivity of $d_n$ to ${\bar \theta}$:  QCD sum rule computations are quite in line with chiral symmetry/NDA expectations. Moreover, a na\"ive scaling of the lattice results with $m_\pi^2$ would imply a value close to our best value\footnote{One should note, however, that the lattice computations to date do not necessarily manifest the expected chiral scaling in other related observables, so the result of na\"ive scaling may be a coincidence.}. Note that since the $\theta$-term does not run, there exists no uncertainty associated with an incomplete analysis of RG evolution, in contrast to several of the dimension six operators. 
\item[(b)] $\lambda_{(0)}$,  the sensitivity of $\gpbz$ to ${\bar\theta}$:  the use of chiral symmetry and lattice results for $(\Delta m_N)_q$ provide a relatively model-independent result. Reduction in the errors on $(\Delta m_N)_q$ and the light quark masses from the lattice will lead to a corresponding narrowing of the theoretical range on this quantity.
\item[(c)] $g_S^{(0)}$, isoscalar scalar form factor that governs in part the sensitivity of atomic and molecular EDMs to the combination of coefficients $\mathrm{Im} (C_{\ell e d q}- C_{\ell equ}^{(1)})$: a model independent value is obtained from $\sigma_{\pi N}$ and the average light quark mass, ${\bar m}$. To the extent that the lattice uncertainties on these quantities are robust, one has a relatively narrow range for the isocscalar scalar form factor.
\item[(d)] ($\beta_n^{q\gamma}$, $\rho_n^q$, $\zeta_n^q$), the sensitivity of $d_n$ to the quark EDMs:  results of explicit computations do not vary considerably from expectations based on either chiral symmetry/NDA or the quark model. It is worth emphasizing, however, that a complete analysis of the RG evolution of the quark EDMs from the weak to hadronic scales, taking into account mixing with the CEDM and four-quark operators, has generally not been carried out.
\end{itemize}

Although further reductions in the uncertainties associated with these quantities would be welcome, we do not consider them to have the greatest urgency. Those seemingly most theoretically fraught are the sensitivities to the CEDM, three-gluon operator, and four-quark operators. 

\begin{itemize}
\item[(e)] ($\beta_n^{qG}$, ${\tilde\rho}_n^q$, ${\tilde\zeta}_n^q$), the sensitivity of $d_n$ to the quark CEDMs: Here we take as best values the average of the existing QCD sum rule results, which in the case of the d-quark CEDM is equal to the chiral/NDA expectation. The ranges here are rather broad, spanning an order of magnitude. Moreover, as in the case of the quark EDMs, a complete implementation of the RG evolution that includes mixing with the four quark operators remains to be performed. We also note that recent studies of the CEDM contribution to the $\rho$-meson EDM using the Dyson-Schwinger approach\cite{Pitschmann:2012by} raise further questions about the reliability of matching of CEDMs onto hadronic quantities. Although the $\rho$-meson EDM is not of experimental interest, the relatively simplicity of the $\rho$-meson bound state makes it a useful \lq\lq laboratory" for various hadronic matrix element computational methods. The results obtained with the Dyson-Schwinger framework imply that the $\rho$-meson EDM is an order of magnitude less sensitive to the CEDMs than one would infer from the corresponding QCD sum rule computation\cite{Pospelov:1999rg}. In contrast, both approaches yield similar sensitivities to the quark EDMs. Should a similar situation emerge for $d_n$, one would need to further inflate the theoretical uncertainty on the ($\beta_n^{qG}$, ${\tilde\rho}_n^q$, ${\tilde\zeta}_n^q$).
\item[(f)] ($\gamma^G_{(i)}$, ${\tilde \omega}_{(i)} $, ${\tilde\eta}_{(i)}$), the sensitivity of the $\gpbi$ to the quark CEDMs: Here the situation is even more uncertain. To our knowledge, only two computations have appeared to date. The ranges quoted in Ref.~\cite{Pospelov:2001ys} is consistent with the magnitude expected from chiral symmetry/NDA, but in the case of the contribution to $\gpbz$ incompasses zero as well. In the case of the $\gpbo$ sensitivity, the range in 
Ref.~\cite{Pospelov:2001ys} is narrower, and we have no present rationale to expand it, but the dearth of analyses and the range for the $\gpbz$ sensitivity should give one pause.
\item[(g)] $\beta_n^{\tilde G}$ and $\gamma_{(i)}^{\tilde G}$, sensitivity of $d_n$ and the $\gpbi$ to the three gluon operator: For this case, we posses a dearth of information. The central value for $\beta_n^{\tilde G}$ given in the QCD sum rule work of Ref.~\cite{Demir:2002gg} that is often quoted elsewhere is an order of magnitude smaller than the chiral symmetry/NDA expectations. Consequently, we take a broad range for this parameter. For the $\gamma_{(i)}^{\tilde G}$, we have only the chiral/NDA expectations, and, thus, employ the \lq\lq factor of a few" criterion for this range. Note that the three-gluon operator is multiplicatively renormalized, so the theoretical uncertainty is associated entirely with the hadronic matching computations. 
\item[(h)] Four quark operators: hadronic matrix element computations for these operators is, if anything, even less advanced than for the three-gluon operator. Apart from issue of RG evolution, explicit computations are few and far between. The factorization computation used to match $Q_{\varphi ud}$ onto $\gpbo$ gives a smaller value for $\gamma_{(1)}^{\varphi ud}$ than the chiral/NDA expectation. Computations for the other operators $Q_{quqd}^{(1,8)}$ have been carried out using a combination of the quark model, factorization, and a relativistic meson loop approach\cite{An:2009zh}. However, the meson loop computation utilized in that work is not consistent with the EFT power counting embodied in HB$\chi$PT and, thus, may overestimate the magnitude of the matrix element by an order of magnitude. Consequently, we are reluctant to use the results in that work for guidance. Instead, we start with the chiral/NDA estimates and give an order of magnitude spread based on the present dearth of consistent computations, implementation of RG running, and comparison with the factorization estimate.
\end{itemize}

We do not include in Table \ref{tab:best} the semileptonic form factors apart from $g_S^{(0,1)}$. The manifestation of the pseudoscalar and tensor form factors in atoms and molecules is suppressed by several factors. The pseudoscalar interactions are higher order in the HB$\chi$PT expansion, while the tensor charge does not receive a nuclear coherent enhancement. We note, however, that the EDM of the  diamagnetic atom  $^{199}$Hg has roughly an order or magnitude greater sensitivity to the tensor $eq$ interaction than it does to the scalar interaction.  The value of $g_T^{(0,1)}$  in this case is, thus, considerably more significant than in the paramagnetic systems.

Two implications should be drawn from our theory uncertainty estimates. First, for BSM scenarios in which the CEDM, three-gluon, and/or four-quark operators have significant Wilson coefficients, it will be particularly important for any global analysis to include the rather sizeable uncertainties at the hadronic level. Second, a concerted effort to refine the hadronic computations and reduce the uncertainties is clearly in order. We hope that our delineation of these best values and ranges will spur future efforts in this direction.

\begin{table}[t]
\centering \renewcommand{\arraystretch}{1.5}
\begin{tabular}{||c||c|c|c||c|c|c||}
\hline\hline
Param & Coeff &  Best Value$^a$ & Range & Coeff &  Best Value$^{b,c}$ & Range$^{b,c}$ \\
\hline
${\bar\theta}$ & $\alpha_n$ & 0.002  & (0.0005-0.004)   &$\lambda_{(0)}$ &0.02 & (0.005-0.04) \\
 & $\alpha_p$ &  &  &$\lambda_{(1)}$ & $2\times 10^{-4}$ & $(0.5-4)\times 10^{-4}$ \\
\hline\hline
$\mathrm{Im}\, C_{q G}$ & $\beta_n^{uG}$ & $4\times 10^{-4}$  & $(1-10)\times 10^{-4}$ &$\gamma_{(0)}^{+G}$ & -0.01 & (-0.03) $-$ 0.03 \\
 & $\beta_n^{dG}$ & $8\times 10^{-4}$  & $(2-18)\times 10^{-4}$  &$\gamma_{(1)}^{-G}$ & -0.02 & (-0.07) $-$ (-0.01) \\
 \hline
 ${\tilde d}_q$ & $e{\tilde\rho}_n^u$ & $-0.35$ & $-(0.09-0.9)$  &${\tilde \omega}_{(0)}$ & 8.8  & (-$25)-25$  \\
 & $e{\tilde\rho}_n^d$ & $-0.7$  & $-(0.2-1.8)$  &${\tilde \omega}_{(1)}$ & 17.7 & $9-62$  \\
 \hline
 ${\tilde\delta}_q$ & $e{\tilde\zeta}_n^u$ & $8.2\times 10^{-9}$ & $(2-20)\times 10^{-9}$  &${\tilde\eta}_{(0)}$ &$-2\times 10^{-7}$ &$($-$6-6)\times 10^{-7}$  \\
& $e{\tilde\zeta}_n^d$ & $16.3\times 10^{-9}$ & $ (4-40)\times 10^{-9}$ &${\tilde\eta}_{(1)}$ &$-4\times 10^{-7}$ &$-(2-14)\times 10^{-7}$  \\
\hline\hline
$\mathrm{Im}\, C_{q \gamma}$ & $\beta_n^{u\gamma}$ & $0.4\times 10^{-3}$ & $(0.2-0.6)\times 10^{-3}$ &$\gamma_{(0)}^{+\gamma}$ &$-$ & $-$ \\
 & $\beta_n^{d\gamma}$ & $-1.6\times 10^{-3}$ &$-(0.8-2.4)\times 10^{-3}$  &$\gamma_{(1)}^{-\gamma}$ &$-$ & $-$ \\
 \hline
 ${ d}_q$ & ${\rho}_n^u$ & $-0.35$   &  $(-0.17)-0.52$ &${\omega}_{(0)}$ &$-$ &$-$  \\
 & ${\rho}_n^d$ & $1.4$ & 0.7-2.1 &${\omega}_{(1)}$ & $-$& $-$ \\
 \hline
 ${\delta}_q$ & $ {\zeta}_n^u$ & $8.2\times 10^{-9}$ & $(4-12)\times 10^{-9}$ &${\eta}_{(0)}$ &$-$ &$-$  \\
& ${\zeta}_n^d$ & $-33\times 10^{-9}$  &  $-(16-50)\times 10^{-9}$&${\eta}_{(1)}$ & $-$& $-$ \\
 \hline\hline
 $C_{\tilde G}$ & $\beta_n^{\tilde G}$ & $2\times 10^{-7}$  & $(0.2-40)\times 10^{-7}$ &$\gamma_{(i)}^{\tilde G}$ & $2\times 10^{-6}$ & $(1-10)\times 10^{-6}$  \\
 \hline\hline
 $\mathrm{Im}\,  C_{\varphi ud }$ & $\beta_n^{\varphi ud}$ & $3\times10^{-8}$  & $(1-10)\times 10^{-8}$ &$\gamma_{(1)}^{\varphi ud}$ & $1\times 10^{-6}$ & $(5-150)\times 10^{-7}$  \\
 $\mathrm{Im}\,  C_{quqd}^{(1,8)}$ & $\beta_n^{quqd}$ & $40\times 10^{-7}$  & $(10-80)\times 10^{-7}$ &$\gamma_{(i)}^{quqd}$ & $2\times 10^{-6}$ & $(1-10)\times 10^{-6}$  \\ 
  \hline\hline
 $\mathrm{Im}\, C_{eq}^{(-)}$ & $g_S^{(0)}$ & $12.7$  & 11-14.5 &  &  &   \\
 $\mathrm{Im}\, C_{eq}^{(+)}$ & $g_S^{(1)}$ & $0.9$  & 0.6-1.2 &  &  &   \\
 \hline\hline
\end{tabular}
\caption{Best values and reasonable ranges for hadronic matrix elements of CPV operators. First column indicates the coefficient of the operator in the CPV Lagrangian, while second column indicates the hadronic matrix element (sensitivity coefficient) governing its manifestation to the neutron EDM. Third and fourth columns give the best values and reasonable ranges for these hadronic coefficients. Firth to seventh columns give corresponding result for contributions to TVPV $\pi NN$ couplings. $^a$ Units are $e$ fm for all but the ${\tilde\rho}_n^q$ and $\rho_n^q$. $^b$ We do not list entries for ($\gamma_{(i)}^{\pm\gamma}$, ${\omega}_{(i)}$, ${\eta}_{(i)}$) as they are suppressed by $\alpha/\pi$ with respect to (${\tilde\gamma}_{(i)}^{\pm\gamma}$, ${\tilde \omega}_{(i)}$, ${\tilde\eta}_{(i)}$) . $^c$ The ${\tilde\omega}_{(0,1)}$ are in units of fm$^{-1}$.
\label{tab:best}}
\end{table}

\section{Nuclear, Atomic, and Molecular Scales}
\label{sec:nuc}

Composite systems are often the most amenable to experiments.  But the EDMs of
composite systems reflect few- or many-body dynamics as well as the
fundamental source of CP violation and QCD.  For some experiments, completely
ionized light nuclei are useful, and for these cases we must employ few-body methods to relate 
the
nuclear EDMs to, {\it e.g.}, $\bar\theta$.  In heavier neutral systems a new
phenomenon is important: the shielding of the EDMs of constituents of one
charge ({\it e.g.} protons in the nucleus) by those of the other (electrons).  
The
transmission of CP violation through a nucleus into an atom must overcome this
shielding and its effectiveness in doing so is expressed by a nuclear Schiff
moment, which we define shortly.

We begin by considering EDMs of light systems, potentially important for
storage-ring experiments, and then move to heavier systems, useful in
experiments on immobilized atoms or molecules. 

\subsection{Light Nuclei}
\label{sec:light-nuclei}


In addition to the continuous improvement in experiments on neutral systems,
a new, exciting prospect is the direct measurement of the EDMs of charged 
particles in storage rings \cite{Onderwater:2012me,Semertzidis:2011qv}.  
When a particle
moves in an electric and/or magnetic field, its spin will precess at a rate
that depends not only on the magnetic dipole moment but also on the EDM.  The
best bound on the muon EDM  \cite{Bennett:2008dy}
comes, in fact, as a by-product on the BNL $g-2$ experiment. 
It can be expected that dedicated experiments in
rings with optimized parameters 
will allow sensitive probing of the EDMs of light nuclei. 
For example, it has been proposed \cite{Onderwater:2012me,Semertzidis:2011qv} 
that for 
$d_p$ a sensitivity of $10^{-16} e$ fm can be achieved.  Similar sensitivity
could be attained
also for the deuteron ($^2$H nucleus) and helion ($^3$He nucleus) EDMs, $d_d$
and $d_h$ respectively.
The triton ($^3$H nucleus) EDM, $d_t$, might be accessible as well.

{}From a theoretical perspective, the EDMs of light nuclei can be calculated 
with
relatively small uncertainty originating in the P-, T-even strong
interactions, as essentially exact calculations are possible.  Moreover, 
with an effective field theory approach based on HB$\chi$PT we can treat the 
nucleon and nuclear EDMs on the same footing, and explore
the sensitivity of nuclear EDMs to different combinations of TVPV hadronic
interactions than that which appears in the nucleon EDM.  
In particular, it has been argued that the
deuteron EDM has some sensitivity to the CPV source 
\cite{Lebedev:2004va,deVries:2011re}, and
that a combined measurement of $d_n$, $d_p$, $d_d$, $d_h$ and $d_t$ could be 
used to disentangle the various sources \cite{deVries:2011an}.  
The reason for this is the different relative strengths
of the various couplings at LO in HB$\chi$PT 
\cite{deVries:2011re,deVries:2011an}, 
which are rooted in the different
chiral symmetry properties of the various sources, as discussed in 
Section \ref{sect:chiralcons}.
Likewise, experimental access to other TVPV moments, such as the deuteron 
magnetic quadrupole moment, 
would be very useful as well for separating sources 
\cite{deVries:2011re,Liu:2012tra},
but it does not look feasible in the near future.

In a nucleus with $A$ nucleons, certain P-, T-even inter-nucleon interactions 
need to be 
resummed in order to produce a bound state and its associated wavefunction
$|\Psi_{A}\rangle$. (For a review of nuclear EFT, see  
Ref. \cite{Bedaque:2002mn}.)
A nuclear EDM $d_{A}$ arises
from the average with such a wavefunction of two TVPV mechanisms: 
{\it i)} the TVPV electromagnetic current
$J_{\slashPTsub}^{0}$, whose one-nucleon component is the 
nucleon EDM;
and {\it ii)} a combination of the TVPV potential $V_{\slashPTsub}$ 
and the P-, T-even electromagnetic current $J_{PT}^{0}$.
We follow here Ref. \cite{deVries:2011an}, which we refer the reader to for 
more details.
Because TVPV interactions are so tiny, we can write in 
first-order perturbation theory
\begin{equation}
d_{A}=
\left\langle \Psi_{A}
\left|\vec{D}_{\slashPTsub}\right|
\Psi_{A}\right\rangle 
+\left( \left\langle \Psi_{A}
\left|\vec{D}_{PT}\right|
\widetilde{\Psi}_{A}\right\rangle 
+ c.c.\right)
\ .
\label{eq:nuclearEDM}
\end{equation}
The electric operators are obtained from the corresponding
time-component electromagnetic current through
$\vec{D} = i\lim_{\vec{q}\rightarrow0}\vec{\nabla}_{\vec{q}}J^{0}(q)$.
At the one-nucleon level
\begin{equation}
\vec{D}_{PT}^{(1)}=
\frac{e}{2}\,\sum_{i=1}^{A}\,\tau_{3}^{(i)}\,\vec{r}_{i}\ ,
\qquad
\vec{D}_{\slashPTsub}^{(1)}= \sum_{i=1}^{A}\,
\left(\bar{d}_{0}+\bar{d}_{1}\,\tau_{3}^{(i)}\right) \,
\vec{\sigma}^{(i)} \ ,
\label{eq:C1_TV_1B}
\end{equation}
in intrinsic coordinates, which obey $\sum_{i=1}^{A}\vec{r}_{i}=0$.
The more complicated two- and more-nucleon
currents are expected to be generically less important,
although this is not always true, as discussed below.
The first term in Eq. \eqref{eq:nuclearEDM} represents
the contribution of the individual nucleons to the nuclear
EDM, as well as the contribution from TVPV many-body currents.
The second term in Eq. \eqref{eq:nuclearEDM} is the contribution
of the parity-admixed wavefunction $|\widetilde{\Psi}_{A}\rangle$,
obtained from the TVPV potential via 
\begin{eqnarray}
(E-H_{PT})|\widetilde{\Psi}_{A}\rangle = V_{\slashPTsub}|\Psi_{A}\rangle 
\qquad{\rm where}\qquad
(E-H_{PT})|\Psi_{A}\rangle =  0 \ ,
\end{eqnarray}
with $H_{PT}$ being the P-, T-even Hamiltonian.

Because of its long range, one-pion exchange (OPE) has long been recognized as
a potentially important component of the TVPV two-nucleon ($NN$) potential
\cite{Haxton:1983dq,Herczeg,herczeg88}, and expressed in terms of the three
non-derivative pion-nucleon couplings in Eq. \eqref{chiPTTV}
\cite{Barton:1961eg}.  In the literature, this potential is sometimes
supplemented by the single exchange of heavier mesons, with the $\eta$
\cite{Gudkov:1992yc}, rho \cite{Towner:1994qe}, and $\omega$ \cite{Towner:1994qe}
being most popular.  Allowing sufficiently many couplings of these mesons to
nucleons one can produce \cite{Liu:2004tq} the most general short-range TVPV
$NN$ interaction with one derivative \cite{H66}.  Although a derivative
expansion is justified on general grounds, the relative importance of terms
with various ranges, spin/isospin structures and number of nucleons depends on
the TVPV source.
The potential obtained from HB$\chi$PT, which accommodates the most important
effect of each source, is discussed in detail in Ref. \cite{Maekawa:2011vs}.
In configuration space, the $NN$ potential corresponding to the Lagrangian
\eqref{chiPTTV} reads
\begin{eqnarray}
V_{\slashPTsub}(\vec{r}_{ij}) & = & 
\frac{g_{A}}{F_{\pi}} 
\left\{\bar{g}_\pi^{(0)}  
\boldtau^{(i)}\cdot\boldtau^{(j)}
\left(\vec{\sigma}^{\,(i)}-\vec{\sigma}^{\,(j)}\right)
\right.
\nonumber \\
& & \left.
+\frac{\bar{g}_\pi^{(1)}}{2}
\left[\left(\tau_{3}^{(i)}+\tau_{3}^{(j)}\right)
\left(\vec{\sigma}^{(i)}-\vec{\sigma}^{(j)}\right)
+\left(\tau_{3}^{(i)}-\tau_{3}^{(j)}\right)
\left(\vec{\sigma}^{(i)}+\vec{\sigma}^{(j)}\right)\right]
\right.
\nonumber \\
& & \left.
+\bar{g}_\pi^{(2)}
\left(3\tau_{3}^{(i)}\tau_{3}^{(j)}-\boldtau^{(i)}\cdot\boldtau^{(j)}\right)
\left(\vec{\sigma}^{(i)}-\vec{\sigma}^{(j)}\right)
\right\}\cdot\left(\vec{\nabla}\, Y(r_{ij})\right)
\nonumber \\
& & +\frac{1}{2}
\left[\bar{C}_{1}+\bar{C}_{2}\boldtau^{(i)}\cdot\boldtau^{(j)}\right]
\left(\vec{\sigma}^{\,(i)}-\vec{\sigma}^{\,(j)}\right)\cdot
\left(\vec{\nabla}\delta^{(3)}(\vec{r}_{ij})\right)
+ \cdots \ ,
\label{eq:PToddV}
\end{eqnarray}
where $\vec{r}_{ij}=\vec r_i-\vec r_j$ is the relative position of the two
interacting nucleons and $Y(r)=\exp(-m_{\pi}r)/4\pi r$ is the usual Yukawa
function, so that \be \vec{\nabla}\, Y(r_{ij}) = -\frac{\vec{r}_{ij}}{4\pi
r_{ij}^3} \left(1+ m_{\pi}r_{ij}\right) \, {\rm exp}(-m_{\pi} r_{ij}) \ .  \ee
The short-range interactions can be thought of as accounting for heavier-meson
exchange, as discussed in Section \ref{subsec:saturation}.  For all sources,
few-body potentials are expected to generate smaller contributions, except for
the left-right four-quark operator \eqref{eq:lreff}, for which the effects of a
three-nucleon potential originating in a LO three-pion vertex
\cite{deVries:2011an} proportional to $\bar{g}_\pi^{(0)}$ remain to be studied.

Equation \eqref{eq:nuclearEDM} has been evaluated in the literature for $A=2,3$,
and the explosive growth in {\it ab initio} methods affords ways to calculate
the EDMs of larger nuclei if needed. 
Most existing work employs TVPV one-meson-exchange potentials, with older
references using simple P-, T-even wavefunctions and single-nucleon currents,
and more recent ones, highly developed phenomenological P-, T-even potentials
and even meson-exchange currents.  In HB$\chi$PT the non-analytic behavior of
the nucleon EDM in $m_\pi$ and the dominance of OPE in nuclear observables can
be accounted for simultaneously, with chiral symmetry playing a central role.
In principle full consistency can be achieved,
but so far calculations are limited to phenomenological P-, T-even interactions
for $H_{PT}$.
At this stage leading contributions, with an uncertainty of roughly $\sim
m_\pi/M_{QCD}\sim 20$\%, are sufficient, and for the most part one can restrict
oneself to the one-body currents and two-body potential described above.  For
the $\theta$-term the situation is more complicated because in nuclei with
equal numbers of protons and neutrons, $N=Z$, the isoscalar component of the
P-, T-odd potential \eqref{eq:PToddV} gives a vanishing contribution in
combination with $\vec{D}_{PT}^{(1)}$ \eqref{eq:C1_TV_1B}
\cite{Haxton:1983dq}.  The latter is an isovector with a conserved third
component, and can only contribute to the EDM if there is a parity-admixed
component of the wavefunction that differs from the ground-state wavefunction
by one unit of isospin.  For dimension-six sources, where the leading $NN$
potential is not expected to be dominantly isoscalar, this is not of particular
consequence.  But, for the $\theta$-term, $\bar{g}_\pi^{(0)}$ is the formally
leading part of the potential, and in $N=Z$ nuclei a non-zero result comes only
from subleading parts of the $NN$ potential as well as two-body currents.

Not all the seven parameters shown explicitly in Eq. \eqref{chiPTTV} are
important for every CPV source.  In fact, as the discussion in Section
\ref{sect:chiralcons} shows, 
$\bar{g}_\pi^{(2)}$ is expected to be small for all sources, and as a
consequence the EDMs of light nuclei should be described at LO in terms of the
{\it six} parameters $\bar{d}_{0,1}$, $\bar{g}_\pi^{(0,1)}$, and
$\bar{C}_{1,2}$ \cite{deVries:2011an}.  Results for the EDMs of light nuclei in
terms of these six LO parameters are reviewed below and summarized in Table
\ref{tab:lightEDMs}.  The first two rows are a reminder that we have absorbed
the loop contributions to the neutron and proton EDMs in $\bar{d}_{0,1}$, as
discussed in Section \ref{sect:chpt}.  The potential-model dependence in the
subsequent rows is not larger than $\sim 25$\%, which is comparable with the LO
HB$\chi$PT error. Exceptions are the short-range contributions from the ${\bar
C}_i$, which can only be considered order-of-magnitude estimates.  As discussed
below, for the tri-nucleon system there are disagreements in the OPE estimates
of about $\pm 50$\% in the values quoted.  After discussing specific results
for the deuteron, helion and triton, we cast them in terms of the $\theta$-term
and dimension-six sources.

%
\begin{table}
\renewcommand{\arraystretch}{1.5}
\begin{center}
\begin{tabular}{||c|cccccc||}
\hline 
LEC & $\bar{d}_{0}$ & $\bar{d}_{1}$ & $\bar{g}_\pi^{(0)}\, e\,\mathrm{fm}$
& $\bar{g}_\pi^{(1)}\, e\,\mathrm{fm}$ & $(F_{\pi}^3\bar{C}_{1})\,
e\,\mathrm{fm}$ & $(F_{\pi}^3\bar{C}_{2})\, e\,\mathrm{fm}$ \tabularnewline
\hline 
$d_{n}$ & $1$ & $-1$ & - & - & - & -\tabularnewline $d_{p}$ & $1$ & $1$ & - & -
& - & -\tabularnewline $d_{d}$ & $2$ & $0$ & $-0.0002+ 0.07\beta_1$ & $0.2$
& - & -\tabularnewline $d_{h}$ & $0.83$ & $-0.93$ & $0.1$ & $0.2$ & $-0.01$
& $0.02$\tabularnewline $d_{t}$ & $0.85$ & $0.95$ & $-0.1$ & $0.2$ & $0.01$ &
$-0.02$\tabularnewline \hline
\end{tabular}
\end{center}
\caption{Dependence of the EDMs of the neutron, proton, deuteron, helion, and
triton on the six relevant TVPV low-energy constants at leading order.  
A ``-'' denotes that the
low-energy constant does not contribute in a model-independent way to the EDM 
at this order.  
For the potential-model
dependence and other uncertainties in the results, see text. 
The P-, T-even isospin-breaking pion-nucleon coupling $\beta_1$
is not well known, 
$\beta_1=(0\pm 9)\cdot 10^{-3}$ \cite{vanKolck:1996rm,vanKolck:1997fu}. 
(Adapted from Ref. \cite{deVries:2011an}.)} 
\label{tab:lightEDMs} 
\end{table}
%

\subsubsection{Deuteron}

The deuteron EDM has been investigated in the meson-exchange picture
\cite{Flambaum:1984fb,Avi85,Khriplovich:1999qr,Liu:2004tq,Afnan:2010xd}, with
various degrees of sophistication in the treatments of the P-, T-even
interaction $H_{PT}$.  Refs. \cite{Liu:2004tq,Afnan:2010xd} found that
differences in the EDM generated by TVPV OPE are rather small among modern
high-quality phenomenological potentials.  Ref. \cite{deVries:2011an} uses the
calculation scheme of Ref. \cite{Liu:2004tq} to obtain wavefunctions
$|\Psi_{2}\rangle$ and $|\widetilde{\Psi}_{2}\rangle$ for the potential
\eqref{eq:PToddV}, in conjunction with HB$\chi$PT currents, for all CPV sources
of dimension up to six.  Ref. \cite{Bsaisou:2012rg} also uses HB$\chi$PT and
phenomenological potentials, but with slight different dimensional estimates
for the various HB$\chi$PT contributions, in the particular case of the
$\theta$ term.  All these references ignore relativistic corrections, which are
absent from the phenomenological potentials they use.  A fully consistent
HB$\chi$PT calculation exists \cite{deVries:2011re}, in which pions are treated
perturbatively --- a good approximation scheme in the loosely bound deuteron.
In this case the LO P-, T- even potential is just a delta function, so this is
an effective field theory extension of Ref. \cite{Khriplovich:1999qr}.  Results
are consistent with Refs. \cite {deVries:2011an,Bsaisou:2012rg}, suggesting
that a fully consistent calculation with non-perturbative pions will not
deviate significantly from the results obtained so far.

The simplest contribution to the deuteron EDM 
originates in the EDMs of its constituents.
Since the deuteron has spin $S=1$ and isospin $I=0$,
the nuclear matrix element 
of $\vec{D}_{\slashPTsub}^{(1)}$ in Eq. \eqref{eq:C1_TV_1B}
gives simply $2\bar{d}_0$.
Since there is no reason to expect a cancellation with other contributions,
$2\bar{d}_0$ serves as a lower-bound estimate for the deuteron EDM.
For the $\theta$-term in particular, 
using the long-range NLO contributions to the 
isoscalar nucleon EDM, Eq. \eqref{d1est}, we expect \cite{deVries:2011re}
\begin{equation}
|d_d|\simge 3\cdot 10^{-4}\bar\theta \, e \, \mathrm{fm} \ .
\end{equation}

In agreement with the more general argument
for $N=Z$ nuclei, the one-body operator $\vec{D}_{PT}^{(1)}$ cannot
bring the deuteron wavefunction back to a (mostly) $^3S_1$ wave
once the isoscalar TVPV potential takes it to $^1P_1$.
In order for $\vec{D}_{PT}^{(1)}$ to yield a non-zero contribution,
the parity-admixed component of the wavefunction has to be in the 
$^3P_1$ state, to which only $\bar{g}_\pi^{(1)}$ contributes.
The corresponding nuclear matrix element has been 
calculated several times in the literature.
As summarized in Ref. \cite {Bsaisou:2012rg},
there is agreement to 
better than 10\% among
modern $NN$ potentials,
and within 30\% between them and a simple delta-function potential.
When OPE through the 
isotensor pion-nucleon coupling $\bar{g}_\pi^{(2)}$ is included,
it gives small contributions \cite{Liu:2004tq}, 
even when estimates about the small magnitude
of $\bar{g}_\pi^{(2)}$ are disregarded.

The expectation from HB$\chi$PT that two- and more-body currents give small
contributions is corroborated by a model calculation \cite{Liu:2004tq}.
They can be neglected for all CPV sources except the $\theta$-term.
For the latter, because the formally LO contribution vanishes,
one has to go to NNLO. This brings in the same dependence on
$\bar{g}_\pi^{(1)}$ as for other sources. Additionally, a dependence
on $\bar{g}_\pi^{(0)}$ emerges through the
subleading potential and two-body currents, 
together with two P-, T-even isospin-breaking parameters:
the quark-mass component of nucleon mass difference, $(\Delta m_N)_q$ 
in Eq. \eqref{eq:nucleonmassL},
which can be estimated from lattice QCD
as we have done in Section \ref{sect:chiralcons}.
and the isospin-breaking pion-nucleon coupling
$(\beta_1/F_\pi) (\partial_\mu \pi_3) \bar{N}S^\mu N$,
with $\beta_1=\mathcal{O}(\epsilon m_{\pi}^{2}/M_{QCD}^{2})$,
for which only the bound   
$\beta_1=(0\pm 9)\cdot 10^{-3}$ \cite{vanKolck:1996rm,vanKolck:1997fu} is known. 

These results are summarized in the third row of Table \ref{tab:lightEDMs}.

\subsubsection{Helion and triton}

There have been fewer calculations of the trinucleon EDMs.  A pioneering
calculation \cite{Avishai:1986dw} of the helion EDM for the $\bar{g}_\pi^{(0)}$
OPE (dominant for the $\theta$-term) used an old phenomenological P-, T-even
potential solved in the adiabatic approximation of the hyperspherical-harmonics
method, and found no nuclear enhancement compared to the neutron EDM.  The era
of modern calculations began with Ref. \cite{Stetcu:2008vt}, when the nuclear
wavefunction was calculated using high-quality P-, T-even potentials including
the Coulomb interaction.  A solution is found with the no-core shell model
(NCSM) method, which employs a model space made from $N_{max}$ properly
antisymmetrized harmonic-oscillator wavefunctions of frequency $\Omega$.  At
large enough $N_{max}$, results for the helion EDM, which are somewhat larger
than Ref. \cite{Avishai:1986dw} where they can be compared, become independent
of $\Omega$. Meson-exchange currents were neglected, as suggested by their
smallness in the deuteron.  For mesonic couplings of equal magnitude, OPE is
found to be dominant over shorter-range interactions.  Ref. \cite{deVries:2011an}
adapted this calculation to the TVPV ingredients from HB$\chi$PT, and
calculated the EDM of triton for the first time.  The two short-range
interactions from $\bar C_{1,2}$, which can be thought of as originating from,
respectively, omega and rho meson exchanges considered in Ref.
\cite{Stetcu:2008vt}, were regulated with Yukawa functions.  Ref.
\cite{Song:2012yh} used similar input but solved Faddeev equations instead.

Calculations with various realistic potentials
agree within $\sim 25\%$ for the nucleon EDM contributions.
They give nuclear matrix elements of roughly
equal magnitude for $\bar d_{0}$ and $\bar d_{1}$, so that, 
as one might have expected, 
the helion (triton) EDM is mostly sensitive to the neutron (proton) EDM,

For the contribution from the TVPV potential, results for triton are very
similar in magnitude to those for helion.  Both Refs. \cite{Stetcu:2008vt} and
\cite{Song:2012yh} find a spread of $\sim 25\%$ between different potentials,
but they disagree by an overall factor of about two in isoscalar and isovector
TVPV terms (and a factor five in the subleading isotensor component).  The reason
for this discrepancy is unclear at present, and it is a priority to resolve
it.  Additionally, in Ref. \cite{deVries:2011an} the short-range contributions
from the ${\bar C}_i$ were found to vary considerably with the explicit
regulator mass from potential to potential (a factor $\sim 5$ in the cases
studied).
There is thus a much stronger potential dependence, and more solid numbers have
to wait for a fully consistent calculation.

These results are summarized in the fourth and fifth rows of Table
\ref{tab:lightEDMs}, where for OPE we took values in between those of Refs.
\cite{Stetcu:2008vt, deVries:2011an} and \cite{Song:2012yh}.  Note that there
is yet no estimate of the effects of a three-nucleon TVPV potential, which
could be significant in the case of the left-right four-quark operator
\cite{deVries:2012ab}.

\subsubsection{Light nuclear EDMs: summary}

Using the results of Section \ref{sec:had}, we can obtain the sensitivity of
each EDM to the underlying CPV sources.  The orders of magnitudes expected for
these EDMs are given in Table \ref{lightOs} for each source.  Recall that the
Weinberg three-gluon operator \eqref{eq:weinbop}
and the two four-quark operators \eqref{eq:hadcpveff} present already at the
electroweak scale are chiral invariant. As a consequence they produce the same
hierarchy of hadronic interactions and cannot be separated at low energies.
For simplicity we use the shorthand notation $\{C_{\tilde G}, \mathrm{Im} \,
C_{quqd}^{(1,8)}\} \to \mathrm{Im}\, C_k $
in Table \ref{lightOs}.

\begin{table}
\renewcommand{\arraystretch}{1.5}
\begin{center}
\begin{tabular}{||c|ccccc||}
\hline 
Source & $\theta$-term & CEDM & quark EDM & chiral-invariant & left-right 
\tabularnewline \hline 
$\Lambda_{\chi} d_{n}/e$ & 
${\cal O}\left(\frac{m_{\pi}^{2}}{\Lambda_{\chi}^{2}}\bar\theta\right)$ & 
${\cal O}\left(\frac{m_{\pi}^{2}}{\Lambda^{2}}{\tilde{\delta}_q}\right)$ & 
${\cal O}\left(\frac{m_{\pi}^{2}}{\Lambda^{2}}\delta_q\right)$ & 
${\cal O}\left(\frac{\Lambda_{\chi}^{2}}{\Lambda^{2}}\mathrm{Im}\, C_k\right)$ &
${\cal O}\left(
\frac{\Lambda_{\chi}^{2}}{(4\pi)^2\Lambda^{2}}\mathrm{Im}\, C_{quqd}^{(1,8)}\right)$ 
\tabularnewline
$d_{p}/d_{n}$ & ${\cal O}\left(1\right)$ & ${\cal O}\left(1\right)$ & 
${\cal O}\left(1\right)$ & ${\cal O}\left(1\right)$ & ${\cal O}\left(1\right)$
\tabularnewline 
$d_{d}/d_{n}$ & ${\cal O}\left(1\right)$ & 
${\cal O}\left(\Lambda_{\chi}^{2}/Q^2\right)$ &
${\cal O}(1)$ & ${\cal O}(1)$ &
${\cal O}\left(\Lambda_{\chi}^{2}/Q^2\right)$
\tabularnewline 
$d_{h}/d_{n}$ & ${\cal O}\left(\Lambda_{\chi}^{2}/Q^2\right)$ & 
${\cal O}\left(\Lambda_{\chi}^{2}/Q^2\right)$ & ${\cal O}(1)$ & 
${\cal O}(1)$& ${\cal O}\left(\Lambda_{\chi}^{2}/Q^2\right)$
\tabularnewline 
$d_{t}/d_{h}$  & ${\cal O}\left(1\right)$ & ${\cal O}\left(1\right)$ & 
${\cal O}\left(1\right)$ & ${\cal O}\left(1\right)$ & ${\cal O}\left(1\right)$
\tabularnewline \hline
\end{tabular}
\end{center}
\caption{Expected orders of magnitude for the neutron EDM (in units of
$e/\Lambda_{\chi}$), and for the EDM ratios proton to neutron, deuteron 
to neutron,
helion to neutron and triton to helion, for the $\theta$-term and
dimension-six sources.  For chiral-invariant sources,
$\mathrm{Im}\, C_k$ stands for $C_{\tilde G}$ and $\mathrm{Im} \, C_{quqd}^{(1,8)}$.
$Q$ represents the low-energy scales $F_\pi$, $m_\pi$,
and $\sqrt{m_NB}$, with $B$ the binding energy.  (Adapted from Ref.
\cite{deVries:2011an}.)}
\label{lightOs} 
\end{table}

The texture of this table underlines the argument 
\cite{Lebedev:2004va,deVries:2011re,deVries:2011an}
that light nuclei act as a ``chiral filter'' for the various CPV sources.
Of course, a measurement of the neutron EDM alone could be due
to a $\theta$-term of just the right magnitude, or to any 
of the dimension-six sources, although if all dimensionless
factors were equal, chiral-invariant sources or the \lq\lq left-right"
four-quark operator \eqref{eq:lreff0} would be favored because 
they require no chirality flip.
Nuclear effects are most significant for the $\theta$-term, CEDM
and left-right four-quark operator.
Just on the basis of orders of magnitude,
we see that a large tri-nucleon EDM compared to a nucleon EDM would point to
them as possibly dominant sources, 
while a large $|d_d|$ compared to $|d_N|$ would be suggestive of
just the CEDM and left-right four-quark operator. 
Bounds on light EDMs would provide tighter bounds on this physics than
comparable bounds on nucleon EDMs.

We can infer more information about CPV sources from Table \ref{tab:lightEDMs}
when we take into account that the relative importance of various pion-nucleon
and short-range interactions is not the same for all sources.  For CEDM and
left-right four-quark operator, the expected dominance of nuclear effects comes
from pion exchange due to both $\bar{g}_\pi^{(0,1)}$ couplings, while only
$\bar{g}_\pi^{(0)}$ is present at LO for the $\theta$-term.  The isoscalar
coupling $\bar{g}_\pi^{(0)}$ approximately cancels in $d_h +d_t$, so while for
the CEDM $d_h +d_t \simeq 3d_d$, for the $\theta$-term $d_h +d_t \simeq 0.8
(d_n+d_p)$.  Effects of the left-right four-quark operator can only be
separated from CEDM if its TVPV three-nucleon potential is significant.  For
the quark EDM, where nuclear effects are much smaller, $d_h +d_t \simeq 0.8
(d_n+d_p)$ also holds but in addition one expects $d_h -d_t \simeq 0.9
(d_n-d_p)$.  The situation is most complicated for chiral-invariant sources,
for which nuclear effects are significant for both $A=2,3$, but they depend in
the deuteron only on $\bar{g}_\pi^{(1)}$ while in the tri-nucleon
$\bar{g}_\pi^{(0)}$ and $\bar{C}_{1,2}$ contribute as well.  In this case $d_h
+d_t \simeq 3d_d -2 (d_n+d_p)$.  By confronting these relations, measurements
of light nuclear EDMs, particularly if they include the triton, could shed
light on the mechanism of CPV \cite{deVries:2011an}.

\subsection{Heavy nuclei, Shielding, and Schiff moments}

Nuclear physics is important in determining the EDMs of neutral atoms.  And the
primary fact from which all other considerations stem is expressed by the
Schiff theorem\cite{Schiff:1963zz}, which states that in the limit that in the limit of a
point-like nucleus and non-relativistic electrons any nuclear EDM is completely
screened by the atomic electrons, so that the net atomic EDM is zero.

We give a brief illustration of this result, following Ref. xxx.  Consider a
system of structureless components (a nucleus and electrons), the $k^\text{th}$
of which has dipole moment $\svec{d}_k$, interacting via the Coulomb force
$V(r)$, so that
\begin{align}
H  &= \sum_k \frac{p_k^2}{2 m_k} + \sum_k V(\svec{r}_k) 
 - \sum_k \svec{d}_k \cdot \svec{E}_k  \nonumber \\
&= H_0  + i \sum_k (1/e_k) \left[ \svec{d}_k \cdot \svec{p}_k, H_0 \right] \,. \nonumber
\end{align}
The perturbing Hamiltonian (the last term above) shifts the unperturbed ground
state $\ket{0}$ to
\begin{align}
\ket{\tilde{0}} & =  \ket{0} + \sum_m \frac{\ket{m}\bra{m} H_d \ket{0} } {E_0
- E_m} 
\quad = \quad \ket{0} + \sum_m \frac{\ket{m}\bra{m}  i \sum_k (1/e_k)
 \svec{d}_k \cdot \svec{p}_k \ket{0} (E_0-E_m)} {E_0 - E_m} \nonumber\\
& = \left( 1+i \sum_k (1/e_k) \svec{d}_k \cdot \svec{p}_k \right) \ket{0}
\end{align}
The induced dipole moment $\svec{d}'$ is then
\begin{align}
\svec{d}' &= \bra{\tilde{0}} \sum_j e_j \svec{r}_j \ket{\tilde{0}} 
 \nonumber \\
&= i \bra{0} \left[ \sum_j e_j \svec{r}_j , \sum_k (1/e_k) \svec{d}_k
\cdot \svec{p}_k \right] \ket{0} 
=  - \sum_k \svec{d}_k \nonumber \\
&= - \svec{d} \,,
\end{align}
so that the net dipole moment of the entire system vanishes. The assumptions
underlying this result are that the constituents are point-like,
non-relativistic, and non-interacting except via the Coulomb force.  In real
systems, none of these assumptions hold fully.  As we shall see immediately
below, the finite nuclear size essentially leads to the replacement the nuclear
dipole operator by the nuclear ``Schiff operator,'' which contains two extra
powers of the nucleon coordinate.  Moments due to finite nuclear size are thus
generically smaller by
$\mathcal{O}\left(R_\text{nucl.}^2/R_\text{atom}^2\right)$ than the unscreened
nuclear EDM.  In diamagnetic atoms, the nuclear physics of which is discussed
next, this suppression is mitigated by relativistic electrons and can be
further mitigated by nuclear octupole deformation.  In paramagnetic atoms,
discussed in the next section, relativistic electrons can lead to a large
enhancement of the atomic EDM.

Further analysis
leads to the result that the post-screening CP-violating
nucleus-electron interaction is
\begin{align}
\label{eq:e-nuc-int}
H&= 4\pi \svec{S}\cdot \svec{\nabla} \delta^3(\svec{r}) \, + \ldots \,,
\end{align}
where the omitted terms come from higher multipoles, e.g.\ the nuclear magnetic quadrupole (M2) and
electric octupole (E3) multipoles. The operator $S$ is the nuclear Schiff operator, defined as
\begin{align}
\label{eq:Ssum}
\svec{S} &= \svec{S}^\text{ch}+\svec{S}^N
\shortintertext{with}
\label{eq:opch}
\svec{S}^\text{ch}&=\frac{e}{10}\,\sum_{p=1}^Z\left(r^2_p-\frac{5}{3}\,
\braket{r^2}_\text{ch}\right) \svec{r}_p \\
\label{eq:opnucl}
\svec{S}^N & = \frac{1}{6} \sum_{j=1}^A \svec{r}_j \, (r_j^2 -
\braket{r^2}_\text{ch}) \nonumber \\
&+\frac{1}{5} \sum_{j=1}^A \left(\svec{r}_j(\svec{r}_j
\cdot \svec{d}_j) - \frac{r^2_j}{3} \svec{d}_j \right) + \ldots \,.
\end{align}
Here $\svec{S}^\text{ch}$ is due to the charge distribution of the nucleus
(usually the dominant piece), $\svec{S}^N$ is due to the EDM of
the nucleon, $e$ is the charge of the proton, $\braket{r^2}_{\rm ch}$ is the
mean squared radius of the nuclear charge distribution, and $\svec{d}_j$ is the
EDM of nucleon $j$. The sum in Eq.\ (\ref{eq:opnucl}) is over all nucleons,
while that in Eq.  (\ref{eq:opch}) is restricted to protons.  Rotational
symmetry lets us express the ground-state matrix elements of the three vector
Schiff operators in terms of a single quantity:
\begin{equation}
\label{eq:Sdef}
S \equiv \bra{\Psi_0}  {S}_z \ket{\Psi_0}  \,,
\end{equation}
where $\ket{\Psi_0}$ is the member of the ground-state multiplet with $J_z=J$.
 

The charge-distribution part of the Schiff moment, $S^\text{ch}$, can only be
induced by an effective $T-$ and $P$-violating inter-nucleon interaction.  Most
studies have been dedicated to the OPE part of the TVPV potential
\eqref{eq:PToddV}.  The moment $S^N$ can have many sources, as we have seen,
and can depend on other quantities besides the $\bar{g}_\pi^{(i)}$. 

Equation (\ref{eq:opch}) is, as mentioned, only approximate.  Corrections come
from nuclear quadrupole deformation (which introduces a term proportional to
the nuclear quadrupole moment), from relativity in electronic wave functions
(which gives terms of order $(Z\alpha)^2$) \cite{flambaum02,flambaum12}, and
more subtle electron-nucleus interactions \cite{Liu:2007zf}, the complete forms of
which are still not entirely settled.  Equation (\ref{eq:PToddV}) is also only
approximate, representing the leading-order part of the chiral effective
potential.  Contact terms and higher-order pieces in effective field theory
(which in heavy systems would be hard to control) or heavier-meson exchange in
older frameworks will modify $V_{\slashPTsub}$.  At present, however,
nuclear-structure theorists have not incorporated any of these corrections save
(occasionally) those of order $(Z\alpha)^2$ into their calculations of Schiff
moments.

Beyond-the-standard-model and hadronic physics, as we have seen, determine the
$\gpbi$ and the nucleon EDMs.  The job of nuclear-structure
theory, within the framework just defined, is to determine the dependence of
the Schiff moment on these quantities.  (Atomic physics in turn determines the
dependence of the atomic EDM on the Schiff moment.)  Here we examine only the
dependence on the $\gpbi$ and $d_{(i)}$; the dependence on the nucleon EDMs can be computed
as well, but is weaker.  Only a few of the calculations cited below (e.g.,
Ref.\ \cite{ban10}) considers this weak dependence.  We can parameterize
the dependence on the $\gpbi$ as follows:
\begin{equation}
\label{eq:coefs}
S = \frac{2m_Ng_A}{F_\pi}
\left( a_0  \, \bar{g}_\pi^{(0)} + a_1 \, g \bar{g}_\pi^{(1)} 
+ a_2 \, \bar{g}_\pi^{(2)} \right) \ .  
\end{equation}
All nuclear structure information is thus encoded in the coefficients $a_i$,
which have units $e$\,fm$^3$.


In what follows we discuss attempts to calculate the $a_i$ in several important
nuclei.  Most take advantage of the weakness of $V_{\slashPTsub}$ compared to nuclear
energies and approximate $S$ in Eq.\ (\ref{eq:Sdef}, essentially perfectly, by
\begin{equation}
\label{eq:Spert}
S  = \sum_{i \neq 0}
\frac{\bra{\Phi_0}  {S}_z \ket{\Phi_i} 
\bra{\Phi_i}  {V}_{\slashPTsub} \ket{\Phi_0} }
              {E_0 - E_i}
         + \text{c.c.} \,,
\end{equation}
where $\ket{\Phi_0}$ is the ``unperturbed'' ground state --- that obtained with
$V_{\slashPTsub}$ turned off --- and the $\ket{\Phi_i}$ are the corresponding excited
nuclear states.

\subsubsection{$^{199}$Hg}

The atom associated with this nucleus has for years had the best limit on its
EDM, and so $^{199}$Hg has received more attention by nuclear-structure
theorists than any other nucleus (though still not nearly enough, as we argue
below).  Calculations range from the extremely schematic to the very
sophisticated.  The table below quotes the results of four, with brief (and
inadequate) phrases signifying the techniques they employ. (A more extensive
table, reporting several of the different estimates in, e.g., Ref.\
\cite{ban10} as well as earlier versions of the $^{225}$Ra numbers presented in
a later table can be found in Ref.\ \cite{ellis11}.)  The first nontrivial
calculation was that of Ref.\ \cite{flambaum86}; it approximated the
unperturbed states in Eq.\ (\ref{eq:Spert}) by the eigenstates of a simple
one-body potential and then treated $V_{\slashPTsub}$ approximately as a
zero-range interaction between the valence nucleon and the $^{198}$ Hg core in
first-order perturbation theory theory.  Ref.\ \cite{dmitriev03} improved this
treatment considerably by using the full finite-range $V_{\slashPTsub}$ and
adding to the perturbative treatment the collective excitation of the core in
the random phase approximation (RPA) by a simplified version of the residual
strong nucleon-nucleon interaction.  The resulting ``core polarization''
decreased the sensitivity of $S$ to $V_{\slashPTsub}$, as the table shows.  Ref.\
\cite{jesus05} also used RPA to treat core polarization but in a diagrammatic
version of self-consistent Skyrme mean-field theory (also known nowadays as
energy-density-functional theory).  The calculation, which contained of a
self-consistent mean-field calculation in $^{198}$Hg before the treatment of
core polarization, employed several state-of-the-art Skyrme energy-density
functionals, giving rise to the range of numbers in the table.  Finally, Ref.\
\cite{ban10} carried out the self-consistent mean-field theory (again with a
number of Skyrme functionals) directly in the odd nucleus $^{199}$Hg,
implicitly including the effects of RPA core polarization by the valence
nucleon.  It also allowed for axially-symmetric nuclear deformation and
included $\mathcal{O}\left( (Z\alpha)^2 \right)$ corrections to the Schiff
moment.  Its result for the coefficient $a_1$ is noticeably different from
those of the other calculations, a fact that is hard to understand because the
methods appear to include much of the same physics, albeit in quite different
ways.

\begin{table}[h]
\centering
\begin{tabular}{||l|l|ccc||}
\hline\hline
Ref.        & Method    & $a_0$ & $a_1$ & $a_2$  \\
\hline
\cite{flambaum86} & Schematic        &  0.087          &   0.087           
& 0.174            \\
\cite{dmitriev03}, \cite{dmitriev05} & Phenomenological RPA     
& 0.00004         &  0.055           & 0.009            \\
\cite{jesus05}    & Skyrme QRPA      & 0.002 -- 0.010 &  0.057 --  0.090  
& 0.011 -- 0.025      \\
\cite{ban10}      & Odd-A Skyrme mean-field theory & 0.009 -- 0.041 
& -0.027 -- +0.005 & 0.009 -- 0.024  \\
\hline\hline
\end{tabular}
\caption{The coefficients $a_i$ 
in $^{199}$Hg from a variety of
nuclear-structure calculations.}
\label{tab:hg}
\end{table}

Which of the calculations is most reliable and what is the uncertainty in our
knowledge of the coefficients $a_i$?  Even if all the calculations included the
same kinds of corrections to the na\"ive Schiff operator in Eq.\ (\ref{eq:opch}),
these questions would be hard to answer.  The calculations agree, more or less,
on the size of $a_0$ and $a_2$, but do not even agree on the sign of $a_1$.
Some possible reasons for the discrepancy between Refs.\ \cite{jesus05} and
\cite{ban10}, which, as mentioned, seem to include essentially the same
many-body effects:
\begin{enumerate}
\item[(a)] One of the calculations is in error.  Ref.\ \cite{ban10} carried out
several internal consistency tests, but did not agree with Ph.D.\ thesis
leading to Ref.\ \cite{jesus05} when repeating one of the calculations there.
The authors suggest as a result that \cite{jesus05} may contain an error.  On
the other hand, the results of \cite{jesus05} agree fairly well with those of
the similar RPA calculations in Ref.\ \cite{dmitriev03}, suggesting that it is
Ref.\ \cite{ban10}, if any, that has problems.
\item[(b)] Some of the mean-field solutions in Ref.\ \cite{ban10} are
metastable, though they are supposed to represent stable ground states.  But
even those that are completely stable use the same Skyrme functional as
\cite{jesus05}, and --- like the solutions in that reference --- correspond to
spherical shapes, disagree with \cite{jesus05}.
\item[(c)] The treatment of core polarization in the two kinds of calculations
are equivalent only up to terms linear in the strong interaction between the
valence nucleon and the core, and only if that interaction is not density
dependent. (It is in fact density dependent in Skyrme functionals.)  But it is
hard to imagine higher-order effects or the density dependence being very
important.  There are a few diagrams in Ref.\ \cite{jesus05} that have no
counterpart in the odd-nucleus mean-field calculation of \cite{ban10}, but
their contributions are apparently small.
\item[(d)] The state of the valence nucleon is represented only approximately
in \cite{jesus05}.  Again, though, the approximation should be reasonably good. 
\end{enumerate}
In short, it is difficult to see how the calculations could disagree so
seriously.  The authors need to revisit their work.

It may, however, that all the calculations just reviewed are missing something
important, and that the spread in results reflects their inadequacy.
$^{199}$Hg is a soft nucleus in which a single mean field, the dominance of
which underlies all of the results obtained thus far, is probably
insufficient.  It is, perhaps, unfortunate that $^{199}$Hg is such a difficult
nucleus, but a better job is not beyond the means of nuclear structure
theorists; techniques to mix mean fields with different properties exist.  We
contrast the state of affairs here here with that in another important and
complex nuclear-structure problem: calculating the matrix elements that govern
neutrinoless double beta decay in complicated nuclei such as $^{76}$Ge.
Theorists believe that they know those matrix elements to within a factor of
two or three, mainly by dint of the number of varied and careful calculations
that have been carried out.  The main theoretical problem with the Schiff
moment in $^{199}$Hg is not the challenging nature of the calculation, but
rather that only a few groups have tried.

$^{129}$Xe presents many of the same problems as $^{199}$Hg.  Fortunately other
nuclei, including the one we discuss next, are better behaved.

\subsubsection{$^{225}$Ra}

This nucleus became the focus of experimental interest after it was shown
\cite{spevak95,spevak97, engel03} that the Schiff moments in nuclei with
asymmetric shapes could be enhanced by two to three orders of magnitude.
$^{225}$Ra is octupole deformed, has favorable atomic physics, and has nuclear
spin 1/2, making the nuclear orientation insensitive to stray quadrupole
fields.  It is thus presents a terrific opportunity to experimentalists.

Shape asymmetry implies parity doubling (see e.g.\ Ref.\ \cite{sheline89}),
i.e.\ the existence in $^{225}$Ra of a very low-energy $|1/2^-\rangle$ state
(55\,keV \cite{helmer87} above the $|\Phi_0\rangle \equiv |1/2^+\rangle$ ground
state, according to measurements).  That low-lying excited state dominates the
sum in Eq.\ (\ref{eq:Spert}) because of the small energy denominator it
introduces.  In the (good) approximation that the shape deformation is rigid,
the ground state and its negative-parity partner in are projections onto good
parity and angular momentum of the same parity-mixed and deformed ``intrinsic
state," which represents the wave function of the nucleus in its own body-fixed
frame.  Equation (\ref{eq:Spert}) reduces in these circumstances to
\cite{spevak97}
\begin{equation}
\label{eq:intr} 
S \approx - \frac{2}{3}\braket{ \hat{S}_z } 
\frac{\braket{\hat{V}_{\slashPTsub}}} { (55\,\mathrm{keV})} \,, 
\end{equation}
where the brackets indicate expectation values in the intrinsic state.  

The results of a couple of Schiff-moment calculations appear in Tab.\
\ref{tab:ra}.  Ref \cite{spevak97}, much like Ref.\ \cite{flambaum86} in
$^{199}$Hg, obtained the intrinsic state by filling single-particle levels in a
phenomenological octupole-deformed potential and using a zero-range
approximation to $V_{\slashPTsub}$, but using Eq.\ (\ref{eq:intr}) instead of
summing over many unperturbed states as in Eq.\ (\ref{eq:Spert}).  Ref.\
\cite{dobaczewski05}, like Ref.\ \cite{ban10} in $^{199}$Hg, treated the
(octupole-deformed) potential in completely self-consistent Skyrme mean-field
theory with several Skyrme functionals, leading to a range of values for the
$a_i$.  It also included (and perhaps exaggerated) the damping effects of
short-range nucleon-nucleon repulsion.  Even so, the octupole deformation makes
the resulting coefficients much larger than in Hg.

As we have already mentioned, these calculations in Ra are almost certainly more
reliable than those in Hg.  The low $1/2^-$ energy implies that the octupole
deformation is strong and rigid, so that a single mean-field shape accurately
represents the intrinsic density.  There is thus little need to go far beyond
mean-field theory here.  Furthermore, experiments promise to increase the
calculations' reliability.  Theoretical work in progress \cite{dobaczewski12}
shows that intrinsic Schiff moments are strongly correlated with E3
transitions, which have been measured in $^{224}$Ra \cite{gaffney11} at ISOLDE
and may be measured in $^{225}$Ra itself.  The resulting data will tightly
constrain the factor $\braket{\hat{S}_z}$ in Eq.\ (\ref{eq:intr}), leaving
$\braket{\hat{V}_{\slashPTsub}}$ as the only real unknown.  Although a reliable
calculation of that quantity is not trivial, it is far easier than calculating
the transition matrix elements of both $\hat{V}_{\slashPTsub}$ and $\hat{S}_z$
to all the excited states of $^{199}$Hg.

\begin{table}[h]
\centering
\begin{tabular}{||l|l|ccc||}
\hline\hline
Ref.\       & Method    & $a_0$ & $a_1$ & $a_2$  \\
\hline
\cite{spevak97}    &  Octupole-deformed Wood-Saxon & -18.6   &  18.6   & -37.2 \\
\cite{dobaczewski05} &  Odd-A Skyrme mean-field theory   & -1.0 -- -4.7 &
 6.0 --  21.5 &
-3.9 -- -11.0        \\
\hline\hline
\end{tabular}
\caption{The coefficients $a_i$ in $^{225}$Ra from two 
nuclear-structure calculations.}
\label{tab:ra}
\end{table}

\subsubsection{Other Nuclei}

Theorists have calculated the Schiff moments of other nuclei as well, though
not with as much care as they have in the nuclei already discussed.  $^{129}$Xe
has and will be the subject of experiments, and so has received some attention;
like $^{199}$Hg, however, it is unfortunately soft.  Researchers have also
examined actinides other than $^{225}$Ra, including some which have no static
octupole deformation; the idea there is that dynamic deformation, i.e.\ octupole
vibrations, may enhance the Schiff moments \cite{auerbach06}.  They have also
considered the spherical nucleus $^{211}$Rn, which is to be examined experimentally
as the first step in a project to work with heavier octupole-deformed Rn
isotopes.  Table \ref{tab:others} lists some of these results.  We have
omitted nuclei that show little prospect of being studied experimentally. 

\begin{table}[h] 
\centering
\begin{tabular}{||l|l|l|ccc||}
\hline\hline
Ref.   & Nucl.     & Method    & $a_0$ & $a_1$ & $a_2$  \\
\hline
\cite{flambaum86} & $^{129}$Xe & Schematic & -0.11 & -0.11 & -0.22 \\
\cite{dmitriev05} & $^{129}$Xe & Phenomenological RPA & -0.008 & -0.006 & -0.009\\
\cite{dmitriev05} & $^{211}$Rn & Phenomenological RPA & 0.019 & -0.061 & 0.053 \\
\cite{ban10} & $^{211}$Rn & Odd-A Skyrme mean-field th. & 0.034 -- 0.042 &
-0.0004 -- -0.018 & 0.064 -- .071 \\
\cite{dmitriev05} & $^{213}$Rn & Phenomenological RPA &  0.012 &  0.021 & 0.016 \\
\cite{spevak97} & $^{223}$Ra & Octupole-def.\ Wood-Saxon & -25 &  25 & - 50 \\
\cite{spevak97} & $^{223}$Rn & Octupole-def.\ Wood-Saxon & -62 &  62 & - 100 \\
\cite{spevak97} & $^{223}$Fr & Octupole-def.\ Wood-Saxon & -31 &  31 & - 62 \\
\cite{auerbach06} & $^{219}$Fr & octupole-quadrupole vibr. & -0.02 & -0.02 &
-0.04  \\ 
\hline\hline
\end{tabular}
\caption{The coefficients $a_i$ in some other nuclei of interest, from several
nuclear-structure calculations.}
\label{tab:others}
\end{table}

\subsubsection{Ranges and Best Values for Schiff Moment}

Table \ref{tab:schiff-ranges} lists best values and ranges for three important
nuclei.  We determined these in a somewhat subjective manner, assessing the
strengths and weaknesses of each calculation.  The entries should be considered
tentative, and we cannot assign a quantitative meaning to our ranges; we simply
consider it likely that the true values lie in them.  For the case of $a_1$ in
$^{198}$Hg, as already discussed, our range includes zero.

\begin{table}[h] 
\centering
\begin{tabular}{||l|ccc|ccc||}
\hline\hline
Nucl. & \multicolumn{3}{|c|}{Best value} & \multicolumn{3}{|c||}{Range} \\
&  $a_0$ & $a_1$ & $a_2$ & $a_0$ & $a_1$ & $a_2$   \\
\hline
$^{199}$Hg & 0.01 & $\pm$ 0.02 & 0.02 & 0.005 -- 0.05 & -0.03 -- +0.09 & 0.01 -- 0.06 \\
$^{129}$Xe & -0.008 & -0.006 & -0.009 & -0.005 -- -0.05 & -0.003 -- -0.05  &
-0.005 -- -0.1 \\
$^{225}$Ra & -1.5 &   6.0 & -4.0 & -1 -- -6 &  4 --- 24 & -3 -- -15 \\
\hline\hline
\end{tabular}
\caption{Best values and ranges for the coefficients $a_i$ in three nuclei used
in experiments.}
\label{tab:schiff-ranges}
\end{table}

\section{CP and T at the Atomic and Molecular Scale}
\label{sec:atom}

As with the physics at the hadronic and nuclear scales, it is convenient to
express the atomic and molecular EDMs in terms of the operators that
characterize physics at shorter distance scales. To that end, we first write
down an expression for a general atomic or molecular EDM $d_A$ in terms of the electron and
nucleon EDMs, the nuclear Schiff moment, and the Wilson coefficients for the
dimension six T- and P-odd electron-quark interactions. We subsequently express
the electron EDM in terms of either the Wilson coefficient $C_{e \gamma}$
or the quantity $\delta_e$. Doing so allows us to express $d_A$ in such a way
as to place all of the fundamental dimension six operators on the same footing,
and in the case of the electron EDM, take into account the additional Yukawa
suppression that accompanies this operator. Thus, we have \bea
\label{eq:datom}
d_A & = & \rho_A^e\, d_e + \sum_{N=p,n} \rho_Z^N d_N +\kappa_S\,  S
+ \left(\frac{v}{\Lambda}\right)^2 \Biggl\{ \left[k_S^{(0)} g_S^{(0)} +k_P^{(1)} g_P^{(1)}\right]\ \mathrm{Im} C_{eq}^{(-)}\\
&& + \left[k_S^{(1)} g_S^{(1)} +k_P^{(0)} g_P^{(0)}\right]\ \mathrm{Im}
C_{eq}^{(+)}
+  \left[k_T^{(0)} g_T^{(0)} +k_T^{(1)} g_T^{(1)}\right]\ \mathrm{Im} C_{\ell equ}^{(3)}\Biggr\}\ \ \ ,
\eea where 
we may
alternately express the electron EDM contribution as \be
\label{eq:rhoArel}
\rho_A^e\, d_e = e\, \zeta_A^e \left(\frac{v}{\Lambda}\right)^2\, \delta_e =
\beta_A^{e\gamma}\,\left(\frac{v}{\Lambda}\right)^2\, \mathrm{Im}\, C_{e\gamma}\ \ \ .  \ee Note that
the $d_N$ and Schiff moment $S$ may then be expressed in terms of
${\bar\theta}$ and the dimension six quark and gluon operator coefficients
using Eqs. (\ref{eq:dNhad}-\ref{eq:etadef}) and (\ref{eq:coefs}), allowing
one to explicitly identify a common factor of $(v/\Lambda)^2$ for all dimension
six operators and, thereby, to place them on a similar footing.

To illustrate the relative sensitivities of various atomic and molecular EDMs to the quantities appearing in Eqs. (\ref{eq:datom}-\ref{eq:rhoArel}) we consider one paramagnetic atom ($^{205}$Tl), one diamagnetic atom ($^{199}$Hg), and one molecule (YbF) for which the most stringent experimental limits have, thus far, been obtained. A summary for other cases is given in Table. As a prelude, we first summarize a few features of the atomic and molecular computations, referring the reader to the extensive reviews in Refs.~\cite{Dzuba:2012bh,Ginges:2003qt} and the recent study in Ref.~\cite{Jung:2013mg} for details. 

The sensitivity of an atom of molecule to the electron EDM is governed by corrections to the Schiff screening as describe above . In contrast to the corrections due to finite nuclear size or higher T- and P-odd nuclear moments, the relevant corrections are relativistic and entail both a first and a second order energy shift proportional to $d_e$: $\Delta E_{(j)}^{(\tilde e)}$ for $j=1,2$. For our purposes, the explicit expressions are not particularly enlightening, and we again refer to Refs.~\cite{Ginges:2003qt,Liu:2007zf} for details. The correction $\Delta E_{(1)}^{(\tilde e)}$ can be expressed alternately in terms of an electronic operator proportional to $\gamma_5$ or $\gamma_0-1$, indicating the dependence on lower components of the electronic wavefunction that vanish in the non-relativistic limit. The second order term $\Delta E_{(2)}^{(\tilde e)}$ entails polarization of the atomic cloud by the presence of $d_e$ (again vanishing in the non-relativistic limit) that is then probed by the external field. For heavy paramagnetic atoms, the polarization correction dominates, growing as $Z^3$. 

The four-fermion, semileptonic interactions in Eq.~(\ref{eq:eNcpv}) lead to an effective atomic Hamiltonian that takes on the following form in the limit of an infinitely heavy nucleus:
\be
{\hat H}_\mathrm{TVPV}^\mathrm{atom} = {\hat H}_\mathrm{S} + {\hat H}_\mathrm{P} +{\hat H}_\mathrm{T}
\ee
where
\bea
{\hat H}_\mathrm{S} & = & \frac{i G_F}{\sqrt{2}}\, \delta({\vec r}) \, \left[(Z+N){C}_S^{(0)}+(Z-N){C}_S^{(1)}\right]\, \gamma_0\gamma_5\\
{\hat H}_\mathrm{T} & = & \frac{2 i G_F}{\sqrt{2}}\, \delta({\vec r}) \sum_N\, \left[{C}_T^{(0)}+{C}_T^{(1)}\tau_3\right]\, \cdot{\vec\sigma}_N\cdot{\vec\gamma} \\
{\hat H}_\mathrm{P} & = & \frac{ i G_F}{4\sqrt{2}m_N}\, \left[{\vec\nabla},\delta({\vec r})\right] \sum_N\, \left[{C}_P^{(0)}+{C}_P^{(1)}\tau_3\right]\, {\vec\sigma}_N \, \gamma_0
\eea
and where the Dirac matrices, $\delta({\vec r})$ and ${\vec\nabla}$ act on the electronic wavefunctions. 

Note that in arriving at the expression for ${\hat H}_S$  we have performed the sum over all nucleons, using the fact that in the non-relativistic limit the operator ${\bar N}N$ just counts the number of nucleons. For ${\hat H}_{T,P}$, in contrast, the nuclear matrix elements of the spin operators is more complicated. The results quoted below for heavy nuclei, which have $N>Z$, were obtained assuming a single unpaired neutron contributes and using a single particle shell model result for the nuclear matrix element of ${\vec\sigma}_n$. Thus, the values for the $k_{T,P}^{(j)}$ quoted below correspond only to the neutron contribution (or the difference $k_{T,P}^{(0)}-k_{T,P}^{(1)}$).  For all paramagnetic atoms, all three interactions ${\hat H}_{S,P,T}$  contribute. For diamagnetic atoms wherein all electrons are paired, ${\hat H}_S$ cannot induce an EDM except in tandem with the hyperfine interaction.

\subsection{Paramagnetic atoms: Thallium}
\label{sec:thallium}

According to the computations of Refs.~\cite{Dzuba:2012bh,Ginges:2003qt} (see also Ref.~\cite{Jung:2013mg} for a recent summary), the EDM of $^{205}$Tl has by far the strongest dependence on the electron EDM and the ${\bar e}i\gamma_5 e {\bar N}N$ interaction of all paramagnetic atoms studied experimentally to date. Compared to the latter, the dependence on tensor and nucleon pseudoscalar operators are suppressed by three and five orders of magnitude, respectively. Numerically, one has\cite{Porsev:2012zx}
\be
\rho_A^e=-573 \pm 20\ \ \ , \qquad \beta_A^{e\gamma} = 0.65 \pm 0.02\ e\ \mathrm{ fm} \ \ \ , \qquad e\zeta_A^e = (1.9\pm 0.07)\times 10^{-6}\ e\ \mathrm{ fm}\ \ \ ,
\ee
while 
\be
k_S^{(0)} = -(7\pm0.3) \times 10^{-5}\ e\ \mathrm{fm}\qquad \mathrm{and}\qquad k_S^{(1)} = 0.2 k_S^{(0)}\ \ \ ,
\ee
and \cite{Ginges:2003qt}
\be
k_P^{(0)} = -k_P^{(1)}  =1.5\times 10^{-10}\ e\ \mathrm{fm}\qquad \mathrm{and}\qquad k_T^{(0)} =  -k_T^{(1)}=0.5\times 10^{-7}\ e\ \mathrm{fm}\ \ \ .
\ee
The numerical dominance of $k_S^{(0)}$ implies that $d_A(^{205}\mathrm{Tl})$ has the greatest sensitivity to $\mathrm{Im}\, C_{eq}^{(-)}$, a somewhat reduced sensitivity to $\mathrm{Im}\, C_{eq}^{(+)}$ and relatively little sensitivity to $\mathrm{Im}\, C_{\ell e qu}^{(3)}$.

It is interesting to compare the relative sensitivity of $d_A(^{205}\mathrm{Tl})$ to $\delta_e$ and $\mathrm{Im}\, C_{eq}^{(-)}$:
\be
\frac{k_S^{(0)}}{e\zeta_A^e} \approx -37\ \ \ .
\ee
To the extent that these two quantities have the same order of magnitude, the four-fermion semileptonic operator would yield a far larger contribution to the thallium EDM than would the electron EDM. 

The corresponding sensitivities for $^{133}$Cs\cite{Das:2008b,Das:2008,Dzuba:2009mw}, $^{85}$Rb \cite{Das:2008}, and $^{210}$Fr\cite{Dzuba99,Dzuba:2011,Das09b,Jung:2013mg}  are also listed in Table \ref{tab:atom1}. We have largely followed Ref.~\cite{Jung:2013mg} in averaging the results for Cesium and in assigning error bars to the results for Francium, for which only the analytic expressions in Ref.~\cite{Dzuba:2011} have been used to obtain $k_S^{(0)}$. 

\begin{table}[t]
\centering
\renewcommand{\arraystretch}{1.5}
\begin{tabular}{||c|c|c|c|c|c|c||}
\hline\hline
Atom  & $\rho_A^e$ & $\beta_A^{e\gamma}$ &$e\zeta_A^e$ & $\rho_p$ & $\rho_n$ & $\kappa_S$ \\
& & $e$ fm &$10^{-8}\, e$ fm   &  $10^{-4}$& $10^{-4}$ & $10^{-4}$ fm$^{-2}$\\
\hline
$^{205}$Tl & $-573(20)$ & 0.65(0.02) & $189 (7) $ &  &  & \\
$^{133}$Cs & $123(4)$ & -0.14(0.005) & $-41 (1.3) $ &  &  & \\
$^{85}$Rb & $25.7(0.8)$ & -0.03(0.0009) & $-8.5 (0.3) $ &  &  & \\
$^{210}$Fr & $903(45)$ & -1.02(0.05) & $ -298(15) $ &  &  & \\
\hline
$^{199}$Hg & 0.01 & $-1.13\times 10^{-5}$ & $-3.3\times 10^{-3}$ & -0.56 & -5.3 & 2.8\\
\hline
\hline
Molecule &  Hz/($e$ fm)   &  Hz & kHz &  & & \\
\hline
YbF & $-(1.1\pm 0.1)\times 10^{12}$ & $(1.2\pm 0.1)\times 10^{9}$ & $(3.6\pm 0.3)$ & & & \\
ThO & $-(4.6\pm 0.4)\times 10^{12}$ & $(5.2\pm 0.5)\times 10^9$ & $(15.2\pm 1.4)$  & & & \\
\hline\hline
\end{tabular}
\caption{Dependence of atomic and molecular EDMs on EDMs of the electron, proton and neutron and on the Schiff moment.
\label{tab:atom1}}
\end{table}

\subsection{Mercury}
\label{sec:mercury}
The diamagnetic nature of $^{199}$Hg makes it far more sensitive to T- and P-odd interactions at the purely hadronic level than those involving electrons as compared to the paramagnetic thallium atom. From Ref.~Refs.~\cite{Dzuba:2012bh,Ginges:2003qt} we first obtain the sensitivity to the nuclear Schiff moment:
\be
\kappa_S = 2.8\times 10^{-4} \ \mathrm{fm}^{-2}
\ee
where the  scale for the Schiff moment is $e$-fm$^3$. While it is possible to include the nucleon EDM contributions in the nuclear Schiff moment, we find it helpful to separate these contributions out explicitly as in Eq.~(\ref{eq:datom}). From Ref.~\cite{Dzuba:2012bh,Ginges:2003qt} we obtain
\be
\rho_A^p = -5.6\times 10^{-5}\qquad \rho_A^n = -5.32\times 10^{-4}\ \ \ ,
\ee
while the sensitivity to the electron EDM is given by 
 \be
 \rho_A^e=0.01\ \ \ , \qquad \beta_A^{e\gamma} = -1.13\times 10^{-5}\ e\ \mathrm{fm} \ \ \ , \qquad e\zeta_A^e = -3.3\times 10^{-11}\ e\ \mathrm{fm}\ \ \ ,
 \ee
essentially five orders of magnitude less sensitive that $^{205}$Tl. The sensitivity to the four-fermion semileptonic interactions ${\bar N}N {\bar e}i\gamma_5 e$ are similarly suppressed with respect to thallium:
\be
\label{eq:ksHg}
k_S^{(0)}[^{199}\mathrm{Hg}] \approx 1.16\times 10^{-4}\, \times\,  k_S^{(0)}[^{205}\mathrm{Tl}] \ \ \ , 
\ee
while the sensitivity to the tensor and nucleon pseudoscalar interactions are somewhat enhanced:
\bea
k_P^{(j)}[^{199}\mathrm{Hg}] & \approx & 4 \times\,  k_P^{(j)}[^{205}\mathrm{Tl}] \ \ \ , \ \ \ j=0,1\\
k_T^{(j)}[^{199}\mathrm{Hg}] & \approx & 4 \times\,  k_T^{(j)}[^{205}\mathrm{Tl}] \ \ \ ,\ \ \  j=0,1 \ \ \ .
\eea
The relatively weak dependence of $d_A(^{199}\mathrm{Hg})$ on the scalar interactions reflects the suppression due to the presence of the atomic hyperfine interaction that must be present to yield a non-vanishing result. Consequently, $d_A(^{199}\mathrm{Hg})$ provides a relatively more effective probe of $\mathrm{Im}\, C_{\ell e qu}^{(3)}$ than does $d_A(^{205}\mathrm{Tl})$. In a scenario where the only T- and P-odd effects arise {\em via} semileptonic interactions, a comparison of results from mercury and thallium could allow one to disentangle between various sources.

Before proceeding with the molecular EDMs, we observe that if a given BSM scenario generated only the dimension six quark EDM operators and not the four-quark, CEDM, or three gluon operators, then the corresponding effect on $d_A(^{199}\mathrm{Hg})$ would be dominated by the induced neutron and proton EDMs. In this case, the present limit on $d_A(^{199}\mathrm{Hg})$ could be interpreted as a limit on $d_n$ at the $10^{-12}$ $e$ fm level, roughly one order of magnitude weaker than the present direct neutron EDM limit. On the other hand, using the latter, one could then infer a bound of roughly $10^{-10}-10^{-11} e$ fm on $d_p$. 

\subsection{Ytterbium Fluoride}
\label{sec:YbF}

In order to assess the sensitivity of the polar molecules to the underlying CPV operators, we first convert to the conventions used in the theoretical literature. Following Ref.~\cite{kozlov98} we write the molecular Hamiltonian as 
\be
\label{eq:amo1}
H_{TVPV}^\mathrm{mol} = \left(W_d d_e + W_S {\bar C}_S\right) {\vec S}\cdot{\hat n}
\ee
where ${\vec S}$ is the electronic spin, ${\hat n}$ is a unit vector along the axis of the YbF molecule, and 
${\bar C}_S= (Z+N) {C}_S^{(0)} +(Z-N) {C}_S^{(1)}$. We do not at present include the dependence on the pseudoscalar or tensor interactions as, to our knowledge, the corresponding evaluations of  molecular sensitivities have not been performed. 

\begin{figure}[tb]
\begin{center}
\begin{minipage}[t]{8 cm}
\epsfig{file=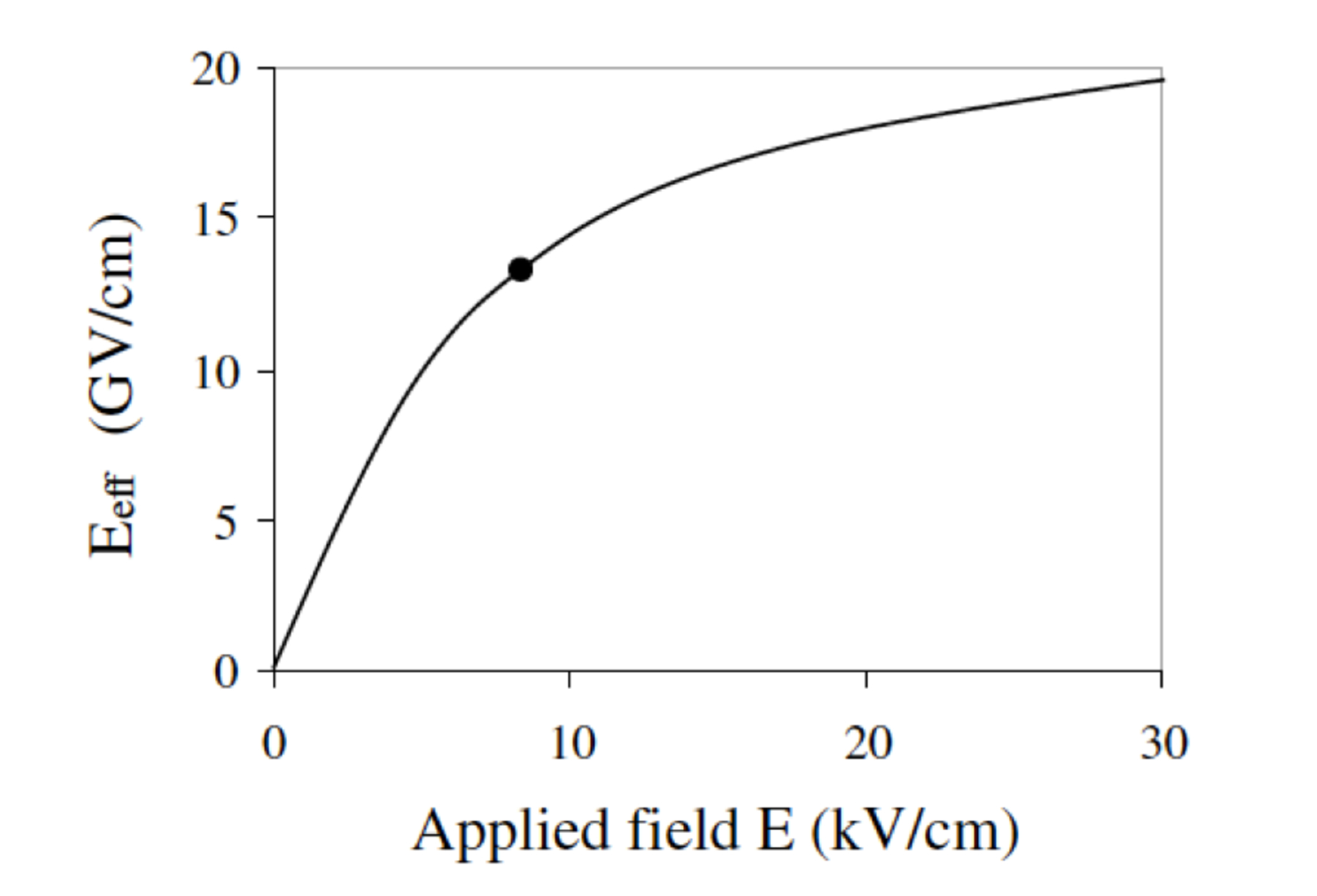,scale=0.5}
\end{minipage}
\begin{minipage}[t]{16.5 cm}
\caption{Dependence of $\eta(E_\mathrm{ext})$ for YbF \cite{Hudson:2002az}. Vertical axis gives $E_\mathrm{eff}$ as a function of the applied field $E_\mathrm{ext}$. Scaling  $E_\mathrm{eff}$ by twice its asymptotic value gives $\eta(E_\mathrm{ext})$ . Figure reprinted with permission from Phys. Rev. Lett. {\bf 89} 023003 (2002).\label{fig:eta}}
\end{minipage}
\end{center}
\end{figure}

Contrary to what one might na\"ively expect, the interaction in Eq.~(\ref{eq:amo1}) contains no dependence on the external electric field, ${\vec E}_\mathrm{ext}$. The experimental observable -- a frequency shift -- depends on the ground state (g.s.) expectation value in the presence of the external field:
\be
\label{eq:amo2} 
\bra{\mathrm{g.s.}} H_{TVPV}^\mathrm{mol} \ket{\mathrm{g.s.}}_{E_\mathrm{ext}} = \left(W_d d_e + W_S {\bar C}_S\right) \, \eta(E_\mathrm{ext})\ \ \ ,
\ee
where 
\be
\label{eq:amo3}
\eta(E_\mathrm{ext}) = \bra{\mathrm{g.s.}} {\vec S}\cdot{\hat n}\ket{\mathrm{g.s.}}_{E_\mathrm{ext}}
\ee
is an effective polarization that increases monotonically with $E_\mathrm{ext}=\vert {\vec E}_\mathrm{ext}\vert$ and has a maximum value of $1/2$. From the first term in Eq.~(\ref{eq:amo1}), then, one may interpret $\eta(E_\mathrm{ext}) W_d$ as the effective internal electric field $E_\mathrm{eff}$ acting on an unpaired electron that is induced by a non-vanishing $E_\mathrm{ext}$.  For YbF, $\eta(E_\mathrm{ext})$ has been reported in Ref.~\cite{Hudson:2002az} and is shown in Fig.  \ref{fig:eta} . The vertical axis gives $E_\mathrm{eff}$ as a function of $E_\mathrm{ext}$. The value of $\eta(E_\mathrm{ext})$ may be obtained by scaling $E_\mathrm{eff}$ by twice its maximum, asymptotic value\footnote{We thank T. Chupp for a helpful discussion of this point.}.

For purposes of this review, it is useful to express $H_{TVPV}^\mathrm{mol}$ in terms of $\delta_e$ and the $\mathrm{Im}\, C_{eq}^{(\pm)}$: 
\be
H_{TVPV}^\mathrm{mol} = \left(\frac{v}{\Lambda}\right)^2\, \left[ e\zeta_A^e \delta_e + g_S^{(0)} k_S^{(0)} \mathrm{Im}\, C_{eq}^{(-)} +  g_S^{(1)} k_S^{(1)} \mathrm{Im}\, C_{eq}^{(+)}\right]\, {\vec S}\cdot{\hat n}\ \ \ ,
\ee
where the quantities $\zeta_A^e$ and $ k_S^{(0,1)}$ are determined by molecular structure. The results for $W_{d}$ and $W_{S}$ are typically quoted in units of Hz$/(e\ \mathrm{cm})$ and cm, respectively . For the sensitivity to the electron EDM, the latest results in the compilation of Ref.~\cite{Jung:2013mg} yields
\be
W_d=-(1.1\pm0.1)\times 10^{12}\ \mathrm{Hz}/(e\ \mathrm{fm})\ ,\qquad \beta_A^{q\gamma}=(1.2\pm 0.1)\times 10^{9}\, \mathrm{Hz}\, ,\qquad e\zeta_A^e = 3.6\pm 0.3\ \mathrm{kHz}
\ee
where we identify $\rho_A^e\equiv W_d$.
For the scalar interactions one has
\be
k_S^{(0)} = 5 k_S^{(1)} = -(92\pm 9)\, \mathrm{kHz}\ \ \ ,
\ee
where we have adopted the ten percent theoretical error suggested in Ref.~\cite{Jung:2013mg}.
To compare with the thallium atom, one has $k_S^{(0)}/e\zeta_A^e\approx -26$, indicating a somewhat stronger relative sensitivity to the electron EDM. However, the sensitivities are sufficiently similar that a combination of the present experimental limits in the two systems does not allow for a significant individual limits on $\delta_e$ and $\mathrm{Im}\,  C_{eq}^{(-)}$. 

Looking to the future, an effort to probe the EDM of ThO is underway.  A value for $W_d$ has been computed in Ref.~\cite{Meyer:2008gc}. The corresponding value for $W_S$ has been inferred from the ratio of $W_d/W_S$ computed analytically in Ref.~\cite{Dzuba:2011}. A conversion to $e\zeta_A^e$ and $k_S^{(0)}$ appears in Tables \ref{tab:atom1} and \ref{tab:atom2}. In both cases, we have arbitrarily assigned a ten percent theoretical uncertainty. We note that the ratio $k_S^{(0)}/e\zeta_A^e$ for ThO lies approximately midway between that of Th and YbF.


\section{Beyond the Standard Model: Examples}
\label{sec:bsm}
The space of BSM scenarios that contain additional sources of CPV is vast, and it is not feasible to provide an exhaustive review here. Instead, we will focus on several representative examples to illustrate the interplay of scales: supersymmetry, models with extended gauge symmetry, and scenarios with extra spacetime dimensions. Before doing so, we first make a few general remarks. Perhaps most importantly, any new source of CPV will generally induce a contribution to the QCD vacuum angle, which we denote as  ${\bar\theta}_\mathrm{BSM}$. In the minimal supersymmetric Standard Model (MSSM) for example, such contributions arise at one-loop order {\em via} corrections to the quark propagators. Given the already severe bounds on ${\bar\theta}$, such contributions to ${\bar\theta}_\mathrm{BSM}$ by themselves imply stringent limits on the CPV phases in the absence of a mechanism to alleviate them. Possibilities include invoking a new symmetry, such as the PQ symmetry or a flavor symmetry that yields a vanishing one-loop result. 

Second, non-observation of atomic, molecular, and neutron EDMs generically
imply that any new CPV phases $\phi_\mathrm{CPV}$ must be quite small if the
BSM mass scale $\Lambda$ is sub-TeV. Conversely, allowing $\sin\phi_\mathrm{CPV}\sim
1$ implies that $\Lambda\gsim$ few TeV. To illustrate, dimensional analysis gives for the
elementary fermion EDM \be
\label{eq:edm0}
d_f\sim e \left(\frac{m_f}{\Lambda^2}\right)\ \frac{\alpha_k}{4\pi}
\sin\phi_\mathrm{CPV}\ \ \ 
\ee 
where $\alpha_k$ is either the fine structure
constant or strong coupling (evaluated at the scale $\Lambda$). For
$\alpha_k=\alpha_\mathrm{em}$ Eq.(\ref{eq:edm0}) gives \be
\label{eq:edm2}
d_f\sim\sin\phi_\mathrm{CPV}\ \left( \frac{m_f}{\mathrm{MeV}}\right)\ \left(
\frac{1\ \mathrm{TeV}}{M}\right)^2 \times 10^{-13}\ e\ \mathrm{fm}\ \ \ .  
\ee The
present limit on the EDM of the electron, $|d_e| < 10.5\times 10^{-15}$
$e\ \mathrm{fm}$~\cite{Hudson:2011zz} obtained from an experiment on the Yb-F molecule,
then implies that \be
\label{eq:edm1}
|\sin\phi_\mathrm{CPV}| \lesssim \left( \Lambda / 2\ \mathrm{TeV}\right)^2\ \
\ .  \ee Thus, for $|\sin\phi_\mathrm{CPV}|\sim 1$ one requires $\Lambda\gsim 2$
TeV.  In order to allow for sub-TeV scale masses and $\mathcal{O}(1)$ CPV
phases while respecting present constraints, one must either invoke
cancellations between different contributions \cite{Ibrahim:1998je} or a mechanism that
suppresses the one-loop EDMs. In the case of the MSSM, for example, taking the
sfermions to have mutli-TeV masses can result in the leading contributions
arising at two-loop order and involving the electroweak gaugino-Higgs/Higgsino
sector with sub-TeV masses\cite{Giudice:2005rz,Li:2008kz}. Given the suppression of an additional loop factor,
the resulting dependence on the CPV phases is weakened and the present
constraints are generally less severe. Alternate strategies can be employed in
other scenarios. 

\subsection{Supersymmetry}
\label{sec:susy}
Supersymmetry (SUSY) remains one of the most strongly motivated BSM scenarios,
providing an elegant solution to the hierarchy problem, candidates for cold
dark matter (the lightest neutralino or gravitino), and copious sources of CPV
that may drive the generation of the baryon asymmetry during the EWSB era. At
the same time, SUSY CPV generically leads to one-loop EDMs that exceed present
experimental bounds, assuming that superpartner masses lie below one TeV [see
Eq.~(\ref{eq:edm1})], leading to the so-called \lq\lq SUSY CP problem". The SUSY
mechanism for solving the hierarchy problem leads one to expect sub-TeV scale
superpartner masses, implying $|\sin\phi_\mathrm{CPV}|\lsim 0.01-0.1$. On the
other hand, one might naturally expect
$\sin\phi_\mathrm{CPV}\sim\mathcal{O}(1)$. Moreover, successful supersymmetric
electroweak baryogenesis typically requires $\mathcal{O}(1)$ phases unless the
relevant portion of the superpartner spectrum is finely-tuned to contain near
degeneracies. 

Several solutions to the SUSY CP problem have been proposed: 
\begin{itemize}
\item[(i)] {\em Heavy sfermions}. It is possible, for example, that the fermion superpartners (sfermions) are considerably heavier than one TeV, leading to a suppression of one-loop EDMs and allowing for $\mathcal{O}(1)$ phases\cite{Giudice:2004tc,Giudice:2005rz,Li:2008kz}.  Null results for superpartner searches at the LHC may be pointing to this \lq\lq split SUSY" scenario, as the generic mass bounds on gluinos and first and second generation squarks are now at the TeV scale\footnote{It is still possible, however, that these strongly interacting superpartners are lighter than one TeV but have a compressed spectrum leading to presently undetectable experimental signatures.}. The electroweak gauge boson and Higgs boson superpartners may still be relatively light, thereby allowing for a viable baryogenesis mechanism (see Ref.~\cite{Morrissey:2012db} and references therein). 
\item[(ii)] {\em CP-conserving SUSY breaking}. It is equally possible that the mechanism of SUSY-breaking that is responsible for both the splitting of SM masses fro those of their superpartners (the \lq\lq soft terms") and the CPV phases suppresses the latter. This possibility has been emphasized in the work of\cite{Kane:2009kv} that considered an M-theory scenario on the G$_2$ manifold in which the only source of CPV at the SUSY-breaking scale is the CKM phase. The resulting effects on low-scale parameters then enters through the RG evolution.
\item[(iii)] {\em Cancellations}. It was proposed some time ago\cite{Ibrahim:1998je} that contributions to EDMs from different CPV phases or those from different dimension-six CPV operators may cancel leading to a suppression that again allows for $\mathcal{O}(1)$ phases and light superpartners.
\end{itemize}

An extensive discussion of (i) and (iii) are given in the reviews of Refs.~\cite{Pospelov:2005pr,RamseyMusolf:2006vr} and the more recent analysis of EDMs in SUSY given in Ref.~\cite{Ellis:1996dg}. Given the comprehensive nature of these articles, we do not attempt to provide an exhaustive review of EDMs in SUSY here. Instead, we summarize several generic features as well as developments that have appeared since publication of these studies. For this purpose, we focus on the minimal supersymmetric Standard Model (MSSM) for which one has the superpotential from which one derives the supersymmetric Lagrangian,
\begin{equation}
W_{\rm MSSM}=\hat{\bar{u}}{\bf y_u} {\hat Q} {\hat H}_u - \hat{\bar{d}}{\bf y_d}\hat{Q} \hat{H}_d -
\hat{\bar{e}}{\bf y_e}{\hat L} {\hat H}_d + \mu \hat{H}_u \cdot \hat{H}_d.
\label{eq:MSSMsuper}
\end{equation}
Here, the hatted quantities $\hat{\bar{f}}$ and $\hat{F}$ are the SU(2)$_L$-singlet and doublet chiral superfields for fermion $F$ while $\hat{H}_{u,d}$ are the two Higgs doublet superfields. The ${\bf y}_f$ are $3\times 3$ Yukawa matrices. For purposes of this discussion, we omit possible R-parity violating terms in the superpotential, which is tantamount to promoting the accidental global  $B-L$ conservation of the SM to a symmetry of the MSSM. Note that superpotential introduces only one new parameter beyond that of the SM, namely, the coefficient of the last term in Eq.~)(\ref{eq:MSSMsuper}). In addition, EWSB allows the two neutral Higgs scalars to have vacuum expectation values, whose ratio defines the angle $\beta$: $\tan\beta = v_u/v_d$.

The soft SUSY-breaking Lagrangian responsible for splitting the SM and superpartner masses is
\begin{eqnarray}
{\cal L}_{\rm
soft}&=&-\frac{1}{2}(M_3 {\tilde{g}}\tilde{g}+M_2 {\tilde{W}}\tilde{W}
+M_1 {\tilde{B}}\tilde{B})+c.c. \nonumber \\
&&-(\tilde{\bar{u}}{\bf a_u}\tilde{Q}H_u-\tilde{\bar{d}}{\bf
a_d}\tilde{Q}H_d -\tilde{\bar{e}}{\bf a_e}\tilde{L}H_d)+c.c. \nonumber
\\ &&-\tilde{Q}^\dagger{\bf m_Q^2}\tilde{Q} -\tilde{L}^\dagger{\bf
m_L^2}\tilde{L} -\tilde{\bar{u}}{\bf
m_{\bar{u}}^2}\tilde{\bar{u}}^\dagger -\tilde{\bar{d}}{\bf
m_{\bar{d}}^2}\tilde{\bar{d}}^\dagger -\tilde{\bar{e}}{\bf
m_{\bar{e}}^2}\tilde{\bar{e}}^\dagger
-m_{H_u}^2H_u^*H_u-m_{H_d}^2H_d^*H_d
\nonumber \\
&&-(bH_uH_d+c.c.)
\label{eq:soft}
\end{eqnarray}
Here, the first line gives the gaugino mass $M_i$, $i=1,2,3$ for the ${\rm
U}(1)_Y$, ${\rm SU}(2)_L$ and ${\rm SU}(3)_C$ gauginos, respectively. 
The second line gives the trilinear ``$A$-term'' that couples Higgs
scalars with left- and right- squarks and sleptons.  The third line
gives the  scalar mass $m_{\tilde{q}_{L,R}}^2$,
$m_{\tilde{l}_{L,R}}^2$,  and $m_{H_{u,d}}^2$ for squarks, sleptons
and Higgs scalars, respectively. As with the Yukawa matrices, the boldfaced quantities indicate matrices in flavor space.  Finally, the last line is the
bilinear $b$-term, which couples up- and down-type Higgs scalar doublets.  It is also important to emphasize that in nearly all SUSY analyses, one takes the ${\bf a}_f$ to be proportional to the corresponding Yukawa couplings, thereby naturally suppressing flavor changing neutral currents:
\be
\label{eq:align}
{\bf a}_f = A_f {\bf y}_f
\ee
where $A_f$ becomes the effective trilinear soft parameter for each fermion species.

The various interactions in ${\cal L}_{\rm soft}$ introduce copious sources of both flavor violation and CPV. Here, we focus on the latter. 
After performing
an appropriate set of field redefinitions, ${\cal L}_{\rm soft}$ --
together with the $\mu$-term in the superpotential -- includes 40
CP-violating phases beyond those of the SM (for a useful discussion, see, {\em e.g.}, 
Ref.~\cite{Dimopoulos:1995ju}). Unlike the CPV phase in the CKM matrix, the effects of these new phases are not suppressed by the
Jarlskog invariant\cite{Jarlskog:1985ht} and light quark Yukawa couplings. Consequently, the
CPV effects need not be suppressed as in the SM, leading to the SUSY CP problem. The new phases can be classified in terms of those that solely enter the gauge-Higgs sector:
\be
 \phi_{i}\equiv {\rm Arg}\, \left(\mu M_i b^\ast \right)\qquad \qquad \phi_{ij}\equiv {\rm Arg}\,  \left(M_i M_j^\ast\right)
\ee
where $i,j$ run over the three gauge groups of the MSSM (leading to a total of three independent phases in this sector which we take to be the $\phi_i$ ), and those involving the sfermions:
\be
\phi_f\equiv {\rm Arg}\, \left(A_f M_i^\ast\right) \qquad \qquad \phi_{ff^\prime}\equiv {\rm Arg}\left(A_f A_{f^\prime}^\ast\right)\ \ \ .
\ee
Given the large number of phases, one often invokes a phase universality assumption:
\bea
\label{eq:universal}
\phi_1=\phi_2=\phi_3&\equiv& \phi_\mu\\
\phi_f=\phi_{f^\prime} & \equiv & \phi_A\ \ \ .
\eea
As we discuss below, this assumption is unlikely to allow for consistency between supersymmetric electroweak baryogenesis and EDM constraints.

\begin{figure}[tb]
\begin{center}
\begin{minipage}[t]{16.5 cm}
\epsfig{file=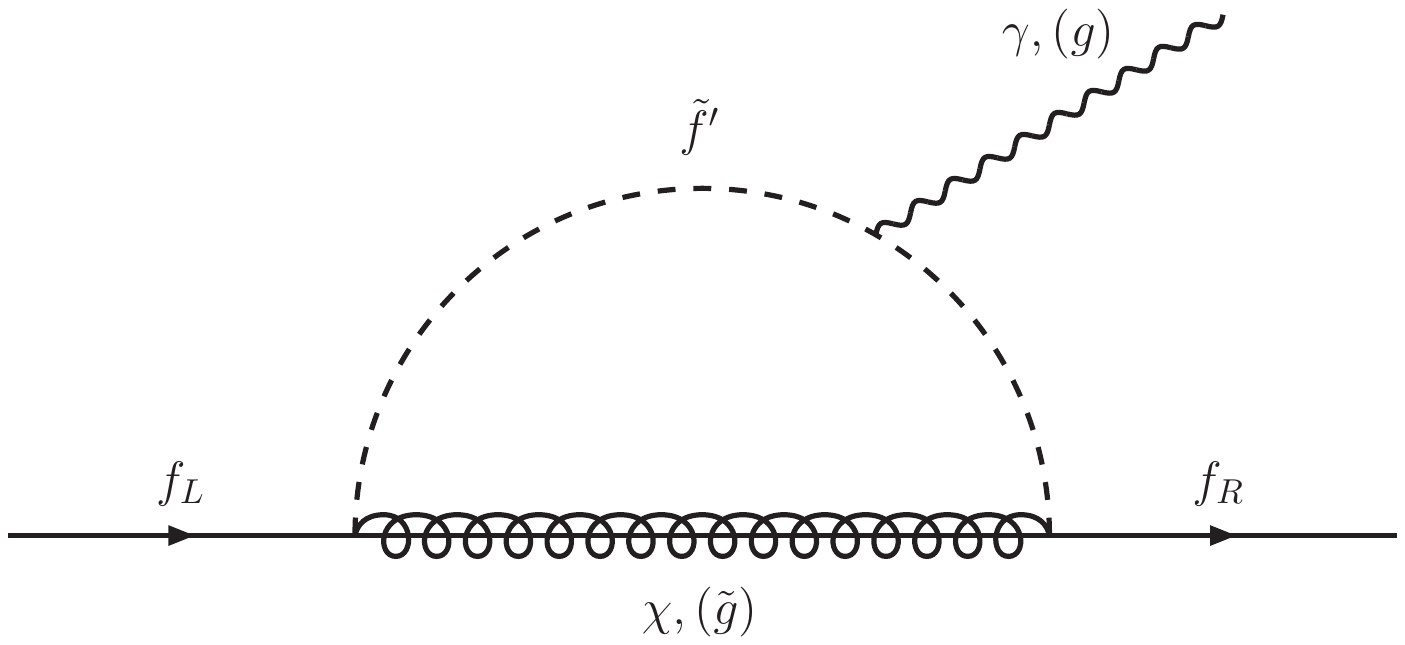,scale=0.4}
\epsfig{file=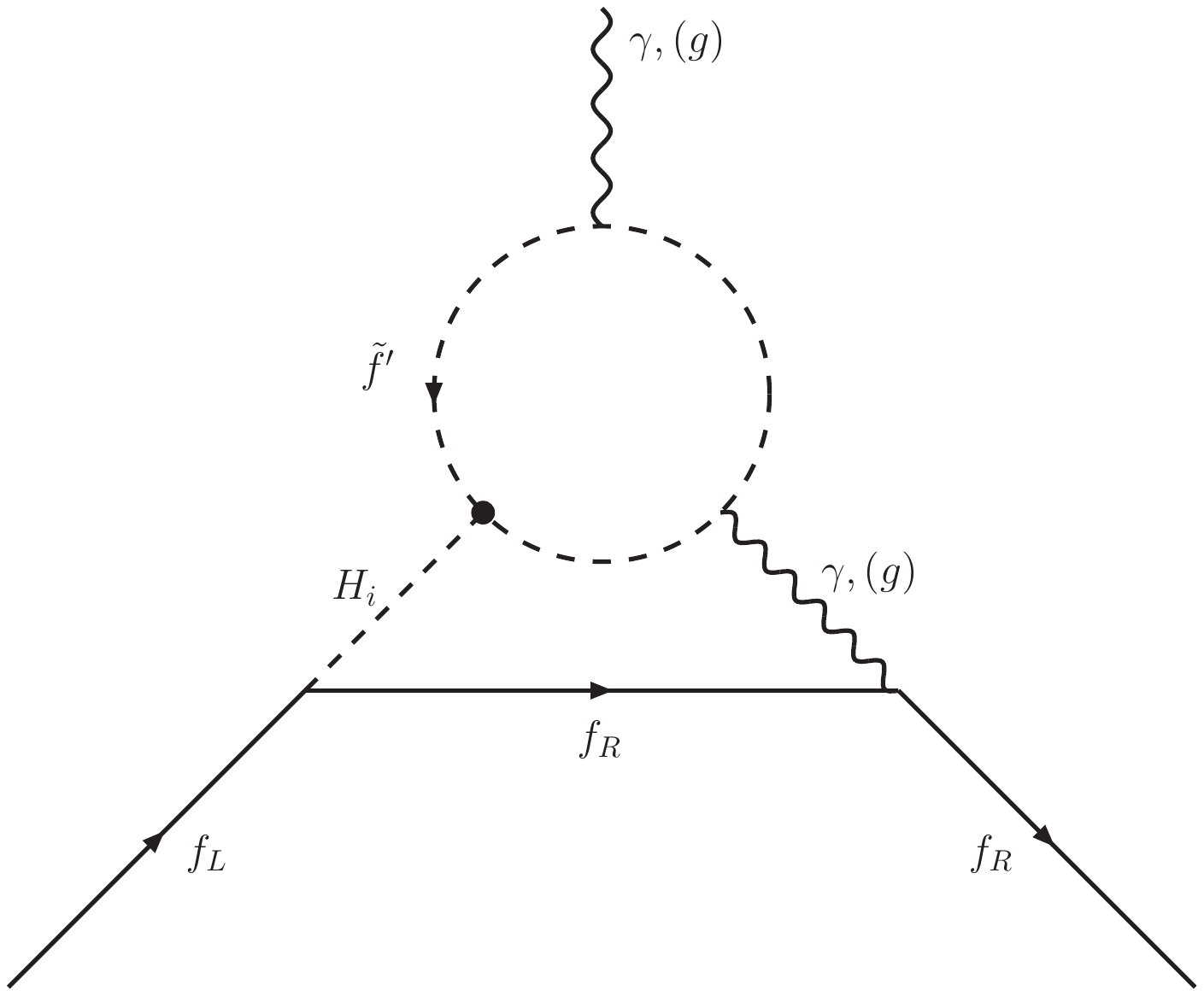,scale=0.4}
\end{minipage}
\begin{minipage}[t]{16.5 cm}
\caption{Illustrative one-loop (left) and two-loop (right) contributions to the fermion EDM and quark CEDM in the MSSM .\label{fig:susy1}}
\end{minipage}
\end{center}
\end{figure}

The CPV interactions in the MSSM give rise to three of the dimension six operators of interest in this article: the EDM and CEDM operators arise at the one-loop level, while the three-gluon operator first enters at two-loop order. The four-fermion operators  are technically dimension eight, but can be enhanced for large $\tan\beta$. Illustrative contributions to the one loop EDM and CEDM arise from the diagrams in Fig.~\ref{fig:susy1}. 

Although not shown explicitly, the insertion of the Higgs fields as needed for electroweak gauge invariance enters in one of two ways:  (a) the mixing of left- and right-handed sfermions and (b) mixing of electroweak gauginos and Higgsinos. The former is characterized by the sfermion mass-squared matrix:
\begin{equation}
{\bf M_f^2} =\left(
\begin{array}{cc}
{\bf M_{LL}^2} & {\bf M_{LR}^2}\\ {\bf M_{LR}^2} & {\bf M_{RR}^2}
\end{array}\right)
\end{equation}
with
\begin{equation}
{\bf{M_{LL}^2}}  =    {\bf m_Q^2}+ {\bf m_q^2 }+{\bf \Delta_f}
\end{equation}
\begin{equation}
{\bf{M_{RR}^2}}  =    {\bf m_{\bar f}^2}+ {\bf m_q^2 }+{\bf
\bar\Delta_f}
\end{equation}
with
\begin{equation}
{\bf \Delta_f} = \left(I^f_3-Q_f\sin^2\theta_W\right)\ \cos 2\beta
M_Z^2
\end{equation}
\begin{equation}
{\bf \bar\Delta_f} = Q_f\sin^2\theta_W \ \cos 2\beta M_Z^2
\end{equation}
and
\begin{equation}
{\bf M_{LR}^2}={\bf M_{RL}^2} =
\begin{cases}
v\left[{\bf a_f} \sin\beta -\mu {\bf y_f} \cos\beta\right]\ , &
{\tilde u}-{\rm type\ sfermion}\\ v\left[{\bf a_f} \cos\beta -\mu {\bf
y_f} \sin\beta\right]\ , & {\tilde d}-{\rm type\ sfermion}
\end{cases}\ \ \ .
\end{equation}
Here ${\bf m_q^2}$ is the mass matrix for the corresponding fermion; 
$I_3^f$ and $Q_f$ are the third component of isospin and fermion charge, respectively; and $v=\sqrt{v_u^2+v_d^2}$. The explicit factor of $v$ in the ${\bf M_{LR}^2}={\bf M_{RL}^2}$ corresponds to the Higgs insertion above the scale of EWSB and leads to mixing between superpartners of the left- and right-handed fermions after EWSB. As a result the incoming and outgoing fermions in Fig.~\ref{fig:susy1} can have opposite handedness as needed for the EDM and CEDM operators. Note that with the assumption of Eq.~(\ref{eq:align}) the left-right mixing is proportional to the fermion Yukawa coupling, implying that the contributions to the EDM and CEDM are as well.

A similar Higgs insertion is implicit in the mixing of the Higgsinos and electroweak gauginos. To illustrate we give the chargino mass matrix for the charged fields 
$\psi^{\pm}=(\tilde{W}^+, \tilde{H}_u^+, \tilde{W}^-, \tilde{H}_d^-)$:
\begin{equation}
{\bf M_{\tilde{C}}}=\left(
\begin{array}{cc}
{\bf 0}&{\bf X^T}\\ {\bf X}&{\bf 0}
\end{array}
\right); \ \ \ {\bf X}=\left(
\begin{array}{cc}
M_2&\sqrt{2}s_{\beta}M_W\\ \sqrt{2}c_{\beta}M_W&\mu
\end{array}\right).
\end{equation}
whose mass eigenstates are the charginos $\chi_i^{\pm}$, $i=1,2$. Note that since $M_W = g v/2$, the off-diagonal terms responsible for gaugino-Higgsino mixing contain an implicit Higgs insertion. Thus, an incoming left-handed fermion in Fig~\ref{fig:susy1} that interacts with the charged SU(2)$_L$ gaugino (the \lq\lq wino") component of the $\chi_i^{\pm}$  can lead to an outgoing right-handed fermion that interacts with the Higgsino component due to this mixing. Since the latter interaction is given by the Yukawa interaction in the superpotential, the corresponding effect on the EDM and CEDM is again proportional to the fermion Yukawa coupling. 

A comprehensive set of expressions for the one-loop contributions to the fermion EDMs and quark CEDMs are given in Ref.~\cite{Ellis:1996dg}, so we do not reproduce them here. Instead, we give an illustrative set of expressions for $\delta_f$ and ${\tilde\delta}_q$ under the universality assumption of Eq.~(\ref{eq:universal}):
\bea
\label{eq:susyEDM1}
\delta_e & = & \frac{Q_e}{32\pi^2} \left[\frac{g_1^2}{12}\sin\phi_A-\left(\frac{5g_2^2}{24}+\frac{g_1^2}{24}\right)\sin\phi_\mu\tan\beta\right] \\
\delta_q & = & -\frac{Q_q}{32\pi^2} \left[\frac{2 g_3^2}{9}\left(\sin\phi_\mu[\tan\beta]^{\pm 1} -\sin\phi_A\right)
+{\cal O}(g_2^2,g_1^2)\right]\\
{\tilde\delta}_q & = & - \frac{1}{32\pi^2} \left[\frac{5 g_3^3}{18}\left(\sin\phi_\mu[\tan\beta]^{\pm 1} -\sin\phi_A\right)
+{\cal O}(g_2^2,g_1^2)\right]
\eea
where $Q_f$ is the fermion charge,
where for simplicity we take SUSY mass parameters to be identical ($\Lambda=|M_j|=|\mu|=|A_f|$), and where the upper (lower) sign corresponds to negatively (positively) charged quarks. 

The expressions in Eqs.~(\ref{eq:susyEDM1}) contain a linear combination of the two universal phases, allowing for the possibility of some cancellation between various contributions. However, as noted in Ref.~\cite{Pospelov:2005pr}, it is unlikely that such a cancellation could allow one to evade all EDM limits since the coefficients of $\sin\phi_\mu$ and $\sin\phi_A$ differ for the various species of fermions as well as between the EDM and CEDMs. Nonetheless, it is in principle possible to obtain a consistent fit to present EDM limits with $\mathcal{O}(1)$ $\sin\phi_\mathrm{CPV}$ if one relaxes the universality assumption, a feature we discuss below.

\begin{figure}[tb]
\begin{minipage}[t]{16.5 cm}
\epsfig{file=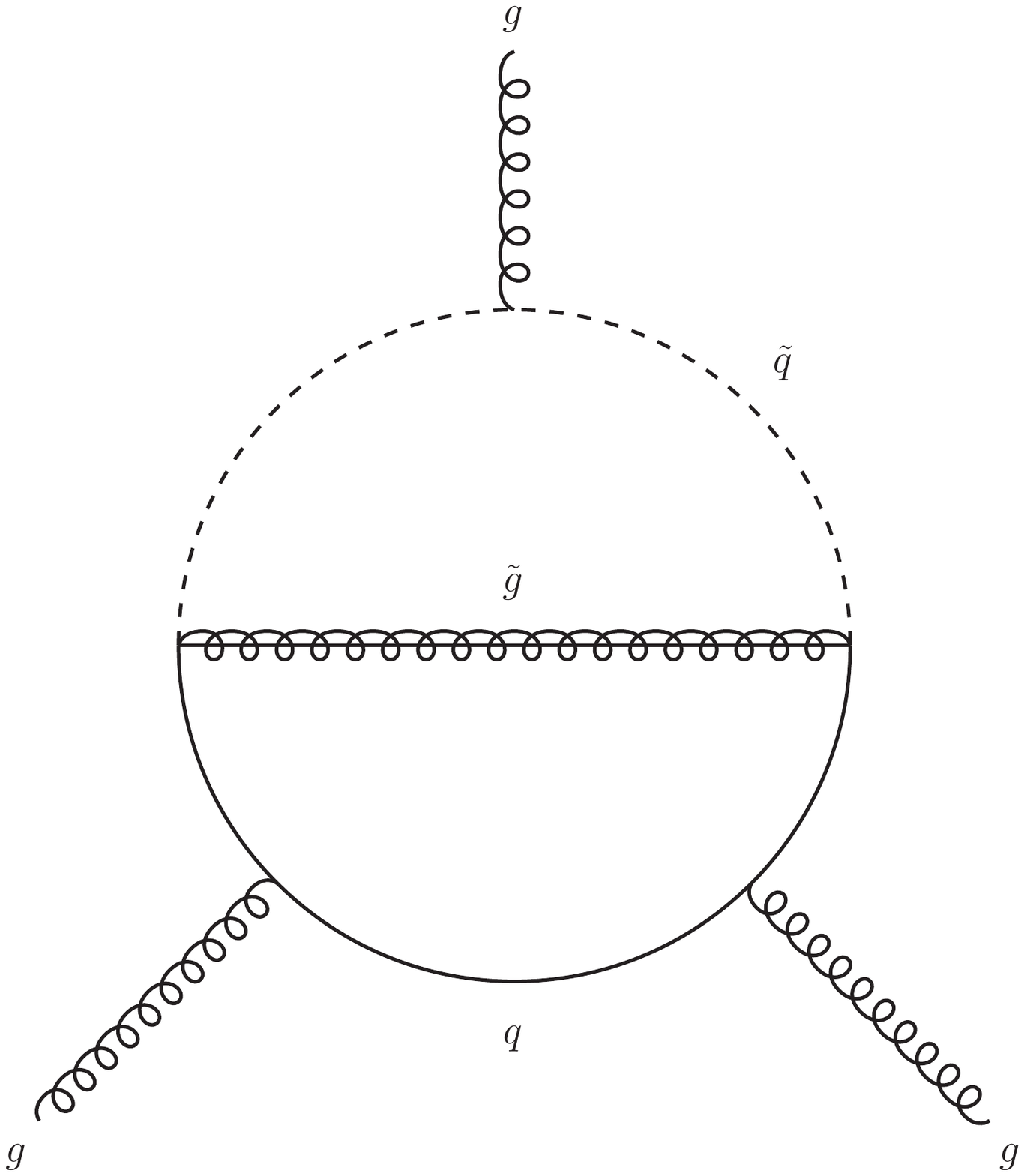,scale=0.4}
\epsfig{file=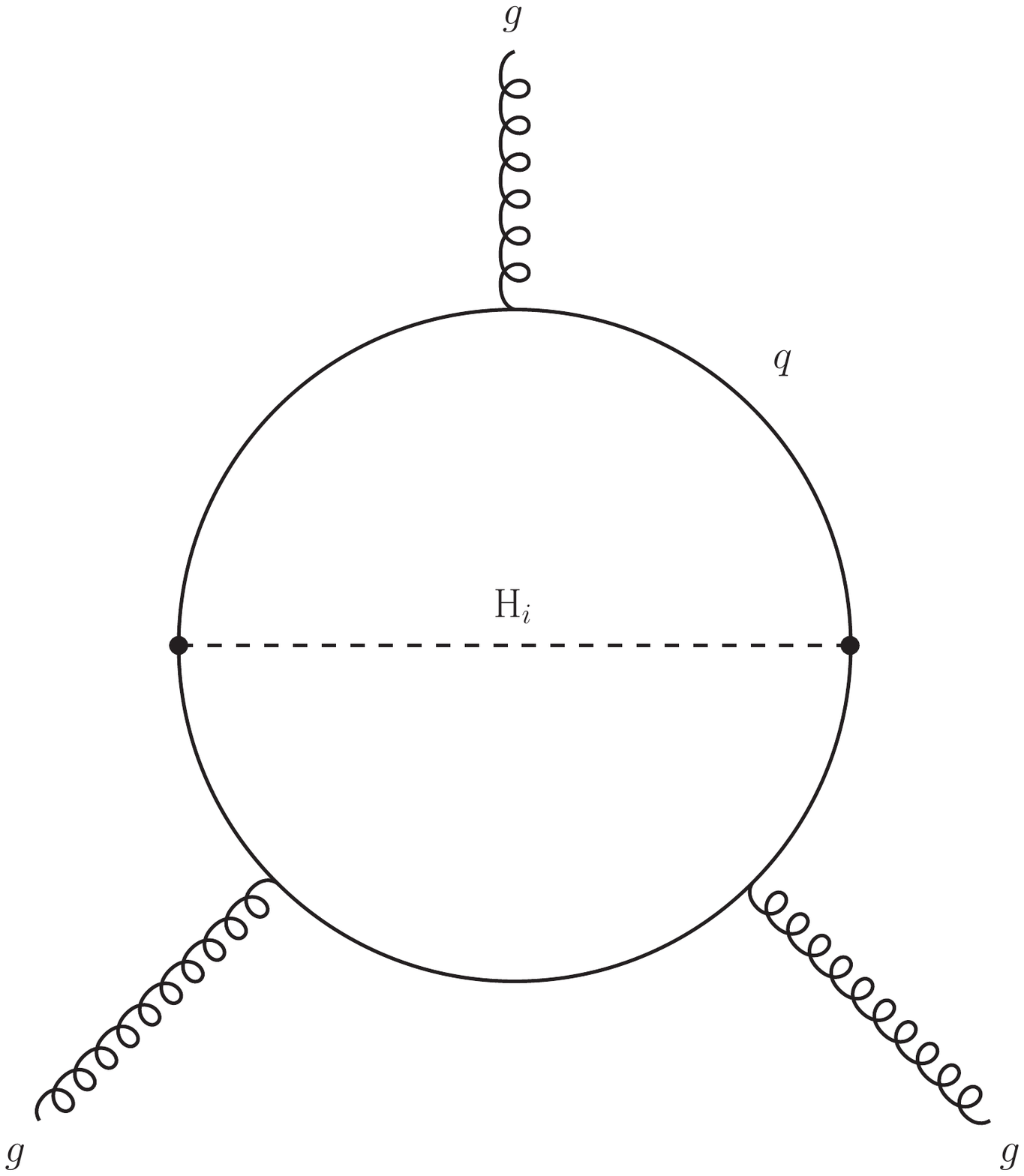,scale=0.4}
\end{minipage}
\begin{center}
\begin{minipage}[t]{16.5 cm}
\caption{Illustrative MSSM contributions to the CPV three-gluon operator $Q_{\tilde G}$ .\label{fig:susy2}}
\end{minipage}
\end{center}
\end{figure}


Going beyond one-loop order, one encounters the first contributions to the three-gluon operator as indicated in Fig. \ref{fig:susy2} as well as the two-loop Barr-Zee graph contributions to the EDM and CEDM operators as indicated by the diagrams in Fig. \ref{fig:susy1}(right). In the limit of heavy sfermions, the Barr-Zee graphs containing electroweak gauginos in the upper loop will give the dominant contribution. Explicit expressions for the contributions generated by the graphs of Fig. \ref{fig:susy2} are given in Ref.~\cite{Ellis:1996dg}, along with those for a subset of the Barr-Zee graphs that contain exchanges of only the lightest CP-even Higgs scalar. In Ref.~\cite{Li:2008kz}, the remaining set of graphs containing electroweak gauginos and exchanges of the charged and CP-odd Higgs as well as other gauge bosons were computed and found in some cases to give the dominant contributions to the fermion EDMs in the heavy-fermion regime. We note that in this regime, even the two-loop CEDM operators are suppressed, since the upper loops in Fig. \ref{fig:susy1}(right) contain only squarks. 

\begin{figure}[tb]
\begin{center}
\begin{minipage}[t]{16.5 cm}
\epsfig{file=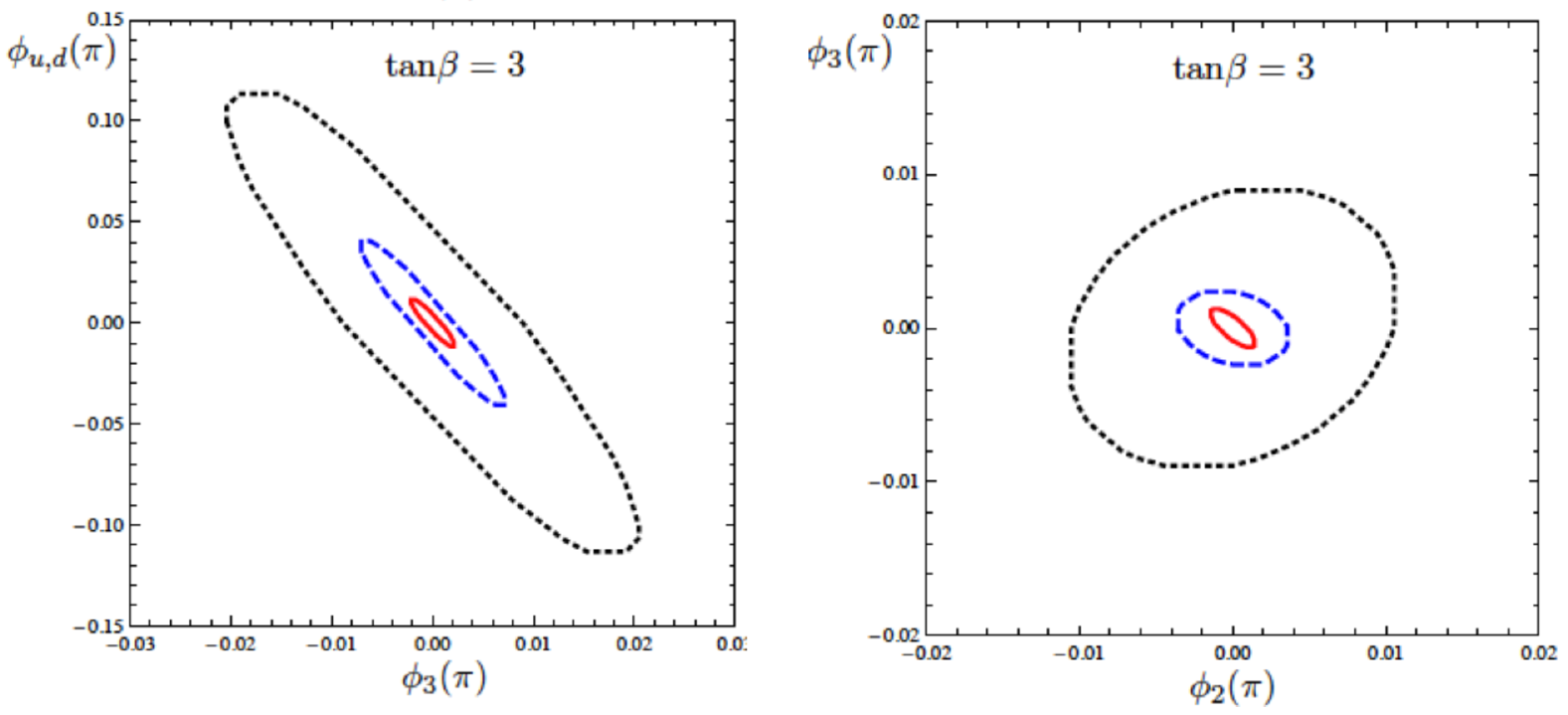,scale=0.8}
\end{minipage}
\begin{minipage}[t]{16.5 cm}
\caption{Constraints on MSSM CPV phases implied by null results for the neutron, Thallium, and Mercury EDMs\cite{Li:2010ax}. Red, blue, and black contours correspond to first generation sfermion masses $(M_{LL,\, RR})_1=200$, $500$, and $1000$ GeV, respectively. 
With kind permission from Springer Science+Business Media: J. High Energy Physics,
\lq\lq A comprehensive analysis of electric dipole moment constraints on CP-violating phases in the MSSM", 08, 2010, p. 062, Y. Li, S. Profumo, M.J. Ramsey-Musolf , Fig. 10 (partial).\label{fig:susy4}}
\end{minipage}
\end{center}
\end{figure}

Based on that work, the authors subsequently performed a global analysis of EDM constraints on CPV phases in the MSSM\cite{Li:2010ax}, up-dating the SuperCPH2.0 code described in Ref.~\cite{Ellis:1996dg}   to include the full set of two-loop graphs. Illustrative results (obtained before publication of the YbF molecular EDM result) are indicated in Fig. \ref{fig:susy4} and Table \ref{tab:current}, based on the use of QCD SR to compute the hadronic matrix elements. In this context, the impact of the three-gluon operators is typically suppressed, as is the effect of the four-fermion operators in the low-to-moderate $\tan\beta$ regime. The results in Table~\ref{tab:current} were obtained assuming three independent phases contribute: $\phi_1$, $\phi_3$, and a common triscalar phase for the first generation squarks: $\phi_u=\phi_d$. The impacts of $\phi_e$ and $\phi_1$  are sufficiently weak that one may omit them from the global analysis, though the sensitivities of on-going and future EDM searches could allow one to probe these phases as well. Fig.~\ref{fig:susy4} shows the relative correlations between pairs of phases, obtained from a fit in each case including only those two phases.  

It is particularly notable that the $d_A(^{199}\mathrm{Hg})$ limit places severe constraints on $\phi_3$ while generating a strong correlation between this phase and $\phi_{u,d}$, both of which enter the CEDM operators at one-loop order. In contrast, the neutron and Thallium EDM limits have a relatively stronger impact on $\phi_2$, though at present the latter constraint is not strongly correlated with any of the other phases. Future measurements with $\sim 100$ times better sensitivity, however, would give rise to such correlations. 

\begin{table}[t]
\caption{Summary of the combined bounds at 95\% c.l. on three phases
($\phi_2$, $\phi_3$, $\phi_{u,d}$) for ${\rm tan}\beta=3$, $60$ and first generation sfermion masses
$(M_{LL,\, RR})_1=200$, $500$, and $1000$ GeV, obtained using current experimental
limits on the neutron, Thallium, and Mercury EDMs\cite{Li:2010ax}.With kind permission from Springer Science+Business Media: J. High Energy Physics,
\lq\lq A comprehensive analysis of electric dipole moment constraints on CP-violating phases in the MSSM", 08, 2010, p. 062, Y. Li, S. Profumo, M.J. Ramsey-Musolf , Table 4.}
\begin{center}
\label{tab:current}
\begin{tabular}{|c|c|c|c|c|c|c|}
\hline ${\rm tan}\beta$ &
\multicolumn{3}{|c|}{3} & \multicolumn{3}{|c|}{60} \\
\hline $(M_{LL,\, RR})_1$ &  200 GeV & 500 GeV & 1000 GeV & 200 GeV &
500 GeV & 1000 GeV \\
\hline $|\phi_2|$ & $<2.1\times 10^{-3}$ & $<5.0\times 10^{-3}$ &
$<1.5\times 10^{-2}$
& $<9.3\times 10^{-5}$ & $<2.5\times 10^{-4}$ & $<6.9\times 10^{-4}$ \\
\hline $|\phi_3|$ & $<2.8\times 10^{-3}$ & $<9.7\times 10^{-3}$ &
$<2.8\times 10^{-2}$
& $<3.1\times 10^{-4}$ & $<4.2\times 10^{-4}$ & $<1.5\times 10^{-3}$ \\
\hline $|\phi_{u,d}|$ & $<1.8\times 10^{-2}$ & $<6.0\times 10^{-2}$
& $<0.17$
& $<1.7\times 10^{-2}$ & $<5.6\times 10^{-2}$ & $<0.21$ \\

\hline
\end{tabular}
\end{center}
\end{table}

\subsection{Extended Gauge Symmetry}
As with SUSY, the embedding of the SM gauge symmetry in a larger gauge group can allow for additional CPV phases in both flavor diagonal and flavor non-diagonal processes at low energies. For purposes of illustration, we consider the well-studied left-right symmetry model (LRSM) with the gauge group SU(2)$_L\times$SU(2)$_R\times$U(1)$_{B-L}$. Symmetry breaking proceeds in two steps, with the first step breaking the left-right symmetry and generating a mass for the right-handed gauge bosons, followed by a second step that breaks the SM gauge symmetry. Implementing this scenario requires augmenting the SM Higgs sector with additional scalar fields: two complex triplets $\Delta_{L,R}$ that transform separately under the SU(2)$_{L,R}$ symmetries and an eight-component bidoublet $\phi$. New CPV phases arise from two sources. The extended gauge symmetry allows for a complex phase $\alpha$ associated with the VEV of $\phi$, corresponding to spontaneous CPV (SCPV):
\be
\langle\phi\rangle =\left(
\begin{array}{cc}
\kappa & 0 \\
0 & \kappa^\prime e^{i\alpha}
\end{array}\right)\ \ \ .
\ee
In addition, new phases can arise in the fermion mixing sector owing to differences between separate rotations of the left- and right-handed fermion fields, $\mathrm{Im} (V_{ij}^L V_{ij}^{R\, \ast})$. Significantly, one requires only two generations of fermions rather than three as in the case of the SM in order to obtain a CPV phase associated with mixing that cannot be removed through field redefinitions. 

The manifestation of these new phases then enters through the mixing of the $W_{L,R}$ gauge bosons into the mass eigenstates $W_{1,2}$:
\bea
W_1^+ & = & \cos\xi\, W_L^+ + \sin\xi\, e^{-i\alpha}\, W_R^+ \\
W_2^+ & = & -\sin\xi\, e^{i\alpha}\, W_L^+ +\cos\xi\, W_R^+
\eea
where the mixing angle is given by 
\be
\tan\xi = -\kappa\kappa^\prime/v_R^2\approx -2\frac{\kappa^\prime}{\kappa}\, \left(\frac{M_1}{M_2}\right)^2
\ee 
with $v_R$ being the vev of the neutral component of the $\Delta_R$ and $M_k$ being the mass of the $W_k^\pm$ boson. 
Interactions of the latter with quarks and leptons can yield a variety of the dimension six effective operators introduced earlier. The resulting contributions to the $d_q$ were first computed in Ref.~\cite{Beall:1981zq} for the two-flavor case. In contrast to the SM, non-vanishing contributions first arise at one-loop order, with the result
\be
\delta_q = \left(\frac{g_L g_R}{96\pi^2}\right)\, \sin\xi\,  A_q\, \left[r_q \cos\theta_L\cos\theta_R\sin\alpha+r_Q\, \sin\theta_L\sin\theta_R\, B_q\right]
\ee
where $\cos\theta_{L,R}$ is the \lq\lq Cabibbo angle" for the left- and right-handed sectors, $g_{L,R}$ are the corresponding gauge couplings for the two sectors, 
\begin{align}
r_u= Y_d/Y_u &\quad r_Q = Y_s/Y_u & A_u=4 &\quad B_q=\sin(\alpha+\delta_R-\delta_L)\\
r_d= Y_u/Y_d &\quad r_Q = Y_c/Y_d & A_d=5 &\quad B_q=\sin(\alpha-\delta_R+\delta_L)
\end{align}
and where we replace
\be
\left(\frac{v}{\Lambda}\right)^2\rightarrow \left(\frac{v}{M_1}\right)^2\left(1-\frac{M_1^2}{M_2^2}\right)
\ee
in the definition of $\delta_q$. Since $M_1$ is the mass of the lightest $W$-boson, one encounters no explicit suppression due to the heavy scale.
Note also that  the EDM for a given quark flavor is proportional to the Yukawa coupling for the quarks having opposite sign third component of $I_{L,R}$. 

The corresponding CEDM operator has been computed in Ref.~\cite{He:1988th} 
One also encounters the four-quark operator of Eq.~(\ref{eq:lreff}) due to the exchange of the $W_{1,2}$ between quarks 
\cite{Ecker:1983dj,He:1988th,Herczeg:1997ei,Zhang:2007da}. Following the notation of Ref.~(\cite{Zhang:2007da}) one obtains
\be
\label{eq:lrsm4q}
\frac{\mathrm{Im}C_{\varphi ud}}{\Lambda^2} = \frac{2\sqrt{2}}{3}\ G_F K^{(-)}\sin\xi \, \mathrm{Im}\left(e^{-i\alpha} V^L_{uq} V^{R\, \ast}_{uq}\right)
\ee
where we have included the contribution from only the exchange of the $W_1$ and have extended the operator $Q_{\varphi ud}$ to include all down-type quarks $q=d,s,b$. The constant $K^{(-)}\approx 3.5$ is a QCD renormalization group factor associated with running from the weak scale to the hadronic scale. Again specifying to the two generation case, we observe that the effect of the phase in the quark mixing matrix will not enter the operators containing only $u$ and $d$ quarks, leaving only a dependence on the SCPV phase. In this case, the contribution to $\gpbo$ will depend solely on this phase and not on $\delta_L-\delta_R$. Including the second generation quarks would then require extending the arguments leading to Eq.~(\ref{eq:lreff4}) to account for the nucleon matrix element of the ${\bar s} s$ and a coupling of the nucleon to the $\eta$ meson. We leave this extension, as well as a consideration of the CEDM and three-gluon operators, to future work. 

Contributions to the neutron EDM in the LRSM have been carried out using a variety of approaches. Ref.~\cite{Beall:1981zq} relied on the quark model result to determine  the dependence of $d_n$ on the $d_q$. The authors of Ref.~\cite{Ecker:1983dj} also performed a quark model evaluation of the contribution from the four quark operator (\ref{eq:lreff}).  Pseudoscalar loops were included in Refs.~\cite{He:1988th,Zhang:2007da}, where one of the pseudoscalar meson-baryon vertices are induced by the underlying CP-violating quark and gluon operators, while Ref.~ \cite{Zhang:2007da} also computed contributions to the nucleon wavefunction due to the CEDM. We note that the pseudoscalar loop results in Refs.~\cite{He:1988th,Zhang:2007da} were not performed using a consistent chiral power counting and are likely to overestimate the corresponding contribution that is proportional to $\gpbo$. 

To illustrate the manifestation of LRSM CPV in EDMs, we consider (a) contributions from the $d_q$ to $d_n$ using the quark model relation \ref{eq:QMedm};  (b) chiral loop contributions to $d_n$ induced by $Q_{\varphi u q}$; (c) contributions to $d_A(^{199}\mathrm{Hg})$ generated by $Q_{\varphi u q}$ {\em via} the nuclear Schiff moment. Starting with the $d_q$, we neglect the heavy quark contributions for simplicity and take $g_L=g_R=e/\sin\theta_\mathrm{W}$, leading to
\be
d_n\sim (1.13\times 10^{-7}e\ \mathrm{fm})\, \left(1-\frac{M_1^2}{M_2^2}\right)\, \sin\xi\, (5 Y_u+4Y_d)\,  \cos\theta_L\cos\theta_R\sin\alpha\ \ \ .
\ee
Noting that $5Y_u+ 4 U_d\sim 2.5\times 10^{-4}$ and that $\vert \sin\xi|\lsim 10^{-3}$ from tests of first row CKM unitarity\cite{Towner:2010zz}, we see that 
\be
|d_n|^\mathrm{d_q}\lsim (3\times 10^{-14}e\ \mathrm{fm})\, \left(1-\frac{M_1^2}{M_2^2}\right)\, \cos\theta_L\cos\theta_R\sin\alpha\ \ \ .
\ee
The present $d_n$ constraint is, thus, not sufficiently stringent to probe this contribution.

A potentially larger contribution may arise from chiral loops involving the isovector TVPV $\pi NN$ interaction. Making the same simplifying assumptions used above and using Eqs.~(\ref{eq:lreff4},\ref{eq:lrsm4q}) one has 
\be
\gpbo\approx -10^{-4}\, \left(1-\frac{M_1^2}{M_2^2}\right)\, \sin\xi\,  \cos\theta_L\cos\theta_R\sin\alpha\ \ \ .
\ee
The corresponding one-loop contribution to $d_n$ is given by 
\be
d_n^\mathrm{chiral} = \frac{e g_A \gpbo}{16\pi^2}\, \frac{\mu_n}{F_\pi}\, F(m_\pi^2/\lamchi^2)
\ee
where $\mu_n=-1.91$ is the neutron anomalous magnetic moment and $F(x)$ is a loop function. An early calculation reported in Ref.~\cite{He:1988th} gave $F(x)=3/2-x+\cdots$, where the \lq\lq $+\cdots$ denote contributions non-analytic in $x$. This computation, however, did not utilize the consistent power counting obtained with HB$\chi$PT and, thus, should be considered unreliable. A consistent HB$\chi$PT computation gives
$F(x)= -x\ln x \approx -0.1$, implying an order of magnitude smaller neutron EDM contribution than one would infer from the computation of Ref.~\cite{He:1988th}. Taking $\vert \sin\xi|\lsim 10^{-3}$ we then obtain
\be
|d_n|^\mathrm{chiral} = (3\times 10^{-10}e\ \mathrm{fm})\, \left(1-\frac{M_1^2}{M_2^2}\right)\, \cos\theta_L\cos\theta_R\sin\alpha\ \ \ ,
\ee
indicating roughly four orders of magnitude greater sensitivity to $\sin\alpha$ than implied by the quark EDM contribution. One may trace this difference to the combination of the quark Yukawa couplings $5Y_u+ 4 U_d\sim 2.5\times 10^{-4}$ that enters the quark EDM contribution  and that does not appear in the chiral loop contribution induced by $Q_{\varphi uq}$. For $\cos\theta_L\approx\cos\theta_R\approx 1$ and $M_1 << M_2$ we then obtain $|\sin\alpha| \lsim 10^{-3}$ from this contribution.

Turning to $d_A(^{199}\mathrm{Hg})$, we use the value of $\kappa_S$ given in Eq.~(\ref{eq:ksHg}) and a representative value for $a_1$ of 0.03 (midpoint of the corresponding range in Table \ref{tab:schiff-ranges})  to obtain 
\be
| d_A(^{199}\mathrm{Hg}) | \lsim  (1.1\times 10^{-11}e\ \mathrm{fm})\, \left(1-\frac{M_1^2}{M_2^2}\right)\, \cos\theta_L\cos\theta_R\sin\alpha\ \ \ ,
\ee
giving an even stronger sensitivity to the SCPV phase than $d_n$, though subject to considerable nuclear theory uncertainties associated with the computation of $a_1$ as discussed above. However, given that the current bound on $d_A(^{199}\mathrm{Hg})$ is three orders of magnitude smaller than the limit on $d_n$, the former is likely to provide the most stringent constraint on the LRSM contribution even allowing for a possibly smaller magnitude for $a_1$ than assumed here. For the benchmark value of $a_1$ used in this example, we would obtain $|\sin\alpha|\lsim 10^{-5}$ for $\cos\theta_L\approx\cos\theta_R\approx 1$ and $M_1 << M_2$.

\begin{figure}[tb]
\begin{minipage}[t]{16.5 cm}
\epsfig{file=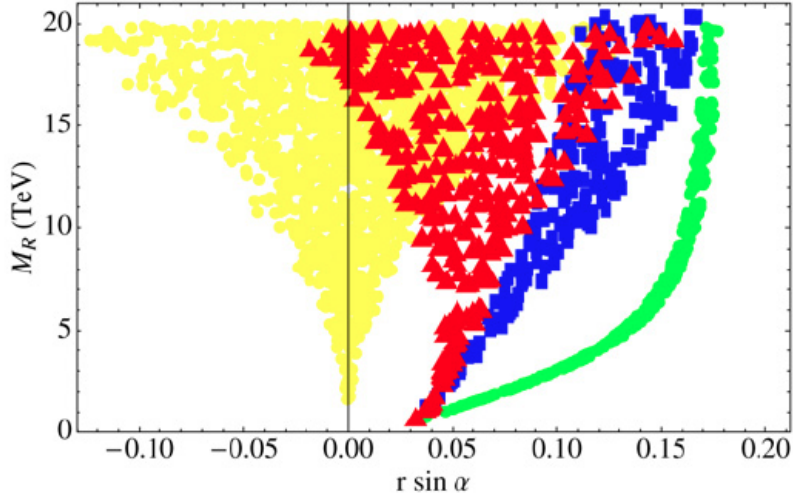,scale=0.55}
\end{minipage}
\begin{center}
\begin{minipage}[t]{16.5 cm}
\caption{Constraints on right-handed $W$-boson mass and CPV parameter $r\sin\alpha$ obtained from present constraints on the neutron EDM (yellow dots) and neutral kaon mixing parameter $\epsilon$ for different representative values of the Higgs mass ($M_H
=\infty$, red triangle; $M_H
= 75$ TeV, blue square; $M_H
= 20$ TeV, large green dots). Reprinted from Nucl. Phys. B., 802, Y. Zhang, H.  An, X. Ji, R. N. Mohapatra, \lq\lq General CP violation in minimal leftÐright symmetric model and constraints on the right-handed scale", p.247., Copyright (2008), with permission from Elsevier .\label{fig:lrsm}}
\end{minipage}
\end{center}
\end{figure}

In the foregoing discussion, we have used the phenomenological constraint on the mixing angle $\xi$ obtained from tests of first row CKM unitarity\cite{Towner:2010zz}. An alternate approach has been followed by the authors of Ref.~\cite{Zhang:2007da}, who observed that one may determine the elements of the right-handed CKM matrix $V^R_{uq}$ in terms of $V^L_{uq}$, the ratio $\kappa^\prime/\kappa$, $\sin\alpha$ and the quark masses by exploiting properties of the LRSM Yukawa matrices, the hierarchy of quark masses, and the Wolfenstein parameterization of $V^L_{uq}$. One then finds that
\be
\sin\xi\, \mathrm{Im}\left(e^{-i\alpha} V^L_{uq} V^{R\, \ast}_{uq}\right)
\ee
can be expressed in terms of $M_1^2/M_2^2$ and $r\sin\alpha$, where $r=(m_t/m_b)(\kappa^\prime/\kappa)$ characterizes the ratio of the two bi-doublet vevs. Illustrative constraints on $M_2\approx M_{W_R}$ and $r\sin\alpha$ are indicated by the yellow points in Fig.~\ref{fig:lrsm}. We note that the yellow points were obtained using the value of $F(x)$ given in Ref.~\cite{He:1988th} that is an order of magnitude larger than the HB$\chi$PT result. Taking into account the latter reduction  and utilizing the bounds on $ d_A(^{199}\mathrm{Hg}) $ we conclude that the region allowed by the $^{199}\mathrm{Hg}$ limit is likely to be considerably narrower than indicated by the yellow points in Fig.~\ref{fig:lrsm}. 

A future improvement in the sensitivity of neutron EDM searches by two orders of magnitude could make $d_n$ a comparably powerful probe of LRSM CPV as $d_A(^{199}\mathrm{Hg})$.

\subsection{Additional Spacetime Dimensions}

The study of EDMs in BSM scenarios involving extra spacetime dimensions, such as the Randall-Sundrum (RS) paradigm for warped extra dimensions or flat but orbifolded extra dimensions, is considerably less advanced than in the case of SUSY or extended gauge symmetries. To our knowledge, EDM estimates have been largely confined to the use of NDA, coupled with an analysis of the flavor and CP structure associated with a given implementation of the RS paradigm. 

For concreteness, we focus on the scenario RS1, wherein SM fields may propagate in the \lq\lq bulk" of the fifth dimension between two branes: the TeV brane and the Planck brane. The Higgs field is localized at the former while gravity lives at the latter, ensuring that the natural scales for the EW and gravitational interactions are the weak and Planck scales, respectively. The dependence on all mass scales in the fifth dimension follows from the \lq\lq warping" associated with the dependence of the metric on the fifth dimensional co-ordinate $z$ as 
\be
(ds)^2= \frac{1}{(kz)^2}\left[ \eta^{\mu\nu} x_\mu x_\nu- (dz)^2\right]
\ee
where $x^\mu$ denotes the usual four dimensional co-ordinate vector and $k$ is a warping factor. 

Flavor structure arises from the $z$-dependence of the fermion wavefunctions rather than from the values of the Yukawa couplings for the 5-dimensional theory. Assuming the latter  to be \lq\lq anarchical", the observed fermion mass hierarchy arises when the light fermions are localized near the Planck brane and the top quark near the TeV brane. Since the Higgs is localized near the TeV brane, its vev gives a significantly larger mass to the top quarks than to the light fermions. The Kaluza-Klein (KK) modes for the light fermions are also localized near the TeV brane, generating a leading order suppression of flavor changing neutral currents (FCNCs).

An early concrete application of this scenario to the flavor and CP problems was carried out in Ref.~\cite{Agashe:2004cp}. The corresponding five-dimensional Lagrangian contains two components of interest: 

\noindent (a) the bulk Lagrangian
\be
\mathcal{L}_\mathrm{fermion} = \sqrt{G}\left\{ i {\bar\psi} \Gamma^M D_M\psi + k  C_{Qud} ({\bar Q} {\bar u} {\bar d}) (Q u d)\right\}
\ee
where all of the SM fields $\psi$ propagate in the warped extra dimension (denoted by a co-ordinate $z$), where $D_M$ is the five dimensional covariant derivative, and where $ C_{Qud}$ are $3\times 3$ Hermitian matrices that determine the 5-D masses. 

\noindent (b) the 5-D Yukawa interaction:
\be
\mathcal{L}_\mathrm{brane} = h \delta(z-z_0) \lambda_{u,d}^{5D} {\bar Q}(u,d)\ \ \ 
\ee
where $ \lambda_{u,d}^{5D}$ are the Yukawa matrices, $h$ is the Higgs field, and $z_0$ indicates the location of the TeV-brane. 

\begin{figure}[tb]
\begin{minipage}[t]{16.5 cm}
\epsfig{file=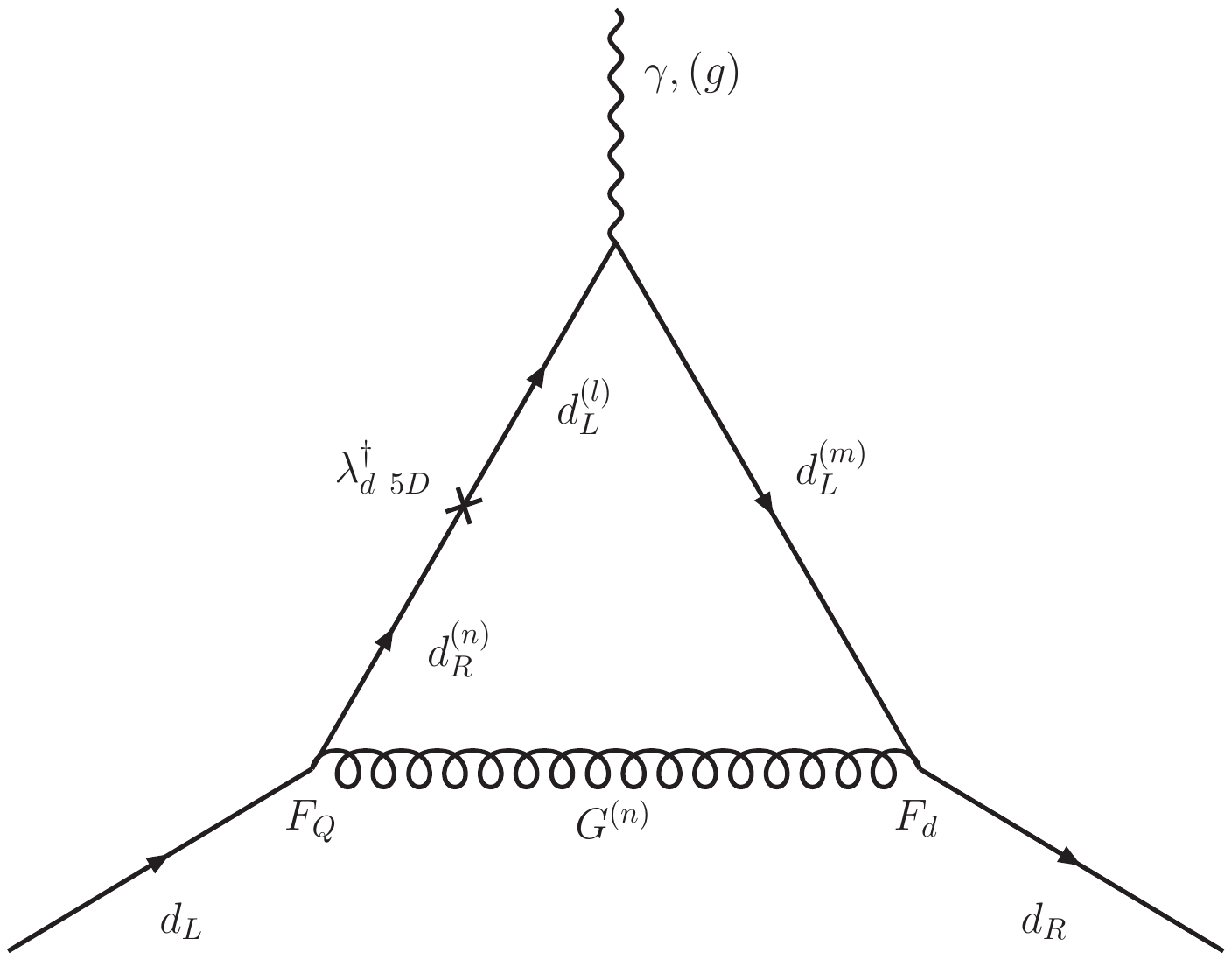,scale=0.4}
\epsfig{file=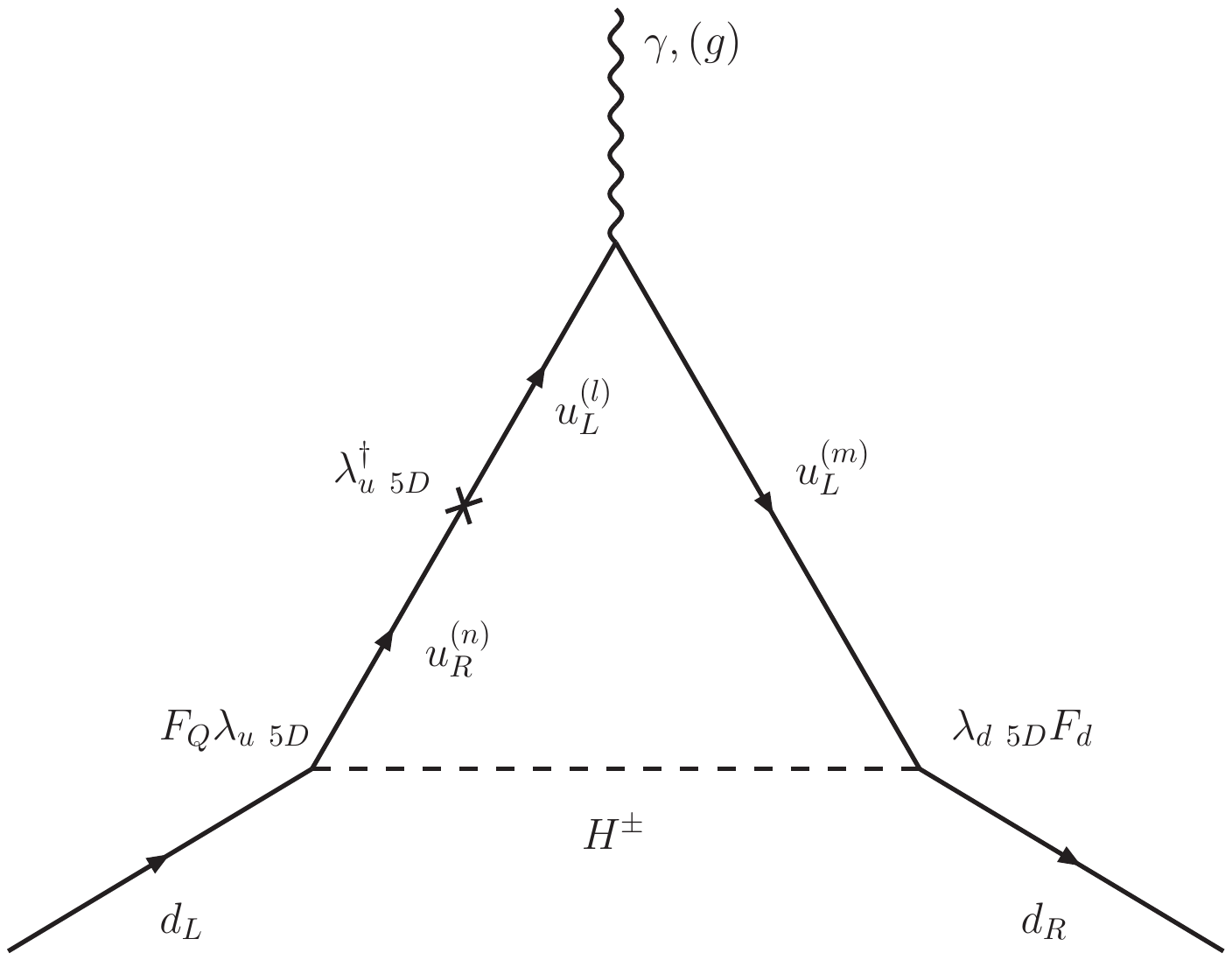,scale=0.4}
\end{minipage}
\begin{center}
\begin{minipage}[t]{16.5 cm}
\caption{Representative contributions to the $d$-quark EDM and CEDM in the Randall-Sundrum scenario.\label{fig:rs}}
\end{minipage}
\end{center}
\end{figure}

Carrying out the Kaluza-Klein (KK) reduction of the 5D theory to an effective 4D theory on the TeV brane yields the SM fields (zero modes) and their KK partners. The quark zero mode masses are then given by $m_q\sim v F_Q \lambda_{u,d}^{5D} F_{u,d}$, where $F_{Q, u, d}$ are the values of the quark wavefunctions on the TeV brane . For purposes of the present discussion, the specific values of the $F_{Q, u, d}$ are not essential. However, due to the different profiles for the light fermion zero mode and KK modes, couplings between the two go as $\lambda_{u,d}^{5D} F_q$. Since these couplings are not aligned with the  quark masses, non-trivial flavor and CPV contributions may be generated at one-loop order. Representative diagrams that generate the quark EDMs are shown in Fig. \ref{fig:rs}. In each case, an odd number of zero mode - KK Yukawa interactions is needed to obtain the chiral structure associated with the EDM. The gauge loops contain only one insertion while the Higgs loops contain three. In the former case for a down quark-gluon loop, one has
\be
d_d [\mathrm{gluon,\, KK}] \sim kv\left( D_L^\dag F_Q \lambda_{d}^{5D} F_d D_R\right)_{11} \sim\left[  \mathrm{diag}(m_d, m_s, m_b)\right]_{23} =0\ \ \ ,
\ee
where $D_{L,R}$ rotate the left- and right-handed down quarks between the flavor and mass bases.  
In contrast, the Higgs loop results are not aligned with the light quark mass matrix and, as shown in Ref.~\cite{Agashe:2004cp}, lead to a non-vanishing EDM contribution:
\be
d_d [\mathrm{Higgs,\, KK}] \sim 2 k^3 v \left[  F_Q \left(    \lambda_u^{5D}\lambda_u^{5D\, \dag} +  \lambda_d^{5D}\lambda_d^{5D\, \dag}\right) \lambda_d^{5D} F_d \right]_{11}\ \ \ .
\ee
Taking the phases that enter this expression to be maximal, the authors of Ref.~\cite{Agashe:2004cp} arrive at the NDA estimate
\be
\label{eq:rsloop1}
d_n[\mathrm{Higgs,\, KK}]  \sim \frac{e}{6} \left(\frac{m_d}{16\pi^2}\right)\, \left( \frac{2k \lambda^{5D}}{m_{KK}^2}\right)
\sim (10^{-11}\, e\ \mathrm{fm})\times\left(\frac{2k \lambda^{5D}}{4}\right)^2\, \left(\frac{ 3\, \mathrm{TeV}}{m_{KK}}\right)^2 
\ee
An earlier analysis by the authors of Ref.~\cite{Chang:2002ww} that included the contribution from the CEDM found a considerably smaller sensitivity to the CPV parameters. In that study, the additional suppression results from a tiny coupling between the first and third generation as well as constraints from the CPV parameter $\epsilon_K$ that enters the neutral kaon system. 

In addition to the loop contribution, one expects contributions from operators that live on the TeV brane. This term is UV-sensitive and, thus, depends on the cutoff of the effective theory on the TeV brane, obtained from the cutoff at the Planck scale by warping: $\Lambda\sim \Lambda_{5D} \exp(-\pi k r_c)$, where $\pi k r_c\sim M_\mathrm{Planck}/\mathrm{TeV}$ to solve the hierarchy problem. From NDA one anticipates
\be
d_n[\mathrm{brane}]  \sim eC_\Lambda\, \left(\frac{m_d}{\Lambda^2}\right) 
\sim (10^{-11}\, e\ \mathrm{fm})\times C_\Lambda\ \left(\frac{2k \lambda^{5D}}{4}\right)^2\, \left(\frac{ 10\, \mathrm{TeV}}{\Lambda}\right)^2 
\ee
The value of $\Lambda$ depends on the specific realization, depending on whether the Higgs is placed on the TeV brane, in the bulk, or in the bulk but localized near the TeV brane. For $\sin\phi_\mathrm{CPV}\sim \mathcal{O}(1)$, the resulting contribution to $d_n$  can be comparable to the present experimental limit or comparable to the considerably larger loop contribution in Eq.~(\ref{eq:rsloop1}).

One should bear in mind that the foregoing results are obtained using NDA and
that loop computations in extra-dimensional scenarios are subject to
theoretical ambiguities. Nonetheless, one thus finds a situation similar to
that in SUSY: current EDM limits imply that either the CPV phases are
suppressed or that the KK mass scale lies well above the TeV scale. A variety
of solutions to the RS CP problem have been proposed. In
Ref.~\cite{Fitzpatrick:2007sa} a variant of RS1 was analyzed under the
assumption of 5D minimal flavor violation, leading to the vanishing of EDMs at
one-loop order. Ref.~\cite{Cheung:2007bu} considered an RS1 scenario with
spontaneous CPV, where the source of CPV was geometrically sequestered from the
TeV brane by placing it in the bulk. The model provides a natural suppression
of ${\bar\theta}$, while the dimension-six EDM operators first appear at
two-loop order. 

These studies notwithstanding, it is evident that there exists considerable
room for further work on EDMs in extra dimensional models. To our knowledge, no
computations of the CEDM, three-gluon, or dimension-six four-fermion operators
has appeared in the literature. In particular, the limits on
$d_A(^{199}\mathrm{Hg})$ may imply more severe constraints on RS CPV than have
been obtained in these earlier studies, given the long-range $\pi$-exchange
contributions to the nuclear Schiff moments sourced by the CEDM operators. Even
with the ambiguities associated with loop computations in $d>4$ dimensions and
with cutoff-dependent TeV brane operators, a study of these additional CPV
effects would be both interesting and potentially significant.

\section{Outlook}
\label{sec:concl}

In the context of fundamental symmetry tests during the LHC era, searches for the permanent electric dipole moments of atoms, molecules, nucleons and nuclei will provide one of the most powerful probes of both BSM physics as well as the remaining, as-yet unseen source of SM CPV -- the QCD $\theta$-term. Improvements in experimental sensitivity are poised to improve by as much as two orders of magnitude in the near term and possibly further on a longer time scale. The observation of a non-zero EDM would constitute a major discovery, pointing to  a non-vanishing ${\bar\theta}$ parameter and/or a new source of CPV associated with new fundamental interactions. Conversely, the non-observation of EDMs at the anticipated sensitivity levels would tighten the already stringent those on ${\bar\theta}$ as well as those on a variety of BSM scenarios. Either way, the implications for the fundamental laws of nature and their consequences for the cosmic baryon asymmetry cannot be overstated.

Theoretically, the challenge is to provide the most robust framework for interpreting the results of EDM searches and delineating their implications. Doing so entails analyzing physics associated with a variety of energy scales, ranging from the short-distance physics of CPV at the elementary particle level to the longer-distance physics at the hadronic, nuclear, and atomic/molecular scales. In this review, we have relied on effective CPV operators at mass dimensions four and six as a bridge between the physics of these various scales. While the use of effective operators is not applicable in all circumstances, such as those in which BSM CPV involves new light degrees of freedom, it  nevertheless provides a  broadly applicable and model-independent context for the interpretation of EDM experiments. Limiting our consideration photons, gluons, and first generation fermions, we encounter thirteen presently undetermined CPV parameters at $d=4$ and $d=6$: ${\bar\theta}$, the fermion EDMs, quark CEDMs, three-gluon operator, and several four-fermion operators. The task for theory, then, is to delineate how these operators may be generated by BSM physics above the weak scale, how they evolve to the hadronic scale, and how they generate the appropriate hadronic, nuclear, atomic and molecular matrix elements that ultimately give rise to EDMs in these systems.

From our review of this theoretical effort, several features emerge:
\begin{itemize}
\item[(i)] The EDMs of paramagnetic atoms and molecules are dominated by two quantities: the electron EDM and one combination of semileptonic, CPV four-fermion operators characterized by the Wilson coefficient $\mathrm{Im}\ C_{eq}^{(-)}$. Moreover, when characterizing the former in terms of the appropriate dimensionless parameter $\delta_e$, one finds that the EDMs of these systems are an order of magnitude more sensitive to $\mathrm{Im}\ C_{eq}^{(-)}$ than to $\delta_e$. The level of theoretical atomic/molecular theory uncertainty in either sensitivity is roughly 10\% or better. For $\mathrm{Im}\ C_{eq}^{(-)}$, the associated hadronic matrix element of the isoscalar scalar density is under reasonable control, given that it can be obtained from the pion-nucleon $\sigma$-term and the average light quark mass. 
\item[(ii)] Diamagnetic atom EDMs are most sensitive to the nuclear Schiff moment, individual nucleon EDMs, and the semileptonic four-fermion tensor operator with Wilson coefficient $\mathrm{Im}\ C_{\ell e q u}^{(3)}$. 
\item[(iii)] Neither system provides a particularly sensitive handle on $\mathrm{Im}\ C_{eq}^{(+)}$, given the relative suppression of the associated isovector scalar or isoscalar pseudoscalar nuclear matrix elements. 
\item[(iv)] There exists considerable room for refinement in computing the nucleon matrix of underlying CPV operators as well as nuclear Schiff moments. While the sensitivity of $\gpbz$ to ${\bar\theta}$ and the dependence of $d_N$ on the quark EDMs is now known fairly reliably, the uncertainties associated with matrix elements of the quark CEDMs, three-gluon operator, and four-quark operators are large. Similar statements apply to the dependence of $d_N$ on ${\bar\theta}$ as well as to the sensitivity of the nuclear Schiff moments to the $\gpbi$. Perhaps, one of the primary challenges facing is now to achieve a more reliable set of hadronic sensitivities. 
\item[(v)] Searches for the EDMs of diamagnetic atoms and nucleons alone is unlikely to disentangle the effects of the quark  CEDMs, three-gluon operator, and four-quark operators. However, the possibility of searching for EDMs of the proton, deuteron, triton and helion in storage rings would offer additional handles on these underlying sources of hadronic CPV based on their chiral transformation properties. 
\item[(vi)] Any global analysis of EDM search results, whether performed at the model-independent level of the effective operators within a given BSM scenario such as supersymmetry, should take into account the rather sizeable theoretical hadronic and nuclear uncertainties associated with the sources of hadronic CPV. 
\end{itemize}

Setting aside the aforementioned uncertainties, current EDM null results imply that any new CPV lies at the TeV scale or that CPV phases are $\mathcal{O}(10^{-2})$ in magnitude or smaller. The next generation of searches will push these sensitivities to $\Lambda\gsim 10$ TeV or equivalently $|\sin\phi_\mathrm{CPV}|\lsim \mathcal{O}(10^{-4})$, putting their reach well beyond that of the LHC. Should the LHC have observed only the SM Higgs boson (a major discovery in its own right) by the end of this decade, then EDM searches will provide one of the most  effective tools for probing the next piece of terrain in the high energy desert. Given these prospects, addressing the open theoretical challenges summarized above becomes all the more important. We hope that our discussion of the rich array of physics associated with EDMs will spur new efforts to take on these challenges.

\section*{Acknowledgments}
The authors are grateful to T. Chupp, J. Donoghue, B. Holstein, J. Hisano, T. Izubuchi, R. Mohapatra, M. Pospelov, A. Ritz, and G. Senjanovic for helpful discussions and comments on the manuscript. We also thank T. Chupp, D. DeMille, B. Filippone, P. Harris, B. Heckel, K. Kirch,  Z.-T. Lu, J. Martin, and Y. Semertzidis for providing future EDM sensitivity goals and H. Guo for invaluable assistance with Figures \ref{fig:susy1}, \ref{fig:susy2}, and \ref{fig:rs}. We also thank E. Shintani for providing Figures \ref{fig:latt1} and \ref{fig:latt2}.
This work was supported in part by the US DOE under contract numbers DE-FG02-08ER41531(MJRM) and DE-FG02-04ER41338
(UvK); by National Science Foundation award 1152096;  and by the Wisconsin Alumni Research Foundation (MJRM).

\appendix

\section{Scalar and pseudoscalar form factors: heavy quark contributions}
\label{sec:heavy}

The $g_S^Q$ for the heavy
flavors can be obtained by using the trace anomaly for the energy momentum 
tensor $\theta_{\mu\nu}$ and nucleon mass and by integrating out the heavy quarks
\cite{Shifman:1978zn,Demir:2003js}. 
We first have 
\be
\label{eq:mntrace}
m_N {\bar \psi_N} \psi_N  =  \bra{N}\theta^\mu_\mu \ket{N} 
=\sum_{q=u,d,s}\bra{N} m_q {\bar q} q\ket{N} 
+ \frac{{\tilde\beta}(\alpha_s)}{2\alpha_s} 
\bra{N}\mathrm{Tr}\left( G_{\mu\nu} G^{\mu\nu}\right)\ket{N}\ ,
\ee
where 
\be
\label{eq:beta1}
{\tilde\beta}(\alpha_s) = -9\alpha_s^2/2\pi
\ee
and where we have used the result that integrating out each heavy quark leads 
to the replacement 
\be
\label{eq:heavyQ1}
\bra{N} m_Q {\bar Q} Q\ket{N} = -\frac{2}{3}\ \frac{\alpha_s}{4\pi}\, 
\bra{N}\mathrm{Tr}\left( G_{\mu\nu} G^{\mu\nu}\right)\ket{N}\  .
\ee
Using Eqs.~(\ref{eq:gs01},\ref{eq:gsstrange})
we may solve for 
$\bra{N} {\bar Q} Q\ket{N}$ and, thus, $g_S^Q$:
\be
-\frac{9\alpha_s}{4\pi} 
\bra{N}\mathrm{Tr}\left( G_{\mu\nu} G^{\mu\nu}\right)\ket{N} = 
\left[m_N-( \bar{m}_N)_q -220\kappa_s\right] {\bar \psi_N} \psi_N \ ,
\ee
so that from Eq.~(\ref{eq:heavyQ1}) we obtain
Eq. \eqref{eq:gsstrange}.

For the $g_P^Q$ of heavy flavors, 
we follow \cite{Shifman:1978zn,Ellis:2008zy,Anselm:1985cf} and exploit the 
U(1$)_A$ anomaly.
Letting
\be
J_{\mu 5} = \sum_{q=u,d} {\bar q} \gamma_\mu\gamma_5 q 
+ \sum_{Q=s,c,b,t} {\bar Q} \gamma_\mu\gamma_5 Q
\ee
and
\be
\bra{N} J_{\mu 5} \ket{N} = g_A^{(0)} 
\bra{N} {\bar N} \gamma_\mu\gamma_5 N\ket{N}\ ,
\ee
we have
\bea
\bra{N} \partial^\mu J_{\mu 5} \ket{N} 
& = & 2 m_N g_A^{(0)} \bra{N} {\bar N} i\gamma_5 N\ket{N}
= 2\sum_{q=u,d} \bra{N} m_q {\bar q} i\gamma_5 q\ket{N}
\\
\nonumber
&& +2\sum_{Q=s,c,b,t} \bra{N} m_Q {\bar Q} i\gamma_5 Q\ket{N}
 + 6 \bra{N} \frac{\alpha_s}{4\pi} \mathrm{Tr} 
\left(G_{\mu\nu}\widetilde{G}^{\mu\nu}\right)\ket{N}\ .
\eea
Using
\be
\bra{N} m_Q {\bar Q} i\gamma_5 Q\ket{N} = 
-\frac{\alpha_s}{8\pi} 
\bra{N} \mathrm{Tr}\left(G_{\mu\nu}\widetilde{G}^{\mu\nu}\right)\ket{N}
\ee
and the expressions for matrix elements of 
${\bar u}i\gamma_5 u \pm {\bar d} i\gamma_5 d$ in terms of $g_S^{(0,1)}$ we obtain
\bea
\bra{N} m_Q {\bar Q} i\gamma_5 Q\ket{N} 
&=& \bar{N} \left\{ \frac{1}{4}\left[ g_A^{(0)} \frac{m_N}{m_Q} 
+ g_P^{(0)}\frac{2{\bar m}}{m_Q}\right]
+\frac{1}{4} g_P^{(1)}\ \frac{\Delta m_q}{m_Q}\tau_3 \right\} i\gamma_5N
\\
&\equiv& \bra{N}\left[ g_P^{Q(0)} +g_P^{Q(1)}\tau_3 \right]i\gamma_5N\ ,
\eea
with the result in Eq. \eqref{eq:gPheavy}.






\section{CPV parameter sensitivities:  a compilation}
\label{sec:compile}

\begin{table}[t]
\centering \renewcommand{\arraystretch}{1.5}
\begin{tabular}{||c|c|c|c|c||}
\hline\hline
CPV Parameter & Coefficient & Method & Value & Remarks \\
\hline
${\bar\theta}$ & $\alpha_n$ & ChPT& $\sim 0.002$ $e$ fm & See Eq.~(\ref{eq:alphanchiral})\\
${\bar\theta}$ & $\alpha_n$ & Lattice QCD\cite{Shintani:2008nt}&  -0.040(28) e-fm& $m_\pi=0.53$ GeV\\
${\bar\theta}$ & $\alpha_p$ & Lattice QCD\cite{Shintani:2008nt}&  0.072(49) e-fm& $m_\pi=0.53$ GeV\\
${\bar\theta}$ & $\alpha_n$ & Lattice QCD\cite{Aoki:2008gv}&  -0.049(5) e-fm& $m_\pi\approx0.61$ GeV\\
${\bar\theta}$ & $\alpha_p$ & Lattice QCD\cite{Aoki:2008gv}&  0.080(10) e-fm& $m_\pi\approx0.61$ GeV\\
${\bar\theta}$ & $\alpha_n$ & QCD Sum Rules\cite{Pospelov:1999ha,Pospelov:1999mv} & ($0.0025\pm 0.0013$) e-fm &$\lambda$ from QCD SR\\
${\bar\theta}$ & $\alpha_n$ & QCD Sum Rules\cite{Hisano:2012sc} & ($0.0004^{+0.0003}_{-0.0002}$) e-fm &$\lambda$ from lattice\\
\hline
${\bar\theta}$ & $\lambda_{(0)}$ & ChPT& $\sim m_\pi^2/ \lamchi F_\pi\sim 0.08$  & See Eq.~(\ref{eq:gpbzchiral3})\\
 & & & $0.017\pm 0.005$ & See Eq.~(\ref{eq:gpbzchiral1}) \\
${\bar\theta}$ & $\lambda_{(1)}$ & ChPT& $\sim m_\pi^4/ \lamchi ^3F_\pi$  & See Eq.~(\ref{eq:lam1theta}) \\
\hline\hline
\end{tabular}
\caption{Dependence of hadronic quantities on ${\bar\theta}$.
\label{tab:hadme1}}
\end{table}

\begin{table}[t]
\centering \renewcommand{\arraystretch}{1.5}
\begin{tabular}{||c|c|c|c|c||}
\hline\hline
CPV Parameter & Coefficient & Method & Value & Remarks \\
\hline
$\mathrm{Im}\, C_{q{ G}}$ & $\beta_n^{uG}$ &ChPT &   $\sim e/v \approx 8\times 10^{-4}$ $e$ fm &   \\
$\mathrm{Im}\, C_{u{G}}$ & $\beta_n^{uG}$ &QCD SR\cite{Pospelov:2000bw} &  -$(0.6\pm 0.3)\times 10^{-3}$ $e$ fm  & PQ assumed  \\
$\mathrm{Im}\, C_{d{ G}}$ & $\beta_n^{dG}$ &QCD SR\cite{Pospelov:2000bw} &  -$(1.2\pm 0.6)\times 10^{-3}$ e-fm & PQ assumed \\
$\mathrm{Im}\, C_{u{ G}}$ & $\beta_n^{uG}$ &QCD SR\cite{Hisano:2012sc} & -$(0.20^{+0.15}_{-0.08})\times 10^{-3}$ e-fm &   PQ assumed \\
$\mathrm{Im}\, C_{d{G}}$ & $\beta_n^{dG}$ &QCD SR\cite{Hisano:2012sc} &   -$(0.40^{+0.31}_{-0.17})\times 10^{-3}$ e-fm &PQ assumed \\
$\mathrm{Im}\, C_{q{G}}$ & $\beta_n^{uG}$ & QM/NDA &  $\sim 1\times 10^{-4}$  $e$ fm &  includes $K_{qG}$  \\
$\mathrm{Im}\, C_{q{G}}$ & $\beta_n^{dG}$ & QM/NDA &  $\sim -4\times 10^{-4}$  $e$ fm &  includes $K_{qG}$ \\
\hline
$ {\tilde d}_q$ & ${\tilde\rho}_N$ &ChPT &  $\sim -0.7 $  &   \\
$ {\tilde d}_q$ & ${\tilde\rho}_N^u$ &QCD SR\cite{Pospelov:2000bw} &  $0.55\pm 2.8$  & PQ assumed  \\
$ {\tilde d}_q$ & ${\tilde\rho}_N^d$ &QCD SR\cite{Pospelov:2000bw} & $1.1\pm 0.55$   &  PQ assumed \\
$ {\tilde d}_u$ & ${\tilde\rho}_N$ &QM/NDA & $\sim -0.09$   &  includes $K_{qG}$ \\
$ {\tilde d}_d$ & ${\tilde\rho}_N$ &QM/NDA & $\sim 0.36$   &  includes $K_{qG}$ \\
\hline
$ {\tilde \delta}_q$ & $e{\tilde\zeta}_N$ &ChPT &  $\sim 5\times 10^{-8}$ $e$ fm &   \\
$ {\tilde \delta}_u$ & $e{\tilde\zeta}_N^u$ &QCD SR\cite{Pospelov:2000bw} &  $-(0.9\pm 0.5)\times 10^{-8}$ $e$ fm & PQ assumed  \\
$ {\tilde \delta}_d$ & $e{\tilde\zeta}_N^d$ &QCD SR\cite{Pospelov:2000bw} & $ (-3.6\pm 1.8)\times 10^{-8}$  $e$ fm & PQ assumed  \\
$ {\tilde \delta}_u$ & $e{\tilde\zeta}_N^u$ &QM/NDA &   $\sim 0.2\times 10^{-8}$ $e$ fm & includes $K_{qG}$  \\
$ {\tilde \delta}_d$ & $e{\tilde\zeta}_N^d$ &QM/NDA &   $\sim -0.8\times 10^{-8}$ $e$ fm & includes $K_{qG}$  \\
\hline\hline
\end{tabular}
\caption{Dependence of nucleon EDM on quark CEDMs expressed in terms of the quantities $\mathrm{Im}\, C_{q G}$, ${\tilde d}_q$, or ${\tilde\delta}_q$.
\label{tab:hadmeCEDMd}}
\end{table}

\begin{table}[t]
\centering \renewcommand{\arraystretch}{1.5}
\begin{tabular}{||c|c|c|c|c||}
\hline\hline
CPV Parameter & Coefficient & Method & Value & Remarks \\
\hline
$\mathrm{Im}\, ( C_{u{ G}}\pm C_{d{ G}}) $ & $\gamma_{(0,1)}^{G}$ & Chiral/NDA & $\sim 0.03$&\\
$\mathrm{Im}\, ( C_{u{ G}}+C_{d{ G}}) $ & $\gamma_{(0)}^{G}$ &QCD SR\cite{Pospelov:2001ys} & $(-1.7\leftrightarrow 0.6)\times 10^{-2}$& PQ assumed \\
$\mathrm{Im}\, (C_{u{ G}}-C_{d{ G}})$ & $\gamma_{(1)}^{G}$ &QCD SR\cite{Pospelov:2001ys} & $-(2.3^{+1.2}_{-4.5}) \times 10^{-2}$&PQ assumed \\
\hline
$ {\tilde d}_q$ & ${\tilde\omega}_{(0,1)}$ &Chiral/NDA & $\sim -26$  &   \\
$ {\tilde d}_u+{\tilde d}_d$ & ${\tilde\omega}_{(0)}$ &QCD SR\cite{Pospelov:2001ys} & $(-5\leftrightarrow 15)$ fm$^{-1}$   &  PQ assumed \\
$ {\tilde d}_u-{\tilde d}_d$ & ${\tilde\omega}_{(1)}$ &QCD SR\cite{Pospelov:2001ys} & $20^{+40}_{-10}$  fm$^{-1}$ & PQ assumed  \\
\hline
$ {\tilde\delta}_q$ & ${\tilde\eta}_{(0,1)}$ &Chiral/NDA & $\sim 1.9\times 10^{-6}$  &   \\
$ {\tilde\delta}_u+{\tilde\delta}_d$ & ${\tilde\eta}_{(0)}$ &QCD SR\cite{Pospelov:2001ys} &  $(-3.5\leftrightarrow 1.2)\times 10^{-7}$ & PQ assumed  \\
$ {\tilde\delta}_u-{\tilde\delta}_d$ & ${\tilde\eta}_{(1)}$ &QCD SR\cite{Pospelov:2001ys} & $(-4.6^{+2.3}_{-9.2})\times 10^{-7}$ & PQ assumed  \\
\hline\hline
\end{tabular}
\caption{Dependence of TVPV $\pi NN$ coupling on quark CEDMs expressed in terms of the quantities $C_{q G}$, ${\tilde d}_q$, or ${\tilde\delta}$.
\label{tab:hadmeCEDMg}}
\end{table}

\begin{table}[t]
\centering \renewcommand{\arraystretch}{1.5}
\begin{tabular}{||c|c|c|c|c||}
\hline\hline
CPV Parameter & Coefficient & Method & Value & Remarks \\
\hline
$C_{q\gamma}$ & $\beta_n^{q\gamma}$ &Chiral/NDA & $\sim e/v \approx 8\times 10^{-4}$ $e$ fm & \\
$C_{u\gamma}$ & $\beta_n^{u\gamma}$ &QCD Sum Rules\cite{Pospelov:2005pr} & $(0.4\pm 0.2)\times 10^{-3}$  $e$  fm & \\
$C_{d\gamma}$ & $\beta_n^{d\gamma}$ &QCD Sum Rules\cite{Pospelov:2005pr} &  -$(1.6\pm 0.8)\times 10^{-3}$  $e$  fm &  \\
$C_{u\gamma}$ & $\beta_n^{u\gamma}$ &QCD Sum Rules\cite{Hisano:2012sc} & $(0.13^{+0.10}_{-0.06})\times 10^{-3}$  $e$  fm & \\
$C_{d\gamma}$ & $\beta_n^{d\gamma}$ &QCD Sum Rules\cite{Hisano:2012sc} &  -$(0.53^{+0.41}_{-0.23})\times 10^{-3}$  $e$  fm &  \\
$C_{u\gamma}$ & $\beta_n^{u\gamma}$ &Quark Model & $0.4 \times 10^{-3}$  $e$  fm & \\
$C_{d\gamma}$ & $\beta_n^{d\gamma}$ &Quark Model &$-1.5 \times 10^{-3}$  $e$  fm  & \\
$C_{u\gamma}$ & $\beta_n^{u\gamma}$ &PQM\cite{Ellis:2008zy,Ellis:1996dg} &$0.6 \times 10^{-3}$  $e$  fm & \\
$C_{d\gamma}$ & $\beta_n^{d\gamma}$ &PQM\cite{Ellis:2008zy,Ellis:1996dg}& $-0.8 \times 10^{-3}$  $e$  fm & \\
\hline
$d_{q}$ & $\rho_N^q$ &Chiral/NDA & $\sim -0.7 $ & \\
$d_{u}$ & $\rho_n^{u}$ &QCD Sum Rules\cite{Pospelov:2005pr} & $(-0.35\pm 0.17)$  & \\
$d_{d}$ & $\rho_n^{d}$ &QCD Sum Rules\cite{Pospelov:2005pr} & $(1.4\pm 0.7)$ &  \\
$d_{u}$ & $\rho_n^{u}$ &QCD Sum Rules\cite{Hisano:2012sc} &  $(-0.11^{+0.09}_{-0.05})$ & \\
$d_{d}$ & $\rho_n^{d}$ &QCD Sum Rules\cite{Hisano:2012sc} & $(0.47^{+0.36}_{-0.2})$ &  \\
$d_{u}$ & $\rho_n^u$ &Quark Model &$-1/3 $ & \\
$d_{d}$ & $\rho_n^d$ &Quark Model & $ 4/3$ & \\
$d_{u}$ & $\rho_n^u$ &PQM\cite{Ellis:2008zy,Ellis:1996dg} &$\left(\Delta u\right)_n = -0.508$ & \\
$d_{d}$ & $\rho_n^d$ &PQM\cite{Ellis:2008zy,Ellis:1996dg} & $\left(\Delta d\right)_n = 0.746$ & \\
$d_{s}$ & $\rho_n^d$ &PQM\cite{Ellis:2008zy,Ellis:1996dg}& $\left(\Delta s\right)_n = -0.226$ & \\
\hline
$\delta_{q}$ & $e\zeta_N^q$ &Chiral/NDA & $\sim 5\times 10^{-8}$ $e$ fm & \\
$\delta_{u}$ & $e\zeta_n^{u}$ &QCD Sum Rules\cite{Pospelov:2005pr} & $ (0.8\pm 0.3)\times 10^{-8}$ $e$ fm & \\
$\delta_{d}$ & $e\zeta_n^{d}$ &QCD Sum Rules\cite{Pospelov:2005pr} & $ (-3.2\pm 2.3)\times 10^{-8}$ $e$  fm &  \\
$\delta_{u}$ & $e\zeta_n^{u}$ &QCD Sum Rules\cite{Hisano:2012sc} & $(0.27^{+0.2}_{-0.1})\times 10^{-8}$ $e$ fm & \\
$\delta_{d}$ & $e\zeta_n^{d}$ &QCD Sum Rules\cite{Hisano:2012sc} & $(-1.1^{+0.8}_{-0.5})\times 10^{-8}$ $e$ fm &  \\
$\delta_{u}$ & $e\zeta_n^u$ &Quark Model & $ 0.8 \times 10^{-8}$ $e$ fm  &  \\
$\delta_{s}$ & $e\zeta_n^s$ &Quark Model & $-3.2 \times 10^{-8}$ $e$ fm  & \\
$\delta_{u}$ & $e\zeta_n^u$ &PQM\cite{Ellis:2008zy,Ellis:1996dg} & $ 1.2 \times 10^{-8}$ $e$ fm  &  \\
$\delta_{s}$ & $e\zeta_n^s$ &PQM\cite{Ellis:2008zy,Ellis:1996dg} & $-1.7 \times 10^{-8}$ $e$ fm  & \\
\hline\hline
\end{tabular}
\caption{Dependence nucleon EDM on quark EDM expressed in terms of the Wilson coefficients $C_{q\gamma}(\lamchi)$, individual quark EDMs $d_q(\lamchi)$, or dimensionless quantities $\delta_q(\lamchi)$. 
\label{tab:hadmeEDMd}}
\end{table}

\begin{table}[t]
\centering \renewcommand{\arraystretch}{1.5}
\begin{tabular}{||c|c|c|c||}
\hline\hline
CPV Parameter & Coefficient & Method & Value  \\
\hline
$\mathrm{Im}\, C_{\tilde G}$ & $\beta_n^{\tilde G}$ &Chiral/NDA & $\sim 40 \times 10^{-7}$ $e$ fm  \\
$\mathrm{Im}\, C_{\tilde G}$ & $\beta_n^{\tilde G}$ &QCD SR\cite{Demir:2002gg} & $2.0\times 10^{-7}$ $e$ fm \\
\hline
$\mathrm{Im}\, C_{\tilde G}$ & $\gamma_{(i)}^{\tilde G}$ & Chiral/NDA & $\sim 2\times 10^{-6}$  \\
\hline\hline
\end{tabular}
\caption{Dependence of hadronic quantities on Weinberg three-gluon operator Wilson coefficient.
\label{tab:hadme3G}}
\end{table}

\begin{table}[t]
\centering \renewcommand{\arraystretch}{1.5}
\begin{tabular}{||c|c|c|c|c||}
\hline\hline
CPV Parameter & Coefficient & Method & Value & Remarks \\
\hline
$\mathrm{Im}\, C_{q u q d}^{(1,8)}$ & $\beta_n^{(1,8)}$ &Chiral/NDA & $\sim 40 \times 10^{-7}$ $e$ fm &  \\
$\mathrm{Im}\, C_{\varphi ud}$ &$\beta_n^{\varphi ud}$ & Chiral/NDA &$\sim 3 \times 10^{-8}$ $e$ fm & \\
$\mathrm{Im}\, C_{\varphi ud}$ &$\beta_n^{\varphi ud}$ & Saturation/ChPT & $1.3\times 10^{-10}$ $e$ fm& NLO\\
\hline\hline
$\mathrm{Im}\, C_{q u q d}^{(1,8)}$ & $\gamma_{(0,1)}^{(1,8)}$ &Chiral/NDA & $\sim 2\times 10^{-6}$   & \\
$\mathrm{Im}\, C_{\varphi ud}$ &$\gamma_{(1)}^{\varphi ud}$ & Chiral/NDA & $\sim  10^{-6}$ & \\
$\mathrm{Im}\, C_{\varphi ud}$ &$\gamma_{(1)}^{\varphi ud}$ & Saturation &$3.3\times  10^{-5}$ & \\
\hline\hline
\end{tabular}
\caption{Dependence of hadronic quantities on CPV four-quark operators. The Saturation result for $\beta_n^{ud}$ has been
obtained by first computing $\gpbo$ and then employing the NLO result from ChPT given in Eqs.~(\ref{d0},\ref{d1}).
\label{tab:hadme4Q}}
\end{table}

\begin{table}[t]
\centering \renewcommand{\arraystretch}{1.5}
\begin{tabular}{||c|c|c|c||}
\hline\hline
Form Factor &  Method & Value & Remarks\\
\hline
 $g_S^{(0)}$ & Lattice QCD & $6.3\pm 0.8$ & Eq.~(\ref{eq:gs01}) and \cite{Aoki:2008sm,Shanahan:2012wh}\\
$g_S^{(1)}$ & Lattice QCD & $0.45\pm 0.15$ & Eq.~(\ref{eq:gs01}) and \cite{Beane:2006fk} \\
$g_S^{(1)}$ & Lattice QCD & 0.4 (2) & Isospin and \cite{Bhattacharya:2011qm} \\
$g_P^{(0)}$ & Chiral &  &\\
$g_P^{(1)}$ & Chiral &  &\\
$g_T^{(0)}$ & Lattice QCD &  &\\
$g_T^{(1)}$ & Lattice QCD & 0.53(18) & Isospin and \cite{Bhattacharya:2011qm}\\
\hline\hline
\end{tabular}
\caption{Form factors entering semileptonic CPV interactions.
\label{tab:hadme4}}
\end{table}

\begin{table}[t] 
\centering
\renewcommand{\arraystretch}{1.5}
\btb{||c|c||c|c||c||} 
\hline
System & Present 90 \% C.L.  & Sensitivity Goal$^b$ & Group & SM CKM ($e$ fm)$^c$ \\
& Limit ($e$ fm)$^a$ & & & \\
\hline \hline
Cs & $1.2\times 10^{-10}$ & & \cite{Murthy:1989zz} & $\sim 10^{-23}$\\
Tl & $9.5\times 10^{-12}$ & &\cite{Regan:2002ta} & $\sim 10^{-22}$\\
YbF$^d$ & $10.5\times 10^{-15}$ & & \cite{Hudson:2011zz} & $\sim 10^{-19}$\\
ThO$^d$ & - & $10^{-15}\to 10^{-17}$ & & \\
\hline
$n$ & $2.7\times 10^{-13}$& & \cite{Baker:2006ts} & $1.6\times 10^{-18}\to 1.4\times 10^{-20}$\\
$n$ & & $(1-3)\times 10^{-14}$ & CryoEDM & \\
$n$ &  & $4\times 10^{-15}$ & nEDM/SNS  & \\
$n$ & & $5\times 10^{-14}$& nEDM/PSI & \\
$n$ & & $5\times 10^{-15}$& n2EDM/PSI & \\
$n$ &  & $2\times 10^{-15}$ & nedm/FRM-II Munich & \\
$n$ & &  $10^{-14}-10^{-15}$  & TRIUMF & \\
\hline
$p$ && $10^{-16}$ & srEDM &\\
\hline
$^{199}$Hg & $2.6\times 10^{-16}$  & $(2.6 - 5)\times 10^{-17}$ & \cite{Griffith:2009zz} &  - \\
$^{225}$Ra &  &  $(10-100)\times 10^{-15}$ & Argonne & - \\
$^{225}$Rn &   &   $1.3\times 10^{-14}$ & TRIUMF & - \\
$^{225}$Rn &     & $2\times 10^{-15}$ & FRIB & - \\
$^{223}$Xe &  $5.5\times 10^{-14}$ & & \cite{Player:1970zz}  & - \\
\hline \hline
\etb
\caption[ . ]{Present EDM limits and  sensitivity goals
for the paramagnetic atoms and molecules (first group); nucleons ($n$,$p$) (second group); and diamagnetic atoms 
(third group). A limit on the electron EDM of $10.5\times 10^{-15}$ $e$ fm (90\% C.L.)  has been derived from the most recent
YbF experiment \cite{Hudson:2011zz} assuming it is the only source of the molecular EDM (see Section \ref{sec:atom} ). Also listed are the expected magnitudes of the SM ``background''
due to the phase in the CKM matrix. $^a$ We thank
T. Chupp for providing the 90\% C.L. limits from existing searches. $^b$ All sensitivity
goals are self-reported by members of the given collaboration. $^c$  We do not quote SM CKM predictions for diamagnetic atoms, due to the incorrect
chiral implementation of chiral symmetry in Ref.~\cite{Flambaum:1984fb} as pointed out in Ref.~\cite{Donoghue:1987dd}. $^d$ Molecular sensitivity expressed in terms of limit on $d_e$ rather than on $d_A$.}
\label{tab:edmexp}
\end{table}

\begin{table}[t]
\centering
\renewcommand{\arraystretch}{1.5}
\begin{tabular}{||c|c|c|c||}
\hline\hline
Atom & $k_S^{(0)}$ &   $k_P^{(0)}$ & $k_T^{(0)}$ \\
 & $10^{-5}$ $e$ fm &   $10^{-10} e$ fm &  $10^{-7}$ $e$ fm \\
\hline
$^{205}$Tl & $-7.0\pm 0.3$ & -1.5 & -0.5   \\
$^{133}$Cs & $-0.78\pm 0.2$ &  -2.2  & -0.92    \\
$^{85}$Rb & $-0.110\pm0.003$ &  &    \\
$^{210}$Fr & $-10.9\pm 1.7$ &   &     \\
\hline
$^{199}$Hg & $-8.1\times 10^{-4}$ &  6 & 4  \\
\hline\hline
Molecule & kHz & kHz & kHz\\
\hline
YbF & $-92\pm 9$ & & \\
ThO & $-564\pm 56$ & & \\
\hline\hline
\end{tabular}
\caption{Dependence of atomic and molecular EDMs on T- and P-odd semileptonic interactions.
\label{tab:atom2}}
\end{table}



\begin{thebibliography}{100}

\bibitem{Morrissey:2012db}
David~E. Morrissey and Michael~J. Ramsey-Musolf.
\newblock {Electroweak Baryogenesis}.
\newblock {\em New J.\ Phys.}, 14:125003, 2012,
  \doi{10.1088/1367-2630/14/12/125003}, \eprint{arXiv}{1206.2942}.

\bibitem{Dine:2003ax}
Michael Dine and Alexander Kusenko.
\newblock {The Origin of the Matter - Antimatter Asymmetry}.
\newblock {\em Rev.\ Mod.\ Phys.}, 76:1, 2003, \doi{10.1103/RevModPhys.76.1},
  \eprint{arXiv}{hep-ph/0303065}.

\bibitem{Riotto:1999yt}
Antonio Riotto and Mark Trodden.
\newblock {Recent Progress in Baryogenesis}.
\newblock {\em Ann.\ Rev.\ Nucl.\ Part.\ Sci.}, 49:35\unskip--\ignorespaces 75,
  1999, \doi{10.1146/annurev.nucl.49.1.35}, \eprint{arXiv}{hep-ph/9901362}.

\bibitem{Kim:2008hd}
Jihn~E. Kim and Gianpaolo Carosi.
\newblock {Axions and the Strong CP Problem}.
\newblock {\em Rev. Mod. Phys.}, 82:557\unskip--\ignorespaces 602, 2010,
  \doi{10.1103/RevModPhys.82.557}, \eprint{arXiv}{0807.3125}.

\bibitem{Shabalin:1978rs}
E.P. Shabalin.
\newblock {Electric Dipole Moment of Quark in a Gauge Theory with Left-Handed
  Currents}.
\newblock {\em Sov.\ J.\ Nucl.\ Phys.}, 28:75, 1978.

\bibitem{Shabalin:1982sg}
E.P. Shabalin.
\newblock {The Electric Dipole Moment of the Neutron in a Gauge Theory}.
\newblock {\em Sov.\ Phys.\ Usp.}, 26:297, 1983.

\bibitem{Bernreuther:1990jx}
Werner Bernreuther and Mahiko Suzuki.
\newblock {The Electric Dipole Moment of the Electron}.
\newblock {\em Rev. Mod. Phys.}, 63:313\unskip--\ignorespaces 340, 1991,
  \doi{10.1103/RevModPhys.63.313}.

\bibitem{Chupp:2008zz}
Tim Chupp.
\newblock {Nuclear and Atomic Electric Dipole Moments}.
\newblock {\em Nucl.\ Phys.\ A}, 827:428c\unskip--\ignorespaces 435c, 2009,
  \doi{10.1016/j.nuclphysa.2009.05.006}.

\bibitem{Chupp:2013}
Tim Chupp and Michael~J. Ramsey-Musolf.
\newblock In preparation.

\bibitem{PPNPIntro}
Vincenzo Cirigliano and Michael Ramsey-Musolf.
\newblock In preparation.

\bibitem{Dzuba:2012bh}
V.A. Dzuba and V.V. Flambaum.
\newblock {Parity Violation and Electric Dipole Moments in Atoms and
  Molecules}.
\newblock {\em Int.\ J.\ Mod.\ Phys.}, E21:1230010, 2012,
  \doi{10.1142/S021830131230010X}, \eprint{arXiv}{1209.2200}.

\bibitem{Ellis:2008zy}
J.~R. Ellis, J.~S. Lee, and A.~Pilaftsis.
\newblock {Electric Dipole Moments in the MSSM Reloaded}.
\newblock {\em JHEP}, 0810:049, 2008.

\bibitem{Ginges:2003qt}
J.S.M. Ginges and V.V. Flambaum.
\newblock {Violations of Fundamental Symmetries in Atoms and Tests of
  Unification Theories of Elementary Particles}.
\newblock {\em Phys. Rept.}, 397:63\unskip--\ignorespaces 154, 2004,
  \doi{10.1016/j.physrep.2004.03.005}, \eprint{arXiv}{physics/0309054}.

\bibitem{Pospelov:2005pr}
Maxim Pospelov and Adam Ritz.
\newblock {Electric Dipole Moments as Probes of New Physics}.
\newblock {\em Ann.\ Phys.}, 318:119\unskip--\ignorespaces 169, 2005,
  \doi{10.1016/j.aop.2005.04.002}, \eprint{arXiv}{hep-ph/0504231}.

\bibitem{Kobayashi:1973fv}
Makoto Kobayashi and Toshihide Maskawa.
\newblock {CP Violation in the Renormalizable Theory of Weak Interaction}.
\newblock {\em Prog.\ Theor.\ Phys.}, 49:652\unskip--\ignorespaces 657, 1973,
  \doi{10.1143/PTP.49.652}.

\bibitem{'tHooft:1976up}
Gerard 't~Hooft.
\newblock {Symmetry Breaking Through Bell-Jackiw Anomalies}.
\newblock {\em Phys.\ Rev.\ Lett.}, 37:8\unskip--\ignorespaces 11, 1976,
  \doi{10.1103/PhysRevLett.37.8}.

\bibitem{Jackiw:1976pf}
R.~Jackiw and C.~Rebbi.
\newblock {Vacuum Periodicity in a Yang-Mills Quantum Theory}.
\newblock {\em Phys.\ Rev.\ Lett.}, 37:172\unskip--\ignorespaces 175, 1976,
  \doi{10.1103/PhysRevLett.37.172}.

\bibitem{Callan:1976je}
Jr. Callan, Curtis~G., R.F. Dashen, and David~J. Gross.
\newblock {The Structure of the Gauge Theory Vacuum}.
\newblock {\em Phys.\ Lett.\ B}, 63:334\unskip--\ignorespaces 340, 1976,
  \doi{10.1016/0370-2693(76)90277-X}.

\bibitem{Grzadkowski:2010es}
B.~Grzadkowski, M.~Iskrzynski, M.~Misiak, and J.~Rosiek.
\newblock {Dimension-Six Terms in the Standard Model Lagrangian}.
\newblock {\em JHEP}, 1010:085, 2010, \doi{10.1007/JHEP10(2010)085},
  \eprint{arXiv}{1008.4884}.

\bibitem{Baluni:1978rf}
Varouzhan Baluni.
\newblock {CP Violating Effects in QCD}.
\newblock {\em Phys.\ Rev.\ D}, 19:2227\unskip--\ignorespaces 2230, 1979,
  \doi{10.1103/PhysRevD.19.2227}.

\bibitem{Beringer:1900zz}
J.~Beringer et~al., Particle Data Group.
\newblock {Review of Particle Physics (RPP)}.
\newblock {\em Phys.\ Rev.\ D}, 86:010001, 2012,
  \doi{10.1103/PhysRevD.86.010001}.

\bibitem{deVries:2012ab}
J.~de~Vries, E.~Mereghetti, R.G.E. Timmermans, and U.~van Kolck.
\newblock {The Effective Chiral Lagrangian From Dimension-Six Parity and
  Time-Reversal Violation}.
\newblock 2012, \eprint{arXiv}{1212.0990}.

\bibitem{Ibrahim:1997gj}
Tarek Ibrahim and Pran Nath.
\newblock {The Neutron and the Electron Electric Dipole Moment in N=1
  Supergravity Unification}.
\newblock {\em Phys.\ Rev.\ D}, 57:478\unskip--\ignorespaces 488, 1998,
  \doi{10.1103/PhysRevD.58.019901, 10.1103/PhysRevD.60.079903,
  10.1103/PhysRevD.60.119901, 10.1103/PhysRevD.57.478},
  \eprint{arXiv}{hep-ph/9708456}.

\bibitem{Ibrahim:1998je}
Tarek Ibrahim and Pran Nath.
\newblock {The Neutron and the Lepton EDMs in MSSM, Large CP Violating Phases,
  and the Cancellation Mechanism}.
\newblock {\em Phys.\ Rev.\ D}, 58:111301, 1998,
  \doi{10.1103/PhysRevD.60.099902, 10.1103/PhysRevD.58.111301},
  \eprint{arXiv}{hep-ph/9807501}.

\bibitem{Weinberg:1989dx}
Steven Weinberg.
\newblock {Larger Higgs Exchange Terms in the Neutron Electric Dipole Moment}.
\newblock {\em Phys.\ Rev.\ Lett.}, 63:2333, 1989,
  \doi{10.1103/PhysRevLett.63.2333}.

\bibitem{RamseyMusolf:2006vr}
M.J. Ramsey-Musolf and S.~Su.
\newblock {Low Energy Precision Test of Supersymmetry}.
\newblock {\em Phys.\ Rept.}, 456:1\unskip--\ignorespaces 88, 2008,
  \doi{10.1016/j.physrep.2007.10.001}, \eprint{arXiv}{hep-ph/0612057}.

\bibitem{Herczeg:1997ei}
P.~Herczeg.
\newblock {Time Reversal Violation in Nuclear Processes}.
\newblock In W.C. Haxton and E.M. Henley, editors, {\em {Symmetries and
  Fundamental Interactions in Nuclei}}, pages 89\unskip--\ignorespaces 125.
  World Scientific, 1997.

\bibitem{Zhang:2007da}
Yue Zhang, Haipeng An, Xiangdong Ji, and Rabindra~N. Mohapatra.
\newblock {General CP Violation in Minimal Left-Right Symmetric Model and
  Constraints on the Right-Handed Scale}.
\newblock {\em Nucl.\ Phys.\ B}, 802:247\unskip--\ignorespaces 279, 2008,
  \doi{10.1016/j.nuclphysb.2008.05.019}, \eprint{arXiv}{0712.4218}.

\bibitem{Xu:2009nt}
Fanrong Xu, Haipeng An, and Xiangdong Ji.
\newblock {Neutron Electric Dipole Moment Constraint on Scale of Minimal Left-
  Right Symmetric Model}.
\newblock {\em JHEP}, 1003:088, 2010, \doi{10.1007/JHEP03(2010)088},
  \eprint{arXiv}{0910.2265}.

\bibitem{Ng:2011ui}
John Ng and Sean Tulin.
\newblock {D versus d: CP Violation in Beta Decay and Electric Dipole Moments}.
\newblock {\em Phys.\ Rev.\ D}, 85:033001, 2012,
  \doi{10.1103/PhysRevD.85.033001}, \eprint{arXiv}{1111.0649}.

\bibitem{Dekens13}
W.~Dekens and J.~de~Vries.
\newblock In preparation.

\bibitem{Leutwyler:2009jg}
H.~Leutwyler.
\newblock {Light Quark Masses}.
\newblock {\em PoS}, CD09:005, 2009, \eprint{arXiv}{0911.1416}.

\bibitem{Cheung:2007bu}
Clifford Cheung, A.~Liam Fitzpatrick, and Lisa Randall.
\newblock {Sequestering CP Violation and GIM-Violation with Warped Extra
  Dimensions}.
\newblock {\em JHEP}, 0801:069, 2008, \doi{10.1088/1126-6708/2008/01/069},
  \eprint{arXiv}{0711.4421}.

\bibitem{Bigi:1991rh}
Ikaros~I.Y. Bigi and N.G. Uraltsev.
\newblock {Effective Gluon Operators and the Dipole Moment of the Neutron}.
\newblock {\em Sov. Phys. JETP}, 73:198\unskip--\ignorespaces 210, 1991.

\bibitem{Bernard:2007zu}
Veronique Bernard.
\newblock {Chiral Perturbation Theory and Baryon Properties}.
\newblock {\em Prog. Part. Nucl. Phys.}, 60:82\unskip--\ignorespaces 160, 2008,
  \doi{10.1016/j.ppnp.2007.07.001}, \eprint{arXiv}{0706.0312}.

\bibitem{Mereghetti:2010tp}
E.~Mereghetti, W.H. Hockings, and U.~van Kolck.
\newblock {The Effective Chiral Lagrangian From the Theta Term}.
\newblock {\em Ann.\ Phys.}, 325:2363\unskip--\ignorespaces 2409, 2010,
  \doi{10.1016/j.aop.2010.03.005}, \eprint{arXiv}{1002.2391}.

\bibitem{Maekawa:2011vs}
C.M. Maekawa, E.~Mereghetti, J.~de~Vries, and U.~van Kolck.
\newblock {The Time-Reversal- and Parity-Violating Nuclear Potential in Chiral
  Effective Theory}.
\newblock {\em Nucl.\ Phys.\ A}, 872:117\unskip--\ignorespaces 160, 2011,
  \doi{10.1016/j.nuclphysa.2011.09.020}, \eprint{arXiv}{1106.6119}.

\bibitem{Barton:1961eg}
G.~Barton.
\newblock {Notes on the Static Parity Nonconserving Internucleon Potential}.
\newblock {\em Nuovo Cim.}, 19:512\unskip--\ignorespaces 527, 1961,
  \doi{10.1007/BF02733247}.

\bibitem{jesus05}
J.H. de~Jesus and J.~Engel.
\newblock {Time-Reversal-Violating Schiff Moment of $^{199}$Hg}.
\newblock {\em Phys.\ Rev.\ C}, 72:045503, 2005.

\bibitem{dobaczewski05}
J.~Dobaczewski and J.~Engel.
\newblock {Nuclear Time-Reversal Violation and the Schiff Moment of
  $^{225}$Ra}.
\newblock {\em Phys.\ Rev.\ Lett.}, 94:232502, 2005.

\bibitem{dmitriev03}
V.~F. Dmitriev and R.~A. Sen'kov.
\newblock {P- and T-Violating Schiff Moment of the Mercury Nucleus}.
\newblock {\em Phys.\ Atom.\ Nucl.}, 66:1940, 2003.

\bibitem{dmitriev05}
V.~F. Dmitriev, R.~A. Sen'kov, and N.~Auerbach.
\newblock {Effects of Core Polarization on the Nuclear Schiff Moment}.
\newblock {\em Phys.\ Rev.\ C}, 71:035501, 2005.

\bibitem{Stoks:1992ja}
Vincent~G.J. Stoks, Rob Timmermans, and J.J. de~Swart.
\newblock {On the Pion - Nucleon Coupling Constant}.
\newblock {\em Phys.\ Rev.\ C}, 47:512\unskip--\ignorespaces 520, 1993,
  \doi{10.1103/PhysRevC.47.512}, \eprint{arXiv}{nucl-th/9211007}.

\bibitem{vanKolck:1996rm}
U.~van Kolck, James~Lewis Friar, and J.~Terrance Goldman.
\newblock {Phenomenological Aspects of Isospin Violation in the Nuclear Force}.
\newblock {\em Phys.\ Lett.\ B}, 371:169\unskip--\ignorespaces 174, 1996,
  \doi{10.1016/0370-2693(96)00009-3}, \eprint{arXiv}{nucl-th/9601009}.

\bibitem{Hisano:2012cc}
Junji Hisano, Koji Tsumura, and Masaki~J.S. Yang.
\newblock {QCD Corrections to Neutron Electric Dipole Moment from Dimension-Six
  Four-Quark Operators}.
\newblock {\em Phys.\ Lett.\ B}, 713:473\unskip--\ignorespaces 480, 2012,
  \doi{10.1016/j.physletb.2012.06.038}, \eprint{arXiv}{1205.2212}.

\bibitem{Morozov:1985ef}
A.~Yu. Morozov.
\newblock {Matrix of Mixing OF Scalar and Vector Mesons oF Dimension D $\leq$ 8
  in QCD. (in Russian)}.
\newblock {\em Sov.\ J.\ Nucl.\ Phys.}, 40:505, 1984.

\bibitem{Chang:1990jv}
D.~Chang, Wai-Yee Keung, C.S. Li, and T.C. Yuan.
\newblock {QCD Corrections to CP Violation from Color Electric Dipole Moment of
  b Quark}.
\newblock {\em Phys.\ Lett.\ B}, 241:589, 1990,
  \doi{10.1016/0370-2693(90)91875-C}.

\bibitem{Arnowitt:1990eh}
Richard~L. Arnowitt, Jorge~L. Lopez, and Dimitri~V. Nanopoulos.
\newblock {Keeping the Demon of SUSY at Bay}.
\newblock {\em Phys.\ Rev.}, D42:2423\unskip--\ignorespaces 2426, 1990,
  \doi{10.1103/PhysRevD.42.2423}.

\bibitem{Braaten:1990gq}
Eric Braaten, Chong-Sheng Li, and Tzu-Chiang Yuan.
\newblock {The Evolution of Weinberg's Gluonic CP Violation Operator}.
\newblock {\em Phys.\ Rev.\ Lett.}, 64:1709, 1990,
  \doi{10.1103/PhysRevLett.64.1709}.

\bibitem{Bsaisou:2012rg}
J.~Bsaisou, C.~Hanhart, S.~Liebig, U.-G. Meissner, A.~Nogga, et~al.
\newblock {The Electric Dipole Moment of the Deuteron from the QCD
  $\theta$-Term}.
\newblock 2012, \eprint{arXiv}{1209.6306}.

\bibitem{Guo:2012vf}
Feng-Kun Guo and Ulf-G. Meissner.
\newblock {Baryon Electric Dipole Moments from Strong CP Violation}.
\newblock {\em JHEP}, 1212:097, 2012, \doi{10.1007/JHEP12(2012)097},
  \eprint{arXiv}{1210.5887}.

\bibitem{Manohar:1983md}
Aneesh Manohar and Howard Georgi.
\newblock {Chiral Quarks and the Nonrelativistic Quark Model}.
\newblock {\em Nucl.\ Phys.\ B}, 234:189, 1984,
  \doi{10.1016/0550-3213(84)90231-1}.

\bibitem{Aoki:2008sm}
S.~Aoki et~al., PACS-CS Collaboration.
\newblock {2+1 Flavor Lattice QCD toward the Physical Point}.
\newblock {\em Phys.\ Rev.\ D}, 79:034503, 2009,
  \doi{10.1103/PhysRevD.79.034503}, \eprint{arXiv}{0807.1661}.

\bibitem{Shanahan:2012wh}
P.E. Shanahan, A.W. Thomas, and R.D. Young.
\newblock {Sigma Terms from an SU(3) Chiral Extrapolation}.
\newblock 2012, \eprint{arXiv}{1205.5365}.

\bibitem{Gasser:1990ce}
J.~Gasser, H.~Leutwyler, and M.E. Sainio.
\newblock {Sigma Term Update}.
\newblock {\em Phys.\ Lett.\ B}, 253:252\unskip--\ignorespaces 259, 1991,
  \doi{10.1016/0370-2693(91)91393-A}.

\bibitem{Baru:2011bw}
V.~Baru, C.~Hanhart, M.~Hoferichter, B.~Kubis, A.~Nogga, et~al.
\newblock {Precision Calculation of Threshold Pi$^-$ d Scattering, Pi N
  Scattering Lengths, and the GMO Sum Rule}.
\newblock {\em Nucl.\ Phys.\ A}, 872:69\unskip--\ignorespaces 116, 2011,
  \doi{10.1016/j.nuclphysa.2011.09.015}, \eprint{arXiv}{1107.5509}.

\bibitem{Hoferichter:2012tu}
Martin Hoferichter, Christoph Ditsche, Bastian Kubis, and Ulf-G. Meissner.
\newblock {Improved Dispersive Analysis of the Scalar Form Factor of the
  Nucleon}.
\newblock 2012, \eprint{arXiv}{1211.1485}.

\bibitem{Beane:2006fk}
Silas~R. Beane, Kostas Orginos, and Martin~J. Savage.
\newblock {Strong-isospin Violation in the Neutron-Proton Mass Difference from
  Fully-Dynamical Lattice QCD and PQQCD}.
\newblock {\em Nucl. Phys. B}, 768:38\unskip--\ignorespaces 50, 2007,
  \doi{10.1016/j.nuclphysb.2006.12.023}, \eprint{arXiv}{hep-lat/0605014}.

\bibitem{WalkerLoud:2010qq}
Andre Walker-Loud.
\newblock {Towards a Direct Lattice Calculation of $m_d - m_u$}.
\newblock {\em PoS}, LATTICE2010:243, 2010, \eprint{arXiv}{1011.4015}.

\bibitem{WalkerLoud:2012bg}
Andre Walker-Loud, Carl~E. Carlson, and Gerald~A. Miller.
\newblock {The Electromagnetic Self-Energy Contribution to $M_p - M_n$ and the
  Isovector Nucleon Magnetic Polarizability}.
\newblock {\em Phys.\ Rev.\ Lett.}, 108:232301, 2012,
  \doi{10.1103/PhysRevLett.108.232301}, \eprint{arXiv}{1203.0254}.

\bibitem{Filin:2009yh}
A.~Filin, V.~Baru, E.~Epelbaum, J.~Haidenbauer, C.~Hanhart, et~al.
\newblock {Extraction of the Strong Neutron-Proton Mass Difference from the
  Charge Symmetry Breaking In pn ---\&gt; d Pi0}.
\newblock {\em Phys. Lett.\ B}, 681:423\unskip--\ignorespaces 427, 2009,
  \doi{10.1016/j.physletb.2009.10.069}, \eprint{arXiv}{0907.4671}.

\bibitem{Anselm:1985cf}
A.~A. Anselm, V.~E. Bunakov, V.~P. Gudkov, and N.~G. Uraltsev.
\newblock {On the Neutron Electric Dipole Moment in the Weinberg CP Violation
  Model}.
\newblock {\em Phys.\ Lett.\ B}, 152:116, 1985.

\bibitem{Bhattacharya:2011qm}
Tanmoy Bhattacharya, Vincenzo Cirigliano, Saul~D. Cohen, Alberto Filipuzzi,
  Martin Gonzalez-Alonso, et~al.
\newblock {Probing Novel Scalar and Tensor Interactions from (Ultra)Cold
  Neutrons to the LHC}.
\newblock {\em Phys.\ Rev.\ D}, 85:054512, 2012,
  \doi{10.1103/PhysRevD.85.054512}, \eprint{arXiv}{1110.6448}.

\bibitem{Crewther:1979pi}
R.J. Crewther, P.~Di~Vecchia, G.~Veneziano, and Edward Witten.
\newblock {Chiral Estimate of the Electric Dipole Moment of the Neutron in
  Quantum Chromodynamics}.
\newblock {\em Phys.\ Lett.\ B}, 88:123, 1979,
  \doi{10.1016/0370-2693(79)90128-X}.

\bibitem{Thomas:1994wi}
Scott~D. Thomas.
\newblock {Electromagnetic Contributions to the Schiff Moment}.
\newblock {\em Phys.\ Rev.\ D}, 51:3955\unskip--\ignorespaces 3957, 1995,
  \doi{10.1103/PhysRevD.51.3955}, \eprint{arXiv}{hep-ph/9402237}.

\bibitem{Hockings:2005cn}
W.H. Hockings and U.~van Kolck.
\newblock {The Electric Dipole Form Factor of the Nucleon}.
\newblock {\em Phys.\ Lett.\ B}, 605:273\unskip--\ignorespaces 278, 2005,
  \doi{10.1016/j.physletb.2004.11.043}, \eprint{arXiv}{nucl-th/0508012}.

\bibitem{deVries:2010ah}
J.~de~Vries, R.G.E. Timmermans, E.~Mereghetti, and U.~van Kolck.
\newblock {The Nucleon Electric Dipole Form Factor From Dimension-Six
  Time-Reversal Violation}.
\newblock {\em Phys.\ Lett.\ B}, 695:268\unskip--\ignorespaces 274, 2011,
  \doi{10.1016/j.physletb.2010.11.042}, \eprint{arXiv}{1006.2304}.

\bibitem{Narison:2008jp}
Stephan Narison.
\newblock {A Fresh Look into the Neutron EDM and Magnetic Susceptibility}.
\newblock {\em Phys.\ Lett.\ B}, 666:455\unskip--\ignorespaces 461, 2008,
  \doi{10.1016/j.physletb.2008.07.083}, \eprint{arXiv}{0806.2618}.

\bibitem{Ottnad:2009jw}
K.~Ottnad, B.~Kubis, U.-G. Meissner, and F.-K. Guo.
\newblock {New Insights into the Neutron Electric Dipole Moment}.
\newblock {\em Phys.\ Lett.\ B}, 687:42\unskip--\ignorespaces 47, 2010,
  \doi{10.1016/j.physletb.2010.03.005}, \eprint{arXiv}{0911.3981}.

\bibitem{Mereghetti:2010kp}
E.~Mereghetti, J.~de~Vries, W.H. Hockings, C.M. Maekawa, and U.~van Kolck.
\newblock {The Electric Dipole Form Factor of the Nucleon in Chiral
  Perturbation Theory to Sub-leading Order}.
\newblock {\em Phys.\ Lett.\ B}, 696:97\unskip--\ignorespaces 102, 2011,
  \doi{10.1016/j.physletb.2010.12.018}, \eprint{arXiv}{1010.4078}.

\bibitem{Cheng:1990pi}
Hai-Yang Cheng.
\newblock {Reanalysis of Strong CP Violating Effects in Chiral Perturbation
  Theory}.
\newblock {\em Phys.\ Rev.\ D}, 44:166\unskip--\ignorespaces 174, 1991,
  \doi{10.1103/PhysRevD.44.166}.

\bibitem{Pich:1991fq}
Antonio Pich and Eduardo de~Rafael.
\newblock {Strong CP Violation in an Effective Chiral Lagrangian Approach}.
\newblock {\em Nucl.\ Phys.\ B}, 367:313\unskip--\ignorespaces 333, 1991,
  \doi{10.1016/0550-3213(91)90019-T}.

\bibitem{Cho:1992rv}
Peter~L. Cho.
\newblock {Chiral Estimates of Strong CP Violation Revisited}.
\newblock {\em Phys.\ Rev.\ D}, 48:3304\unskip--\ignorespaces 3309, 1993,
  \doi{10.1103/PhysRevD.48.3304}, \eprint{arXiv}{hep-ph/9212274}.

\bibitem{Borasoy:2000pq}
B.~Borasoy.
\newblock {The Electric Dipole Moment of the Neutron in Chiral Perturbation
  Theory}.
\newblock {\em Phys.\ Rev.\ D}, 61:114017, 2000,
  \doi{10.1103/PhysRevD.61.114017}, \eprint{arXiv}{hep-ph/0004011}.

\bibitem{Aoki:1989rx}
S.~Aoki and A.~Gocksch.
\newblock {The Neutron Electric Dipole Moment in Lattice QCD}.
\newblock {\em Phys.\ Rev.\ Lett.}, 63:1125, 1989.

\bibitem{Shintani:2005xg}
E.~Shintani {\it et al.}
\newblock {Neutron Electric Dipole Moment from Lattice QCD}.
\newblock {\em Phys.\ Rev.\ D}, 72:014504, 2005.

\bibitem{Berruto:2005hg}
F.~Berruto, T.~Blum, K.~Orginos, and A.~Soni.
\newblock {Calculation of the Neutron Electric Dipole Moment with Two Dynamical
  Flavors of Domain Wall Fermions}.
\newblock {\em Phys.\ Rev.\ D}, 73:054509, 2006.

\bibitem{Shintani:2006xr}
E.~Shintani {\it et al.}
\newblock {Neutron Electric Dipole Moment with External Electric Field Method
  in Lattice QCD}.
\newblock {\em Phys.\ Rev.\ D}, 75:034507, 2007.

\bibitem{Shintani:2008nt}
E.~Shintani, S.~Aoki, and Y.~Kuramashi.
\newblock {Full QCD Calculation of Neutron Electric Dipole Moment with the
  External Electric Field Method}.
\newblock {\em Phys.\ Rev.\ D}, 78:014503, 2008.

\bibitem{Aoki:2008gv}
S.~Aoki {\it et al.}
\newblock {The Electric Dipole Moment of the Nucleon from Simulations at
  Imaginary Vacuum Angle Theta}, \eprint{}{arXiv:0808.1428 [hep-lat]}.

\bibitem{Shintani:2009}
E.~Shintani, S.~Aoki, and Y.~Kuramashi.

\bibitem{Hisano:2012sc}
J.~Hisano, J.~Y. Lee, N.~Nagata, and Y.~Shimizu.
\newblock {Reevaluation of Neutron Electric Dipole Moment with QCD Sum Rules}.
\newblock {\em Phys.\ Rev.\ D}, 85:114044, 2012, \eprint{arXiv}{1204.2653}.

\bibitem{Pospelov:1999mv}
Maxim Pospelov and Adam Ritz.
\newblock {Theta Vacua, QCD Sum Rules, and the Neutron Electric Dipole Moment}.
\newblock {\em Nucl.\ Phys.\ B}, 573:177\unskip--\ignorespaces 200, 2000,
  \doi{10.1016/S0550-3213(99)00817-2}, \eprint{arXiv}{hep-ph/9908508}.

\bibitem{Pospelov:1999ha}
Maxim Pospelov and Adam Ritz.
\newblock {Theta Induced Electric Dipole Moment of the Neutron via QCD Sum
  Rules}.
\newblock {\em Phys.\ Rev.\ Lett.}, 83:2526\unskip--\ignorespaces 2529, 1999,
  \doi{10.1103/PhysRevLett.83.2526}, \eprint{arXiv}{hep-ph/9904483}.

\bibitem{Pospelov:2000bw}
Maxim Pospelov and Adam Ritz.
\newblock {Neutron EDM from Electric and Chromoelectric Dipole Moments of
  Quarks}.
\newblock {\em Phys.\ Rev.\ D}, 63:073015, 2001,
  \doi{10.1103/PhysRevD.63.073015}, \eprint{arXiv}{hep-ph/0010037}.

\bibitem{Pospelov:2001ys}
Maxim Pospelov.
\newblock {Best Values for the CP Odd Meson Nucleon Couplings from
  Supersymmetry}.
\newblock {\em Phys.\ Lett.\ B}, 530:123\unskip--\ignorespaces 128, 2002,
  \doi{10.1016/S0370-2693(02)01263-7}, \eprint{arXiv}{hep-ph/0109044}.

\bibitem{Demir:2002gg}
Durmus~A. Demir, Maxim Pospelov, and Adam Ritz.
\newblock {Hadronic EDMs, the Weinberg Operator, and Light Gluinos}.
\newblock {\em Phys.\ Rev.\ D}, 67:015007, 2003,
  \doi{10.1103/PhysRevD.67.015007}, \eprint{arXiv}{hep-ph/0208257}.

\bibitem{Ellis:1996dg}
John~R. Ellis and Ricardo~A. Flores.
\newblock {Implications of the Strange Spin of the Nucleon for the Neutron
  Electric Dipole Moment in Supersymmetric Theories}.
\newblock {\em Phys.\ Lett.\ B}, 377:83\unskip--\ignorespaces 88, 1996,
  \doi{10.1016/0370-2693(96)00312-7}, \eprint{arXiv}{hep-ph/9602211}.

\bibitem{GHM93}
V.P. Gudkov, X.-G. He, and B.H.J. McKellar.
\newblock {CP-odd Nucleon Potential}.
\newblock {\em Phys.\ Rev.\ C}, 47:2365, 1993.

\bibitem{TH94}
I.S. Towner and A.C. Hayes.
\newblock {P,T-Violating Nuclear Matrix Elements in the One-Meson Exchange
  Approximation}.
\newblock {\em Phys.\ Rev.\ C}, 49:2391, 1994.

\bibitem{Tim+04}
C.-P. Liu and R.G.E. Timmermans.
\newblock {P- and T-Odd Two-Nucleon Interaction and the Deuteron Electric
  Dipole Moment}.
\newblock {\em Phys.\ Rev.\ C}, 70:055501, 2004.

\bibitem{Stoks:1994wp}
V.G.J. Stoks, R.A.M. Klomp, C.P.F. Terheggen, and J.J. de~Swart.
\newblock {Construction of High Quality N N Potential Models}.
\newblock {\em Phys.\ Rev.\ C}, 49:2950\unskip--\ignorespaces 2962, 1994,
  \doi{10.1103/PhysRevC.49.2950}, \eprint{arXiv}{nucl-th/9406039}.

\bibitem{Machleidt:2000ge}
R.~Machleidt.
\newblock {The High Precision, Charge Dependent Bonn Nucleon-Nucleon Potential
  (CD-Bonn)}.
\newblock {\em Phys.\ Rev.\ C}, 63:024001, 2001,
  \doi{10.1103/PhysRevC.63.024001}, \eprint{arXiv}{nucl-th/0006014}.

\bibitem{Tiator:1994et}
L.~Tiator, C.~Bennhold, and S.S. Kamalov.
\newblock {The Eta N N Coupling in Eta Photoproduction}.
\newblock {\em Nucl.\ Phys.\ A}, 580:455\unskip--\ignorespaces 474, 1994,
  \doi{10.1016/0375-9474(94)90909-1}, \eprint{arXiv}{nucl-th/9404013}.

\bibitem{Pitschmann:2012by}
Mario Pitschmann, Chien-Yeah Seng, Michael~J. Ramsey-Musolf, Craig~D. Roberts,
  Sebastian~M. Schmidt, et~al.
\newblock {Electric Dipole Moment of the Rho-Meson}.
\newblock {\em Phys.\ Rev.\ C}, 87:015205, 2013,
  \doi{10.1103/PhysRevC.87.015205}, \eprint{arXiv}{1209.4352}.

\bibitem{Pospelov:1999rg}
Maxim Pospelov and Adam Ritz.
\newblock {Hadron Electric Dipole Moments from CP Odd Operators of Dimension
  Five via QCD Sum Rules: The Vector Meson}.
\newblock {\em Phys.\ Lett.\ B}, 471:388\unskip--\ignorespaces 395, 2000,
  \doi{10.1016/S0370-2693(99)01343-X}, \eprint{arXiv}{hep-ph/9910273}.

\bibitem{An:2009zh}
Haipeng An, Xiangdong Ji, and Fanrong Xu.
\newblock {P-odd and CP-odd Four-Quark Contributions to Neutron EDM}.
\newblock {\em JHEP}, 1002:043, 2010, \doi{10.1007/JHEP02(2010)043},
  \eprint{arXiv}{0908.2420}.

\bibitem{Onderwater:2012me}
C.J.G. Onderwater.
\newblock {Search for Electric Dipole Moments at Storage Rings}.
\newblock 2012, \doi{10.1007/s10751-012-0584-9}, \eprint{arXiv}{1204.2512}.

\bibitem{Semertzidis:2011qv}
Yannis~K. Semertzidis, Storage Ring EDM Collaboration.
\newblock {A Storage Ring Proton Electric Dipole Moment Experiment: Most
  Sensitive Experiment to CP-Violation Beyond the Standard Model}.
\newblock 2011, \eprint{arXiv}{1110.3378}.

\bibitem{Bennett:2008dy}
G.W. Bennett et~al., Muon (g-2) Collaboration.
\newblock {An Improved Limit on the Muon Electric Dipole Moment}.
\newblock {\em Phys.\ Rev.\ D}, 80:052008, 2009,
  \doi{10.1103/PhysRevD.80.052008}, \eprint{arXiv}{0811.1207}.

\bibitem{Lebedev:2004va}
Oleg Lebedev, Keith~A. Olive, Maxim Pospelov, and Adam Ritz.
\newblock {Probing CP violation with the Deuteron Electric Dipole Moment}.
\newblock {\em Phys.\ Rev.\ D}, 70:016003, 2004,
  \doi{10.1103/PhysRevD.70.016003}, \eprint{arXiv}{hep-ph/0402023}.

\bibitem{deVries:2011re}
J.~de~Vries, E.~Mereghetti, R.G.E. Timmermans, and U.~van Kolck.
\newblock {Parity- and Time-Reversal-Violating Form Factors of the Deuteron}.
\newblock {\em Phys.\ Rev.\ Lett.}, 107:091804, 2011,
  \doi{10.1103/PhysRevLett.107.091804}, \eprint{arXiv}{1102.4068}.

\bibitem{deVries:2011an}
J.~de~Vries, R.~Higa, C.-P. Liu, E.~Mereghetti, I.~Stetcu, et~al.
\newblock {Electric Dipole Moments of Light Nuclei From Chiral Effective Field
  Theory}.
\newblock {\em Phys.\ Rev.\ C}, 84:065501, 2011,
  \doi{10.1103/PhysRevC.84.065501}, \eprint{arXiv}{1109.3604}.

\bibitem{Liu:2012tra}
C.-P. Liu, J.~de~Vries, E.~Mereghetti, R.G.E. Timmermans, and U.~van Kolck.
\newblock {Deuteron Magnetic Quadrupole Moment From Chiral Effective Field
  Theory}.
\newblock {\em Phys.\ Lett.\ B}, 713:447\unskip--\ignorespaces 452, 2012,
  \doi{10.1016/j.physletb.2012.06.024}, \eprint{arXiv}{1203.1157}.

\bibitem{Bedaque:2002mn}
Paulo~F. Bedaque and Ubirajara van Kolck.
\newblock {Effective Field Theory for Few Nucleon Systems}.
\newblock {\em Ann.\ Rev.\ Nucl.\ Part.\ Sci.}, 52:339\unskip--\ignorespaces
  396, 2002, \doi{10.1146/annurev.nucl.52.050102.090637},
  \eprint{arXiv}{nucl-th/0203055}.

\bibitem{Haxton:1983dq}
W.C. Haxton and E.M. Henley.
\newblock {Enhanced T Violating Nuclear Moments}.
\newblock {\em Phys.\ Rev.\ Lett.}, 51:1937, 1983,
  \doi{10.1103/PhysRevLett.51.1937}.

\bibitem{Herczeg}
P.~Herczeg.
\newblock {T Violating Effects in Neutron Physics and CP Violation in Gauge
  Models}.
\newblock In N.R. Robertson, C.R. Gould, and J.D. Bowman, editors, {\em {Tests
  of Time-Reversal Invariance in Neutron Physics}}, Singapore, 1987. World
  Scientific.

\bibitem{herczeg88}
P.~Herczeg.
\newblock {T-Violation in Nuclear Interactions ¿ An Overview}.
\newblock {\em Hyperfine Interactions}, 43:77, 1988.

\bibitem{Gudkov:1992yc}
Vladimir~P. Gudkov, Xiao-Gang He, and Bruce~H.J. McKellar.
\newblock {On the CP Odd Nucleon Potential}.
\newblock {\em Phys.\ Rev.\ C}, 47:2365\unskip--\ignorespaces 2368, 1993,
  \doi{10.1103/PhysRevC.47.2365}, \eprint{arXiv}{hep-ph/9212207}.

\bibitem{Towner:1994qe}
I.S. Towner and A.C. Hayes.
\newblock {P, T violating Nuclear Matrix Elements in the One Meson Exchange
  Approximation}.
\newblock {\em Phys.\ Rev.\ C}, 49:2391\unskip--\ignorespaces 2397, 1994,
  \doi{10.1103/PhysRevC.49.2391}, \eprint{arXiv}{nucl-th/9402026}.

\bibitem{Liu:2004tq}
C.-P. Liu and R.G.E. Timmermans.
\newblock {P- and T-odd Two-Nucleon Interaction and the Deuteron Electric
  Dipole Moment}.
\newblock {\em Phys.\ Rev.\ C}, 70:055501, 2004,
  \doi{10.1103/PhysRevC.70.055501}, \eprint{arXiv}{nucl-th/0408060}.

\bibitem{H66}
P.~Herczeg.
\newblock {The General Form of the Time-Reversal Non-Invariant Internucleon
  Potential}.
\newblock {\em Nucl.\ Phys.}, 75:665, 1966.

\bibitem{vanKolck:1997fu}
U.~van Kolck, M.C.M. Rentmeester, James~Lewis Friar, J.~Terrance Goldman, and
  J.J. de~Swart.
\newblock {Electromagnetic Corrections to the One Pion Exchange Potential}.
\newblock {\em Phys.\ Rev.\ Lett.}, 80:4386\unskip--\ignorespaces 4389, 1998,
  \doi{10.1103/PhysRevLett.80.4386}, \eprint{arXiv}{nucl-th/9710067}.

\bibitem{Flambaum:1984fb}
V.V. Flambaum, I.B. Khriplovich, and O.P. Sushkov.
\newblock {On The Possibility to Study P Odd and T Odd Nuclear Forces in Atomic
  and Molecular Experiments}.
\newblock {\em Sov.\ Phys.\ JETP}, 60:873, 1984.

\bibitem{Avi85}
Y.~Avishai.
\newblock {Electric Dipole Moment of the Deuteron}.
\newblock {\em Phys.\ Rev.\ D}, 32:314, 1985.

\bibitem{Khriplovich:1999qr}
I.B. Khriplovich and R.A. Korkin.
\newblock {P and T Odd Electromagnetic Moments of Deuteron in Chiral Limit}.
\newblock {\em Nucl.\ Phys.\ A}, 665:365\unskip--\ignorespaces 373, 2000,
  \doi{10.1016/S0375-9474(99)00403-0}, \eprint{arXiv}{nucl-th/9904081}.

\bibitem{Afnan:2010xd}
I.R. Afnan and B.F. Gibson.
\newblock {Model Dependence of the 2H Electric Dipole Moment}.
\newblock {\em Phys.\ Rev.\ C}, 82:064002, 2010,
  \doi{10.1103/PhysRevC.82.064002}, \eprint{arXiv}{1011.4968}.

\bibitem{Avishai:1986dw}
Y.~Avishai and M.~Fabre De La~Ripelle.
\newblock {Electric Dipole Moment of He-3}.
\newblock {\em Phys.\ Rev.\ Lett.}, 56:2121\unskip--\ignorespaces 2123, 1986,
  \doi{10.1103/PhysRevLett.56.2121}.

\bibitem{Stetcu:2008vt}
I.~Stetcu, C.-P. Liu, James~Lewis Friar, A.C. Hayes, and P.~Navratil.
\newblock {Nuclear Electric Dipole Moment of He-3}.
\newblock {\em Phys.\ Lett.\ B}, 665:168\unskip--\ignorespaces 172, 2008,
  \doi{10.1016/j.physletb.2008.06.019}, \eprint{arXiv}{0804.3815}.

\bibitem{Song:2012yh}
Young-Ho Song, Rimantas Lazauskas, and Vladimir Gudkov.
\newblock {Nuclear Electric Dipole Moment of Three-Body System}.
\newblock {\em Phys.\ Rev. C}, 87:015501, 2013,
  \doi{10.1103/PhysRevC.87.015501}, \eprint{arXiv}{1211.3762}.

\bibitem{Schiff:1963zz}
L.I. Schiff.
\newblock {Measurability of Nuclear Electric Dipole Moments}.
\newblock {\em Phys.\ Rev.}, 132:2194\unskip--\ignorespaces 2200, 1963,
  \doi{10.1103/PhysRev.132.2194}.

\bibitem{flambaum02}
V.~V. Flambaum and J.~S.~M. Ginges.
\newblock {Nuclear Schiff Moment and Time-Invariance Violation in Atoms}.
\newblock {\em Phys.\ Rev.\ A}, 65:032113, 2002.

\bibitem{flambaum12}
V.~V. Flambaum and A.~Kozlov.
\newblock {Screening and Finite-Size Corrections to Octupole and Schiff
  Moments}.
\newblock {\em Phys.\ Rev.\ C}, 85:068502, 2012.

\bibitem{Liu:2007zf}
C.-P. Liu, M.J. Ramsey-Musolf, W.C. Haxton, R.G.E. Timmermans, and A.E.L.
  Dieperink.
\newblock {Atomic Electric Dipole Moments: The Schiff Theorem and Its
  Corrections}.
\newblock {\em Phys.\ Rev.\ C}, 76:035503, 2007,
  \doi{10.1103/PhysRevC.76.035503}, \eprint{arXiv}{0705.1681}.

\bibitem{ban10}
Shufang Ban, Jonathan Engel, Jacek Dobaczewski, and A.~Shukla.
\newblock {Fully Self-Consistent Calculations of Nuclear Schiff Moments}.
\newblock {\em Phys.\ Rev.\ C}, 82:015501, 2010.

\bibitem{ellis11}
John Ellis, Jae~Sik Lee, and Aportolos Piliaftsis.
\newblock {Maximal Electric Dipole Moments of Nuclei with Enhanced Schiff
  Moments}.
\newblock {\em J.\ High Energy Phys.}, 2011(2):45, 2011.

\bibitem{flambaum86}
V.~V. Flambaum, I.~B. Khriplovich, and O.~P. Sushkov.
\newblock {On the P- and T-Nonconserving Nuclear Moments}.
\newblock {\em Nucl.\ Phys.\ A}, 449:750, 1986.

\bibitem{spevak95}
V.~Spevak and N.~Auerbach.
\newblock {Parity Mixing and Time Reversal Violation in Nuclei with Octupole
  Deformations}.
\newblock {\em Phys.\ Lett.\ B}, 359:254, 1995.

\bibitem{spevak97}
V.Spevak, N.~Auerbach, and V.~Flambaum.
\newblock {Enhanced T-odd P-odd Electromagnetic Moments in Reflection
  Asymmetric Nuclei}.
\newblock {\em Phys.\ Rev.\ C}, 56:1357, 1997.

\bibitem{engel03}
J.~Engel, M.~Bender, J.~Dobaczewski, J.~H. de~Jesus, and P.~Olbratowski.
\newblock {Time-Reversal Violating Schiff Moment of $^{225}$Ra}.
\newblock {\em Phys.\ Rev.\ C}, 68:025501, 2003.

\bibitem{sheline89}
R.~K. Sheline, A.~K. Jain, K.~Jain, and I.~Ragnarsson.
\newblock {Reflection Asymmetric and Symmetric Shapes in $^{225}$Ra.
  Polarization Effects of the Odd Particle}.
\newblock {\em Phys.\ Lett.\ B}, 219:47, 1989.

\bibitem{helmer87}
R.~G. Helmer, M.~A. Lee, C.~W. Reich, and I.~Ahmad.
\newblock {Intrinsic Reflection Asymmetry in $^{225}$Ra: Additional Information
  from a Study of the Alpha-Decay Scheme of $^{229}$Th}.
\newblock {\em Nucl.\ Phys.\ A}, 474:77, 1987.

\bibitem{dobaczewski12}
J.~Dobaczewski and J.~Engel.
\newblock In preparation.

\bibitem{gaffney11}
Liam~P. Gaffney.
\newblock {Octupole collectivity: Coulomb excitation of $^{224}$Ra}, 2011.
\newblock Talk at ARIS-2011, Advances in Radioactive Isotope Science.

\bibitem{auerbach06}
N.~Auerbach, V.F. Dmitriev, V.V. Flambaum~A. Lisetskiy, R.A. Sen'kov, and V.G.
  Zelevinsky.
\newblock {Nuclear Schiff Moment in Nuclei with Soft Octupole and Quadrupole
  Vibrations }.
\newblock {\em Phys.\ Rev.\ C}, 74:025502, 2006.

\bibitem{Jung:2013mg}
Martin Jung.
\newblock {A Robust Limit for the Electric Dipole Moment of the Electron}.
\newblock 2013, \eprint{arXiv}{1301.1681}.

\bibitem{Porsev:2012zx}
S.G. Porsev, M.S. Safronova, and M.G. Kozlov.
\newblock {Electric Dipole Moment Enhancement Factor of Thallium}.
\newblock 2012, \eprint{arXiv}{1201.5615}.

\bibitem{Das:2008b}
H.S. Nataraj, Sahoo, B.P. B.K, Das, and D.~Mukherjee.
\newblock {Electric Dipole Moment of the Electron in the YbF Molecule}.
\newblock {\em Phys.\ Rev.\ A}, A78:010501, 2008.

\bibitem{Das:2008}
Sahoo, B.P. B.K, Das, R.~Choudhuri, D.~Mukherjee, and E.~Venugopal.
\newblock {Electric Dipole Moment of the Electron in the YbF Molecule}.
\newblock {\em Phys.\ Rev.\ Lett.}, 101:033002, 2008.

\bibitem{Dzuba:2009mw}
V.A. Dzuba and V.V. Flambaum.
\newblock {Calculation of the (T,P)-odd Electric Dipole Moment of Thallium}.
\newblock {\em Phys.\ Rev.\ A}, 80:062509, 2009,
  \doi{10.1103/PhysRevA.80.062509}, \eprint{arXiv}{0909.0308}.

\bibitem{Dzuba99}
T.~M.~R Byrnes, Dzuba~V. A., V.~V. Flambaum, and D.~W. Murray.
\newblock {P-odd and CP-odd Four-Quark Contributions to Neutron EDM}.
\newblock {\em Phys.\ Rev.\ A}, 59:3082, 1999, \doi{10.1103/PhysRevA.59.3082.},
  \eprint{arXiv}{0908.2420}.

\bibitem{Dzuba:2011}
V.A. Dzuba and V.V. Flambaum.
\newblock {Calculation of the (T,P)-odd Electric Dipole Moment of Thallium}.
\newblock {\em Phys.\ Rev.\ A}, 84:052108, 2011,
  \doi{10.1103/PhysRevA.80.062509}, \eprint{arXiv}{1109.6082}.

\bibitem{Das09b}
D.~Mukherjee, B.K Sahoo, H.S. Nataraj, and B.~P Das.
\newblock {P-odd and CP-odd Four-Quark Contributions to Neutron EDM}.
\newblock {\em J.\ Phys.\ Chem.\ A}, 113:12549, 2009,
  \doi{doi/abs/10.1021/jp904020s.}

\bibitem{kozlov98}
N.S Mosyagin, M.G. Kozlov, and A.V. Titov.
\newblock {Electric Dipole Moment of the Electron in the YbF Molecule}.
\newblock {\em J.\ Phys.\ B}, 31:L763, 1998.

\bibitem{Hudson:2002az}
J.J Hudson, B.E. Sauer, M.R. Tarbutt, and E.A. Hinds.
\newblock {Measurement of the Electron Electric Dipole Moment Using YbF
  Molecules}.
\newblock {\em Phys.\ Rev.\ Lett.}, 89:023003, 2002,
  \doi{10.1103/PhysRevLett.89.023003}, \eprint{arXiv}{hep-ex/0202014}.

\bibitem{Meyer:2008gc}
Edmund~R. Meyer and John~L. Bohn.
\newblock {Prospects for an Electron Electric Dipole Moment Search in
  Metastable ThO and ThF}.
\newblock {\em Phys.\ Rev.\ A}, 78:010502(R), 2008, \eprint{arXiv}{0805.0161}.

\bibitem{Hudson:2011zz}
J.J Hudson, D.M. Kara, I.J. Smallman, B.E. Sauer, M.R. Tarbutt, et~al.
\newblock {Improved Measurement of the Shape of the Electron}.
\newblock {\em Nature}, 473:493\unskip--\ignorespaces 496, 2011,
  \doi{10.1038/nature10104}.

\bibitem{Giudice:2005rz}
G.F. Giudice and A.~Romanino.
\newblock {Electric Dipole Moments in Split Supersymmetry}.
\newblock {\em Phys.\ Lett.\ B}, 634:307\unskip--\ignorespaces 314, 2006,
  \doi{10.1016/j.physletb.2006.01.027}, \eprint{arXiv}{hep-ph/0510197}.

\bibitem{Li:2008kz}
Yingchuan Li, Stefano Profumo, and Michael Ramsey-Musolf.
\newblock {Higgs-Higgsino-Gaugino Induced Two Loop Electric Dipole Moments}.
\newblock {\em Phys.\ Rev.\ D}, 78:075009, 2008,
  \doi{10.1103/PhysRevD.78.075009}, \eprint{arXiv}{0806.2693}.

\bibitem{Giudice:2004tc}
G.F. Giudice and A.~Romanino.
\newblock {Split Supersymmetry}.
\newblock {\em Nucl.\ Phys.\ B}, 699:65\unskip--\ignorespaces 89, 2004,
  \doi{10.1016/j.nuclphysb.2004.11.048}, \eprint{arXiv}{hep-ph/0406088}.

\bibitem{Kane:2009kv}
Gordon Kane, Piyush Kumar, and Jing Shao.
\newblock {CP-violating Phases in M-theory and Implications for EDMs}.
\newblock {\em Phys.\ Rev.\ D}, 82:055005, 2010,
  \doi{10.1103/PhysRevD.82.055005}, \eprint{arXiv}{0905.2986}.

\bibitem{Dimopoulos:1995ju}
Savas Dimopoulos and David~W. Sutter.
\newblock {The Supersymmetric Flavor Problem}.
\newblock {\em Nucl.\ Phys.\ B}, 452:496\unskip--\ignorespaces 512, 1995,
  \doi{10.1016/0550-3213(95)00421-N}, \eprint{arXiv}{hep-ph/9504415}.

\bibitem{Jarlskog:1985ht}
C.~Jarlskog.
\newblock {Commutator of the Quark Mass Matrices in the Standard Electroweak
  Model and a Measure of Maximal CP Violation}.
\newblock {\em Phys.\ Rev.\ Lett.}, 55:1039, 1985,
  \doi{10.1103/PhysRevLett.55.1039}.

\bibitem{Li:2010ax}
Yingchuan Li, Stefano Profumo, and Michael Ramsey-Musolf.
\newblock {A Comprehensive Analysis of Electric Dipole Moment Constraints on
  CP-violating Phases in the MSSM}.
\newblock {\em JHEP}, 1008:062, 2010, \doi{10.1007/JHEP08(2010)062},
  \eprint{arXiv}{1006.1440}.

\bibitem{Beall:1981zq}
G.~Beall and A.~Soni.
\newblock {Electric Dipole Moment of the Neutron in a Left-Right Symmetric
  Theory of CP Violation}.
\newblock {\em Phys.\ Rev.\ Lett.}, 47:552, 1981,
  \doi{10.1103/PhysRevLett.47.552}.

\bibitem{He:1988th}
Xiao-Gang He, Bruce~H.J. McKellar, and Sandip Pakvasa.
\newblock {Epsilon-Prime / Epsilon and the Electric Dipole Moment of The
  Neutron in Left-Right Symmetric Models}.
\newblock {\em Phys.\ Rev.\ Lett.}, 61:1267\unskip--\ignorespaces 1270, 1988,
  \doi{10.1103/PhysRevLett.61.1267}.

\bibitem{Ecker:1983dj}
G.~Ecker, W.~Grimus, and H.~Neufeld.
\newblock {The Neutron Electric Dipole Moment in Left-Right Symmetric Gauge
  Models}.
\newblock {\em Nucl.\ Phys.\ B}, 229:421, 1983,
  \doi{10.1016/0550-3213(83)90341-3}.

\bibitem{Towner:2010zz}
I.S. Towner and J.C. Hardy.
\newblock {The Evaluation of V(ud) and its Impact on the Unitarity of the
  Cabibbo-Kobayashi-Maskawa Quark-Mixing Matrix}.
\newblock {\em Rept.\ Prog.\ Phys.}, 73:046301, 2010,
  \doi{10.1088/0034-4885/73/4/046301}.

\bibitem{Agashe:2004cp}
Kaustubh Agashe, Gilad Perez, and Amarjit Soni.
\newblock {Flavor Structure of Warped Extra Dimension Models}.
\newblock {\em Phys.\ Rev.\ D}, 71:016002, 2005,
  \doi{10.1103/PhysRevD.71.016002}, \eprint{arXiv}{hep-ph/0408134}.

\bibitem{Chang:2002ww}
We-Fu Chang and John~N. Ng.
\newblock {CP violation in 5-D split Fermions Scenario}.
\newblock {\em JHEP}, 0212:077, 2002, \eprint{arXiv}{hep-ph/0210414}.

\bibitem{Fitzpatrick:2007sa}
A.~Liam Fitzpatrick, Gilad Perez, and Lisa Randall.
\newblock {Flavor Anarchy in a Randall-Sundrum Model with 5D Minimal Flavor
  Violation and a Low Kaluza-Klein Scale}.
\newblock {\em Phys.\ Rev.\ Lett.}, 100:171604, 2008,
  \doi{10.1103/PhysRevLett.100.171604}, \eprint{arXiv}{0710.1869}.

\bibitem{Shifman:1978zn}
A.~I.~Vainshtein M.~A.~Shifman and V.~I. Zakharov.
\newblock {Remarks on Higgs Boson Interactions with Nucleons}.
\newblock {\em Phys.\ Lett.\ B}, 78:443, 1978.

\bibitem{Demir:2003js}
D.~A. Demir, OLebedev, K.~A. Olive, M.~Pospelov, and A.~Ritz.
\newblock {Electric Dipole Moments in the MSSM at Large Tan Beta}.
\newblock {\em Nucl.\ Phys.\ B}, 680:339, 2004.

\bibitem{Murthy:1989zz}
S.A. Murthy, D.~Krause, Z.L. Li, and L.R. Hunter.
\newblock {New Limits on the Electron Electric Dipole Moment from Cesium}.
\newblock {\em Phys.\ Rev.\ Lett.}, 63:965\unskip--\ignorespaces 968, 1989,
  \doi{10.1103/PhysRevLett.63.965}.

\bibitem{Regan:2002ta}
B.C. Regan, E.D. Commins, C.J. Schmidt, and D.~DeMille.
\newblock {New Limit on the Electron Electric Dipole Moment}.
\newblock {\em Phys.\ Rev.\ Lett.}, 88:071805, 2002,
  \doi{10.1103/PhysRevLett.88.071805}.

\bibitem{Baker:2006ts}
C.A. Baker, D.D. Doyle, P.~Geltenbort, K.~Green, M.G.D. van~der Grinten, et~al.
\newblock {An Improved Experimental Limit on the Electric Dipole Moment of the
  Neutron}.
\newblock {\em Phys.\ Rev.\ Lett.}, 97:131801, 2006,
  \doi{10.1103/PhysRevLett.97.131801}, \eprint{arXiv}{hep-ex/0602020}.

\bibitem{Griffith:2009zz}
W.C. Griffith, M.D. Swallows, T.H. Loftus, M.V. Romalis, B.R. Heckel, et~al.
\newblock {Improved Limit on the Permanent Electric Dipole Moment of Hg-199}.
\newblock {\em Phys.\ Rev.\ Lett.}, 102:101601, 2009,
  \doi{10.1103/PhysRevLett.102.101601}.

\bibitem{Player:1970zz}
M.A. Player and P.G.H. Sandars.
\newblock {An Experiment to Search for an Electric Dipole Moment in the P(2)-3
  Metastable State of Xenon}.
\newblock {\em J.\ Phys. B}, 3:1620\unskip--\ignorespaces 1635, 1970.

\bibitem{Donoghue:1987dd}
John~F. Donoghue, Barry~R. Holstein, and M.J. Musolf.
\newblock {Electric Dipole Moments of Nuclei}.
\newblock {\em Phys.\ Lett.}, B196:196, 1987,
  \doi{10.1016/0370-2693(87)90603-4}.

\end{thebibliography}


\end{document}